%% file: thesis.tex
\documentclass[oneside,10pt]{report}
\usepackage{setspace}
\usepackage{amssymb}
\usepackage{amsmath}
\usepackage{cite}
\usepackage{fancyhdr}
\usepackage{a4}
\usepackage{epsfig}
\usepackage{graphicx}
\usepackage{subfigure}
\usepackage{wrapfig}
\usepackage{sectionappendix}
\newcommand{\lam}{\lambda}
\newcommand{\la}[1]{\label{#1}}
 
\newcommand{\be}{\begin{equation}}
\newcommand{\ee}{\end{equation}}
\newcommand{\ba}{\begin{eqnarray}}
\newcommand{\ea}{\end{eqnarray}}
\newcommand{\bi}{\begin{itemize}}
\newcommand{\ei}{\end{itemize}}

\newcommand{\tr}{{\rm Tr\,}}
\newcommand{\re}{\mathop{\rm Re}}

\newcommand{\im}{\mathop{\rm Im}}

\newcommand{\<}{\langle} 
\renewcommand{\>}{\rangle}  

\renewcommand{\vec}[1]{{\bf #1}}

\newcommand{\RR}{{\rm I\kern -.2em  R}}

\newcommand{\half}{{\frac{1}{2}}}

\def\lsi{\raise0.3ex\hbox{$<$\kern-0.75em\raise-1.1ex\hbox{$\sim$}}}
\def\gsi{\raise0.3ex\hbox{$>$\kern-0.75em\raise-1.1ex\hbox{$\sim$}}}

\begin{document}


\newpage
\pagenumbering{arabic}
\setcounter{page}{1} \pagestyle{fancy}
\renewcommand{\chaptermark}[1]{\markboth{\chaptername%
\ \thechapter:\,\ #1}{}}
\renewcommand{\sectionmark}[1]{\markright{\thesection\,\ #1}}

\newpage
\include{front}

\newpage

\addtolength{\headheight}{3pt}    
\fancyhead{}


\fancyhead[RE]{\sl\leftmark}
\fancyhead[RO,LE]{\rm\thepage}
\fancyhead[LO]{\sl\rightmark}

\fancyfoot[C,L,E]{}

\pagenumbering{arabic}
\pagestyle{fancy}
\onehalfspacing    
\include{chapter1}
\include{chapter2}
\include{chapter3}
\include{chapter4}
\include{chapter5}
\include{chapter6}

\include{chapter7}
\include{chapter8}
\include{appendix}

\onehalfspacing
\bibliographystyle{h-physrev2} 
\bibliography{hmbib}

\end{document}

%% file: front.tex
\newpage
\thispagestyle{empty}
\singlespacing

\vspace*{1cm}
\begin{center}
{\Huge\bf Glueball Regge trajectories \\}
\vspace{2cm}
{\Large
Harvey Byron Meyer\\
\vspace{0.35cm}
Lincoln College, Oxford\\
\vspace{0.35cm}}
\end{center}

\begin{center}
\vspace{2.5cm}
\end{center}
\begin{center}
{\large 
\vspace{0.2cm}
Rudolf Peierls Centre for Theoretical Physics\\
Department of Physics, University of Oxford}
\end{center}
\vspace{7cm}
\begin{center}
{\large  
Thesis submitted for the
degree of \\
Doctor of Philosophy at the University of Oxford\\
\vspace{0.3cm}
$\cdot$ Trinity term, 2004 $\cdot$}
\end{center}

%
%
%
%
%

\newpage
\thispagestyle{empty}
\vspace*{1cm}
\begin{center}
{\huge\bf Glueball Regge trajectories \\}
\vspace{0.5cm}
{\large 
Harvey Byron Meyer \\
Lincoln College \\
\vspace{0.3cm}
Thesis submitted for the
degree of \\
Doctor of Philosophy at the University of Oxford\\ 
Trinity Term, 2004}
\end{center}
\vspace{2cm}
\begin{center}
\large{\bf Abstract}
\end{center}
\noindent 
We investigate the  spectrum of glueballs in $SU(N_c)$ gauge theories.
Our motivation is to determine whether the states lie on straight Regge trajectories.
It has been  conjectured for a long time that glueballs are the physical states
lying on the pomeron, the trajectory responsible for the slowly rising hadronic
cross-sections at large centre-of-mass energy.

\vspace{0.2cm}\noindent
After a review of Regge phenomenology, we show that string models of glueballs
predict states to lie on linear trajectories with definite sequences of quantum 
numbers. We then move on to the lattice formulation of gauge theory.
Because the lattice regularisation breaks rotational symmetry, there is an ambiguity
in the assignment of the spin of lattice states. We develop numerical methods to 
resolve these ambiguities in the continuum limit, in particular how to extract high
spin glueballs from the lattice. We also devise a multi-level algorithm that reduces
the variance on Euclidean correlation functions from which glueball masses are extracted.

\vspace{0.2cm}\noindent
In 2+1 dimensions, we determine the $SU(2)$ spectrum up to spin 6, and relabel
a previously published $SU(3)$ spectrum with the correct spin quantum numbers.
We find well-defined Regge trajectories, but the leading trajectory goes through
the lightest scalar glueball and has an intercept close to (-1).

\vspace{0.2cm}\noindent
In 3+1 dimensions, we perform a detailed survey of the $SU(3)$ spectrum. 
A comparison to the low-lying $SU(8)$ spectrum, that we also compute, 
indicates that these gauge theories are `close' to $SU(\infty)$. 
Although the spectrum is more complex than
in two space dimensions, we can clearly identify the leading trajectory: it goes
through the lightest $2^{++}$ and $4^{++}$ states, has slope $\alpha'=0.28(2)$
in units of the mesonic slope and intercept $\alpha_0=0.93(24)$, 
in remarkable agreement with phenomenological values. We conclude with some implications
of these results.

\newpage

\thispagestyle{empty}

\vspace*{3in}
\begin{center}
{\Large \textit{To Anne $\&$ R\"udiger}}
		
\end{center}

\newpage

\pagestyle{empty}
\pagenumbering{roman}
\setcounter{page}{1} \pagestyle{plain}

\tableofcontents


\chapter*{Acknowledgements} 
\addcontentsline{toc}{chapter} 
		 {\protect\numberline{Acknowledgements\hspace{-96pt}}} 

First and foremost I would like to thank my supervisor Dr.~Michael~Teper
for sharing his experience and providing guidance and support.
Most of the ideas and methods presented in this thesis resulted from the
enjoyable discussions we had throughout the course of my D. Phil. 
I~would also like to express my gratitude to every member of the Rudolf Peierls 
Centre for Theoretical Physics, for creating a friendly environment and making
the time I spent in the Centre a very rewarding experience; 
I have profited greatly from discussions with many of its members.
In particular, Dr.~Biagio~Lucini and Dr.~Urs~Wenger often
provided me with useful suggestions concerning physics and programming issues. 
The numerical calculations presented in this thesis were performed partly on
Compaq Alpha workstations, partly on `Hydra', the cluster of 80 Xeon processors
of the Theoretical Physics Centre. These machines were partly funded by EPSRC and 
PPARC grants. I warmly thank Lory Rice and Jonathan Patterson for their efficient
help and support.

\vspace{0.2cm}\noindent
 I am indebted to Prof.~Martin~L\"uscher for 
introducing me to the subject of lattice gauge theory, and
to Prof.~Kari~Rummukainen, who introduced me to the art
of Monte-Carlo simulations and continued offering his advice by email.
I would also like to use this opportunity to extend my gratitude to
Prof.~Mikhail~Shaposhnikov who helped me a lot in obtaining 
a position as a graduate student 
at Oxford University and whose lasting influence has accompanied me
over the past three years. 

\vspace{0.2cm}\noindent
Last but not least I thank my family, especially my parents, 
Anne~and~R\"udiger~Walter~Meyer, 
for their sustained and unconditional support throughout my studies of physics.

\vspace{1in}\noindent
This work was supported by the Berrow Scholarship of Lincoln College, 
the ORS Award Scheme (UK) and
the Bourse de Perfectionnement et de Recherche of the University of Lausanne.
\vspace{0.5in}
\section*{Publications}
A substantial part of this thesis is published work. 
Chapter~\ref{ch:hspin} is a condensed version of~\cite{Meyer:2002mk},
Chapter~\ref{ch:mla} is a contraction of papers~\cite{Meyer:2002cd} 
and~\cite{Meyer:2003hy} and Chapter~\ref{ch:regge_2d} corresponds 
to~\cite{Meyer:2003wx}.
\newpage


%% file: chapter1.tex
\chapter{High energy hadronic reactions}
\label{ch:hehr}
\section{Regge theory}
The paradigm for the quantum relativistic description of interactions
between particles is the exchange of a bosonic particle. Its
coupling to `matter' particles and its propagator determine the 
force between them. A classic example is the Yukawa potential between
nucleons resulting from the exchange of a pion. In particular,
for a scattering process where $s$ is the square centre-of-mass energy
and $t$ the square momentum transfer, the usual Mandelstam variables, 
a pole appears in the scattering matrix when $t$ goes through the value
of the square mass of the exchanged particle.
In the following   we give a bird's view of $S$-matrix theory, based on
\cite{Spearman:1970cr, Forshaw:1997dc}.

The \emph{Lorentz invariance} of the scattering matrix $S$ implies that it can be 
taken to be a function of the Lorentz invariants $s$ and $t$. The 
conservation of probability expresses itself in the \emph{unitarity} of $S$,
$S^\dagger S=1$. An equivalent expression of this property is the set of  Cutkovsky 
rules, which allow us to determine the imaginary part of an amplitude by 
considering the scattering amplitudes of the incoming and outgoing states
into all possible intermediate states. Defining the scattering amplitude $A_{ab}$
through
\be
S_{ab}=\delta_{ab}+ i(2\pi)^4\delta^4\left(\sum_i p_i\right)A_{ab},
\ee
unitarity implies
\be
2\im~A_{ab}=(2\pi)^4\delta^4\left(\sum_i p_i\right)\sum_c A_{ac}A^\dagger_{cb}.
\ee

A special case of these rules is the optical theorem, which relates the 
imaginary part of the forward (elastic) amplitude to the total cross-section for 
the scattering of two particles:
\be
2\im~A_{aa}=F\sigma_{\rm tot},\la{eq:opt_thm}
\ee
where at high energies the flux factor $F$ tends to $2s$.

The requirement of causality of the theory, namely that two regions at space-like
separation do not influence each other, leads to the property of
\emph{analyticity} of $A(s,t)$, with only those singularities required
by unitarity. For instance, below the two-particle threshold, the imaginary
part of the amplitude can be chosen to vanish on the real $s$-axis. 
The Schwarz reflection principle then implies $A(s,t)^*=A(s^*,t)$
throughout the domain of analyticity. Since the imaginary part of the amplitude
is non-zero above threshold, there must be a cut along the real $s$-axis starting
at the branch point of the threshold energy. The imaginary part of the physical 
amplitude can be defined as 
\be
\im A(s,t)=\frac{1}{2i}(A(s+i\epsilon,t)-A(s-i\epsilon,t)).
\ee

A further consequence of analyticity is crossing symmetry. While in the $s$-channel
$s>0$ and $t<0$, the amplitude may be analytically continued to the region
$t>0$ and $s<0$. Thus we can use the same amplitude to describe
 the crossed-channel process:
\be
A_{a+\bar c\rightarrow \bar b+d}(s,t)=A_{a+b\rightarrow c+d}(t,s).
\ee
The argument about the existence of a cut along the real $s$-axis above
threshold can be repeated in the crossed channel, leading to the conclusion that
there is also a cut along the real $s$-axis running from $-\infty$ to $-t$.
\emph{Dispersion relations} allow us to reconstruct the real part 
of an amplitude from its imaginary part.
 By integrating along a contour around the cuts~\cite{Walker:1970}, 
one learns that
\be
A(s,t)=\frac{1}{\pi}\int_{s^+}^\infty \frac{\im A(s',t)}{s'-s} ds' + 
\frac{1}{\pi}\int_{-\infty}^{s_-} \frac{\im A(s',t)}{s'-s} ds'~+~{\rm poles},
\ee
which in particular gives us $\re A(s,t)$ for $s$ on the real axis.

The $s$-channel amplitude can be written as a partial wave expansion:
\be
A(s,t) = \sum_{\ell=0}^\infty~ (2\ell+1)~a_\ell(s)~P_\ell(\cos{\theta}),
\la{eq:part_wave}
\ee
where $\cos{\theta}=1+2t/s$. This expansion is very useful at low energies, where
a classical argument shows that partial waves with $\ell>pb$ are exponentially 
suppressed, where $b$ is the transverse size of the target particle. At high energies, 
the expression does not seem to be very useful, 
given that more and more partial waves contributing to the amplitude must be determined
and that the whole series must be resummed in order to get the asymptotic behaviour.
Nevertheless, from the same classical argument, a bound can be inferred on the 
amplitude, called the Froissart bound, 
which can be expressed as a unitarity constraint on the total cross-section:
\be
\sigma_{\rm tot} \leq\frac{\pi}{m^2}\log^2\left(\frac{s}{m^2}\right),
\qquad m={\rm mass~gap}. \la{eq:froissart}
\ee
However,  we can analytically 
continue the partial wave amplitudes to negative values of their argument;
after the interchange $s\leftrightarrow t$, by crossing symmetry they correspond
to the $t$-channel partial wave amplitudes $a_\ell(t)$.
Furthermore, following the ideas of Regge, we consider the analytic continuation 
in the complex angular momentum plane $a(\ell,t)\equiv a_\ell(t) $.
Now 
a Sommerfeld-Watson transform~\cite{Walker:1970} may be performed, 
which expresses the partial wave expansion as a contour integral in the complex 
angular momentum plane:
\be
A(s,t)=\frac{1}{2i}\int_C d\ell~ \frac{2\ell+1}{\sin{\pi \ell}}~
\sum_{\eta=\pm}~ \frac{\eta+e^{-i\pi\ell}}{2}~a^{(\eta)}(\ell,t)~ 
P\left(\ell,1+\frac{2s}{t}\right)
\ee
In the process, even and odd `signature' partial waves $a^\eta$ had to be introduced,
with $a(\ell,t)\equiv a^{(-1)^\ell}(\ell,t)$. The point of this transformation
becomes clear when the contour is deformed to a large half circle with its diameter along
the $\re \ell=-\frac{1}{2}$ axis. For instance, each time a pole of $a^\eta(\ell,t)$ enters
the contour at position $\ell=\alpha(t)$, a new term must be added to the expression. 
$\alpha(t)$ is called a Regge trajectory; 
when $t$ goes through $m^2$, the square mass of a physical state, $\alpha(t)$
is equal to its spin. Because of the asymptotic behaviour of
the Legendre functions $P(\ell,1+2s/t)$, at high $s$-channel energies
 $s\gg |t|$ the amplitude 
is dominated by the rightmost singularity in the complex $\ell$ plane:
\be
A(s,t)~\sim~ \frac{\eta+e^{-i\pi\alpha(t)}}{2\sin{\pi\alpha(t)}} 
~ \pi(2\alpha(t)+1)\beta(t)~   s^{\alpha(t)}
\qquad({\rm simple~pole}),
\ee 
where $\beta$ is the residue of the pole.  The amplitude
behaves as if a single object, called the \emph{reggeon}, was being exchanged:
it may be interpreted as the superposition of amplitudes for the exchanges of 
a whole family of particles in the $t$-channel.  In particular, $\beta(t)$ contains
the information on the coupling of the reggeon to the particles that are scattering.
This coupling depends only on $t$ and obeys the \emph{factorisation} property.
Through the optical theorem, it is seen that the total cross-section
behaves at high energy $\sqrt{s}$ as
\be
\sigma_{\rm tot}\propto s^{\alpha(0)-1}. \la{eq:sigma_tot}
\ee

We shall also encounter examples of more complicated singularities below.
%
%
%
\section{Regge phenomenology}
%
\subsection{The soft pomeron}
The data on hadronic {\bf total cross-sections} exhibits a universal behaviour  
at high energy: they are almost constant, in fact they even slightly increase. 
The object responsible for this non-trivial behaviour is by definition called the pomeron.
From Eqn.~(\ref{eq:sigma_tot}), it is seen that the simplest explanation is
 that the pomeron is a Regge pole with intercept close to one~\cite{Donnachie:1992ny}:
\be
 \alpha_0=1+\epsilon_0, \qquad {\rm with}\quad \epsilon_0\simeq 0.08 \la{eq:spom_a0}.
\ee
The coefficients in front of the power of $s$ depend on the process. In particular,
it is well-known that 
\be
\frac{\sigma(\pi p)}{\sigma(p p)} \simeq \frac{2}{3}\la{eq:add_quark_rule}, 
\ee
which suggests an \emph{`additive quark rule'}: it seems that the pomeron couples
 to the individual valence quarks inside hadrons. 
The Pomeranchuk theorem (1956) states that
 any scattering process in which there is charge exchange vanishes asymptotically.
Thus the pomeron must have vacuum quantum numbers and positive signature.

If $\epsilon_0$ is strictly positive, Eqn.~(\ref{eq:spom_a0})
 eventually leads to a violation of
the Froissart bound~(\ref{eq:froissart}). However, the exchange of two pomerons
leads to a cut in the complex angular momentum plane:
\be
\alpha_{PP}(t)=1+2\epsilon+\frac{\alpha't}{2}.
\ee
Two pomerons produce an asymptotic cross-section
behaving as $s^{2\epsilon_0}/f(\log{s})$, where logarithms of $s$ appear
and the proportionality coefficient has the opposite sign of the single-pomeron 
amplitude. The superposition of single and double pomeron exchange leads
to an effective power law $\sigma_{\rm tot}\sim s^\epsilon$, with $\epsilon<\epsilon_0$
decreasing with $s$. This eventually leads to the \emph{unitarisation} of the
scattering amplitude. There has been a controversy in the literature
\cite{Landshoff:1996ab,Capella:1994cr} concerning the importance of the mixing. 
The small-mixing version accords more naturally with the additive-quark rule,
because two-pomeron exchange would spoil the factorisation property. On the other hand,
the strong-mixing version can perhaps explain deep inelastic scattering data more
economically (see below).

The data on the differential {\bf elastic cross-sections} contains information on the 
trajectory of the pomeron. Donnachie and Landshoff~\cite{Donnachie:1986iz}
 used the proton form factor
from $ep$ elastic scattering to obtain the prediction
\be
\frac{d\sigma}{dt}={\rm const.}\times F_1(t)^4 ~(\alpha's)^{2(\alpha(t)-1)}. 
\ee
It turns out that a linear trajectory $\alpha(t)=\alpha_0+\alpha't$ with 
\be
\alpha'=0.25{\rm GeV}^{-2}
\ee
can be fitted to the CERN ISR data~\cite{Breakstone:1984te} 
at small $t$; at larger $t$, this ansatz
still matches the data well, which is a non-trivial check on the functional 
form used for $F_1(t)$. It is not understood why the form factor corresponding
to the photon ($C=-$) also works for the pomeron ($C=+$).

A further type of data where the pomeron phenomenon shows up is {\bf diffractive 
dissociation}. In such a process, a projectile ($p,~\gamma$) only carries
off a small fraction $\xi\ll1$ of a target proton (which remains intact). The projectile 
is then dissociated into a number of products $X$. The experimental signature for
such an event is a large rapidity gap, and $\xi$ is measured as $M_X^2/s$.
Using the factorisation property, one may write~\cite{Landshoff:1996ab}
\be
\frac{d^2\sigma}{dtd\xi} = F_{Pp}(\xi,t)~\sigma_{PA}(M_X^2,t)
\ee
In the special case of an off-shell photon ($A=\gamma^*$) (the `very-fast-proton' events
at HERA), a single-pomeron exchange gives a factorising contribution to the proton
 structure function.

Finally, {\bf exclusive electroproduction of vector mesons} (e.g. $\gamma^*p\rightarrow \rho p$)
is another standard process where the soft pomeron is seen. Whilst 
it describes the data well  up to $Q^2<25$GeV$^2$ 
when used in conjunction with the additive quark rule, it fails
to describe the increase in charm production $\gamma p \rightarrow J/\psi p$ with
the centre-of-mass energy $W$ of the system. This brings us to the 
more recent subject of the `hard' pomeron.
\subsection{The hard pomeron}
The HERA and ZEUS experiments on deep inelastic scattering (DIS)
at DESY gathered a wealth of new data throughout the
nineties. Two (related) discoveries came as surprises. 

Firstly, the proton structure function
$F_2(x,Q^2)$ was found to rise sharply at small $x\equiv \frac{Q^2}{Q^2+W^2}$
($W$ is the centre-of-mass energy of the $\gamma^*p$ system).
The stronger rise at $x<0.01$ therefore suggests the presence of
a  `harder' singularity with a higher intercept $\alpha_1(0)=1+\epsilon_1$.
 The experimentally determined value of $\epsilon_1$ is then~\cite{Donnachie:2001xx}
\be
\epsilon_1\simeq0.44.\la{eq:hpom_a0}
\ee
Two interpretations have been proposed. Donnachie and Landshoff~\cite{Donnachie:2001xx}
 postulate 
the existence of a new, `hard' pomeron with the intercept given above. Thus they write
the structure function as
\be
F_2(x,Q^2)= f_0(Q^2)x^{-\epsilon_0} + f_1(Q^2)x^{-\epsilon_1}
\ee
with $\epsilon_{0,1}$ fixed and given by (\ref{eq:spom_a0}) and (\ref{eq:hpom_a0}).
Another interpretation~\cite{Capella:1994cr} 
is that the large value of the effective intercept comes
from the perturbative evolution of a unique pomeron. The intercept thus acquires 
a dependence on $Q^2$:
\be
F_2(x,Q^2)= f(Q^2) x^{-\epsilon(Q^2)}.
\ee
Clearly it is hard to distinguish between these two forms through fits to experimental
data~\cite{Landshoff:2000mu}. 
We must look at other processes to choose between the two interpretations.

A second surprise came in the data on charm production $\gamma p \rightarrow J/\psi p$.
The differential cross-section rises with a similar `hard' power of the centre-of-mass
energy as the proton structure function. Assuming an amplitude which is the superposition
of the original `soft' pomeron and a new `hard' pomeron with a linear trajectory, 
Donnachie and Landshoff were able to fit both the total cross-section and 
the $t$ dependence of the differential  cross-section. They find 
\be
\alpha_h'=0.1{\rm GeV}^{-2}.
\ee
It seems however that the HERA data can also be accommodated 
within the second approach mentioned above~\cite{Merino:1999qf}.

If we accept the two-pomeron interpretation  for the moment,
a key question is whether the hard pomeron is already contributing in on-shell processes.
Recently it was claimed~\cite{Cudell:2003ci, Donnachie:2004pi} that a  combined fit 
to several total cross-sections and elastic amplitudes indicates the presence of a hard
pomeron compatible with that observed in DIS. The hard component would have been missed
previously, because in its bare form it leads to too strong a rise of the $pp$ and 
$\bar pp$ cross-section; however, the best overall fit is obtained 
when an interpolation between the power-law behaviour and the unitarised 
logarithmic behaviour at asymptotically large energies such as
$
\sigma_{\rm tot}\propto \log{\left[1+\left(\frac{s}{s_o}\right)^{\epsilon_1}\right]}
$
is used. Cudell finds that the ratio of the hard pomeron coupling to the soft one varies
from $0.2\%$ in $pp$ to $1\%$ in $\pi p$ and $K p$ and remarks that the coupling 
mechanism of the hard pomeron must be very different from that of the soft pomeron.
%
\subsection{The odderon\la{sec:odde}}
The elastic $pp$ differential cross-section famously exhibits a dip: for instance, it
is situated at $\sqrt{|t|}\simeq1.2$GeV at $\sqrt{s}=23$GeV (data from the 
CHHAV collaboration~\cite{Nagy:1979iw}).
On the other hand, no such dip is seen in the $\bar p p$ case.
While the interference between single and double pomeron exchange is destructive,
an additional contribution, odd under charge conjugation, must be invoked to 
explain the asymmetry between the $pp$ and $\bar p p$ processes. 
This $C=-$ object is called the odderon~\cite{Lukaszuk:1973nt}.
It is thus probable~\cite{Landshoff:2000mu} that the dominant exchange is $C=+$ at 
small $|t|$  and $C=-$ beyond the dip.
The status of this phenomenon remains unclear however, because it has not been observed
in other processes. We shall come back to this point in Section~(\ref{sec:p_odde}).
\section{The perturbative-QCD pomeron and odderon \la{sec:qcd_pom_odd}}
%
\subsection{The Low-Nussinov pomeron\la{sec:low-nussinov}}
It is natural to ask whether the pomeron phenomenon can be addressed within 
perturbative QCD. In the following we shall keep track of colour factors for
a general number of colours $N_c$. The smallest number of gluons that can lead
to colour-singlet exchange is two. Therefore the leading contribution to the 
cross-section is of order $\alpha_s^2$. Another point can be made prior
to any calculation: if a constant cross-section (up to logarithms) is to be obtained,
the only scale available to give the cross-section its unit of area is the transverse
size of hadrons. Incidentally, this is also what is suggested by the classical 
`black-disk' picture, where the two objects interact whenever their impact parameter 
is smaller than their diameter. Let us now see how these ideas show up in explicit
calculations.

It was noted in the early days of QCD~\cite{Low:1975sv}
 that the box diagram describing two-gluon exchange between quarks is the dominant one
at high energies and leads to 
a constant total cross-section:
\be
\sigma_0=G_o\alpha_s^2\int_{\Lambda}^\infty d^2k_T ~\frac{1}{k_T^4}.
\la{eq:low_pom}
\ee
where  $G_o$ is the colour factor $G_o\equiv \frac{N_c^2-1}{N_c^2}$.
However it is obvious that the expression diverges quadratically 
in the infrared if the cutoff $\Lambda$ is removed. Therefore, the result
is sensitive to the infrared region of the theory. Naturally in reality the 
quarks are embedded in hadrons, for which an \emph{impact factor} must be introduced. 
The impact factor gives a distribution in off-shellness of the quarks; effectively 
these momenta provide the infrared cutoff for~(\ref{eq:low_pom}). An alternative, 
equivalent description is obtained in impact parameter space $\vec b$: for heavy
meson-heavy meson scattering for instance, a wave function in the valence 
$q\bar q$ dipole size can be introduced. The corresponding formula 
(see e.g.~\cite{Mueller:1999yb})
 for the dipole-dipole cross-section is given by a modification of 
Eqn.~(\ref{eq:low_pom}):
\ba
&&\sigma_{\rm dd}(d,d')= \nonumber\\
&&G_o\alpha_s^2\int \frac{d\hat n}{2\pi} 
\int \frac{d\hat n'}{2\pi}   \int \frac{d^2k_T}{(k_T^2)^2}
\left(2-e^{ik_T\cdot d\hat n}-e^{-ik_T\cdot d\hat n}\right) 
  \left(2-e^{ik_T\cdot d'\hat n'}-e^{-ik_T\cdot d' \hat n'}\right)  \nonumber  \\
&&=  2\pi~G_o~\alpha_s^2 d_<^2\left( 1 + \log{\frac{d_>}{d_<}}  \right),\qquad\qquad
d_<={\rm min}(d,d').
\la{eq:sig_dipole_4d}
\ea
Here $d,d'$ are the sizes of the two dipoles and
 $\hat n, \hat n'$ are their orientations, 
over which we have averaged. The result is manifestly finite; one may now 
introduce a weighted average over the sizes of the dipoles, as described by mesonic
wave functions. The main characteristics of the result are already clear at this stage
though: the scale for the cross-section is provided by the size of hadrons (i.e. the 
confinement scale), with logarithmic corrections depending on the details of QCD
dynamics. This is an indication that the scattering process in the Regge limit is 
dominated by small momentum scales.

Another crucial point is that because diagrams leading to the renormalisation
of the coupling constant are a subleading effect at large $s$ -- they are not enhanced
by a factor of log$s$ -- there is an ambiguity in choosing a scale at which to evaluate 
the (running) coupling $\alpha_s$. This arbitrariness can only be lifted by going to 
next-to-leading order calculations; more on  this below.
%
%
\subsection{The BFKL pomeron}
The two-gluon exchange diagram is only the first diagram of an infinite series, each
term of which carries an extra factor of $\alpha_s \log{s}$; 
this series is a subset of the full perturbative series. The approach of Balitsky,
Fadin, Kuraev and Lipatov (BFKL)~\cite{Kuraev:1977fs}
was to take the limit $\alpha_s\rightarrow0$,
$\log{s}\rightarrow\infty$, whilst keeping their product fixed; that is, they resummed
the perturbative series, keeping only the leading-logarithmic terms.
The most effective method to carry out the resummation is to write down an  equation
describing the evolution in energy. More precisely, it describes the evolution in
longitudinal momentum of the real gluons produced in the scattering process (the `rungs'
of the `ladder'). The equation does however not include the effects of evolution in the 
virtuality of the gluons along the $t$-channel exchange (along the `ladder').
Therefore the calculation can only strictly apply to processes dominated by a single 
hard momentum scale. An idealised case is the scattering
of two small dipoles (providing the hard scale), 
the `heavy-onium' collision 
considered above. Experimentally, the closest processes to the ideal situation are
forward jets in $pp$, $\bar pp$ scattering~\cite{Mueller:1987ey},
hard forward jets in DIS and $\gamma^*\gamma^*$ collisions~\cite{Ewerz:2004rc}.

A gluon is exchanged in the $t$-channel from which a number of real gluons can be emitted
in the $s$-channel. Virtual corrections lead to the `reggeisation' of the $t$-channel
gluon.
This means that the $t$-channel gluon becomes a collective excitation of the gluon field
-- the exchange of which leads to a scattering amplitude of the type $s^{\alpha(t)}$ --,
rather than a simple perturbative gluon. The gluons along the ladder are strongly ordered
in longitudinal momentum fraction -- this will lead to the log$s$ enhancement of the 
amplitude. The transverse momenta of the $s$-channel gluons on the other hand are not
ordered. In fact it can be shown that the gluon emissions along the ladder lead to a
random
walk in $\log{{\bf k}^2}$. Thus the probability distribution of momenta along the ladder
resembles a diffusion process. With increasing energy $\sqrt{s}$, 
the distribution widens and the lower part of the distribution dangerously approaches the
non-perturbative region. If the coupling is assumed to run as a function of the transverse
gluon momenta along the ladder, emissions with smaller momenta become even more likely
as $\alpha_s$ is larger at smaller momenta.

It is also worth noting that in the leading logarithmic approximation (LLA), 
the gluons emitted from the $t$-channel reggeised gluon can be produced without any cost
in energy. At large energies, the energy conservation constraint at the vertex is a 
subleading effect. With Monte-Carlo methods~\cite{Brodsky:1998kn,Schmidt:1999mz}
it is possible to study the corrections introduced by the energy constraint.
It is found that the growth of the cross-section at high energies is tamed.

The BFKL evolution is normally expressed as an integral equation, where
the kernel describes the emission of a real gluon. Famously, the solution of the BFKL
equation can be found analytically for the quark-quark scattering amplitude; 
in particular~\cite{Forshaw:1997dc}, 
\be
\sigma_{\rm qq}(s)=\alpha_s^2\frac{N_c^2-1}{N_c^2}\int \frac{d^2{\bf k_1}}{{\bf k_1}^2}
\frac{d^2{\bf k_2}}{{\bf k_2}^2}~F(s,{\bf k_1},{\bf k_2},0)
\la{eq:bfkl_sig}
\ee
with
\be
F(s,{\bf k_1},{\bf k_2},0)=\frac{1}{|{\bf k_1}|~|{\bf k_2}|}~
\frac{1}{2\pi a} \exp{\left(-\frac{\log^2({\bf k_1^2}/{\bf k_2^2})}
{4a^2\log(s/{\bf k^2})}\right)}~
\frac{\left(\frac{s}{\bf k}\right)^{\epsilon_0}}{\sqrt{\pi \log(s/{\bf k}^2)}}
\la{eq:bfkl_F}
\ee
where $a=14\frac{\alpha_s N_c}{\pi}\zeta(3)$ and 
\be
\epsilon_0 = 4\frac{\alpha_s N_c}{\pi}\log2 \la{eq:bfkl_exp}
\ee
is the famous BFKL exponent. Impact factors are introduced as functions 
$\Phi({\bf k_{1,2}})$ in the integrals over transverse momenta in 
Eqn.~(\ref{eq:bfkl_sig}) and lead to an infrared-finite expression.

Thus leading logarithm perturbation theory gives a cut
rather than a simple pole. $\epsilon_0$ evaluates to $\sim0.5$ for a choice
$\alpha_s=0.2$.
As noted above, the value of $\alpha_s$ can only be fixed by a next-to-leading order
calculation that includes the effects of the running coupling. Nevertheless the definite
prediction of the calculation is a strong rise of the cross-section at high energies.
Unfortunately, the recent $\gamma^*\gamma^*$ data from the L3 experiment in LEP2 runs 
show no sign of such a strong rise~(see for instance~\cite{Bartels:2000sk}). 
In~\cite{DelDuca:2002qt} it was found that
a calculation at fixed order in $\alpha_s$, but including the next-to-leading diagrams
with respect to the $\log{s}$ expansion, produces a better description of the data, 
although the prediction is somewhat too low~\cite{Ewerz:2004rc}.

In 1998, the next-to-leading order (NLO) evolution equation was found independently 
by two groups~\cite{Fadin:1998py}.
The exponent of $s$ in the cross-section expression 
finds itself being reduced. Together with renormalisation group improvement of the BFKL 
equation~\cite{Brodsky:1998kn} resumming additional large logarithms of the 
transverse momentum,
the NLO equations produce an exponent for the energy dependence of about 0.2-0.3 (the 
precise value is somewhat scheme dependent). The ultimate test, namely to compare this 
improved prediction to the L3 data, has not been carried out yet.
%

Meanwhile, in most other processes such as small-$x$ DIS, one must expect  
the perturbative pomeron to get convoluted with `soft physics'; it may be that such a
`convolution' corresponds to the phenomenological pomeron observed in HERA 
data~\cite{Donnachie:2001xx}. Time will tell.
\subsection{The perturbative odderon\la{sec:p_odde}}
The simplest PQCD diagram that can produce a colour singlet
$C=-1$ trajectory is the 3-gluon exchange diagram. Analogously to the BFKL generalisation
of the 2-gluon exchange diagram, there exists an integral equation
\cite{Bartels:1980pe} that resums
the leading logarithms of the energy $\sqrt{s}$ for three reggeised gluons in 
the $t$-channel. This `Bartels-Kwiecinski-Praszalowicz' (BKP) equation is analytically
solvable. Several classes of solutions were found. The Janik-Wosiek 
solution~\cite{Janik:1998xj} has an intercept 
$\alpha_0 = 1 - 0.24717\frac{\alpha_sN_c}{\pi}$,
while the Bartels-Lipatov-Vacca solution (BLV)~\cite{Bartels:1999yt} has an 
intercept that is 
exactly 1. Due to their different couplings, it is likely~\cite{Ewerz:2004rc} 
that the BLV 
solution gives the leading contribution to most processes. Thus perturbative QCD 
firmly predicts an important contribution of the odderon to cross-sections.

Phenomenologically, the odderon remains largely a mystery, due to the difficulty
of disentangling it from the other reggeon contributions and its strong
dependence on $\alpha_s$~\cite{Ewerz:2004rc}. 
Interest has shifted to exclusive processes, 
for instance $pp\rightarrow J/\psi$~\cite{Schafer:1991na} or 
 $\gamma^* p\rightarrow pM_{PS}$, both requiring an odderon exchange. 
Another promising idea is to use the pomeron-odderon 
interference~\cite{Merino:1999yn}.
\section{Unitarisation\la{sec:unitarisation}}
As mentioned in the previous section, processes where the BFKL amplitude is expected to 
be valid are rather rare. At the theoretical level, it is interesting to investigate how 
far one can go with perturbation theory in a way to describe the effects of unitarisation.

It is clear that at asymptotic energies, the unitarisation (and hence the saturation of 
the Froissart bound) has to come from non-perturbative
effects, as can be seen from the following intuitive argument (originally due to Heisenberg
and reported e.g. in~\cite{Kovner:2002xa}). In a theory with a mass gap, the
distribution
of matter density in a target must decay exponentially at the periphery, 
$\rho(b)\sim\exp{(-mb)}$. For a projectile to scatter inelastically on the target, at least one
particle must be produced. Therefore the overlap of the probe and the target must contain
an energy at least equal to the mass gap $m$. In the `infinite momentum frame', where
all the energy $E=s/m$ of the reaction is stored in the target, the target energy density
is $E\rho(b)$. Thus the maximal impact parameter at which the scattering can occur
is given by $E\exp{(-mb)}=m$; hence $b_{\rm max}=\frac{1}{m}\log{s/m^2}$, which leads to the
Froissart bound. Conversely, a power growth with energy of the cross-section implies
a power-law distribution of matter rather than an exponential one. Indeed with 
$\rho(b)\propto b^{-\lambda}$ one obtains $ b_{\rm max}\propto s^{1/\lambda}$.
The fact that hadronic cross-sections are still well fitted by power-laws in the energy 
ranges where they have been measured was taken as a hint that the true asymptotic regime
has not been observed yet~\cite{Kovner:2002xa}, 
and that the currently available data may perhaps
be understood within the perturbative framework.

Let us consider again the ideal case of the scattering of two heavy `onia'.
There are two equivalent ways to
interpret the BFKL ladder with respect to the simple two-gluon exchange, 
depending on the reference frame that is chosen. The original point of view
held by its authors, 
expressed in the centre-of-mass frame, is that the BFKL ladder resums the 
leading-logarithmic exchanges of gluons between \emph{one} constituent of the 
left-moving onium with \emph{one} constituent of the right-moving onium.
A different perspective was taken by A.H.~Mueller; let us 
choose for instance the target rest-frame. The BFKL evolution equation can now
be interpreted as the evolution with energy $\sqrt{s}$ of the projectile's 
gluon content, each constituent of which then simply scatters via two gluon-exchange on the 
gluon field of the target. Alternatively, of course, one could describe the evolution
of the target wave function in the rest frame of the projectile.

This change of point of view, together with some simplifications due to the large-$N_c$
formalism, leads to Mueller's colour dipole picture (CDP) 
of high-energy scattering~\cite{Mueller:1994rr}.
Indeed, the large-$N_c$ limit allows one to treat a gluon
diagrammatically like a $\bar q q$ pair. The emission of a gluon by the primary 
dipole (the valence quarks of the onium) is interpreted as its splitting into two:
each new dipole is made of the (anti-)quark component of the primary dipole and the 
(anti-)quark component of the emitted gluon. The iteration of this process leads
to an `evolved' wavefunction description as a system of dipoles. The large-$N_c$ limit
implies that one can neglect the interference between emissions from different dipoles:
they emit independently, resulting in a tree of dipoles. The linear approximation 
means that the interaction among dipoles is neglected, implying that the dipole number
density evolves according to the (linear) BFKL equation. Incidentally, this approximation
puts a high-energy limit on the applicability of the dipole picture; indeed, 
non-linear effects, such as dipole recombination, become important at energies such that
the dipole density  $\exp{(y' \epsilon_0)} $ compensates for the weakness of the
dipole-dipole 
interaction $\sim\alpha_s^2$; $y'\sim \log{s}$ is the rapidity of one of the onia. 
This is the so-called \emph{saturation} of the onium wave function, 
an effect which is difficult to describe in the dipole formalism. 

On the other hand, Kovchegov~\cite{Kovchegov:1999yj}, following the 
work at general $N_c$ of~\cite{Balitsky:1996ub},
was able to write an evolution equation\footnote{It is called the BK equation.} 
taking into account the \emph{multi-scatterings} 
of \emph{different} dipoles of one onium on \emph{different} dipoles of the
other\footnote{Note that the probability of a single 
dipole undergoing multiple scattering is still suppressed (it is of order $\alpha_s^2
e^{\epsilon_0 y/2}$).}. 
A key question is which of the two effects, saturation or multi-scattering, 
is the dominant sub-leading correction to the linear BFKL evolution.
Depending on which reference frame one chooses to calculate the wave functions and 
the multi-scattering processes, the rapidities of the two onia are different.
Thus in general, one of them has a dipole density $e^{\epsilon_0 y'}$ and the other
$e^{\epsilon_0 (y-y')}$. While the effects of multi-scattering become important when
$\alpha_s^2 e^{\epsilon_0 y}={\cal O}(1)$, saturation has to be taken into account
when $\alpha_s^2 e^{\epsilon_0{\rm max}(y-y',y')}={\cal O}(1)$. 
In the centre-of-mass frame, 
$y'=y/2$, the effects of saturation are maximally suppressed, hence it is the 
frame of choice in this formalism. The distinction between saturation 
and unitarisation is frame-dependent; the Lorentz invariance of the scattering amplitude
allows us to make the most convenient choice of reference frame.

The  BK  equation thus takes care of the dominant sub-leading corrections to the 
BFKL evolution, which are perturbative. It is a non-linear rapidity-evolution equation 
of the dipole scattering probability $N$, and the non-linearity describes the onset of
unitarisation. Although the equation cannot be solved exactly, the qualitative behaviour
of the solution is given by~\cite{Golec-Biernat:1999qd} 
\be
N(x,y)\simeq 1-\exp{\left[-(x-y)^2Q_s(y)^2\right]},
\ee
where $(x,y)$ are the transverse coordinates of the `legs' of the dipoles and $Q_s$
is the saturation momentum, which has a qualitative dependence on the rapidity of the type
$\sim \exp{(\lambda \alpha_s y)}$, $\lambda={\cal O}(1)$; it also represents the
centre of the distribution in transverse momentum of the gluons.
 Thus the equation does yield
a unitary evolution, and moreover it can be shown that the width of the latter
distribution is roughly independent of $s$, thus avoiding
the diffusion into the infrared region that occurs in the BFKL evolution.

The validity of the BK equation, as discussed above,
is limited in rapidity to the regime $\alpha_s^2e^{\epsilon_0 y/2}\ll 1$. At higher 
energies recombination of dipoles in each wave function 
must be taken into account --- presumably an evaluation of loops of BFKL ladders becomes
necessary. This task has not yet been completed~(see for instance~\cite{Navelet:2002zz}).
In the mean time however, a different formalism was developed to describe the effects 
of saturation: the colour glass condensate formalism (see the review~\cite{Iancu:2003xm} 
and references therein).

In this formalism, the small-$x$ partons (the `wee partons') are viewed as radiation 
products of faster-moving colour charges; the latter's internal dynamics is frozen by 
Lorentz time dilation. Since they are produced by partons with a large spread in 
momentum fraction $x$, the
distribution in colour in the transverse space becomes random~\cite{Iancu:2003xm}.
A hadron's wave function is fully specified by giving the probability
law for the spatial distribution of the colour charges. The framework is still 
perturbative QCD, but  
the coupling between the quantum fluctuations and the classical colour field radiated
by the faster sources includes non-linear effects. These processes are described by 
a functional renormalisation group equation, called the JIMWLK equation\footnote{The authors
involved are Jalilian-Marian, Iancu, McLerran, Weigert, Leonidov, Kovner.}. 
In the limit of weak colour field the equation reproduces the BFKL 
evolution~\cite{Jalilian-Marian:1997jx}. At high energies however, 
the equation describes the formation of a highly dense gluonic state -- the colour glass
condensate, characterised by the saturation momentum $Q_s$ and large occupation numbers
${\cal O}(1/\alpha_s)$ for the modes with momentum less than or equal to $Q_s$.
It was shown in~\cite{Iancu:2003uh} that the colour glass condensate formalism yields 
the same answer, in the weak-field regime and at large $N_c$, 
as  the colour dipole formalism for onium-onium scattering.
The formalism can potentially  give a universal description of scattering processes
through the description of a new `state' of QCD matter at high-energies.
The way forward to deal with the complexity of the JIMWLK equation seems to be numerical 
techniques~\cite{Rummukainen:2003ns}.
\section{Non-perturbative models of the soft pomeron \la{sec:np_spom}}
As discussed in the previous section, unitarisation of the scattering amplitude
eventually requires non-perturbative input at high enough energies. Several approaches
have been attempted to give a semi-quantitative description of high-energy scattering
invoking non-perturbative effects. We  only mention a small number of them.

A general formula for the high energy quark -- anti-quark scattering amplitude 
in the eikonal approximation was worked out by Nachtmann~\cite{Nachtmann:1991ua}.
It relates
the amplitude to the correlator of two Wilson lines running along the light-cone, 
$x^\pm=$cst, where $x^\pm = (t\pm z)/\sqrt{2}$. This formula is powerful because it 
allows one to use any formalism of choice to
evaluate the correlator of Wilson lines. In the colour-glass condensate formalism
for instance, onium-onium scattering 
in an asymmetric reference frame is described as 
 the scattering process of
 a dipole $ ({\bf x},{\bf y})$ by a colour glass \cite{Iancu:2003uh}. 
Then the scattering matrix simply reads
\be
S({\bf x},{\bf y})=\frac{1}{N_c}\langle \tr\{ V_{\bf x}^\dagger V_{\bf y} \}\rangle~,
\qquad\quad
V^\dagger_{\bf x}[A]= \tr\left[{\rm P} ~ \exp{\int_{-\infty}^\infty
 dx^- A_-(x^-,{\bf x})}\right].
\ee
The correlator now takes the meaning of a weighted average over the gluon field
of the colour glass.

Herman and Erik Verlinde~\cite{Verlinde:1993te} developed an effective high-energy 
Lagrangian 
formalism. In the high-energy limit, the longitudinal degrees of freedom can be integrated
out and the effective action becomes two-dimensional. In that context, the correlator
is evaluated as a path integral where the distribution of fields
 is given by the Boltzmann weight associated with the effective action.
While a perturbative treatment of the effective action reproduces 
the ordinary perturbative results, this formalism offers the possibility of 
a non-perturbative treatment.

Landshoff and Nachtmann proposed one of the first non-perturbative model for the soft 
pomeron~\cite{Landshoff:1987yj}. 
It was based on ideas of the stochastic vacuum~\cite{Olesen:1982zp}
and of the QCD sum rules~\cite{Shifman:1979bx}. In particular, it was shown 
that if the size of hadrons ($\sim1$fm) happened to be numerically larger than 
the typical size $a$ of a domain of constant magnetic flux in the non-perturbative vacuum,
then the additive-quark-rule naturally followed. Indeed, $a$ acts as the correlation length
of the gluon field in the vacuum, and therefore two quarks can only exchange one such 
`massive' gluon if they cross each other at an impact parameter smaller than $a$.
Therefore the two gluons necessary to ensure colour-singlet exchange will predominantly 
couple to the same quark. The cutoff for the transverse momentum integration 
in the expression of the cross-section~(\ref{eq:low_pom}) is then given by $a^{-1}$.
Below that momentum, a non-perturbative ansatz has to be made for the gluon propagator, 
whose behaviour at $k^2=0$ is assumed to be finite and determined by the scale of the
gluon 
condensate, $M_c^4$. Most of the contribution to the cross-section comes from this
non-perturbative region. By using phenomenological values of hadronic 
cross-sections, Donnachie and Landshoff inferred~\cite{Donnachie:1989nj} estimates
 for $a\simeq 0.3$fm  and $M_c\simeq 1.2-1.6$GeV.

More recently, a picture of the soft pomeron 
where the rungs of the BFKL ladder couple to fluctuations
 of the non-perturbative vacuum  was proposed~\cite{Kharzeev:1999vh}.
Attempts have been made to describe an evolution from the `hard', BFKL ladder
 to the `soft' pomeron using this picture~\cite{Bondarenko:2003jv}.
%
%
%
\section{Conclusion}
The review presented above (hopefully) gives an impression of 
the wealth of ideas developed in the subject
of high-energy hadronic reactions. In our view, the importance of the subject stems
from the fact that these processes probe the dynamics of the theory at all scales,
thus providing an opportunity to study the cross-over from the perturbative to the 
non-perturbative regime. 

There is one aspect which has been left aside.
According to Regge theory, at positive $t$, one should find physical states
lying on the pomeron trajectory. Since in QCD the pomeron is thought to correspond
to the exchange of excitations of the gluon field, these states should be bound states
of gluons, the `glueballs'. The relation between glueballs and the pomeron was 
investigated within a constituent gluon model in 
\cite{Kaidalov:1999yd,Llanes-Estrada:2000jw}. 
In the former article the leading glueball trajectory was found to be 
$\alpha_P(t) = \alpha_0 + \alpha't$, $\alpha'=\frac{1}{2\pi\sigma_a}$,
where $\sigma_a$ is the adjoint string tension and $\alpha_0\simeq0.5$. 
In this model it is the mixing of gluonic with $\bar qq$
states which must account for the intercept $\sim1$ of the phenomenological pomeron.
A similar trajectory is expected to correspond to the odderon
 \cite{Kaidalov:1999yd,Llanes-Estrada:2000jw}.  In the constituent gluon
model, these states have to be formed of three gluons at least. The $3^{--}$ state
is found to be around 3.6GeV, and assuming the same slope for the $C=-$ as for the $C=+$,
this leads to a negative intercept.

In the next chapter, we discuss string models of glueballs in more detail
and work out the qualitative features of the Regge trajectories they predict.

%% file: chapter2.tex
\chapter{String models of glueballs}
\label{ch:string}
If we assume that the fundamental degrees of freedom of low-energy QCD
are those of the QCD string, or `flux-tube'~\cite{Isgur:1985bm}, 
then it seems natural to associate the glueball spectrum
with the spectrum of the bosonic  
string~\cite{Huang:1970iq, Goddard:1973qh}. 
However,  famously the bosonic string must live in 26 dimensions in order 
for its spectrum to preserve Lorentz invariance; moreover
 the fundamental state of the string is a tachyon, and it has 
a massless spin 2 mode~\cite{Green:1987sp}.
Secondly, it is only in the case of a stretched open string 
-- whose length $L\gg \sigma^{-1/2}$ ensures
 that the excitations of the effective string action are much 
lower-lying than the intrinsic excitations of the flux-tube --
that the leading correction to the 
energy of the string assumes a universal form~\cite{Luscher:1981ac};
it is then calculable by semi-classical methods and  depends only on the
 central charge of the string action.
The closed-string configuration, on the other hand,
naturally takes a size of order $\sigma^{-1/2}$. In such a situation, 
one cannot expect to find a universal spectrum, independent of the 
internal properties of the flux-tube. 
And yet universality could be regained at large angular momentum $J\gg \hbar$. 
It is well-known that the semi-classical Bohr-Sommerfeld model
 of the hydrogen atom works well at large angular momentum.
The presence of the large parameter $J$ allows us to treat quantum mechanical
effects as small corrections to the classical result. Another 
 way to understand this is that for a given  angular momentum, the string
will try and minimise its moment of inertia. In fact one finds that the 
square length of the string increases proportionally to $J$. Eventually this 
brings us back to the situation of the stretched string, where universality
should manifest itself.
The correction $\alpha_0$ 
to the classical Regge trajectories $\alpha(t)=\alpha't$ is the analog
of the L\"uscher correction to the energy of a long string.
In general a high-spin glueball
would decay very rapidly into lighter glueballs; 
unless we take the limit where hadrons are  stable, that is, 
the planar limit $N_c\rightarrow \infty$.

Hence, there is a strong theoretical motivation for investigating
string models of glueballs:
if they can be thought of as spinning and vibrating configurations of an
effective string, then their spectrum should be universally calculable
in the planar limit and in the large angular momentum regime.
Static-potential~\cite{Luscher:2002qv} and torelon-mass\cite{Lucini:2001nv} 
calculations provide strong numerical
 evidence that the universality class of the QCD string is bosonic.
We therefore expect to find the same bosonic class for the string
 configurations corresponding to glueballs.
Establishing the large angular momentum glueball spectrum is thus part
of the long-standing program of relating gauge theories to string theories.
%
\section{Two string  models of glueballs}
\label{sec:models}
\noindent 
In the standard valence quark picture, a high spin meson
will consist of a $q$ and $\bar{q}$ rotating rapidly around
their common centre of mass. For large angular momentum $J$ they
 will be far apart and the chromoelectric
flux between them will be localised in a flux tube which
also rotates rapidly, and so contributes to $J$. In a
generic model of such a system, a simple calculation shows
that the spin and mass are related linearly, 
$J=\alpha_0 + \alpha^\prime M^2$, and that the slope
is related to the tension $\sigma_f$ of the
confining string\footnote{The subscript $f$ indicates that the charges and
flux are in the fundamental representation. We will often 
follow convention and use $\sigma\equiv\sigma_f$ instead.}
 as  $\alpha^\prime = 1/2\pi\sigma_f$.
If one uses a phenomenologically sensible value for
$\sigma_f$ one obtains a value of $\alpha^\prime$
very similar to that which is experimentally observed 
for meson trajectories. This picture might well
become exact in the large-$N_c$ limit where the
fundamental string will not break and all the
mesons are stable. 

This picture can be generalised directly to glueballs.
We have two rotating gluons joined by a rotating flux tube 
that contains flux in the adjoint rather than fundamental 
representation. This is the first model we consider below. 
However for glueballs there is another possibility that is
equally natural: the glueballs may be composed of closed
loops of fundamental flux. This is the second model
we consider. The first  is natural in a 
valence gluon approach, while the second arises naturally
in a string theory. They are not exclusive; both may contribute
to the glueball spectrum. Indeed if there are two classes of 
glueball states, each with its own leading Regge trajectory,
one might for instance speculate that they correspond to
the phenomenological  `hard' pomeron and  `soft' pomeron
(see Chapter~\ref{ch:hehr}). 
In our view the connexion between glueball Regge trajectories
and high-energy scattering constitutes another strong motivation for
studying string models of glueballs, since the latter
 naturally lead to Regge trajectories.
Both above-mentioned models can be motivated as easily
 in 2+1 as in 3+1 dimensions and in both cases predict linear glueball
trajectories with some pomeron-like properties.
%
\subsection{The adjoint-string model\label{kaidalov}}
In this model~\cite{Kaidalov:1999yd}, 
the glueball  is modelled as an adjoint string binding 
together two adjoint sources, the constituent gluons. 
It is a direct extension
to glueballs of the usual model for high-$J$ $q\bar{q}$ mesons
where the $q$ and $\bar{q}$ are joined by a `string' in the 
fundamental representation. The adjoint string is of course unstable, 
once it is long enough (as it will be at high $J$), but this is also 
true of the fundamental string in QCD. The implication is  that 
glueballs cannot strictly  have a definite number of constituent gluons.
What is important for the model to make sense is that the decay width should
be sufficiently small -- essentially that the lifetime of the adjoint 
string should be much longer than the period of rotation.
In $SU(N_c)$ gauge theories, both the
adjoint and fundamental strings become completely stable as
$N_c\to\infty$. So if we are close to that limit the model can
make sense. Since adjoint string breaking in $SU(N_c)$ occurs at
$O(1/N_c^2)$ while fundamental string breaking in $QCD_{N_c}$ occurs 
at $O(1/N_c)$, one would expect the instability to be less of
a problem in the former case. Moreover there is now considerable
evidence
\cite{Lucini:2001ej,Lucini:2002ku,DelDebbio:2002xa}
from lattice calculations that the $D=3+1$ $SU(3)$ 
gauge theory is indeed `close' to $SU(\infty)$, and that this 
is also the case for $D=2+1$ $SU(N_c)$ gauge theories for $N_c\geq2$
\cite{Teper:1998te}.

The calculation of the  $J$ dependence of the  glueball mass $M$
 is exactly as for the  $q\bar{q}$ case
\cite{Perkins:1982xb}.
That is to say, we consider the string joining the two gluons as a 
rigid segment of length $2r_0$, rotating with  angular momentum $J$ 
(the contribution of the valence gluons being negligible at high
enough $J$). The local velocity at a point along the segment is thus
$v(r)=r/r_0$ (one maximises $J$ at given $M$ if the end-points move 
with the speed of light), so that
\ba
M&=&2\int_0^{r_0} \frac{\sigma_a dr}{\sqrt{1-v^2(r)}}=\sigma_a\pi r_0\\ 
J&=&2\int_0^{r_0} \frac{\sigma_a r v(r)dr}{\sqrt{1-v^2(r)}}=
\frac{\pi}{2}\sigma_a r_0^2,
\ea
and, eliminating $r_0$, 
%
%
we obtain a linear Regge trajectory 
$J=\frac{M^2}{2\pi\sigma_a}$ where $\sigma_a$ is the adjoint
string tension. So this model predicts that the slope of the leading
glueball trajectory is smaller than that of the leading meson
trajectory by a factor $\sigma_f/\sigma_a$.
If Casimir scaling is used~\cite{Deldar:1999vi},
$\frac{\sigma_a}{\sigma}\simeq
 \frac{2N_c^2}{N_c^2-1}$, the predicted Regge slope is $4/9$ of the mesonic 
trajectories in $SU(3)$. Thus the leading glueball trajectory will have a slope
$\alpha_{AS}' \sim 0.88/2.25 \sim 0.39 ~\mathrm{GeV}^{-2}$ 
if we input the usual Regge slope of about 
$\alpha_{R}'=\frac{1}{2\pi\sigma_f} \simeq 0.88~\mathrm{GeV}^{-2}$.
This is only a little larger
than the actual slope of the `soft' pomeron. Thus to this extent the model
is consistent with the idea that the pomeron is the leading glueball
trajectory, perhaps modified by mixing with the flavour-singlet 
meson Regge trajectory.  In the planar 
limit, the adjoint string becomes stable and the ratio of string tensions
approaches 2; the Regge slope is then $(4\pi\sigma)^{-1}$. Interestingly, 
this is also the result for the collapsed,  segment-like configuration
of the closed flux-tube discussed below.

Since in this model the rotating glueball lies entirely within 
a plane, the calculation is identical for $D=2+1$ and  $D=3+1$.
Thus it is also a plausible model for the leading Regge trajectory
in $D=2+1$ $SU(N_c)$ gauge theories. 

In~\cite{Kaidalov:1999yd} the adjoint string model was taken beyond the 
classical limit just presented. The full spectrum obtained is well 
approximated by~\cite{Kaidalov:1999yd}
\be
\frac{M^2}{2\pi\sigma_a} = J + 2n_r + c_1,
\ee
where $c_1$ is a number of order 1, and $n_r$ is a radial quantum number.
The leading trajectory contains $PC=++$, even spin states.
The `einbein' formalism  to deal with the relativistic Hamiltonian 
is reviewed in~\cite{Kalashnikova:2001ig}; it would be very interesting to also apply it 
to the flux-tube model described below
in view of obtaining its relativistic corrections.

The semi-classical corrections to the classical trajectory
were calculated at large $J$ in~\cite{Baker:2002km} in the context of 
mesonic trajectories. 
The action is expanded to quadratic order in the fluctuations around
the classical solution that we considered. We quote the result:
\be
\alpha(t)=\frac{t}{2\pi\sigma_a}+\frac{(D-2)\pi}{24}+{\cal O}
\left(\frac{\sigma_a}{t}\right).
\ee
%
\subsection{The flux-tube model\label{sec:isgur}}
An `open' string model of the kind described above, is
essentially forced upon us if we wish to describe high-$J$
mesons within the usual valence quark picture. For
glueballs, however, there is no experimental or theoretical
support for a valence gluon picture. A plausible alternative
is to see a glueball as being composed of a closed loop
of fundamental flux.  A 
simple first-quantised model of glueballs as closed
flux tubes was formulated some time ago
\cite{Isgur:1985bm}
and has been tested with some success 
\cite{Moretto:1993dc,Johnson:2000qz}
against the mass spectrum of D=2+1 $SU(N_c)$ gauge theories as 
obtained on the lattice
\cite{Teper:1998te}.

In this model the essential component is a circular
closed string (flux tube)  of radius $\rho$. There are 
phonon-like excitations of this closed string
which move around it clockwise or anticlockwise and
contribute to both its energy and its angular momentum.
The system is  quantised so that we can calculate, 
from a Schr\"odinger-like wave equation, the amplitude for
finding a loop in a particular radius interval. The phonon
excitations are regarded as `fast' so that they contribute
to the potential energy term of the equation and
the phonon occupation number is a quantum number
labelling the wave-function.

Let us be more specific and consider the model in 3+1 dimensions. 
The fundamental configuration is a circular ring. 
Small fluctuations of the loop in the radial direction, parametrised by
\[ \delta\rho(\varphi) = \sum_{|m|\geq2} \alpha_m\sin{m\varphi} + 
    \beta_m\cos{m\varphi}, \]
and similar fluctuations in the $\hat z$ direction orthogonal to the string 
plane are expected to have a harmonic oscillator 
Hamiltonian~\cite{Moretto:1993dc,Isgur:1985bm}
\ba
H_{\rm phon} &=& \frac{1}{2\pi\sigma\rho}\sum_{n\geq2} 
( p^{(z)2}_{\alpha_n} + p^{(z)2}_{\beta_n} +
 p^{(\rho)2}_{\alpha_n} + p^{(\rho)2}_{\beta_n})\nonumber \\
&&+\frac{\pi\sigma}{2\rho}\sum_{n\geq2} n^2
(\alpha_n^{(\rho)2} + \beta_n^{(\rho)2} + \alpha_n^{(z)2} 
+ \beta_n^{(z)2}),   \la{eq:phon_hamil}
\ea
where the $p$'s are their conjugate momenta. The quantised normal modes, 
or \emph{phonons}, carry  the following eigenvalues of  energy $M/\rho$
 and angular momentum $\Lambda$:
\ba
M &=& \sum_{m\geq 2} m\left( n^{(\rho)}_m + n^{(\rho)}_{-m} +
	 n^{(z)}_m + n^{(z)}_{-m}  \right) \label{phonon_e}\\
\Lambda &=&  \sum_{m\geq 2} m\left( n^{(\rho)}_m - n^{(\rho)}_{-m} +
	 n^{(z)}_m - n^{(z)}_{-m}  \right) \label{phonon_j}
\ea
The $m=1$ phonons would correspond to translation or rotation and are 
therefore  spurious degrees of freedom. The $m=0$ phonon, 
or `breathing mode', describes the dynamics of the radius
$\rho$ of the circle. 
The loop is of course classically unstable, but, just as 
it happens for the hydrogen atom, quantising the radial variable stabilises 
it via the uncertainty principle.

The `collective' motion, namely the rotation of the whole ring around an
axis lying inside its plane, can also contribute to the angular momentum:
$\mathbf{J} = \mathbf{L} + \mathbf{S}$,
where $\mathbf{L}$ is the orbital angular momentum and 
$\mathbf{S}$ the phonon angular momentum.
However, in the spirit of the collective models in nuclear 
physics~(see \cite{Feshbach:1974zp}, Chapter~VI), 
the `Coriolis effect' $\mathbf{L}\cdot\mathbf{S}$ 
of the collective motion on the `internal', 
phononic modes is neglected. The change in mass and moment of inertia
due to the phonons is neglected in the treatment of the collective motion
in $\rho,\theta,\phi$: the time-scale of the radial and orbital
modes is supposed to be much larger than that of the phonons - this is 
sometimes referred to as the adiabatic approximation. These approximations
work well  in nuclear physics, where `rotational
bands' build up on each of the `intrinsic' states. In that physical situation,
the energy gap between the latter is empirically found to be much larger than
those within a rotational band; in the present case, given that we are dealing
with a one-scale problem, such a separation of scales is in general 
a crude approximation. In particular,
the model does not describe the deformation of the circular loop into an 
ellipse which has larger moment of inertia and therefore allows for a lower
energy at fixed angular momentum.

Under these simplifying assumptions, the Hamiltonian describing the 
collective and phononic degrees of freedom separates. The phonons have a 
harmonic oscillator spectrum (Eqn.~\ref{phonon_e}). 
The Hamiltonian, restricted to the Hilbert subspace with quantum
numbers $J,\Lambda,M$, is 
\be
H=\pi\sigma \rho \dot\rho^2 + 2\pi\sigma \rho + \frac{M-\gamma}{\rho}
+\frac{J(J+1)-\Lambda^2}{2\pi \sigma \rho^3}. \la{eq:ft_hamil}
\ee
A parameter $\gamma$ has been introduced:
the zero-point fluctuations of the phonons renormalise the string tension
and produce a `L\"uscher term' proportional to $1/\rho$. However since
we do not expect this coefficient to be universal, we keep it as a free 
parameter of the model (following~\cite{Isgur:1985bm}).

We note again that the fundamental state, $\Lambda=M=J=0$, is 
classically unstable. Although it is stabilised quantum mechanically,
 its energy becomes very low. Partly for that reason,
 a `fudge factor' $(1-e^{-f'\sqrt{\sigma}\rho})$ multiplying the $1/\rho$ 
term was introduced in the original article by Isgur and 
Paton~\cite{Isgur:1985bm}
 that prevents the ring to shrink to zero radius at the classical level. 
On the other hand, as soon as the string is excited the states are 
classically stable without any need for a `fudge factor'. We shall ignore 
such a factor for the moment.
\paragraph{Regge trajectories in  the flux-tube model  spectrum}
In  Appendix~\ref{ap:ft}, 
we show that two types of straight Regge trajectories are
 obtained from the Hamiltonian~(\ref{eq:ft_hamil}) at large angular momentum: 
one of them is due to the phonon dynamics and is given by
\be
\alpha(t)=\alpha' t+\alpha_0+{\cal O}(1/t)
\ee
with
\be
\alpha' =\frac{1}{4}\left(\frac{1}{2\pi\sigma}\right),\qquad
\alpha_0=\gamma-1-\frac{1}{\sqrt{2}}.   \la{eq:regge_phon}
\ee
The second is associated with the orbital motion. It has 
\be
\alpha' =\frac{3\sqrt{3}}{16}\left(\frac{1}{2\pi\sigma}\right),\qquad
\alpha_0=\frac{1}{2}\left(\sqrt{3}(\gamma-1)-1 \right)
 \la{eq:regge_orbi}
\ee
The unusual slope is associated with the moment of inertia of the ring. 
If the model described its collapse to a segment in the plane 
orthogonal to the axis of rotation, 
the slope would be $\frac{1}{4\pi\sigma}$.

The fully quantum mechanical trajectories can be obtained numerically.
Fig.~(\ref{fig:traj}) shows the corresponding Chew-Frautschi plot and 
compares them to the  semi-classical
 predictions, for the value $\gamma=13/6$ (the value obtained by 
summing up the zero-point fluctuations of the phonons~\cite{Moretto:1993dc}).

We remark that for $SU(N_c>3)$ gauge theories,
the fundamental string is no longer the only one that
is absolutely stable, and closed loops of these 
higher representation strings provide an equally
good model for glueballs
\cite{Johnson:2000qz,Lucini:2001nv}.
These extra glueballs
will however be heavier and, to the extent that
we are only interested in the leading Regge trajectory,
will not be relevant here. 
\begin{figure}[t!]
\vspace{-0.7cm}
\centerline{\begin{minipage}[c]{9cm}
\psfig{file=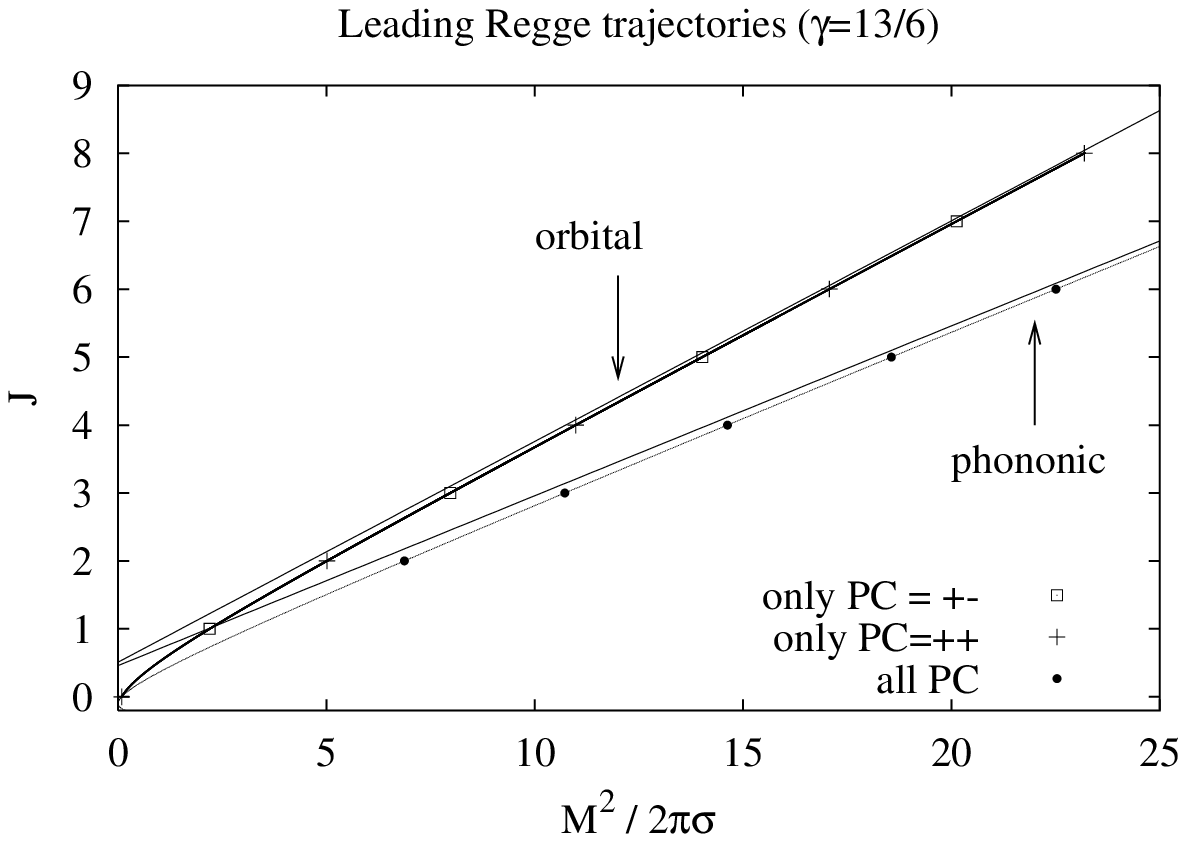,angle=0,width=9cm}
\psfig{file=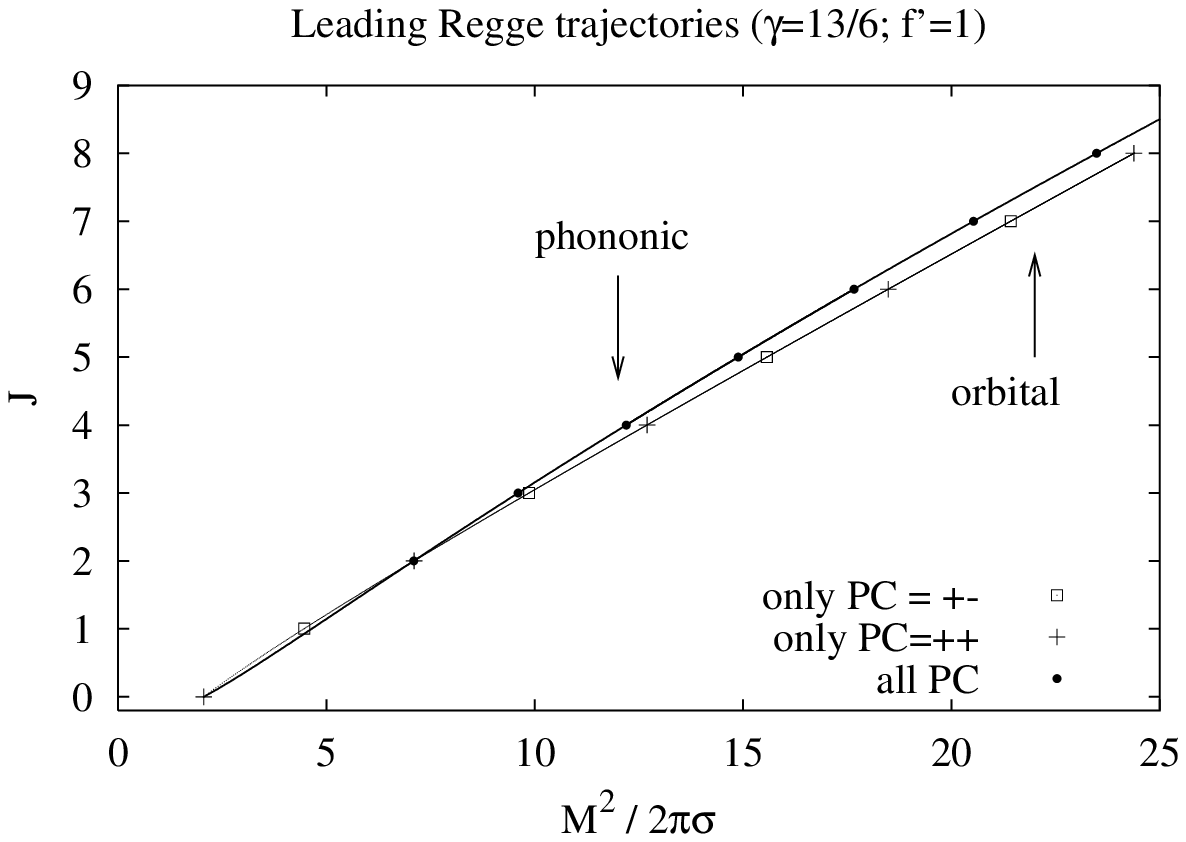,angle=0,width=9cm}
	    \end{minipage}}

\caption[a]{The leading phononic and orbital Regge trajectories in the 
flux-tube model in 3+1D. Top: without `fudge factor'; 
the straight lines are the 
semi-classical trajectories (Eqn.~(\ref{eq:regge_phon})
and~(\ref{eq:regge_orbi})). Bottom: with fudge factor, $f'=1$. 
Crosses, circles and 
squares indicate the position of physical states with the corresponding
quantum numbers.}
\la{fig:traj}
\end{figure}
%
\paragraph{Quantum numbers of the Regge trajectories}
 For the orbital trajectory, the geometry of the circular loop 
automatically gives it positive parity $P=+$. Furthermore, 
the mere fact that an oriented planar loop is spinning around an axis
 contained in its plane implies that the charge conjugation is determined 
by the spin:
\be
P~=~+,\qquad C~=~(-1)^{J},\qquad J=0,1,2\dots\qquad({\rm orbital~trajectory})
\ee
For the leading phononic trajectory, the most obvious feature is the 
absence of a $J=1$ state, because there is no $m=1$ phonon. Secondly, 
for a planar loop, parity has the same effect as a $\pi$-rotation around
an axis orthogonal to its plane. Therefore:
\be
P~=~(-1)^J, \qquad C=\pm
\qquad J=0,2,3,4,\dots\qquad({\rm leading~}\hat\rho~{\rm phononic~trajectory})
\ee
On the other hand, $P=(-1)^{J+1}$ for the $\hat z$ phonons:
\be
P~=~(-1)^{J+1},\qquad C=\pm
\qquad J=0,2,3,4,\dots\qquad({\rm leading~}\hat z~{\rm phononic~trajectory})
\ee
\paragraph{Other topologies}
It is conceivable that for those quantum numbers
for which the simple flux-tube model predicts a very large mass,
other topologies of the string provide ways
to construct a lighter fundamental state. A new pattern of quantum numbers 
arises 
if the oriented closed string adopts a twisted, `8' type configuration, 
whilst remaining planar. The parity of such an object is automatically locked to
the charge conjugation quantum number, $P=C$. The orbital trajectory
 built  on such a configuration leads to a sequence of states
\be
0^{++},~1^{--},~2^{++},~3^{--},~4^{++},~\dots\qquad
({\rm twisted~orbital~trajectory})\la{eq:tw_traj}
\ee
More exotic topologies of the string 
have been advocated in~\cite{Niemi:2003hb}, but they
presumably lead to more massive states. Such objects are at best relevant 
to the large $N_c$ limit, where they will not decay.
%
\paragraph{The 2+1-dimensional case}
In the case of two space dimensions, the Hamiltonian does not contain any
orbital term (the last term in Eqn.~\ref{eq:ft_hamil}), and the $\hat z$
phonons are absent. This does not modify the slope of the phononic 
trajectory, but
modifies the intercept to $\alpha_0(D_s=2)=\gamma-\frac{1}{\sqrt{2}}$.
Moreover, the parameter $\gamma$ is expected to be smaller in 2 space
dimensions, because the number of transverse dimensions to the string is 
smaller (for the bosonic string, it is 13/12~\cite{Moretto:1993dc}). 
The quantum numbers for the leading $C=+$ and $C=-$ phononic trajectories are
\ba
J^{PC}&=&0^{++},~2^{P+},~3^{P+},~4^{P+},~\dots \quad
C=+~({\rm leading~phononic~trajectory})\nonumber \\
J^{PC}&=&0^{--},~2^{P-},~3^{P-},~4^{P-},~\dots \quad
C=-~({\rm leading~phononic~trajectory}),
\ea
where $P$ is arbitrary (and corresponds to the trivial parity doubling 
of non-zero spin states in two space-dimensions).
In the simplest form of the model, the two trajectories are degenerate.
An orbital trajectory is only present if the string can  acquire a 
`permanent deformation', as heavy nuclei can do. The largest possible slope
is obtained in the extreme case of the collapse to a segment, 
when the slope is  $\frac{1}{4\pi\sigma}$. The twisted orbital trajectory
carries the states $0^{++}$, $1^{P-}$, $2^{P+}$, $3^{P-}$, \dots
($P$ arbitrary).
%
%
\paragraph{The  Hagedorn temperature in the flux-tube model}
The degeneracy $P$ of an energy operator such as~(\ref{phonon_e}) is given
for large eigenvalue $M$ in~\cite{Huang:1970iq} (we have four sets of 
creation/annihilation operators in four dimensions, and $2(D-2)$ for a 
general number of space-time dimensions $D$):
\be
P \propto \exp{\left[ 2\pi\sqrt{\frac{(D-2)M}{3}} \right]}.
\ee
We have just seen that the phonons lead to Regge trajectories, 
$\alpha' m^2\simeq M$. Therefore the density of states grows as
\be
N(m)dm \propto e^{\beta_H m} dm, 
\qquad \beta_H=2\pi\sqrt{\frac{(D-2)\alpha'}{3}}.
\ee
Since we obtained $\alpha'=(8\pi\sigma)^{-1}$, we find for the Hagedorn
temperature, where the partition function diverges, 
\be
\frac{T_H}{\sqrt{\sigma}}  = \sqrt{\frac{6}{\pi(D-2)}}~,\qquad\qquad
\left.\frac{T_H}{\sqrt{\sigma}}\right|_{D=4} \simeq 0.98.
\la{eq:t_h}
\ee
In this context it is interesting to note that lattice 
calculations~\cite{Lucini:2003zr}
find  the deconfining temperature of 4D pure gauge theories to be 
\be
T_c/\sqrt{\sigma} = 0.596(4) + 0.453(30)/N_c^2
\ee
and that the transition is first order for $N_c\geq3$.
The Hagedorn phase transition at $T=T_H$ is second 
order~\cite{Polyakov:1978vu}, with the effective string tension $\sigma(T)$
vanishing at the critical point. The order parameter
for the transition is the mass of the string mode winding around the 
`temperature cycle' of length $1/T$. In fact
if the L\"uscher expression for the Polyakov loop mass 
$m_T=\sigma /T -\frac{\pi(D-2)T}{6}$ is equated to zero, we recover
Eqn.~\ref{eq:t_h}.
It turns out that the winding modes condense before
they become massless~\cite{Atick:1988si}, leading to a first order phase
transition which occults the original Hagedorn 
transition; in the large-$N_c$ gauge theory,
the latent heat grows with $N_c^2$~\cite{Atick:1988si}. 
It is therefore not surprising that we find a Hagedorn temperature associated
with the phonons of the closed flux-tube to be larger than the actual 
deconfinement phase transition.
\section{Glueball spectra from gravity\label{sec:brower}}
A calculation of the glueball spectrum based on the correspondence
between supergravity on an $AdS^7\times S^4$ manifold and the large $N_c$ 
supersymmetric gauge theory living on the boundary of this space
was presented in~\cite{Brower:2000rp}. The order of the states in terms of their 
quantum numbers matches the lattice data of the pure gauge theory (at 
finite $N_c=3$). 
Pushing the comparison further probably has little significance, 
given that the classical gravity equations correspond to strong t'Hooft 
coupling  on the boundary.
The success of model calculations is best measured
by comparison with a `default'  model.
As the authors note, it has been known for a long
 time~\cite{Jaffe:1986qp} that
the ordering of the low-lying states obtained from the 
lattice can be understood very  economically
in terms of the minimal dimensionality of local gauge-invariant operators
 that carry the quantum numbers of the various glueballs. 
 
It is interesting to compare the spectrum obtained from the gravity side
to strong coupling expansions~\cite{Drouffe:1983fv} based on 
the Wilson lattice action.
In the latter case the three lightest states are at leading order
$m_{0^{++}} = m_{2^{++}} = m_{1^{+-}}$. The fact that some of the 
dimension-four interpolating operators are absent both 
on the supergravity side and in the lattice strong-coupling expansion 
(e.g. the operator having quantum numbers $2^{-+}$),
while the $1^{+-}$ (which has $d=6$) is present, is an intriguing similarity
between the two approaches. 

Finally, we note that 
spinning strings in gravity backgrounds were investigated by semi-classical
methods in~\cite{PandoZayas:2003yb}; the result is a glueball spectrum
on the boundary which extends to higher angular momenta than $J=2$.
\section{The flux-tube model from the Nambu-Goto action}
In this section we show how the flux-tube model derives from the Nambu-Goto
action for closed strings. The action is expanded around the circular 
configuration and the non-relativistic limit is taken. One benefit of the 
derivation is a better understanding of the approximations made.
It also paves the way to compute the leading corrections to the model, which
are entirely determined by the string action and require no further input
parameters.
\subsection{Generalities\la{sec:nr}}
We shall start with the Nambu-Goto action, where we consider only the
simplest topology of the closed string: a single, oriented loop
of radius $\rho$ with the world-sheet coordinates
$\tau_{\rm ws}=t$ and $\sigma_{\rm ws}=r(\varphi)\varphi$. 
Then the Lagrangian takes the form
\be
L_{NG}= -\sigma \int d\varphi~\frac{d\ell(\varphi)}{d\varphi}~
\sqrt{1-v^2(\varphi)},
\ee
where $\sigma$ is the string tension. As an aside, 
if the string had an intrinsic stiffness, we would replace $\sigma d\ell$
by $(\sigma+\frac{\kappa}{R^2_{\rm curv}})d\ell$.
The theory of elasticity~\cite{elasticity}
 relates $\kappa$ to the Young modulus $Y$ and 
the half-width of the string $R_o$ through
 $\kappa=\frac{\pi}{4}YR_o^4$. For a planar loop 
for instance, the `mass' of an element of string $d\ell$ is given by
\be
\sqrt{r^2+r'^2}
 ~\left[\sigma ~ + ~\kappa~\frac{
\left(2r'^2+r^2-rr''
\right)^2}
{\left(r^2+r'^2\right)^3}\right]~d\varphi
\ee
where the prime denotes differentiation with respect to $\varphi$.
For the non-relativistic circular loop, the main effect of 
the curvature term is to add the expression $+\frac{2\pi\kappa}{\rho}$ 
to the potential~\cite{Johnson:2000qz}. 
If $2\pi\kappa> \gamma$, the curvature term wins over
the attractive L\"uscher term
 $-\frac{\gamma}{\rho}$. The loop is then classically
stable, with an energy 
$E_{\rm cl}=4\pi\sqrt{\sigma\kappa'}$, where
$\kappa'$ is now the effective curvature coefficient.
The new term mostly affects the low-angular momentum states, 
which will generically become heavier.

We now return to the pure Nambu-Goto action.
Let us first consider the case of a perfectly circular loop, 
$\frac{d\ell}{d\varphi}=\rho$, with no orbital motion, $v=\dot \rho$.
The conjugate momentum $p_\rho$ takes the expression
\be
p_\rho \equiv \frac{\partial L}{\partial \dot \rho} = 
 \frac{(2\pi\sigma \rho) \dot \rho}{\sqrt{1-\dot \rho^2}},
\ee
and the Hamiltonian reads
\be
H\equiv \dot \rho p_\rho -L =\frac{2\pi \sigma \rho}{\sqrt{1-\dot \rho^2}}.
\ee
Note that $\dot \rho = \frac{p_\rho}{H}$. Therefore, under quantisation
\be
p_\rho \rightarrow -i\frac{1}{\rho}~\frac{\partial}{\partial \rho}~\rho,
\ee
the `Schr\"odinger' equation
\be
\{ p_\rho^2 + (2\pi\sigma \rho)^2\}\psi(\rho)  = E^2 \psi(\rho),
\ee
after substitution $\xi(\rho)=\psi(\rho)\rho$,
corresponds to a one-dimensional harmonic oscillator. Only the odd solutions 
are acceptable if $\psi$ is to be normalisable. Therefore the spectrum is 
\be
E^2 = 2\pi \sigma (4n + 3),\qquad n=0,1,2\dots
\ee
This corresponds to a straight trajectory in the radial quantum number, with 
the same slope as the phononic trajectory of the flux-tube model. 
Incidentally, the prediction for the lightest glueball mass is
$m_{0^{++}}/\sqrt{\sigma} = 4.34$, which is not too bad (see 
Chapter~\ref{ch:regge_3d}).

It is instructive to consider the non-relativistic case for comparison.
The Nambu-Goto Lagrangian then takes the form
\be
L_{NR} = \sigma \int d\varphi ~\frac{d\ell}{d\varphi}~\left(\frac{v^2}{2}-1\right).
\la{eq:ng_nr}
\ee
For the circular loop,
\be
L = 2\pi\sigma \rho \frac{\dot\rho^2}{2} - 2\pi\sigma \rho.
\ee
After substitution $x=(\sqrt{\sigma}\rho)^{3/2}$~\cite{Isgur:1985bm} and 
quantisation  $p_x\rightarrow -i\frac{d}{dx}$, 
the Schr\"odinger equation in the $x$ variable reads
\be
\{-\frac{9}{16\pi}\frac{d^2}{dx^2} + 
2\pi x^{2/3} \} \psi(x) = \left(\frac{E}{\sqrt{\sigma}}\right)\psi(x).
\ee
It is well-known that this particular 
power-law potential leads to straight Regge trajectories. 
Brau~\cite{Brau:2000st} obtains an approximate energy formula by applying the 
Bohr-Sommerfeld quantisation prescription - which becomes exact at large 
quantum number $n$. In this case it leads to
\be
E^2 \simeq 2\sqrt{2}\sigma(4n+3), \qquad n=0,1,2\dots
\ee
We see that the non-relativistic levels are almost 
identical to the relativistic ones, 
up to an overall rescaling transformation. This will serve as a justification
for using the non-relativistic approximation~(\ref{eq:ng_nr}) from now on.
Our aim will be mainly to identify the relevant degrees of freedom 
and physical  effects that determine the 
global features of the string spectrum. For that purpose,
we assume that the non-relativistic approximation is good enough.
\subsection{The vibrating closed string in 2+1 dimensions}
We now consider fluctuations around the circular configuration:
\be
r^2(\varphi) = \rho^2 + \rho\sum_{|m|\geq 2} d_m ~e^{im\varphi}, 
\qquad d_{-m}=d_m^*.\la{eq:d_m}
\ee
This parametrisation  assumes that the function $\rho(\varphi)$ remains 
single-valued, that is, there is no topology change. We shall expand 
the action to quadratic order in the `deformations' $d_m$. We now have
\be
\frac{d\ell}{d\varphi} = \sqrt{r^2+r'^2},\qquad r'\equiv \frac{dr}{d\varphi}.
\ee
Plugging the expansion~(\ref{eq:d_m}) into this expression and expanding to
quadratic order, we get
\be
\sqrt{r^2+r'^2} \simeq
\rho + \frac{1}{2}\sum_{|m|\geq 2} d_m ~e^{im\varphi}
-\frac{1}{8\rho}\sum_{|m|,|n|\geq2} d_m d_n (mn+1) ~e^{i(m+n)\varphi}
\la{eq:line_elem}
\ee
The expression for the `velocity' becomes
\ba
\dot r^2 &\simeq& \dot \rho^2 ~+~ 
\dot\rho\sum_{|m|\geq 2}  \dot d_m ~e^{im\varphi} 
~-~\frac{\dot \rho}{2\rho} \sum_{|m|\geq 2}  d_m ~e^{im\varphi}
 \sum_{|n|\geq 2} \dot d_n ~e^{in\varphi} \nonumber \\
&&+\left(\frac{\dot \rho}{2\rho}\right)^2 
\left(\sum_{|m|\geq 2} d_m ~e^{im\varphi}\right)^2
~+~\frac{1}{4}\left(\sum_{|m|\geq 2} \dot d_m ~e^{im\varphi}\right)^2.
\la{eq:2d_veloc}
\ea
Plugging $\frac{d\ell}{d\varphi}$ and $\dot r^2$ into~(\ref{eq:ng_nr}), 
we obtain
\ba
L &=& 2\pi\rho\sigma \frac{\dot\rho^2}{2}
	\left[ 1 + \frac{1}{4\rho^2} \sum_{n\geq2} |d_n|^2(n^2+1) \right]
	\nonumber\\
	&&+~ \frac{\pi\sigma\rho}{2}\sum_{n\geq2}|\dot d_n|^2
	~-~ \frac{\pi\sigma}{2\rho} \sum_{n\geq2} |d_n|^2 (n^2-1) 
	~-~ 2\pi\sigma\rho
\ea
It is worth noting that the parametrisation~(\ref{eq:d_m}) was chosen 
so that no cross kinetic term $\dot\rho \dot d_m$ appears in the Lagrangian.
The Hamiltonian is therefore obtained in a straightforward way. In terms of 
 real components $d_n\equiv \alpha_n + i \beta_n$, it reads
\ba
H &=& \frac{p_\rho^2}{2}\frac{1}{2\pi\rho\sigma}
\left[1-\frac{1}{4\rho^2}\sum_{n\geq2} (n^2+1) (\alpha_n^2+\beta_n^2) \right]
~+~ \frac{1}{2\pi\sigma\rho} \sum_{n\geq2} (p_{\alpha_n}^2+p_{\beta_n}^2)
\nonumber \\
&&+~\frac{\pi\sigma}{2\rho} \sum_{n\geq2} (n^2-1) (\alpha_n^2+\beta_n^2)
~+~2\pi\sigma\rho
\ea
An important point is that the phonon occupation numbers are conserved (this corresponds
to the Lagrangian symmetry $d_n\rightarrow d_n e^{i\alpha}$, $n$ fixed). 
A straightforward implication is  that energy eigenstates can be written as
$|\psi\>_\rho\otimes|\xi\>_{\rm phon}$, where $|\xi\>_{\rm phon}$ has definite occupation
numbers.
If we ignore the correction to the kinetic term $p_\rho^2$ for the moment, 
the fluctuations $\alpha_n,~\beta_n$ have the same harmonic-oscillator
Hamiltonian  as~(\ref{eq:phon_hamil}), except that the frequencies
 $\omega_n$ are now given by $\omega_n=\frac{\sqrt{n^2-1}}{\rho}$. This 
will also affect the evaluation of the zero-point energy $\gamma/\rho$.
The other difference with respect to the flux-tube model Hamiltonian is
that the `phonons' modify the weight of the kinetic term for $\rho$.
For large quantum numbers, the expectation value of 
$\rho$ becomes large and therefore the kinetic term becomes small, 
thus justifying the adiabatic approximation; but
for the low-lying states it is 
in general not so. First order perturbation theory tells us that 
the leading correction to the energy levels 
obtained in the adiabatic approximation is given by the expectation
value of the new term on the factorised `wave function' 
$|\psi\>_\rho\otimes|\xi\>_{\rm phon}$.
\subsection{The spinning and vibrating closed string in 3+1 dimensions}
There are two complications when moving from 3 to 4 dimensions.
The first is benign: fluctuations in the $\hat z$ direction are now possible.
If we use the parametrisation
\be
z(\varphi) = \frac{1}{2}~\sum_{|n|\geq2} z_n e^{in\varphi}, \la{eq:z_m}
\ee
the line element $\frac{d\ell}{d\varphi}$ is modified 
from~(\ref{eq:line_elem}) to 
\be
\sqrt{r^2+r'^2+
z'^2} \simeq 
\sqrt{r^2+r'^2} -\frac{1}{8\rho}
\left(\sum_{|n|\geq2}nz_ne^{in\varphi}\right)^2
\ee
in cylindrical coordinates.

The second complication is that a `collective' orbital motion becomes
possible. It can be parametrised by  two angles
characterising the orientation of the loop-plane.
The description of the fluctuating loop is now associated with a 
non-inertial, `body-fixed' frame. 
Non-relativistically, the velocity ${\bf v}_{\rm bf}$ of a point 
in the body-fixed frame is related to 
the velocity ${\bf v}_i$ in the inertial frame by 
\be
{\bf v}_i = {\bf v}_{\rm bf} + {\bf\omega}_{\rm bf}\times {\bf x}
\ee
where
\be
{\bf x} = r {\bf\hat r} + z{\bf\hat z}
\ee
If we first set ${\bf\omega}=0$, the velocity is 
\be
{\bf v} \equiv {\bf v}_i = \dot r {\bf\hat r} + \dot z {\bf\hat z},
\qquad v^2=\dot r^2+\dot z^2.
\ee
In terms of the expansions~(\ref{eq:d_m}) and~(\ref{eq:z_m}) , the 
expression~(\ref{eq:2d_veloc}) becomes
\be
v^2|_{4d,{\bf\omega=0}} ~~=~~ v^2|_{3d} ~+~ 
\frac{1}{4}\left(\sum_{|n|\geq2}\dot z_n e^{in\varphi}\right)^2.
\ee
Therefore, if the orbital degree of freedom $\bf \omega$ is switched off,
the Lagrangian reads
\ba
L_{0} &=& 2\pi\rho\sigma \frac{\dot\rho^2}{2}
	\left[ 1 + \frac{1}{4\rho^2} \sum_{n\geq2} |d_n|^2(n^2+1) 
	+ n^2 |z_n|^2\right]
	\nonumber\\
	&&+~ \frac{\pi\sigma\rho}{2}\sum_{n\geq2}|\dot d_n|^2 + |\dot z_n|^2
\nonumber \\
	&&- 2\pi\sigma\rho ~-~
	 \frac{\pi\sigma}{2\rho} \sum_{n\geq2} |d_n|^2 (n^2-1) +|z_n|^2n^2~.
\la{eq:L_w=0}
\ea

We will use the Eulerian angles $(\phi,\theta,\psi)$ 
to parametrise the orientation of the 
body-fixed frame (see~\cite{Marion:1970zp}; $\phi$ is not to be mixed up 
with $\varphi$, the parametrisation of the string!). 
However, because the string is `immaterial', there 
is no dynamical degree of freedom associated with the rotation of the ring
around its symmetry axis (although there is a freedom in the choice of 
parametrisation). As a consequence, the component $\omega_z$ of the angular 
velocity in the body-fixed frame is arbitrary -- its sole effect is to modify
the parametrisation of the internal degrees of freedom of the string.
We choose the third Euler angle $\psi$ to vanish identically. 
The Cartesian coordinates 
of ${\bf \omega}$ in the body-fixed frame are now~\cite{Marion:1970zp}
\be
\omega_{x}=\dot\theta,\qquad 
\omega_{y}=\dot\phi\sin{\theta},\qquad 
\omega_{z}=\dot\phi\cos{\theta},
\ee
while ${\bf x}$ has coordinates
\be
x = r\cos\varphi,\qquad 
y = r\sin\varphi,\qquad 
z = z.
\ee
Therefore 
\ba 
v^2&=&\dot r^2+\dot z^2 ~+~ \omega^2(r^2+z^2)-({\bf\omega}\cdot {\bf x})^2
~+~2(\dot r z - \dot z r) {\bf \hat r}\cdot({\bf\omega}\times {\bf \hat z})
\nonumber \\
&=&\dot r^2+\dot z^2 ~+~
	r^2 \left[ \dot\phi^2(1-\sin^2\theta\sin^2\varphi)
	-\dot\theta\dot\phi\sin\theta\sin2\varphi + 
	\dot\theta^2\sin^2\varphi \right]
\nonumber \\
&& + z^2(\dot\theta^2 + \dot\phi^2\sin^2\theta)
\nonumber \\
&& - zr(2\dot\phi\dot\theta\cos\theta\cos\varphi 
	+ \dot\phi^2 \sin2\theta\sin\varphi)
\nonumber \\
&& +2(\dot r z - \dot z r)
	(\dot\phi\sin\theta\cos\varphi-\dot\theta\sin\varphi)~.
\ea 
Thus the Lagrangian can be written as  $L ~=~ L_{0} ~+~L_{\bf \omega}$,
\ba 
\frac{L_{\bf \omega}}{\pi\sigma}&=& \rho^3 
\left(\dot\phi^2\left(1-\frac{\sin^2\theta}{2}\right)
	+\frac{\dot\theta^2}{2}\right)
	\left[ 1+\frac{1}{4\rho^2}
	\sum_{n\geq2} (n^2+3)|d_n|^2 + n^2|z_n|^2 \right] 
\nonumber\\
&& +\frac{\rho}{2}~(\dot\phi^2\sin^2\theta+\dot\theta^2)~\sum_{n\geq2}|z_n|^2
\nonumber\\
&& +\frac{3\rho^2}{4} \left[ 2\dot\phi\dot\theta \im d_2 
	+ (\dot\phi^2\sin^2\theta-\dot\theta^2) \re d_2 \right]
\nonumber\\
&& +\frac{\rho}{8} \left[ (\dot\phi^2\sin^2\theta-\dot\theta^2)
~\re\sum_{n\geq2} d_{n+2}^*d_n(n^2+2n+3) + n(n+2) z_{n+2}^*z_n\right.
\nonumber\\
&& \qquad\left. -2\dot\phi\dot\theta\sin\theta~\im\sum_{n\geq2}
d_{n+2}^*d_n(n^2+2n+3) + n(n+2) z_{n+2}^*z_n\right]
\nonumber\\
&&-\frac{\rho}{2} \left[ 2\dot\phi\dot\theta\cos\theta
	~\re\sum_{n\geq2} (z_{n+1}^*d_n+d_{n+1}^*z_n)
	+\dot\phi^2\sin2\theta ~\im\sum_{n\geq2} (z_{n+1}^*d_n+d_{n+1}^*z_n)
	\right]
\nonumber\\
&&-\rho\left[\dot\phi\sin\theta~\re\sum_{n\geq2} 
	(\dot z_{n+1}^*d_n + d_{n+1}^* \dot z_n)
	-\dot \theta ~\im \sum_{n\geq2} 
	(\dot z_{n+1}^*d_n + d_{n+1}^* \dot z_n)	\right]
\nonumber\\
&&+\frac{\rho}{2}\left[\dot\phi\sin\theta
	~\re\sum_{n\geq2} (\dot d_{n+1}^*z_n + z_{n+1}^* \dot d_n)
	-\dot \theta ~\im \sum_{n\geq2} 
	(\dot d_{n+1}^*z_n + z_{n+1}^* \dot d_n)	\right]
\nonumber\\
&&+\frac{\dot\rho}{2}\left[\dot\phi\sin\theta~\re\sum_{n\geq2} 
	( z_{n+1}^*d_n + d_{n+1}^*  z_n)
	-\dot \theta ~\im \sum_{n\geq2} 
	( z_{n+1}^*d_n + d_{n+1}^*  z_n)	\right].\la{eq:Lo}
\ea 
The first term describes the rotational kinetic energy associated with
the ring, which in the absence of phonons has moment of inertia $\pi\rho^3\sigma$. 
The moment of inertia is  modified by the phonons in a similar way to the 
`mass' associated with the breathing mode (Eqn.~\ref{eq:L_w=0}).
All the following terms specify the intertwining of the orbital motion with
the internal degrees of freedom: the `spin-orbit' interactions 
imply that the phonon occupation numbers are no longer conserved.
However the matrix representing the Hamiltonian in the phononic basis
is band-diagonal, that is to say, the transitions in phonon occupation numbers
are `local' in that basis. Also, the spin-orbit terms allow
the $z$ and $\rho$ phonons to mix.
A special role is played by the $m=2$ phonons, since their dynamics
 are affected by the orbital motions at linear order. 

We leave the quantisation procedure
and the computation of the spectrum for the future.
\section{Conclusion}
String models of glueballs are particularly attractive in the pure gauge 
theories, where the stability of the `flux-tube' makes it a natural
object to describe the low-energy dynamics of the theory. 
In the context of large $N_c$ gauge theories, the adjoint string, which can 
be thought of as two weakly interacting fundamental strings, is equally
 natural. The geometry of the closed, oriented string leads
to definite predictions on the quantum numbers of the states corresponding
to spinning and vibrating configurations. At large angular momentum, the 
`orbital' and `phononic' Regge trajectories have semi-classically calculable
slopes $\alpha'$ and offsets $\alpha_0$. The intercept and slope at $J=0$
are however deeply quantum mechanical quantities which in general depend
 on the mixing between the different trajectories and the details of the 
underlying gauge theory. The lightest glueball could well be 
an intricate superposition of many different topologies of the closed string.

We note that the bag model for glueballs was revisited in recent 
years~\cite{Kuti:1998rh} and that good agreement was claimed to be found
  between the predicted spectrum and the low-lying lattice spectrum.
It would be interesting to see whether such a model can lead to linear Regge
trajectories at large angular momentum. As long as
the cavity is spherically symmetric, this seems impossible. For instance, 
the 2+1D spectrum based on harmonic modes inside a disk~\cite{Karl:1999sz}
predicts $J\propto M^{3/2}$. On the other hand, 
if the `bag' gets elongated by the angular momentum
of the constituent gluons, then the adjoint string can form between them.
An increase in the spin as fast as $M^2$
appears to be  possible only for an object whose moment of inertia grows 
with angular momentum.

We did not discuss the issue of
glueball decays. An attempt to model the decays in the flux-tube model 
context in presence of quarks was made in~\cite{Iwasaki:2003cr}. The mechanism
employed was the so-called Schwinger mechanism (string$\rightarrow q\bar q
\rightarrow$ hadrons) and the result is that the width is proportional to 
the mass of the glueballs.

Although models can give insight into the qualitative features
of the glueball spectrum, well-established methods are available in numerical 
lattice gauge theory that allow us to compute \emph{ab initio}
 the low-lying glueball spectrum with remarkable numerical accuracy. 
We shall describe these methods in the next chapter.

%% file: chapter3.tex
\chapter{Lattice gauge theory\la{ch:lgt}}
Lattice gauge theory~\cite{Wilson:1974sk} 
is one of the only known non-perturbative regularisations of QCD.
Several introductory texts to the subject are available, the most recent 
being~\cite{Rothe:1997kp} and~\cite{Smit:2002ug}.
\section{Generalities}
We will be using the path integral formalism, partly because it is a powerful tool of 
quantum field theory, and partly because it provides a natural framework to perform
numerical simulations. 
Quarks are mathematically represented by Dirac spinors.
We work in the Euclidean formulation of the theory; 
the Euclidean Dirac operator is $D=\gamma_\mu\partial_\mu+m$, 
with $\{\gamma_\mu,\gamma_\nu\}=2\delta_{\mu\nu}$ and $\gamma_\mu^\dagger=\gamma_\mu$.
The anti-commuting nature of the fermionic field, 
$\{\psi_\alpha(x),\psi_\beta(y)\}_{x_0=y_0}=\delta_{\alpha\beta}\delta({\bf x}-{\bf
y})$, requires that Grassmann variables be used in the path integral.

Let us define a lattice version of the free-fermion theory. 
Consider a space-time lattice,
$x=a(n_0,n_1,n_2,n_3)$, $n_\mu\in {\bf Z}$, where $a$ is the lattice spacing.
Although more symmetric  4D lattices exist~\cite{Celmaster:1982ht}, 
the regular hyper-cubic lattice is by far the most commonly used.
The lattice Dirac field is the assignment of a Dirac spinor $\psi(x)$ to each 
point on the lattice. The lattice spacing provides a momentum cutoff of order $1/a$:
momenta are restricted to the Brillouin zone $[-\frac{\pi}{a},\frac{\pi}{a}]$.

Next we consider a   gauge group $SU(N_c)$ with group generators $T^a$, $a=1,\dots, N_c^2-1$, 
a basis of hermitian, traceless $N_c\times N_c$ complex matrices satisfying
\be
[T^a,T^b]=if^{abc}T^c\qquad \tr\{T^a T^b\}= \frac{\delta^{ab}}{2}
\ee
where the structure constants $f^{abc}$ are real and totally antisymmetric in their
indices. In the $SU(2)$ case, $T^a=\frac{\tau^a}{2}$, where $\tau^a$ are the Pauli 
matrices. The continuum gauge field $A_\mu^a(x)$ is defined as 
\be
A_\mu(x)=A_\mu^a(x)T^a
\ee
and similarly
\be
F_{\mu\nu} = \partial_\mu A_\nu - \partial_\nu A_\mu -i [A_\mu,A_\nu]
 = \{\partial_\mu A_\nu^a-\partial_\nu A_\mu^a+f^{abc}A_\mu^bA_\nu^c\}T^a.
\ee

A gauge transformation of the quark field is defined by a field of $SU(N_c)$ matrices
$\Lambda(x)$:
\be
\psi(x)\rightarrow \Lambda(x)\psi(x),\qquad 
\bar\psi(x)\rightarrow \bar \psi(x)\Lambda(x)^{-1}
\ee
The fermion action is only invariant under
such a local transformation if the partial derivatives in the Dirac operator are 
supplemented by a `gauge field', $D_\mu=\partial_\mu - iA_\mu(x)$,
 and the transformation law
\be
A_\mu(x) \rightarrow
\Lambda(x)A_\mu(x)\Lambda^{-1}~+~i\Lambda(x)\partial_\mu\Lambda(x)^{-1}.
\la{eq:cont_gauge_transf}
\ee
The field tensor then transforms covariantly 
\be
F_{\mu\nu}(x)\rightarrow \Lambda(x)F_{\mu\nu}(x)\Lambda(x)^{-1}
\ee
and the following gauge field action is therefore also gauge invariant:
\be
S_G[A] = \frac{1}{2g_o^2}\int d^4x \tr\{F_{\mu\nu}(x)F_{\mu\nu}(x)\},\qquad
g_o= {\rm bare~coupling.}
\ee

Since differential operators are replaced by difference operators on the lattice, 
the `purpose' of the gauge field will be to ensure that the `covariant difference operator'
acting on the fermionic field transforms in the same way as the original field.
If $U_\mu(x)\in SU(N_c)$ transforms as
\be
U_\mu(x) \rightarrow \Lambda(x)U_\mu(x)\Lambda(x+a\hat\mu)^{-1},
\la{eq:lat_gauge_transf}
\ee
we indeed find  that
\be
\nabla_\mu\psi(x)\equiv\frac{1}{a}\{U_\mu(x)\psi(x+a\hat\mu)-\psi(x)\}
\ee
transforms like $\psi(x)$. Therefore the lattice gauge field is the assignment
 of a gauge group element $U_\mu(x)$ to each point $x$ and direction $\mu$, 
and can be pictured as living on the `links' between site $x$ and $x+a\hat\mu$.
The explicit relation between the lattice and the continuum field is provided by the 
Wilson line. For a given continuum field $A_\mu(x)$, the corresponding lattice gauge
field is given by 
\be
U_\mu(x) = {\rm P}~ \exp{\left(i a\int_0^1 d\xi A_\mu(x+a\xi\hat\mu)\right)}.
\ee
The symbol P means that the exponentiation  maintains the ordering of the matrices
along the path.  It is a short exercise to check that the transformation
property~(\ref{eq:lat_gauge_transf})
corresponds to the gauge transformation~(\ref{eq:cont_gauge_transf}) of the continuum
field.
For that reason, the lattice formulation is said to preserve the gauge symmetry
exactly.
Its elegance and naturalness remains a permanent source of delight for lattice
practitioners 
around the world.

It is now obvious that any product of gauge links along a closed path transforms
covariantly;
in particular its trace is gauge invariant. The simplest of non-trivial closed paths is the
`plaquette'
\be
P_{\mu\nu}(x)=U_\mu(x)U_\nu(x+a\hat\mu)U_\mu(x+a\hat\nu)^{-1}U_\nu(x)^{-1}.
\ee
In the classical continuum limit, its trace can be expanded in a power series in $a$, 
where the coefficients of the series are gauge invariant operators in $A_\mu$ and its
derivatives:
\be
\re\tr\{1-P_{\mu\nu}\}=\frac{a^4}{2}~\tr\{F_{\mu\nu}^2\}-\frac{a^6}{24}
\tr\left[F_{\mu\nu}(D_\mu^2+D_\nu^2)F_{\mu\nu}\right]+{\rm total~deriv.}+
{\cal O}(a^8),
\ee
where $D_\mu=\partial_\mu+[A_\mu,.]$.
Thus the Wilson  action for the lattice gauge field is given by:
\ba
S_G[U] &=& \frac{1}{g_o^2}\sum_{x,\mu,\nu}\re\tr(1-P_{\mu\nu}(x))\la{eq:wilson_action}
\ea
Finally, to complete the definition of  the quantum theory, we must specify the integration
measure  over the gauge fields,
\be
D[U] = \prod_{x,\mu}dU_\mu(x).
\ee
 Usually the only property that is needed  is 
\be
\int D[U] f[UV] = \int D[U] f[VU] = \int D[U] f[U].
\ee

In addition to the ultraviolet cutoff $a$, 
an infrared cutoff can be introduced: the system described by the 
action~(\ref{eq:wilson_action}) is often considered on a finite volume, 
$V=L_0\times L_1\times L_2\times L_3$, $L_\mu=a\hat L_\mu$,
and specific boundary conditions must be 
imposed. In order not to lose translational invariance, they
are chosen periodic. Although in some cases `twisted' 
boundary conditions~\cite{'tHooft:1981sz}
can be of interest, by far the most commonly used are the plain 
\be
U_\mu(x+L_\nu\hat\nu)=U_\mu(x)\qquad \forall\mu, ~\nu.  \la{eq:periodic_bc} 
\ee
In the following, we consider $L_0=L_1=\dots= L$. As a consequence of the finite 
volume, the lowest non-zero momentum has a magnitude $2\pi/L$. The path integral is 
now simply a multi-dimensional integral, and can be evaluated by numerical means. 
Naturally one wishes to eventually remove both the ultraviolet and infrared cutoffs.
These operations are respectively referred to as the continuum and the 
infinite-volume limit.
\section{The continuum limit\la{sec:cont_lim}}
Let us assume that there is a mass gap in the lattice theory defined 
by~(\ref{eq:wilson_action}); if it is to describe QCD, this had better be the case!
Since the physical mass $m$ of the corresponding continuum field theory must stay finite,
the mass measured in lattice units $\hat m$ must vanish in the continuum limit. In
other
words, a correlation length must diverge, and the continuum limit is associated with 
a second order phase transition. 
In the theory without quarks, there is only
one parameter in the action, the bare coupling $g_o$. 
In the four-dimensional theory, it is dimensionless.
The study of a statistical mechanics 
system near a phase transition requires a tuning of parameters, in this case
$g_o=g_o(a)$.
This dependence becomes apparent when it is noted that $a^{-1}\hat m(g_o)$ must
converge
toward a fixed $m$, the physical value.

A physically intuitive way to derive the dependence of the lattice spacing on the bare
coupling is based on the computation of the 
static quark potential at order $g_o^4$ 
in lattice regularisation~\cite{Rothe:1997kp}:
\be
V(R,g_o,a)=-\frac{C_F}{4\pi R}\left[g_o^2 + 2\gamma g_o^4 
\log{\left(\frac{R}{a}\right)} + {\cal O}(g_o^6) \right]. \la{eq:stat_pot_1loop}
\ee
where $C_F=4/3$ and $\gamma=\frac{11}{(4\pi)^2}$ for the gauge group $SU(3)$.
If the left-hand side is to converge to a finite value, we must require 
\ba
\left[ a\frac{\partial}{\partial a} - 
\beta(g_o)\frac{\partial}{\partial g_o}\right]V(R,g_o,a)&=&0, \la{eq:stat_pot_rg}\\
\qquad \beta(g_o) \equiv -a\frac{\partial g_o}{\partial a}. \la{eq:beta_func}
\ea
Combining Eqn.~(\ref{eq:stat_pot_1loop}) and~(\ref{eq:stat_pot_rg}) leads to 
$\beta(g_o)\simeq-\gamma ~g_o^3$. The solution of~(\ref{eq:beta_func}) is then
\be
a = \frac{1}{\Lambda_L} \exp{\left( -\frac{1}{2\gamma g_o^2}\right)},
\ee
where $\Lambda_L$ is an arbitrary integration constant.
This shows that the critical point is reached at $g_o=0$, which is also
the fixed point of the renormalisation group equation~(\ref{eq:beta_func}).
The arbitrariness of $\Lambda_L$, which corresponds to the arbitrariness in the choice
of 
energy unit,  means that lattice simulations can only predict
dimensionless quantities. In general, a number of physical observables equal to the
number 
of bare parameters in the action must be measured before any prediction can be made. 
Working in the pure gauge theory, we shall choose the string tension $\sigma$ to set
the scale (see Section~\ref{sec:2pt_func}). 
Another popular choice is the Sommer scale $r_o$~\cite{Sommer:1994ce}.
Because at the classical level, the action has only ${\cal O}(a^2)$ corrections, we
expect such ratios to behave as
\be
\left.\frac{m}{\sqrt{\sigma}}\right|_{\rm cont}= 
 \frac{m}{\sqrt{\sigma}}(a) + {\cal O}(\sigma a^2). \la{eq:cont_extrapol}
\ee
\section{Monte-Carlo simulations}
In the Euclidean path integral formalism, expectation values of field operators are 
evaluated as ensemble averages:
\be
\langle {\cal O}\rangle = \frac{\int D[\phi]~{\cal O}[\phi]~e^{-S[\phi]}}
{\int D[\phi]~ e^{-S[\phi]}}.
\ee
The problem of numerical lattice gauge theory thus amounts to performing a large number
of integrals. It is well-known that beyond a handful of integrals, it is much more
efficient
to use statistical techniques. In fact, given that the lattice action is real, the 
exponential of the action can be interpreted as the Boltzmann weight of a statistical 
mechanics system whose Hamiltonian in units of the temperature is given by the 
action~(\ref{eq:wilson_action}). Now, importance sampling is based on the idea that
only a small fraction of the configurations has an appreciable probability to appear
in the statistical system. The ensemble average is thus estimated through
\be
\langle {\cal O} \rangle\simeq \frac{1}{N_{\rm config}}~~\sum_{\rm config} {\cal
O}[\phi_i]
\la{eq:mc_aver}
\ee
where the configurations $\phi_i$ have been generated with a probability distribution
given by the Boltzmann weight $e^{-S}$. This estimate now has a statistical error, 
which decreases as $1/\sqrt{N_{\rm config}}$ if the configurations are statistically 
independent.

The problem of Monte-Carlo simulations is thus to produce an ensemble of configurations
with the correct distribution function. The general scheme of a simulation run is 
the following. Given a `rule' for generating a sequence of 
configurations from a given configuration, one first `updates' the configurations a 
sufficiently large number of times until `thermalisation' is achieved. Then the 
configuration is updated a (much smaller) number of times, to produce a Markov chain
of configurations (for a clear introduction, see~\cite{Rothe:1997kp}). 
The observables are measured on this sequence of configurations. Under very general 
assumptions on the Markov chain, it can be shown that the average of these measurements
converges according to formula~(\ref{eq:mc_aver}). 
The only requirement the `rule' must satisfy is detailed balance:
\be
e^{-S[C]}~{\rm Prob.}(C\rightarrow C') = e^{-S[C']}~
{\rm Prob.}(C'\rightarrow C) \la{eq:detailed_balance},
\ee
where $C$ and $C'$ are two configurations.

The number of 
steps required depends on the algorithm used, the observable  and the
value of the input, bare parameters. In particular, close to the continuum limit 
it becomes increasingly hard to generate a thermalised ensemble, due to the critical
behaviour; the number of steps required grows as a power of the correlation length
(this is called critical slowing down). Once the system is thermalised, it should have
lost all memory from the starting configuration. Typically, the starting configuration
is either `cold' (a classical solution of the equations of motion), or `hot'
(the variables taking random values, independent of each other). There can be exceptions
to the uniqueness of the ensemble achieved, if a `bulk phase transition' separates 
the strong coupling from the weak coupling regime of the theory (as is the case for 
$SU(N_c)$ gauge theory with $N_c\geq4$~\cite{Lucini:2001ej}). 

The quality of the algorithm is measured in terms
of the `speed' (in Monte-Carlo time) at which the system travels through its
configuration
space, so as to ensure statistical independence of the configurations, versus
its computational cost. The Metropolis algorithm~\cite{Metropolis:1953am} is 
applicable to any action. However, more specifically adapted algorithms
usually perform better. In that respect, there is a qualitative difference between
local
and non-local actions. The full Wilson action is of course local; however the fermionic 
fields are represented by Grassmann variables. This would ordinarily require an amount
of memory which grows exponentially with the number of degrees of
freedom~\cite{Creutz:1998ee}. However, the fact that the
fermionic action $S_F$ is quadratic in the fermionic fields allows one to integrate
them
out by hand, and this yields the determinant of the Dirac operator in the numerator 
of the path integral. The effective
action is now non-local in the gauge fields. The most efficient known algorithms 
in this case  are those
of the `hybrid Monte-Carlo' type~\cite{Gottlieb:1987mq}.

Here we shall only be dealing with the pure gauge action $S_G$;
in this case, the action is local in the link variables, and each of them can be
updated 
in turn during a Monte-Carlo `sweep' through the lattice. 
In numerical simulations the bare coupling is conventionally parametrised as
\be
S=\beta\sum_{\rm plaq} (1-\frac{1}{N_c}\re\tr P_{\mu\nu}), \qquad \beta=\frac{2N_c}{g_o^2},
\ee
where the sum extends over all `plaquettes' of the lattice; the continuum limit is thus
obtained as $\beta\rightarrow \infty$. For the gauge group $SU(2)$, a `heat-bath' algorithm
is known (its first version is due to Creutz~\cite{Creutz:1983cr}; the version of 
Kennedy and Pendleton~\cite{Kennedy:1985nu}, 
more efficient close to the continuum limit, is explained in
appendix~\ref{ap:NRLGT}). Heat-bath algorithms are generally more efficient than the 
Metropolis algorithm because the updated variable does not depend on its previous
value.
For general $SU(N_c)$ groups, the Cabbibo-Marinari algorithm (see
appendix~\ref{ap:NRLGT}),
which updates a covering set of $SU(2)$ subgroups using one of 
the known heat-bath algorithms,
partially reduces the dependence of the updated variable on its previous value, and is
the most widely used in the pure gauge case. One should be aware that updating the 
configurations by local changes can lead to a critical slowing down of observables that
are not affected by local changes of the configuration in the continuum,
such as the topological charge $Q$~\cite{DelDebbio:2004xh}. This means that the simulation 
can get `stuck' in the fixed topological charge sectors of the theory. The effect of this 
critical slowing-down on the spectrum is however suppressed by the space-time volume: 
assuming a $\theta$-angle dependence 
$M(\theta)=M|_{\theta=0}+\frac{1}{2}M''|_{\theta=0}\theta^2+\dots$, 
it was estimated in~\cite{Brower:2003yx} that a mass evaluated in the sector of topological 
charge $Q$ is given by
\be
M_Q = M|_{\theta=0}+\frac{M''|_{\theta=0}}{2L^4
\chi_t}\left(1-\frac{Q^2}{\chi_t L^4}\right)+\dots,
\la{eq:m_q}
\ee
where $\chi_t\equiv \frac{\langle Q^2\rangle}{L^4}$ is the topological susceptibility.

Experience in numerical simulations has shown that over-relaxation 
steps~\cite{Adler:1981sn} (appendix~\ref{ap:NRLGT}) can help  reduce the correlations
 of a sequence of configurations.
It amounts to making the maximal change of a link variable that does not change the
action.
It is therefore not ergodic (only a subset of measure zero of the phase space
 is explored), and must be used in conjunction with a heat-bath
algorithm. Since an over-relaxation step is faster than a heat-bath step, a ratio
1:3 or 1:4 is usually chosen between the number of heat-bath and over-relaxation
sweeps.
Such a sequence of four or five sequences is sometimes called a `compound sweep'.
We shall use this `hybrid algorithm' throughout this work. The number of thermalisation
sweeps is typically chosen 3000-5000, while the number of compound
sweeps between measurements is one or two, depending on the lattice spacing.

As we shall see in Chapter~\ref{ch:mla}, the locality of the action can be further
exploited to reduce the variance of the statistical estimates.
\subsection{Extracting the spectrum from two-point functions\la{sec:2pt_func}}
The mass of a particle is given by the position of the pole in its propagator.
In Euclidean field theory, by Fourier transform, this corresponds to the large-`time'
decay rate of the two-point function of an interpolating field, 
in our case a functional of the gauge field.
A physically appealing interpretation of such correlation functions is provided by the
transfer matrix formalism (see~\cite{Luscher:1988sd}). It pictures the system
as a 3-dimensional quantum mechanical system with a Hilbert space $\cal H$ of physical
states, a Hamilton operator $H$ and linear operators $\hat {\cal O}$ corresponding to the
Euclidean functionals $\cal O$. The transfer matrix T can be defined
explicitly~\cite{Luscher:1988sd} as an operator acting on $\cal H$, 
with the fundamental identity $Z=\tr\{{\rm T}^{\hat L_0}\}$ 
and from which the Hamiltonian is defined through T$=e^{-aH}$.
Thus the two-point function of an operator ${\cal O}$ localised in a `time-slice' reads
\be
\langle {\cal O}^*(x_0) {\cal O}(x_0=0)\rangle 
~=~ \langle \Omega | \hat{\cal O}^\dagger ~e^{-Hx_0}~\hat{\cal O}| \Omega \rangle
~=~ \sum_n |\langle  n | \hat{\cal O}|\Omega\rangle|^2~ e^{-E_nx_0},\la{eq:tranf_matrix}
\ee
where we have inserted a complete set of energy eigenstates $|n\rangle$. The
interpretation 
is particularly simple: the operator $\hat{\cal O}$ `creates' the states
$|n\rangle$
with amplitudes $\langle n| \hat{\cal O}|\Omega\rangle$ at time $t=0$, and 
$\hat{\cal O}^\dagger$ `annihilates' them at time $t=x_0$. The basic principle of hadron
spectroscopy 
measurements is that at large time separations $x_0\rightarrow\infty$, the correlator
is dominated by the lightest state whose `overlap' with $\hat {\cal O}$,
$c_n\equiv |\langle  n | \hat{\cal O}|\Omega\rangle|^2$, does not vanish.

In fact, if the lattice extent is finite in the time direction, fluctuations 
propagating over the separation $L_0-x_0$ have to be taken into account\footnote{
We assume that the time-extent is large enough so that the partition function $Z$
 is dominated by the vacuum.}:
\ba
&&\langle {\cal O}^*(x_0) {\cal O}(x_0=0)\rangle_{L_0}
= \frac{1}{Z} \tr\{ e^{-(L_0-x_0)H}~\hat {\cal O}^\dagger ~e^{-Hx_0}~\hat{\cal O} \}
 \nonumber \\
&&= \sum_{n,m} ~\left|\langle n |\hat{\cal O}|m\rangle\right|^2~
 e^{-(E_n+E_m)\frac{L_0}{2}}~\cosh\left[(E_n-E_m)\left(\frac{L_0}{2}-x_0\right)\right]
\ea
The leading correction to~(\ref{eq:tranf_matrix}) for $(E_1-E_0)x_0\gg 1$ 
and $x_0\ll L_0$ is thus given by
\be
\langle {\cal O}^*(x_0) {\cal O}(x_0=0)\rangle_{L_0}\simeq
2c_0~e^{-E_0L_0/2} ~\cosh{\left[E_0(\frac{L_0}{2}-x_0)\right]}. \la{eq:cosh_form}
\ee
This is the form we shall use to fit the time dependence of two-point functions 
(see appendix~\ref{ap:NRLGT}).

What operators are capable of `creating' glueballs when acting on the vacuum?
Certainly, they have to be gauge-invariant.
In fact, we shall be using operators that do not depend on the time-like links; 
they are traces of `magnetic' closed Wilson loops, which span the full Hilbert
space of states~\cite{Kogut:1975ag}.
Since the system with periodic boundary conditions has translational invariance, 
the operators can always be chosen to have definite momenta, ${\cal O} = \sum_{\bf x} 
{\cal O}({\bf x})e^{i{\bf p\cdot x}}$. In the following we restrict ourselves to the
case of 
zero-momentum:
\be
{\cal O}({\bf p}=0) = \sum_{\bf x} {\cal O}({\bf x})\la{eq:0mom}.
\ee

The operation consisting in replacing the links by their hermitian conjugates, 
$U\rightarrow U^\dagger$, is a symmetry of the action. Consider the trace of a closed
Wilson loop;
its real part is invariant under this operation while the imaginary part changes sign.
Therefore the real and imaginary parts create states lying in orthogonal sub-spaces
of the full Hilbert space.
Time-like Wilson lines  transform as $W\rightarrow W^\dagger$.
 Since they are the propagators of static quarks (see below), this symmetry
is nothing but the charge conjugation operation $C$.

As explained above~(Section~\ref{sec:cont_lim}), we also need 
to compute the string tension $\sigma$ to set the scale.
 This observable is defined as the decay
constant of the Wilson loop as a function of its area.
 For an $R\times T,~R,T\rightarrow \infty$ rectangle,
\be
\tr\{W(R,T)\}=\tr\left[{\rm P} ~ \exp{\left(i\int_{C(R,T)}A_\mu(x)
dx^\mu\right)}\right]
 ~\propto ~e^{-\sigma RT}.
\ee
It can be shown~(see e.g.~\cite{Rothe:1997kp}) 
that $\tr \{W(R,T\rightarrow\infty)\}$ has the interpretation $ e^{-V(R)T}$, 
where $V(R)$ is the static potential of two infinitely massive quarks
located at distance $R$ of each other. 
The time-like Wilson lines of the rectangle are
 the propagators of the static quarks.

There are alternative ways to measure the string tension 
\cite{Luscher:2002qv,Lucini:2001nv}. We shall follow~\cite{Lucini:2001nv},
whose approach we now briefly describe. 
Firstly, note that for a generic direction $\hat x$, and 
for a fixed coordinate $\bar x$, the transformation 
\be
U_{\hat x}(\bar x,y,z)\rightarrow z_k ~U_{\hat x}(\bar x,y,z),
 \qquad z_k=e^{i2\pi k/N_c} \in {\bf Z}_{N_c}; ~\forall(y,z)
\la{eq:cen_sym}
\ee
is a symmetry of the action (${\bf Z}_{N_c}$ 
is the centre of the gauge group $SU(N_c)$ and the 
property is called the centre symmetry). Suppose $L_x=L$ is finite; 
while local gauge-invariant operators are also left invariant, the Polyakov loop 
\be
P=\tr\left\{\prod_{j=1}^{\hat L_x} ~U_{\hat x}(x=ja,\dots)\right\}
\ee
transforms according to $P\rightarrow z_k P$. In particular this implies that its
expectation
value has to vanish\dots as long as we are in the confined phase! 
Indeed, it is the famous order parameter
for the deconfinement phase transition (in that context, 
$\hat x$ corresponds to the `temperature' direction).
For $\hat x$ spatial, this operator, being gauge-invariant, creates eigenstates of the 
finite-volume Hamiltonian. These states are called `torelons', 
because they wind around a cycle of the hypertorus on which the theory is defined.
They belong to a sub-space of the full Hilbert space which is orthogonal to 
the sub-space of glueballs, and which decouples in the infinite volume
limit. 
The energy of the fundamental state of this sub-space is asymptotically proportional to
$L$:
\be
\frac{m_T}{L}\rightarrow \sigma.
\ee
If the dynamics of the torelon is described by an effective string theory, then
the leading correction to this linear growth of the mass is given by 
\be
\frac{m_T}{L}= \sigma-\frac{\gamma}{L^2}+ {\cal O}\left(\frac{1}{L^3}\right).
\ee
This is the so-called `L\"uscher correction'~\cite{Luscher:1980fr}. The coefficient
$\gamma$
depends only on the universality class of the string~\cite{Luscher:1981ac};
 for the bosonic string in $D$ space-time dimensions, 
$\gamma=\frac{\pi}{6}(D-2)\la{eq:bosonic_string}$.
There is growing numerical evidence~\cite{Luscher:2002qv,Lucini:2001nv} that
the `flux-tube' energy in non-Abelian gauge theories indeed admits the L\"uscher term as 
leading correction to the linear growth with $L$. In this work we assume the validity
of this correction from $\sigma L^2\simeq 10$,
where its represents $\sim 10\%$ of the lineic mass of the torelon. 

\subsection{Glueball spectrum calculations\la{sec:glueball_intro}}
The numerical techniques to compute  the spectrum of pure gauge theories are well
developed (Ref.~\cite{Teper:1998te} gives a lot of useful details). In particular, the
method of
construction of operators with large overlaps on the lightest states has been perfected
over many years of numerical experimenting~(see for instance
\cite{Lucini:2004my}).
It was realised a long time ago that using `bare' plaquettes to compute the mass gap
becomes increasingly inefficient as the continuum is approached, because the physical 
size of the plaquette shrinks, and it is thus dominated by ultra-violet fluctuations.
Generally speaking, the `fuzzing' techniques, such as smearing~\cite{Albanese:1987ds} and 
blocking~\cite{Teper:1987wt}, provide ways of maintaining the spatial extent of operators at
a constant physical size. In the following chapters
we shall describe the details of the fuzzing algorithm
we used for each series of simulations.

Another extremely useful technique is the variational method~\cite{Berg:1983kp,Luscher:1990ck}.
Indeed, the large-time behaviour of two-point functions always tells us the mass of the
lightest state with non-vanishing overlap on the operator. To extract the excited
spectrum,
we need to measure the full correlation matrix $C(t)$ of a set of $N_o$ operators:
\be
C_{ij}(t) = \langle {\cal O}_i(0) {\cal O}_j(t) \rangle~.
\ee
While the method is also used in mesonic and baryonic measurements, 
here the operators ${\cal O}_i$ are either the real or imaginary parts 
of closed Wilson loop traces. Thus $C(t)$ is a real, symmetric $N_o\times N_o$ matrix.
The spectral decomposition of $C_{ij}(t)$ reads
\be
C_{ij}(t) = \sum_{n=1}^\infty c_n^{(i)} c_n^{(j)} e^{-E_nt}
          = \sum_{n=1}^{N_o} c_n^{(i)} c_n^{(j)} e^{-E_nt} + 
            {\cal O}\left(\exp{\left(-E_{N_o+1}t\right)}\right)
\ee
where $c_n^{(i)}=\langle  n | \hat{\cal O}_i |\Omega \rangle$. 
For large enough $t$, the remaining terms
can be neglected. Then $C(t)$ can be considered as a scalar product expressed
in the basis $\{c^{(i)}\}_{i=1}^{N_o}$. In the canonical basis the scalar product
is simply diag$(\exp-{(E_1t)},\dots,\exp-{(E_{N_o}t)})$. Thus diagonalising $C(t)$
at large enough $t$ can in principle yield the $N_o$ lightest states of the spectrum.
Because statistical noise can make the diagonalisation unstable at large $t$, 
an alternative method is usually considered preferable~\cite{Luscher:1990ck}. 
It amounts to solving the generalised eigenvalue problem:
\be
C(t)\psi = \lambda(t,t_o)C(t_o) \psi
\ee
The eigenvalues $\lambda$ are such that the determinant of 
$D(t,t_o)\equiv C(t)-\lambda C(t_o)$ vanishes. $D(t,t_o)$ is the bilinear form which
in the canonical basis is represented by a diagonal matrix whose $n^{\rm th}$
diagonal element is given by $\exp{(-E_nt)}-\lambda \exp{(-E_nt_o)}$. Thus it is clear that
the solutions to the generalised eigenvalue problem are 
\be
\lambda_n(t,t_o)=e^{-E_n(t-t_o)}
\ee
and the eigenvectors $\psi_n$ can be interpreted as the `wave functions' of the states
of energy $E_n$ (since the latter correspond to the canonical basis in the linear
algebra),
expressed in the initial basis of operators ${\cal O}_i$.  
Appendix~\ref{ap:NRLGT}
describes how the generalised eigenvalue problem is solved in practice.
 
A word of caution is in order. The prescription we have just described
will~\emph{always}
produce a set of orthogonal operators and accompanying energy levels. However, in
general
only the first few states are actually stable one-particle states; the higher states
are scattering states of two or more glueballs. The spectrum of one-particle states
admits
only exponentially small corrections in a finite volume, while two-particle states have
discrete levels with both strong and complicated dependence on $L$~\cite{Luscher:1988sd}.
L\"uscher showed~\cite{Luscher:1991cf}
that from a careful study of the spectrum of scattering states in a finite volume, 
one can extract the scattering matrix in the elastic regime at discrete values of 
momentum; in principle, by fitting the data with a Breit-Wigner formula 
one can extract the energy and width of resonances. However, in the pure gauge theory,
where the lightest state has a mass $\sim 1.5$GeV, the widths of the first resonances
are expected to be rather small compared to their mass. Moreover, the single-trace
operators
that we use naturally couple to the states with energy around the resonance mass. 
In fact, 
in the planar limit $N_c\rightarrow \infty$, single-trace operators are expected to
create 
only one-glueball states. The latter all become stable in that limit; 
more precisely, the width of glueball
resonances is of order $1/N_c^2$~\cite{Witten:1979kh}, while it has been demonstrated 
numerically that the masses are of order 
$N_c^0$ with $1/N_c^2$ corrections~\cite{Lucini:2001ej}. Furthermore,
the non-zero orbital angular momentum decay of resonances is suppressed
close to the two-particle threshold, $A(k)\propto  k^\ell$.

For all these reasons,  we shall sometimes extract the energies of states 
that are above the two-particle threshold. A finite-volume study gives
us some control over the volume dependence of the energy level: if the dependence is
weak, the resonance is almost certainly very narrow.

\section{Outlook}
The study of pure gauge theories on the lattice at zero temperature has 
been for a few years in  an era of precision `numerical experiments'. 
Highlights  include
the determination of the low-lying glueball spectrum in 2+1~\cite{Teper:1998te}
and 3+1~\cite{Morningstar:1999rf} dimensions, 
the confining string spectrum~\cite{Luscher:2002qv,Juge:2003ge}
 as well as ratios of stable string tensions in $SU(N_c)$ gauge 
theories~\cite{DelDebbio:2001sj,Lucini:2001nv}. 
The reasons for this progress lie
both in the increase in computing power and in the development of new 
numerical techniques, such as the fuzzing procedures 
\cite{Albanese:1987ds,Teper:1987wt}, improved actions 
(e.g.~\cite{Morningstar:1997ff}) and 
multi-level algorithms (MLA).

The MLA idea dates back to the `multihit method'~\cite{Parisi:1983hm}, 
which   consists in replacing the link
variables by their average under fixed nearest-neighbour links, when
computing a Wilson or Polyakov loop. It is a realisation of the
real-space renormalisation group transformation. More recently, 
a version adapted to the stochastic computation of quark propagators
  was developed in~\cite{Michael:1998sg}.
 Thanks to the locality property~\cite{Luscher:2001up},
  nested averages can be performed under fixed boundary conditions (BC). 
Multi-level algorithms are particularly powerful in theories with 
a mass gap, where distant regions of the lattice are almost uncorrelated.
An  impressive increase in performance with respect to
 the ordinary 1-level algorithm was achieved in the Polyakov loop 
correlator~\cite{Luscher:2001up}.
A generalisation was proposed in~\cite{Meyer:2002cd}, where the 
MLA was considered for any factorisable functional of the links - including fuzzy 
operators. Indeed the factors need not even be
 gauge-invariant. The efficiency of the
algorithm is based on the fact that the UV fluctuations can be averaged out
separately for each factor, effectively achieving $n^{n_f}$ measurements
by only actually computing $n$ of them, 
where $n$ is the number of measurements done at the lower level of the 
algorithm and $n_f$ the number of factors. The choice of the factorisation
is thus dictated by a competition between having as many factors as possible
and each factor being as independent of the BC as possible.
Chapter~\ref{ch:mla} describes in detail how these ideas are implemented in practice.

%% file: chapter4.tex
\chapter{High-spin glueballs from the lattice\label{ch:hspin}}
\section{Introduction}
\noindent 
The  standard method to measure the spectrum of a gauge theory 
is to evaluate the correlation function of a gauge-invariant 
operator in Euclidean space (Eqn.~\ref{eq:tranf_matrix}).
In order to correctly  
label the energy eigenstates $|n\rangle$ with the quantum numbers of the
rotation group, one has to construct operators that project out the undesired
states.  In continuum Euclidean space, it is possible to construct
operators belonging to an irreducible representation of the rotation 
group; that is, with a definite spin. In
two space dimensions, the case we consider in this chapter, 
the construction amounts to 
\begin{equation}
{\cal O}_J=\int \frac{d\phi}{2\pi} ~ e^{iJ\phi}~ {\cal O}_\phi \la{opj}
\end{equation}
where $J$ is the spin and  ${\cal O}_\phi$ is obtained 
by rotating the operator ${\cal O}_{\phi=0}$ by an angle $\phi$. 
If $|j;n_j\rangle$ is a state of spin $j$ and other quantum numbers
$n_j$, we have
\begin{equation}
\langle \Omega| \hat{\cal O}_J|j;n_j\rangle=\delta_{Jj}~
~\langle \Omega| \hat{\cal O}_{\phi=0}|j;n_j\rangle~.
\end{equation}
The definite-spin subspaces in which the Hamiltonian can separately be
diagonalised  are still infinite-dimensional, so that extracting
their lightest states requires an additional piece of strategy.
The most commonly used is the variational method 
\cite{Berg:1983kp,Luscher:1990ck}. 

However, on a  lattice, rotation symmetry is broken and only a 
handful of rotations leave the lattice invariant. Therefore physical 
states can only be classified into irreducible representations of the
lattice point group. This is a much less thorough classification, in
the sense that each of the diagonal blocks of the lattice Hamiltonian
contains a whole tower of smaller blocks of the continuum Hamiltonian. 
For instance, the trivial square lattice irreducible representation 
contains all multiple-of-4 spin states of the continuum, because
$\exp\{iJ\phi\}$ is unchanged if $J \to J+4n, \ n\in {\bf Z}$, for 
any symmetry rotation of the lattice, 
i.e. for $\phi=n^\prime\pi/2, \ n^\prime\in {\bf Z}$.

When extracting glueball masses from lattice calculations, it has been 
customary to label the obtained states with the lowest spin falling into
the lattice irreducible representation; in the example above,
it would be `spin 0'. Not only does
the procedure used till now lack the capability of extracting the higher
spin spectrum, but this labelling could very well be wrong, especially for 
excited states.  Suffice it to think of 
the hydrogen atom, where the degeneracy in $\ell$ implies, for example, 
that the $n=1, \ell=1$ is lighter than the $n=2,\ell=0$ state. In
the case of D=2+1 gauge theories, the simple flux tube model
predicts~\cite{Johnson:2000qz} that the lightest $J^{PC}=0^{-+}$ 
state is much heavier than the lightest $4^{-+}$ state, and the mass that 
it predicts for the latter agrees with the value obtained on the 
lattice~\cite{Teper:1998te} for the state labelled as $0^{-+}$. 
This suggests a possible misidentification, 
as emphasised in~\cite{Johnson:2000qz}. 

In the case of the two-dimensional square lattice, the only states one can 
distinguish in the `traditional' fashion are those of spin 0, 1 and 2. 
Here we attempt to go beyond that apparent limitation of lattice 
calculations: we want to check the correctness of the conventional labelling
of states and  extract the lowest lying states carrying spin higher 
than 2. It is
clear  that a necessary ingredient in the realisation of this program 
is a reliable way to construct operators that are rotated with respect to each
other by angles smaller than $\frac{\pi}{2}$ but that are otherwise
(nearly) identical.
Indeed one expects that as the lattice spacing becomes very much 
smaller than the dynamical length scale, this becomes possible
due to the recovery of the continuum symmetries.
In fact it has been known for a while that rotation
symmetry is restored dynamically rather early in the approach to the 
continuum. An early piece of evidence came from the ($D=3+1$) $SU(2)$ static
potential measured off the lattice axis~\cite{Lang:1982tj}. Later it was 
shown~\cite{deForcrand:1986ww} that the ($D=3+1$) $SU(3)$ glueball  dispersion 
relation $E(\vec{p})$ to a good approximation depends only on $|\vec{p}|$ already
at $\beta=5.7$. Also, the detailed glueball spectra obtained more recently in
(2+1)~\cite{Teper:1998te} and (3+1)~\cite{Morningstar:1999rf} dimensions exhibit 
the degeneracies between states belonging to different lattice irreducible representations
that are expected in the continuum limit.

The outline of this chapter is as follows. We begin by 
discussing techniques for constructing arbitrary rotations of a 
given operator. 
We design systematic tests to evaluate how well these methods 
perform in $D=2+1$ $SU(2)$ gauge theory\footnote{The generalisation
to $SU(N_c)$ is trivial.}. We then discuss how to use 
these techniques to extract the high spin spectrum from lattice 
simulations. As an example, we analyse the case
of the lightest $4^-$ and $0^-$ glueballs. We find that the state
conventionally labelled as a $0^-$ is in fact a $4^-$, as was
suggested by calculations of the spectrum within the flux tube 
model~\cite{Johnson:2000qz}. We also find that the
$J=3$ ground state is lighter than the $J=1$ --- again,
as suggested by the flux tube model. 
%
\section{Two methods of operator construction}
Lattice glueball operators are usually constructed out
of space-like Wilson loops. For these to project onto arbitrary spins
we need linear combinations of Wilson loops rotated by arbitrary 
angles. Since we are working on a cubic lattice, 
such a rotated loop will usually only be an 
approximate copy of the initial loop. The better the
approximation, the less ambiguous the spin assignment. Now, a general
Wilson loop consists of a number of sites connected by products of 
links. To obtain a good projection onto the lightest states in
any given sector, these links need to be `smeared' or `blocked'
so that they are smooth on physical rather than just ultraviolet
length scales\footnote{See, for example, \cite{Teper:1998te} 
and references therein.}. 
To construct arbitrary  rotations of various Wilson loops,  
it is clear that we need to be able to construct `smeared' parallel
transporters between arbitrary sites in a given time-slice of the
lattice. We now describe two techniques to do this. We begin
by reviewing and elaborating on a method~\cite{Teper:1988ea} 
that has recently been used~\cite{Johnson:2002qt} in an attempt to address 
the $0^-/4^-$ ambiguity referred to in the Introduction. This method 
is however  computationally too expensive to allow a realistic
continuum extrapolation with our resources, and without such an 
extrapolation one has very little control over the restoration of 
rotational invariance. We therefore develop a second 
method that is much less expensive and provides a practical approach 
to the problem.
%
\subsection{The matrix method}
\label{subsection_matrix}
%
Consider the two spatial dimensions of a given time-slice of size
$N\equiv L\times L$ with periodic boundary conditions. 
Each point $p$ is parametrised by integers $(m,n)$ representing its 
Cartesian coordinates in lattice units. For $i=1,\dots N$, 
we choose a mapping  $i \rightarrow (m(i), n(i))\equiv \varphi(i)$ 
and  define an $N \times N$ matrix $M$  by its elements
\begin{equation}
M_{ij}\equiv U(\varphi(i);\varphi(j))
\end{equation}
where $U(p;q)$ is the link matrix joining points $p$ to $q$ if they are
nearest neighbours, and vanishes otherwise. Since
$U(p,q)= (U(q,p))^\dagger$, $M$ is hermitian. For notational simplicity,
if $\varphi(i)$ is the origin and $\varphi(j)=(m,n)$, 
we shall write $M_{ij}\equiv M[m,n]$.

It is straightforward to see that $\left(M^\ell\right)_{ij}$ 
contains all paths of length $\ell$ going from $\varphi(i)$ to 
$\varphi(j)$.  One can construct a `superlink'
connecting these two points by adding up paths of all lengths, weighted by
a damping factor that ensures the convergence of the series:
\begin{equation}
K=\sum_\ell c(\ell) M^\ell
\end{equation}
In general it is very costly to calculate such a  power series
numerically. For $c(\ell) = \alpha^{\ell}$,  the
series can be resummed:
\begin{equation}
K=\sum_{\ell\geq 0} \alpha^\ell M^\ell=\left(1-\alpha M\right)^{-1}, \la{kdef}
\end{equation}
and the calculation of the geometric series can be reduced to the
inversion of a matrix. For instance, one can now compute a triangular 
`fuzzy' Wilson loop by simply multiplying together three elements 
of $K$
\begin{equation}
W = \tr \{ K_{ij} K_{jk}  K_{ki} \},
\label{eqn_triangle}
\end{equation}
where $i,j,k$ are the vertices of the triangle. It is clear that
as we increase $\alpha$, the longer paths are less suppressed,
i.e. increasing $\alpha$ increases the smearing of the `superlink'.
Thus by inverting a single matrix we obtain fuzzy parallel 
transporters between all pairs of sites in the given time-slice. We can 
now use these to construct any Wilson loops we wish.

While the above construction is valid for an infinite volume, one
must be careful if the volume is finite. Consider, for example, the
triangular Wilson loop defined in Eqn.~(\ref{eqn_triangle}). On a finite
spatial torus, the sum of paths contributing to $K_{ij}$ contains not
only the `direct' paths from $i$ to $j$ but also paths that go
the `long' way around  the torus between these two points.
That is to say, the Wilson loop defined in  Eqn.~(\ref{eqn_triangle})
is not necessarily a contractible triangle; some of the 
contributions to $W$ are non-contractible closed paths that wind 
once around the torus. In the confining phase such an operator
projects onto flux loops that wind once around the torus. These
states are orthogonal to glueballs. Moreover, in the kind of
volume with which one typically works, this loop will be much lighter
than any of the excited glueballs. Such effects induce an infra-red
breaking of rotation symmetry that, unlike lattice spacing effects, 
will survive the continuum limit. Fortunately it is simple
to modify our matrix procedure so as to explicitly suppress 
such contributions and we now describe two ways of doing so.

We label the link matrix emanating from the site $(x,y)$ in the direction
$\mu$  by $U_{\mu}(x,y)$ and we consider for simplicity the $SU(2)$ gauge 
group. We define the matrix $M^{(x)}$,
obtained from $M$ by a ${\bf Z}_2$ transformation, 
as in Eqn.~(\ref{eq:cen_sym}):
\begin{equation}
U_{x}(x=L,y) \to -1 \times U_{x}(x=L,y) \ \ \ \ \ \forall \ y.
\label{eqn_Mx}
\end{equation}
The corresponding matrix $K^{(x)}$ will produce
`superlinks' that are identical to those from $K$ except that
the contribution of paths that wind once (or an odd number of times)
around the $x$-cycle will come in with a relative minus sign. 
Thus if we replace $K$ by $K + K^{(x)}$ in Eqn.~(\ref{eqn_triangle}) 
the contribution of all the non-contractible paths winding 
around the  $x$-torus will cancel. In the same way we can
define a matrix  $M^{(y)}$ from  $M$ by
\begin{equation}
U_{y}(x,y=L) \to -1 \times U_{y}(x,y=L) \ \ \ \ \ \forall \ x
\label{eqn_My}
\end{equation}
and a matrix $M^{(xy)}$ which includes both the modifications
in equations (\ref{eqn_Mx}) and (\ref{eqn_My}). It is easy to
see that the sum of the corresponding inverse matrices,
$K + K^{(x)} + K^{(y)} + K^{(xy)}$, will produce superlinks that
have no contributions from non-contractible paths that
wind around the $x$-torus or the $y$-torus or simultaneously
around both tori. This is a simple and effective modification
although it would appear to suffer from the fact that it quadruples 
the length of the calculation. However it is easy to see that
one can considerably reduce this cost. We start by noting that on
a lattice with an even number of sites in both $x$ and $y$ directions, even
and odd powers of $M$ connect a given point (say, the origin) to two disjoint
sets of lattice sites. This implies a partitioned structure for
 $K_e\equiv(1-\alpha^2M^2)^{-1}=\sum_{n\geq0}(\alpha M)^2$ and  
allows us to store any 
polynomial in the matrix $M^2$ in two matrices of size $N/2 \times N/2$.
If standard inversion algorithms are applied to compute $K_e$ (for which
the CPU time scales as $N^3$), this represents a reduction in CPU time
 by a factor four. If $L_x$ and $L_y$ are odd, this trick cannot be used,
but paths joining two points by going around the world have an opposite-parity
weighting in powers of $\alpha$ to those connecting them directly. Therefore
in that case we can proceed as follows: use 
$K_e$ to propagate by an even number of lattice links and 
$K_o\equiv \alpha M K_e$ for an odd number of 
steps\footnote{$K_o$ is very fast to obtain from $K_e$, given the sparsity
of the matrix $M$.}. 
Thus paths with an odd winding number are excluded by 
construction. This way of proceeding has the additional advantage that one
can truly propagate by a distance larger than $L_{x,y}/2$, which is not the 
case with $L_{x,y}$ even.

In summary, obtaining the superlinks free of odd-winding-number paths
requires either of the following computations:
if the lattice has an odd number of sites, it is sufficient
to perform one full matrix inversion, plus two multiplications by $M$; 
if $L_{x,y}$ are even, we have to use the four ${\bf Z}_2$ transformations,
as discussed above, but we can compute and store the superlinks in
matrices smaller by a factor two.
\paragraph{Interpretation as a propagator}
There is an interesting interpretation to the matrix construction in
an infinite volume.
If we choose the mapping such that $\varphi^{-1}(m,n) =L\cdot m+n$, then
in the frozen configuration (all link variables set to unity),
$M$ coincides with the matrix used in 
discretising partial differential equations in the finite difference scheme;
more precisely, $M-4$ is exactly the expression of a discretised
Laplacian operator on a torus. For that reason, the Klein-Gordon equation
$-\nabla^2 F+m^2 F=0$ translates into
\begin{equation}
[(am)^2+4-M]_{ij}F_{j}=0,\quad \forall i=1,\dots,N
\end{equation} 
where $F$ is now a column vector containing the approximate values of the 
function $F$ on the lattice sites. If we introduce a point-like source 
$v$ on the RHS, that is $v^{(k)}_i=\delta_{ik}$, we obtain the 
interpretation that $[((am)^2+4-M)^{-1}]_{ij}$ is the $2d$ 
lattice propagator of a massive scalar field from point
$\varphi(i)$ to point $\varphi(j)$. For a scalar field 
minimally coupled to the gauge field at finite $\beta$, the 
ordinary derivatives are simply replaced by covariant derivatives.
If the scalar field is in the fundamental representation 
of the gauge group, then
our matrix $1-\alpha M$ provides a discretisation of the kinetic
term where the parameter $\alpha$ in Eqn.~(\ref{kdef}) corresponds to
\begin{equation}
\alpha=\frac{1}{(am_0)^2+4}
\label{eqn_alphatom}
\end{equation} 
and $m_0$ is the tree level mass of the scalar particle.
Setting $\alpha$ to $\frac{1}{4}$ corresponds to $m_0=0$. The propagator
calculation amounts to introducing a scalar particle 
in the configuration, a `test-charge' that does not modify 
the configuration, but the closed
paths of which reveal gauge-invariant information on the background gauge
field. In other words, this method is analogous to performing
a quenched simulation of the gauge theory minimally coupled to a scalar
field --- with the latter confined to a time-slice. The lighter the mass
of the scalar particle, the greater the transverse distance that
it explores as it propagates between two sites. 
We recover our earlier conclusion: as $\alpha$ increases, 
the propagator becomes increasingly fuzzy.

We expect on general grounds that if we calculate a propagator
over some physical length scale for a mass $am_0$ that is fixed in
physical units, then the lattice corrections to continuum rotational
invariance will be ${\cal O}(a^2)$. This provides a general theoretical
argument that our matrix method will yield `rotationally invariant'
superlinks if one chooses the parameter $\alpha$ suitably. 
Of course, since there is no
symmetry to protect against mass counter-terms, there will be both
an additive and a multiplicative renormalisation of the mass.
That is to say, choosing the mass involves a `fine-tuning'
problem that is very similar to the one that one encounters when
using Wilson fermions. 

The recovery of rotational invariance at large distances and
at weak coupling can, in general, only be seen numerically 
(see below). However, in the special case of a frozen configuration, 
it can be studied analytically~\cite{Meyer:2002mk}:
\be
M^{2k+m+n}[m,n]= \left(\begin{array}{c} 2k+m+n\\ k+n\end{array}\right) 
\left(\begin{array}{c} 2k+m+n\\ k\end{array}\right);\quad
\left(\begin{array}{c} p \\ q \end{array}\right) \equiv \frac{p!}{q!(p-q)!}
;\quad k\geq 0. 
\la{eq:M2k+m+n}
\ee
This study leads to the following conclusions: if $m^2+n^2=d^2$, for
$1\ll d \ll (am_0)^{-1}$ the propagation from  $(0,0)$ to $(m,n)$
results from a Brownian motion and the length of the dominating 
paths is of the order $d^2$. In this regime rotational invariance is 
recovered\footnote{For $d\gg (am_0)^{-1}$, an unphysical range of distances,
the superlinks become more directed and rotational invariance is lost.}: indeed,
expression~\ref{eq:M2k+m+n} has the asymptotic behaviour for $k\gg m,n$
\[
\left[\frac{2^{2k+m+n}}{\sqrt{k}}\left(1-\frac{(m-n)^2}{4k}\right)\right]
\left[\frac{2^{2k+m+n}}{\sqrt{k}}\left(1-\frac{(m+n)^2}{4k}\right)\right]
=\frac{4^{2k+m+n}}{k}~\left(1-\frac{m^2+n^2}{2k}+\dots\right).
\]
\paragraph{Cost}
The matrix method is a simple and powerful tool for obtaining
fuzzy `superlinks' between all pairs of sites in a given
time-slice. However even in D=2+1 SU(2) and at moderate
$\beta$ values the matrix is large and the inversion expensive.
While a calculation with modest statistics on say a $16^3$ lattice 
may be readily performed  on a workstation, this is no longer the 
case for the $24^3$ and $32^3$ lattices that would be needed
for even a minimal attempt at a continuum extrapolation.
To circumvent this problem we develop a much faster alternative 
method in the next subsection. 
\paragraph{Generalisations}
We have noted that the elements of our inverse matrix 
$K=\left(1-\alpha M\right)^{-1}$ 
are nothing but the propagators of a minimally
coupled scalar particle in the fundamental representation,
whose bare mass is determined by the parameter $\alpha$.
In principle we are free to consider propagators of other
particles: these should provide equally good `superlinks'.
Consider then a fermion in the fundamental representation and
suppose we discretise it as a two-dimensional staggered lattice 
fermion. The propagators are obtained by inverting a
matrix which is obtained from our matrix $M$  
by multiplying the elements of $M$
by position-dependent factors of -1 \cite{Kogut:1975ag}.
This lattice discretisation maintains a chiral symmetry
which protects the massless fermions from an additive
renormalisation. This removes the fine-tuning problem we
referred to earlier: a first advantage. Moreover we expect the 
long-distance physics to be encoded in the lowest eigenvalues,
and corresponding eigenvectors, of our discretised Dirac operator.
Now we recall that the fermion propagator can be expressed 
in terms of all the eigenvectors and eigenvalues of the Dirac
operator. If we truncate this sum to include only some
suitably chosen set of these lowest eigenvalues and eigenvectors,
then this should provide us with an approximation to the propagators
that maintains the long-distance physics; for example the
restoration of full rotational invariance. That is to say,
they can be used as `superlinks' for our purposes.
As we approach the continuum limit we do not need to
increase the number of these eigenvectors, as long as the
volume is fixed in physical units, so the computational cost
scales in a way that is far better than that of the
full Dirac operator inversion -- a second, major, advantage. 
As an added bonus we note that we can
expect the chiral symmetry to be spontaneously broken. This
implies a non-zero density of modes near zero which generates
the chiral condensate \cite{Banks:1980yr}, and the choice
of what are `small' modes then becomes unambiguous.
This is of course only an outline of a strategy; its
practical application is something we do not attempt here.
%
\subsection{The path-finder method}
%
We turn now to a simpler, more direct and, above all, faster alternative
method for constructing superlinks between any two sites.
In order to define a path from site $A$ to site $B$, we first
introduce a `d-link' in the diagonal direction of the lattice:
\begin{equation}
U_{\mu\nu}(x)={\cal U}\left(U_\mu(x)U_\nu(x+a\hat{\mu})+
	U_\nu(x)U_\mu(x+a\hat{\nu})\right),
\end{equation}
where $\cal{U}$ represents a unitarisation 
procedure\footnote{In $SU(2)$, the 
operation amounts to dividing the matrix by the square root of 
its determinant; see Appendix~\ref{ap:NRLGT}.}.
From a point $x$, there are now 8 directions available.
It is easy to write an algorithm that finds the path following 
the straight line from $A$ to $B$ as closely as possible.
 Indeed, at each step, it is 
sufficient to try all directions by adding the corresponding vector
to the current state of the path and select the result that has maximal
 projection on the
$\overrightarrow{AB}$ vector. As this can lead to a path that is not invariant
under a $\pi$-rotation, one can average with the opposite path 
 obtained with the same algorithm by starting from $B$
and inverting it at the end.
In practice, before starting to calculate the path, we may
 smear the ordinary links to reduce short-wavelength fluctuations and
achieve better overlap onto physical states. 
%
\subsection{A test for the operator construction methods}
%
We  now calculate appropriate Wilson loops using the two kinds 
of superlinks introduced above,  so as to test the extent to which
rotation symmetry is violated. The result is obviously determined
by the dynamics of the lattice gauge theory and the operator being measured.
Therefore, to obtain rotational invariance to a good accuracy, 
 the two following conditions must be satisfied: 
\begin{equation}
aL\gg \sigma^{-1/2}\qquad {\rm and } \qquad \sigma^{-1/2}\gg a,
\label{eqn_conditions} 
\end{equation}
where $L$ is the size of the loop and $\sigma$ the string tension.
It is our task, when constructing rotated copies of operators, to ensure that
these conditions are also sufficient. 

We will refer to the 
method using the propagator as `method I', to the path-finder
as `method II'. In both cases,
in order to have a gauge-invariant operator, we must form a closed path.
Since a great number of paths contribute to the superlinks in method I, 
an operator of the type 
\begin{equation}
{\cal O}(x,y)=\mathrm{Tr}~ \{ K(y,x)\cdot K(x,y) \}
\end{equation}
is a perfectly acceptable operator characterising the direction $\phi$
determined by the points $x,y$. We call it a `segment' operator; it has an
interesting physical interpretation which we shall elaborate upon below.
We can choose pairs of points $x,y$ that have approximately 
the same length -- up to the percent level -- and which
are rotated by an approximately constant angle. These approximations
mean that there is an intrinsic limitation to the rotational invariance 
that we can expect to observe at a fixed length $|x-y|$. 
One can finesse this problem by plotting the values of 
${\cal O}(x,y)$ for all points  $x,y$ and seeing to what extent 
they fall on a single smooth curve. For our purposes an alternative
two-part strategy is more illuminating. In the first step, working 
at a fixed value of $|x-y|$, we ask if the 
violations in rotation symmetry are of the same order of magnitude
as the differences in the lengths. In the second step, as
a more  direct test of rotational invariance, 
we first normalise the operators  ${\cal O}={\cal O}(\phi)$ to a common
value and then calculate the  correlation between rescaled segment 
operators at different angles. 

We perform these tests  on a $16^3$ lattice at $\beta=6$, where we 
know~\cite{Teper:1998te} that the string tension is
$a\surd\sigma \simeq 0.254$ so that the second condition of
Eqn.~(\ref{eqn_conditions}) would appear to be satisfied. 
We choose $\alpha=0.24$ and there is no preliminary smearing of the links. 
Using the tree-level relation in Eqn.~(\ref{eqn_alphatom}) 
and the above value of $a\surd\sigma$ 
we see that this value of $\alpha$ corresponds to a mass for the
scalar particle of $m_0\simeq 1.6\surd\sigma$, i.e. a physical rather  
than an ultra-violet scale. The segment we use is of length $7a$, 
and is rotated by multiples of
approximately $\frac{\pi}{12}$ or $\frac{\pi}{16}$ angles. 
To illustrate the necessity of the 
torelon-suppression procedure, we shall present our results with and 
without implementation of the latter.

We present in Fig.~\ref{meaI} the results of the first step of
our test. The points
$x,y$ that we use lie on an (approximate) semicircle and are joined by 
solid lines for clarity. We do not label the $x$ and $y$ axes, but the
on-axis distance from the origin of the (semi)circle is 7 in lattice
units and this sets the (separate) scales for the $x$ and $y$ axes. 
Using this information, the Euclidean length $R(\phi)$ of the segment  
in each direction $\phi$ can be read directly off this polar plot;
the x-axis of the plot corresponds to the lattice x-axis. 
For each point $x,y$  we plot the average value of the segment operator,
as a point along the same  direction, with the distance to the
origin representing its value. For clarity these points have been
joined up by dashed lines. Both sets of points have
been rescaled so that they can be plotted on the same graph,
and there are separate plots with and without torelon removal.

With method II, the superlink from  a point $A$ to a point $B$  is a
unitary matrix (or a sum of two such matrices).
Therefore we cannot use segment operators since
they would be trivial. Instead we use long rectangular Wilson loops, 
typically $7\times1$; they each characterise a specific direction $\phi$.
We present the results in Fig.~\ref{meaI} in a polar plot similar 
to that used for method I: in each direction $\phi$, the 
Euclidean length $R(\phi)$ of the segment is given, as well as the average
value of the rectangle operator pointing in that direction.
In this case there are no torelon contributions that need to be subtracted.
%
\begin{figure}[tb]
\vspace{-0.5cm}
\centerline{\begin{minipage}[c]{14.7cm}
    \psfig{file=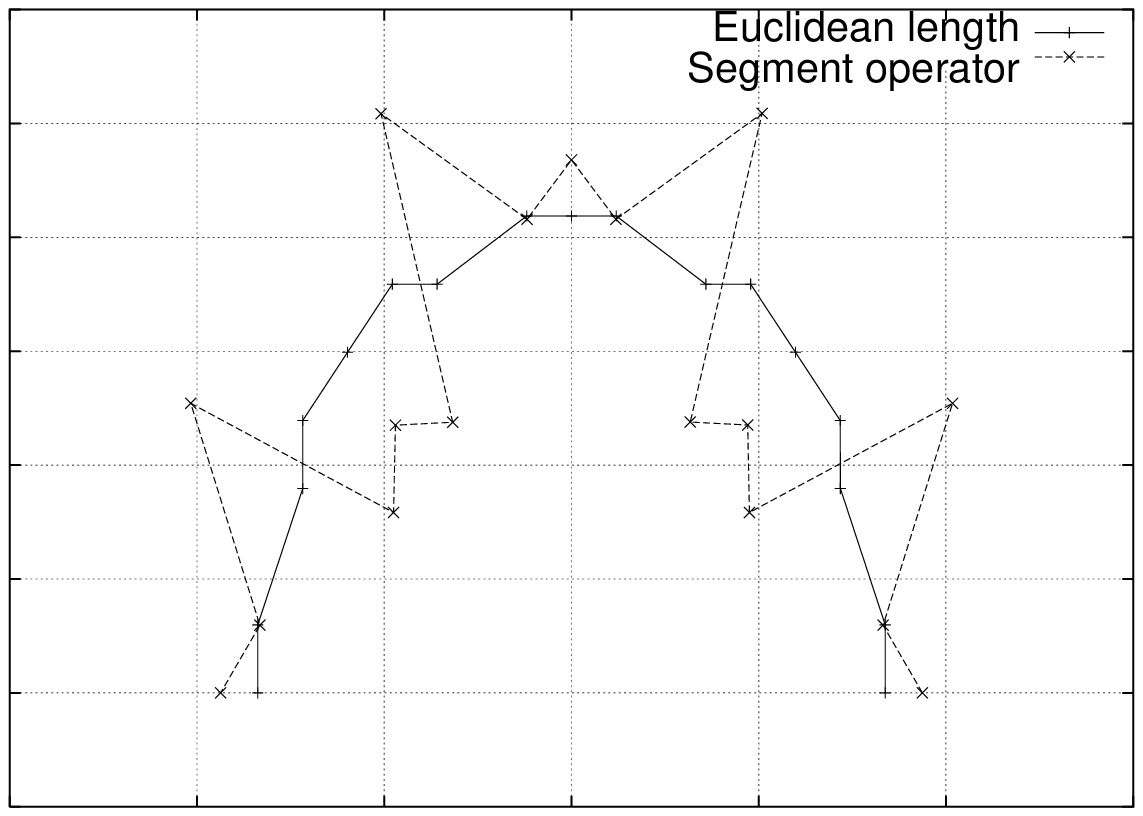,angle=0,width=4.8cm, height=3cm}
    \psfig{file=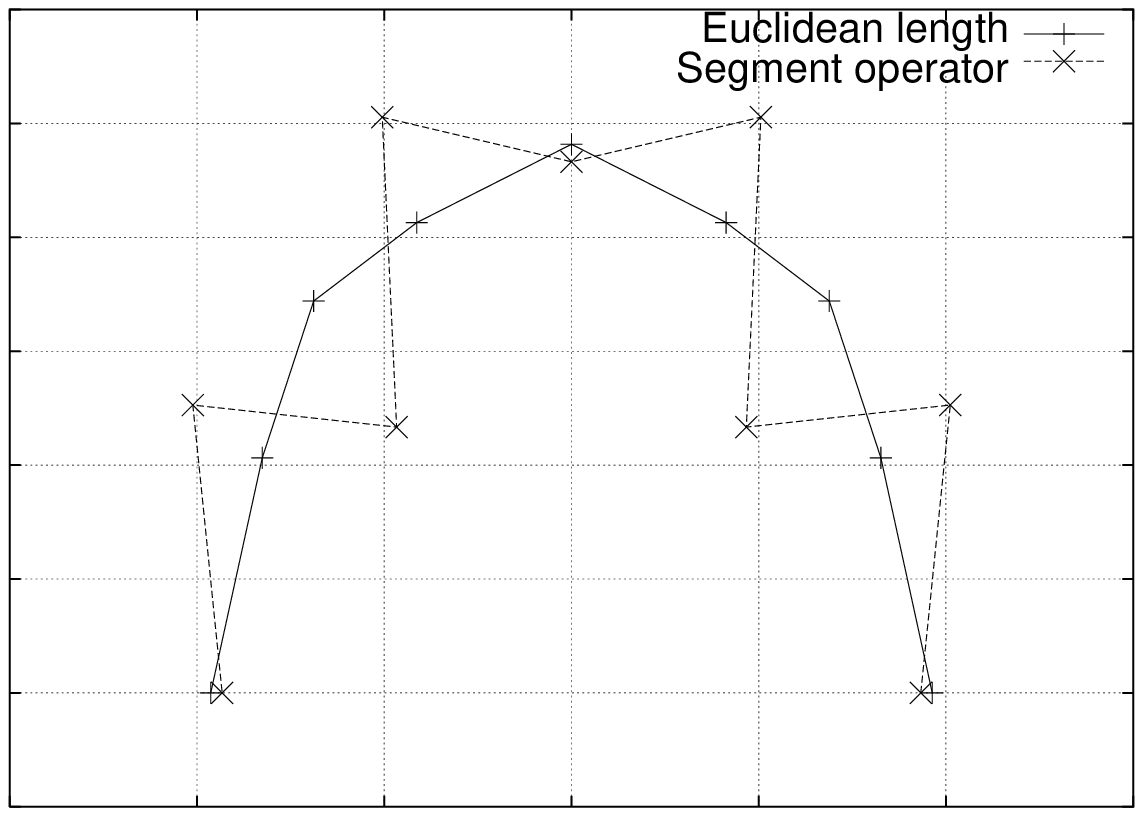,angle=0,width=4.8cm, height=3cm}
    \psfig{file=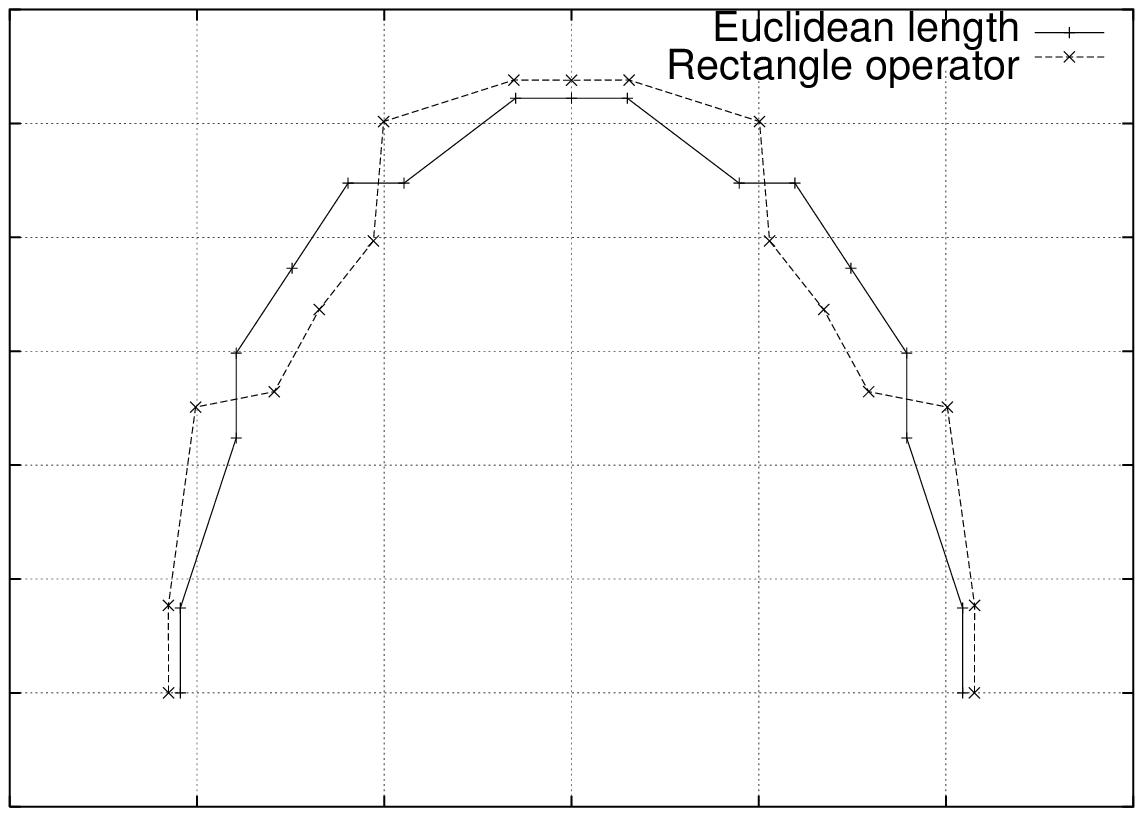,angle=0,width=4.8cm, height=3cm}
    \end{minipage}}
\caption[a]{In polar coordinates: the Euclidean length and the average
 value of segment operators in different directions $\phi$. Left: 
method I without torelon-suppression. Middle: method I with 
torelon-suppression. Right: method II.}
\la{meaI}
\end{figure}
\begin{figure}[tb]
\vspace{-0.5cm}
\centerline{\begin{minipage}[c]{14.7cm}
    \psfig{file=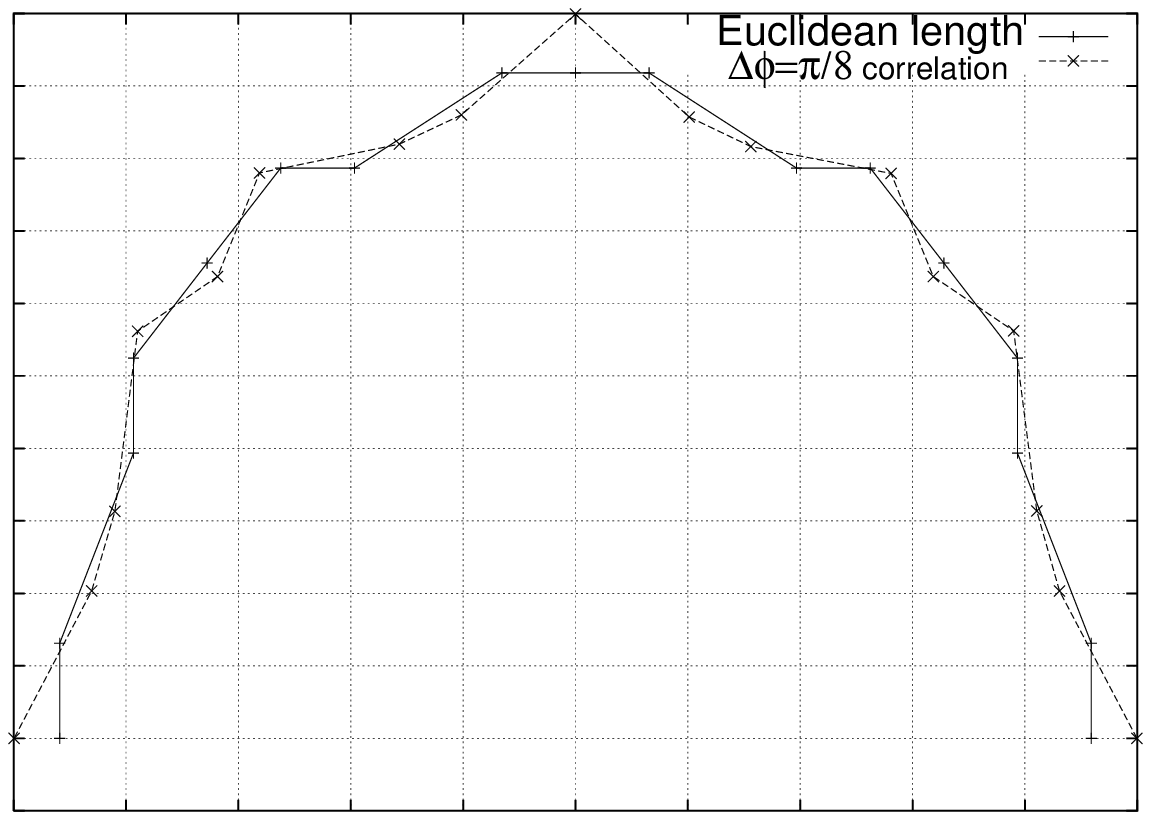,angle=0,width=4.8cm, height=3cm}
    \psfig{file=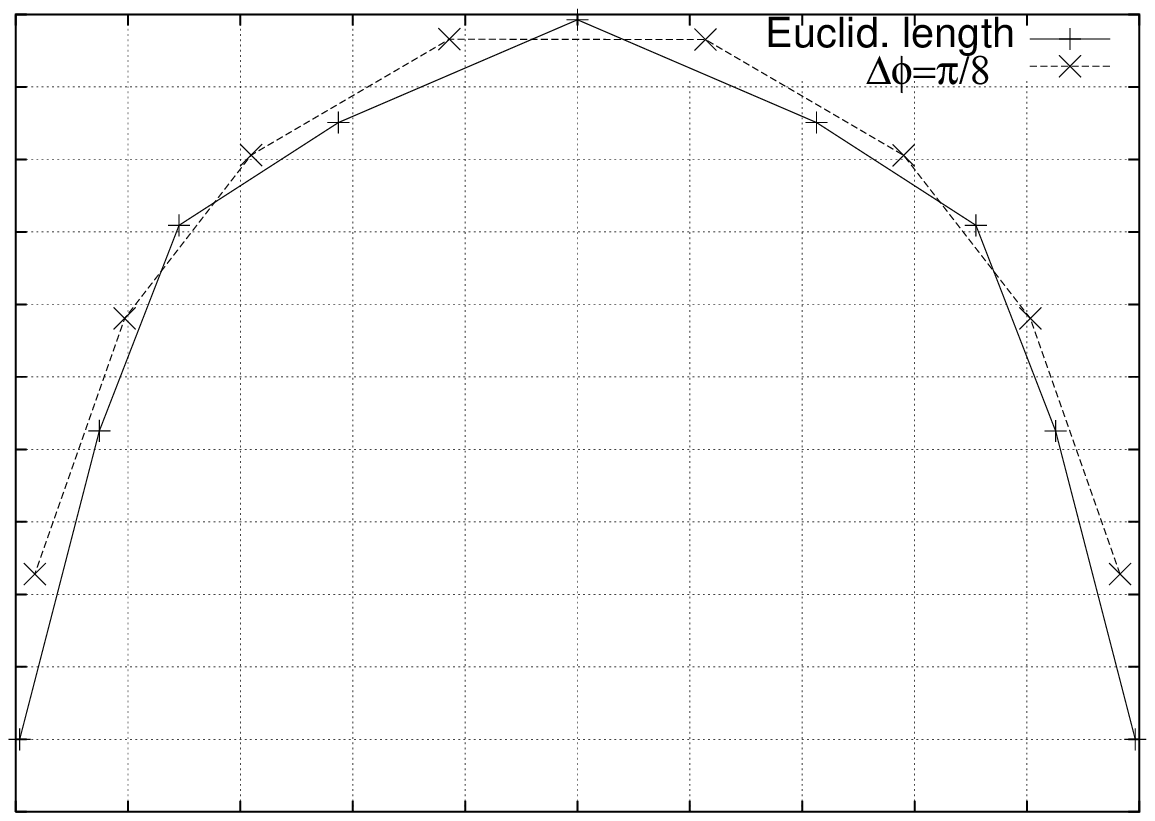,angle=0,width=4.8cm, height=3cm}
    \psfig{file=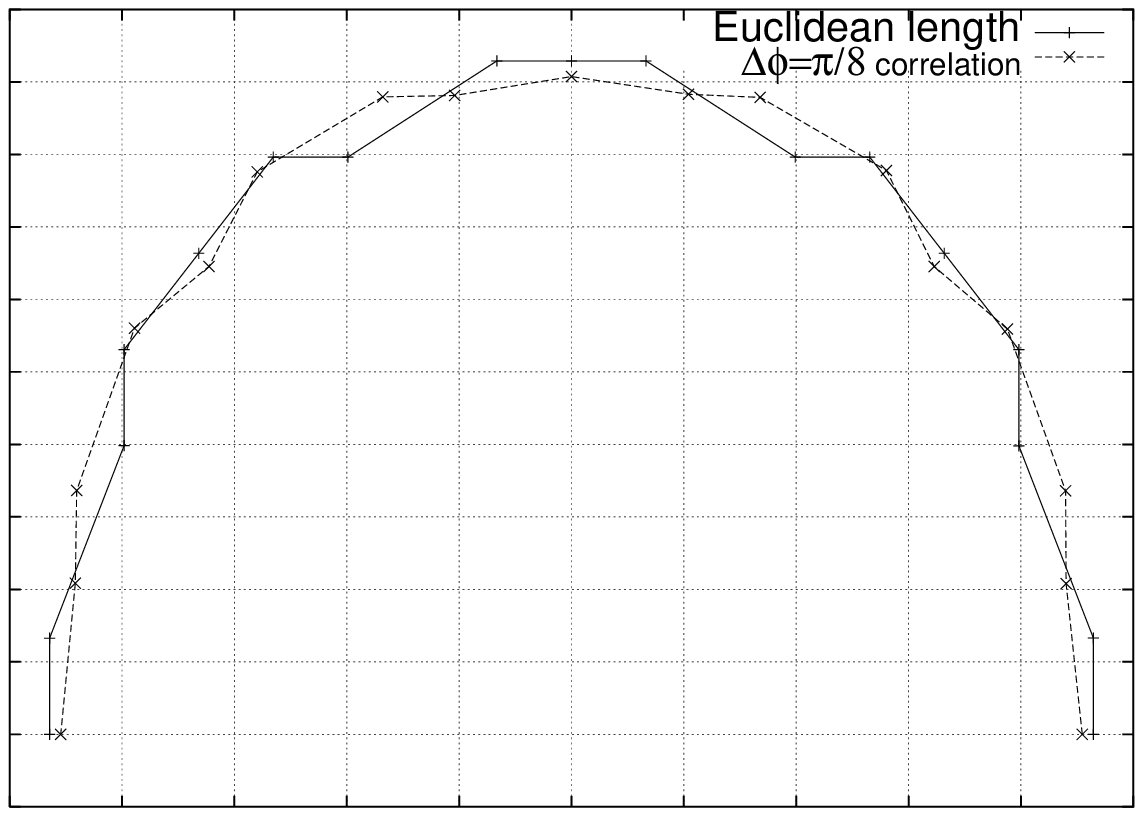,angle=0,width=4.8cm, height=3cm}
    \end{minipage}}
 \centerline{
 \begin{minipage}[c]{14.7cm}
    \psfig{file=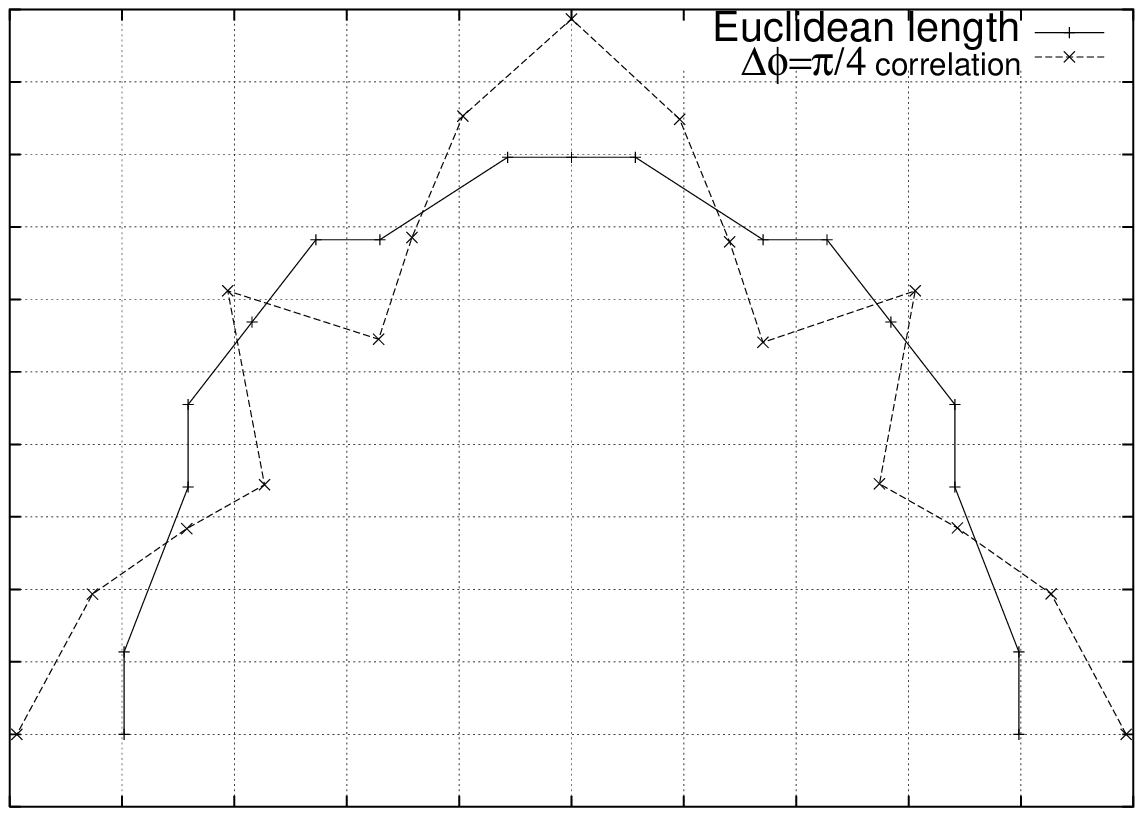,angle=0,width=4.8cm, height=3cm}
    \psfig{file=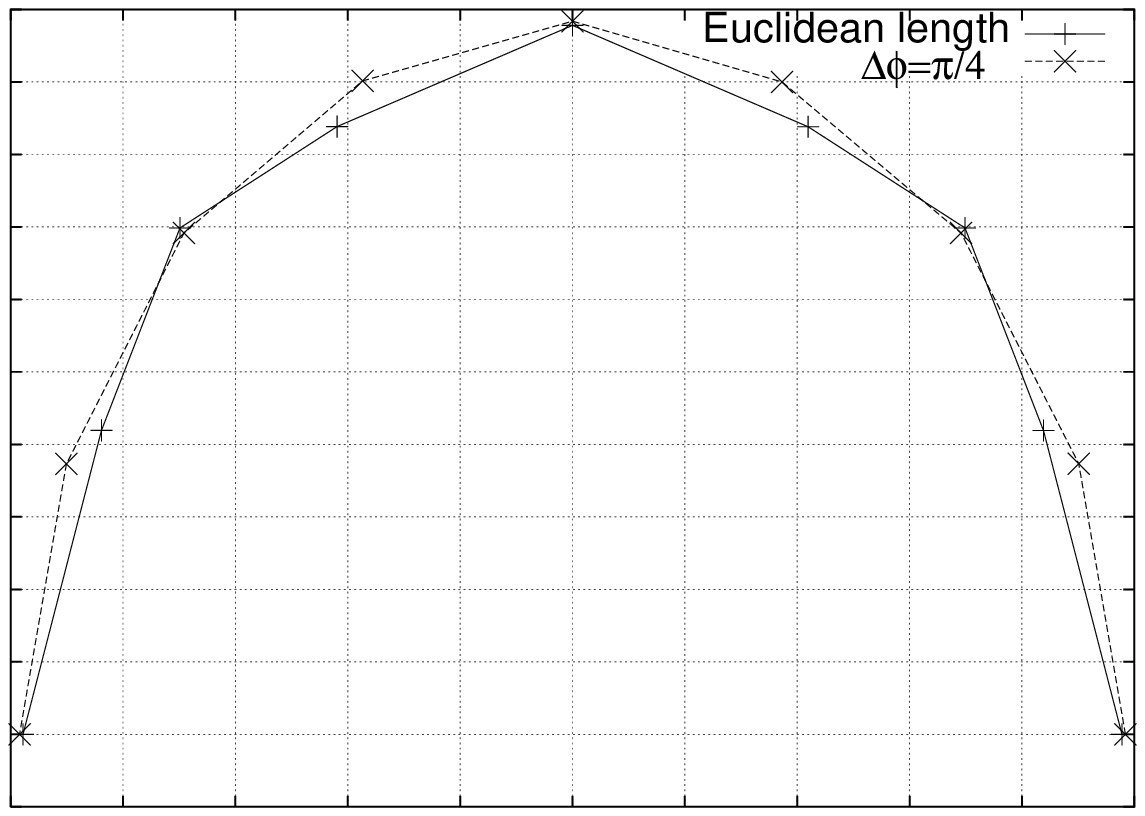,angle=0,width=4.8cm, height=3cm}
    \psfig{file=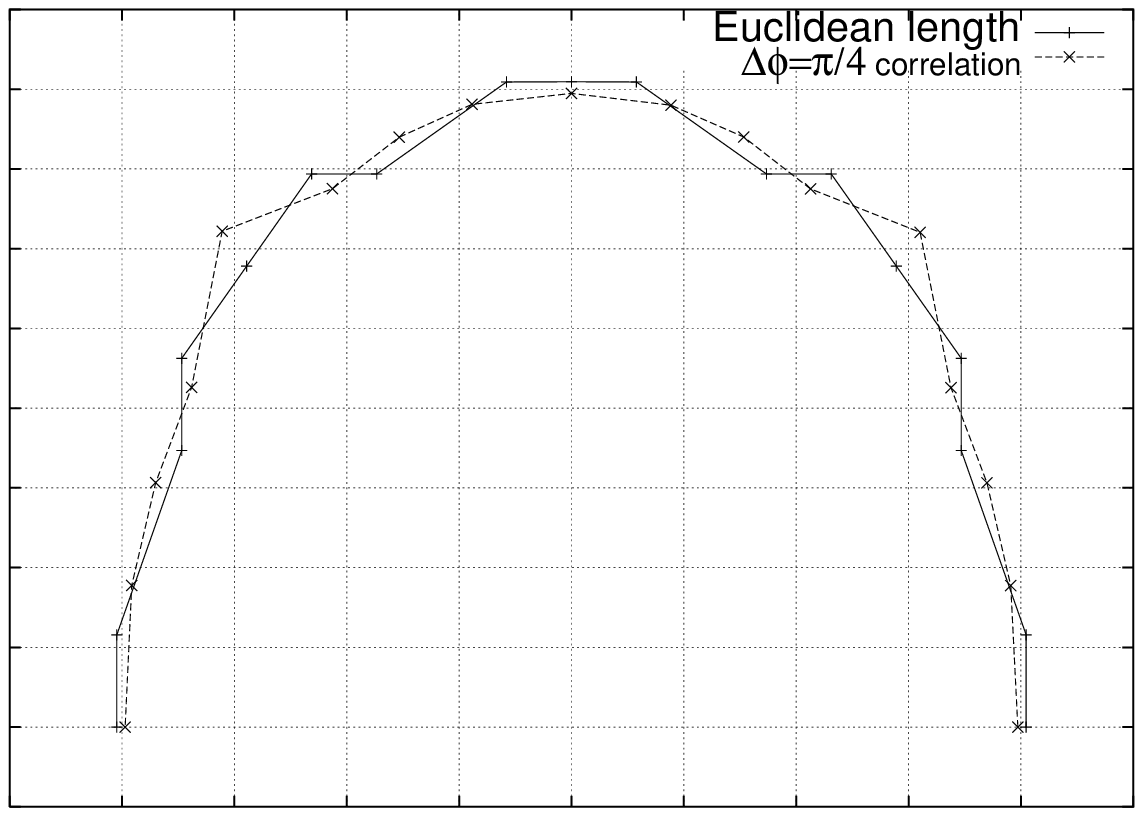,angle=0,width=4.8cm, height=3cm}
    \end{minipage}}
 \centerline{
 \begin{minipage}[c]{14.7cm}
    \psfig{file=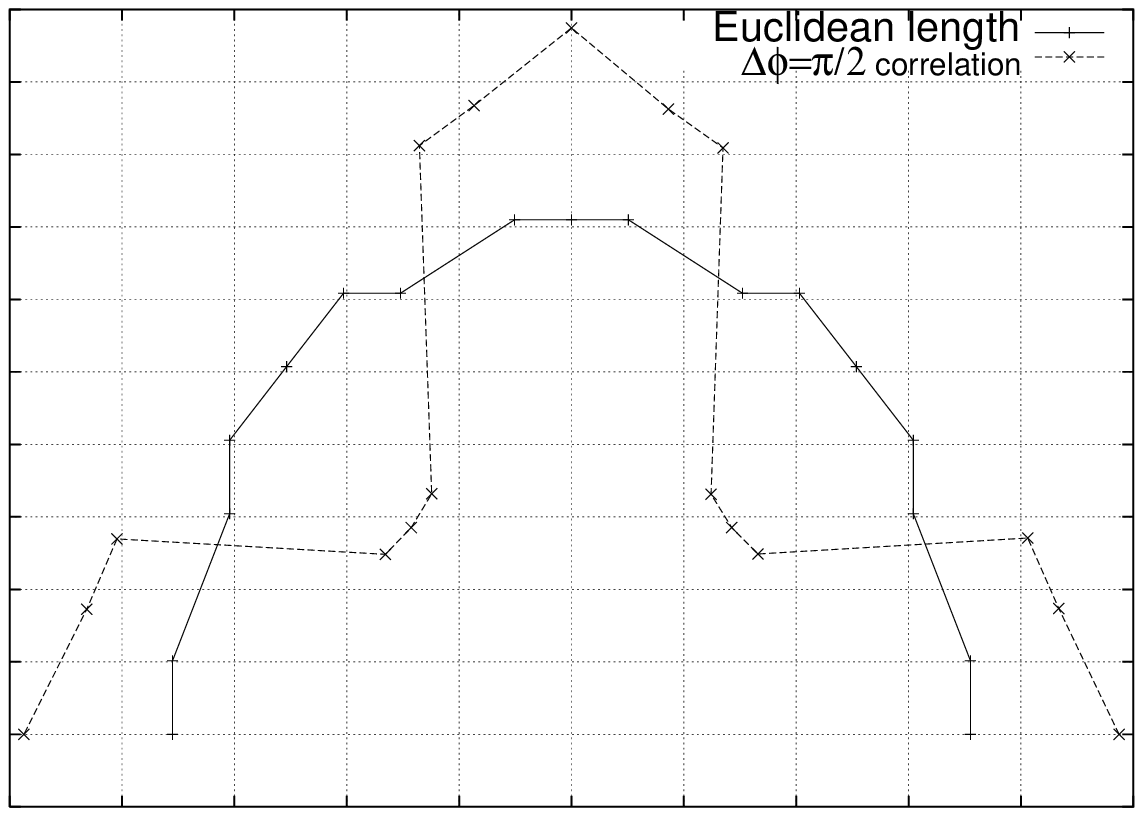,angle=0,width=4.8cm, height=3cm}
    \psfig{file=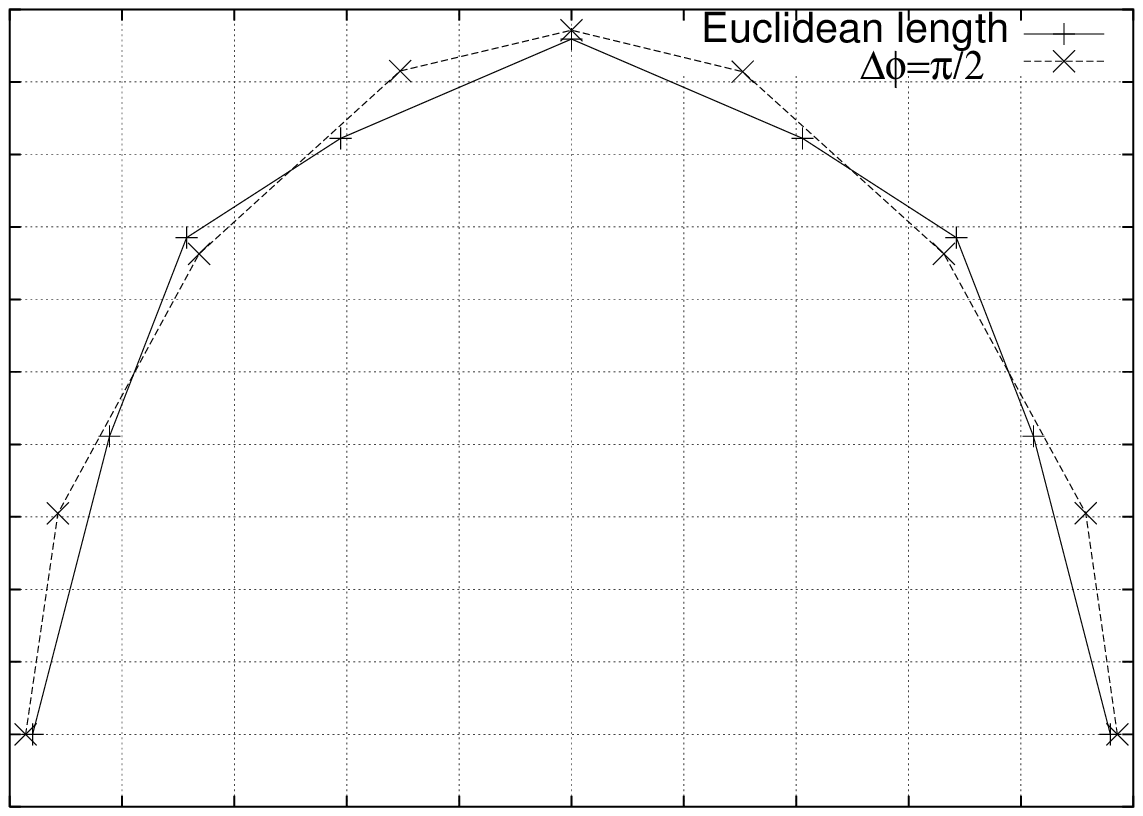,angle=0,width=4.8cm, height=3cm}
    \psfig{file=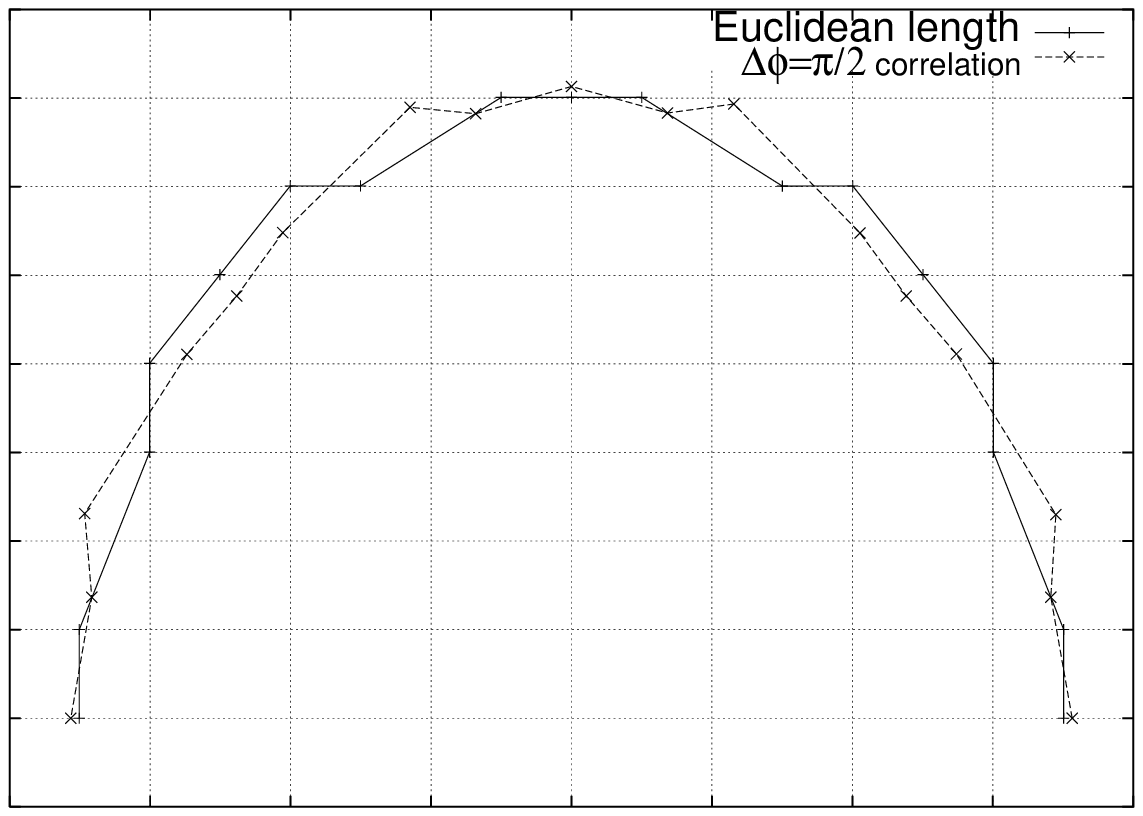,angle=0,width=4.8cm, height=3cm}
    \end{minipage}}
\caption[a]{The Euclidean length and the correlation
 function (\ref{corphi}) in different directions $\phi$, 
for $\Delta\phi=\frac{\pi}{8},~\frac{\pi}{4},~\frac{\pi}{2}$ (from top to
bottom). Left: method I without torelon-suppression. Middle:  method I with 
torelon-suppression. Right: method II.}
\la{cor_standI}
\end{figure}

\underline{Method I}: The average operators have a significantly larger
vacuum expectation value (VEV) in the $\phi=\frac{\pi}{8}$ direction, 
and a smaller VEV in the $\frac{\pi}{4}$ direction. Notice that the data
is symmetric around the $\frac{\pi}{4}$ direction. 
The observed distortions are not due to  winding paths, since the data 
with the torelon suppression implemented shows the same pattern.

\underline{Method II}:  Although there are still variations of the operators'
VEVs along the semicircle, they are of the same order as the geometric
distortions. It must be said that the right angles of the rectangles also
get distorted, so that in  general, the rectangle becomes a parallelogram
at an arbitrary angle $\phi$.

The above comparison provides a first hint as to the rotational
properties of these operators, although it is not as direct as
it might be because the Wilson loops used for method I and method II 
are somewhat different. In 
the case of the segment operator used with method I, there is a 
simple and useful physical interpretation. As we saw earlier, the
superlink $K(x,y)$ is the propagator of a scalar particle
in the fundamental representation of the gauge group in
two Euclidean dimensions. Its tree-level mass  $m_0$ is given 
by Eqn.~(\ref{eqn_alphatom}). Thus the segment operator in
${\cal O}(x,y)$ is the quenched propagator of
`mesons' composed of a scalar and its antiparticle. Such a
point to point propagator includes contributions from 
all allowed energies and from excited as well as ground state 
masses. Since our value of $\alpha$ corresponds to $2am_0 \simeq 0.8$,
the propagator will vary rapidly with distance. In addition, the
short-distance part of the propagator will certainly vary strongly
with the angle $\phi$. Note that if we were to use in the
case of method I
the same rectangular loops as we used for method II, then the physical
interpretation would remain the same, except that the `meson'
wavefunctional would now be smeared, extending over roughly
one lattice spacing, which one would expect to favour the
contribution to the propagator of the lighter intermediate states,
leading to a weaker dependence on distance.

In any case it is clear from the above that if we want
to construct trial wave-functionals with definite rotational
properties, then we should  renormalise the individual operators  
${\cal O}(\phi)$ in such a way that they have exactly the same VEV. 
Doing so we shall now investigate how far we can restore 
rotational invariance by looking at the correlation between 
rescaled segment operators at different angles.
\paragraph{Correlation of rotated operators}
Using the values of the  segment operators calculated above,
we calculate the correlation function
\begin{equation}
\langle \bar{\cal O}(\phi-\frac{\Delta \phi}{2})\bar{\cal O}
(\phi+\frac{\Delta \phi}{2}) \rangle
\label{corphi}
\end{equation}
for a fixed $\Delta \phi$ (the bar indicates that the operators are now 
rescaled so that $\<\bar{\cal O}(\phi)\>=1$). This quantity is plotted 
in the direction $\phi$ in the  polar plots in Fig.~\ref{cor_standI}.
Ideally it should be independent of $\phi$.

\underline{Method I}: without the torelon suppression, we see that for a small
 angle between the operators 
(first graph, $\Delta\phi=\frac{\pi}{8}$), the correlation function has
only small variations, of the order of the errors and 
geometric length and angle variations. However for a larger angle (second 
graph,  $\Delta\phi=\frac{\pi}{4}$), the variations are significantly larger.
   The worst case is the difference between the
 $\Delta\phi=\frac{\pi}{2}$ correlation of two segment operators along 
the lattice main directions or along the diagonals; here there is no
geometric error on $\Delta\phi$ and hardly any ($\sim 1\%$) on the length:
yet the two correlation functions differ by a factor 2.5.
This can be easily understood in terms of paths winding around the torus. 
Once the latter are suppressed (middle column in Fig.~\ref{cor_standI}), 
the rotational invariance is restored to a good approximation.

\underline{Method II}:
 We notice that the correlation function is independent of $\phi$, at the 
level of a few percent. In particular, the variations are practically
the same at the three values $\Delta\phi=\frac{\pi}{8},~\frac{\pi}{4},~
\frac{\pi}{2}$ and are of the same order as the geometric distortions.
\paragraph{Conclusion}
The fact that we observe correlations between Wilson loops that are 
approximately independent of their orientation with respect to the lattice
axes constitutes evidence for a dynamical restoration of rotational invariance.
From these correlation functions of operators 
constructed with two different algorithms, 
we conclude that both methods are suitable
to calculate operators rotated by angles smaller than $\frac{\pi}{2}$. 
Since method II is very much faster
than method I, we shall use the former in practical calculations. 
\section{High spin states on the lattice?}
We now suppose that we have a reliable way of constructing operators  
rotated by angles of the type $\frac{2\pi}{n}$ at our disposal.
How can we use this tool to resolve the spin quantum numbers of the physical
states in the  continuum limit?
%
%
\subsection{Lattice vs. continuum symmetry group}
%
The symmetry group of the square lattice contains two 
 rotations by $\frac{\pi}{2}$, one rotation
by $\pi$ and two types of symmetry axes  
($x$ and $y$ axes, and $y=\pm x$ axes). For convenience, the character table
of  this group is given in Appendix~\ref{ap:group}. 
There are four one-dimensional 
representations, plus one two-dimensional representation. 

The continuum rotation group only has one-dimensional irreducible 
representations (IRs), due to
the commutativity of rotations in the plane. However, because parity does not
commute with rotations (see Appendix~\ref{ap:group}), 
we may also want to consider two-dimensional representations, 
which are irreducible under the full symmetry group (rotations + parity). 

The continuum two-dimensional representations are in general reducible 
with respect to the square group and  can be decomposed into 
the irreducible representations of the square. For instance,
the spin $4^\pm$ representation $D_4$ decomposes into two one-dimensional 
irreducible representations of the lattice group: 
\begin{equation}
D_4=A_1\oplus A_2
\end{equation} 
This tells us in what lattice IRs to look in order to extract information
on the $D_4$ states. 

We shall make use of the notation
\ba
|~\{m\}~\> &=& \frac{1}{\sqrt{2}}\left(~|m\> ~+~ |-m\>~\right) \nonumber \\
|~[m]~\> &=& \frac{1}{\sqrt{2}}\left(~|m\> ~-~ |-m\>~\right).
\ea
The phases are chosen such that $\<\phi | \pm m\> = e^{\pm im\phi}$, where 
the $\phi=0$ direction coincides with one of the lattice axes.

The most general state belonging to $A_1$ can be written as a linear combination
\begin{equation}
|\psi^{(A_1)}\> = \sum_{m=0,4,8,\dots}~ |\psi^{(A_1)}_m \> 
~|\{m\}\> 
\label{eq:se}
\end{equation} 
and correspondingly  for $A_2$
\begin{equation}
|\psi^{(A_2)}\>=\sum_{m=0,4,8,\dots}~ |\psi^{(A_2)}_m \>~ |[m]\>. 
\end{equation} 
The same notation is used for different Hilbert spaces; however 
it should be clear from the context which one is meant. 
The $|\psi^{(A_{1,2})}_m \>$ are vectors of a Hilbert space describing 
for instance the `radial' part of the wave function; their norm 
 represents the quantum mechanical amplitude for an $A_{1,2}$ state to be found
 with a  definite spin $m$. We introduce the notation
\be
c_m = ||\psi^{(A_1)}_m||^2 =  \<\psi^{(A_1)}_m | \psi^{(A_1)}_m\>\la{eq:c_m}
\ee
and $c_m'$ correspondingly for $A_2$.
As we evolve from a small lattice spacing $a\ll \xi\equiv 1/\sqrt{\sigma}$
to coarser and coarser lattices, we imagine the following scenario in terms
of the coefficients $c_m$:
\bi
\item close to the continuum, for any particular Hamiltonian eigenstate
$\psi^{(A_1)}$, 
the $\{c_m\}$ are close to $\delta_{mn}$ for some $n$. If for instance $n=4$, 
these states  `remember' that their wave function changes sign under 
approximate $\frac{\pi}{4}$ rotations that are available on the lattice 
at length scales much greater than $a$. Moreover, the state in $A_2$ with 
$c_m'\simeq \delta_{nm}$ is almost degenerate with  $\psi^{(A_1)}$.

\item as we move away from the continuum, the sharp 
dominance of one particular $c_m$ in the series becomes smoother, and we can 
think of the angular wave functions as having a `fundamental mode' $m_f$, plus
some fluctuations due to `higher modes'. It is as if we started with a sound
of pure frequency, and the effect of the lattice is to add
 contributions from higher harmonics, giving the sound a richer timbre.
 The degeneracies between the states in $A_1$ and $A_2$ 
are broken more and more badly. This is due to the non-equivalence of the
two classes of parity transformations available on the lattice.

\item in general, more and more terms contribute to the series in 
Eqn.~(\ref{eq:se}). 
Thus it seems that the angular dependence of a general state 
in $A_1$ or $A_2$ becomes very intricate. 
 However as $a\rightarrow \xi$, higher modes in the expansion  must 
become irrelevant, because there are no lattice points to support
 their fluctuations on the length scale of the theory.
 We know from the strong coupling expansion 
that the lowest lying states have a simple behaviour as $\beta\rightarrow 0$:
the wave function of the fundamental state is simply a plaquette.
\ei

In fact, we have ignored a possible complication. We have assumed that
 no phase or roughening transition occurs, and that 
we can define smooth trajectories of the states in an energy vs. $a$ plane. 
However, in general we must expect crossings of states to occur. 
For any given range of energies, $E\le E_0$,
there will exist a lattice spacing $a_0$ such that for $a<a_0$, 
there are no more crossings until the continuum is reached\footnote{There is 
a possible exception to that: we know 
that states come as parity doublets, which means that pairs of trajectories
must converge and could possibly cross many times in doing so. This is not
a problem for the present discussion.}. At $a_0$, 
the states  represent the continuum spectrum faithfully, with only
small numerical deviations on their energies. We now follow the trajectory of
one particular state as the lattice spacing is increased. 
Suppose we  meet another trajectory at $a=a_1$. At that particular 
lattice spacing, there will seem to be an `accidental' degeneracy. 
Nearly-degenerate states will mix with the mixing driven by the matrix
element of the lattice Hamiltonian between the `unperturbed' 
eigenstates, i.e. $\langle 1 | H(a) | 2\rangle$. Near the continuum
limit the unperturbed states will be close to continuum spin
eigenstates, $H(a)$ will be close to the continuum Hamiltonian and
so the mixing parameter $\langle 1 | H(a) | 2\rangle$ will be close to
zero. Nonetheless sufficiently close to the crossing, the states
will mix completely and so will the angular Fourier components of the
state. That is to say, the Fourier components need not have a simple
 behaviour with $a$ as $a\to 0$, and care must be taken to
identify any near-degeneracies in following the Fourier components
toward the continuum. 
\subsection{Two strategies}
\label{subsec_twostrat}
With these ideas in mind on the evolution of the rotational properties of the 
physical states as functions of $a$, 
at least two (related) strategies are available in order to extract the
high spin continuum spectrum numerically:

{\bf I.~} If we can afford to work close to the continuum, we can 
construct operators with an approximate continuum wave function 
$e^{i J \phi}$, using the operator construction technique presented earlier.
Ideally this kind of operator belongs to one of the 
irreducible representations of 
the lattice symmetry group. But because the perturbation of such an operator 
off  the continuum wave function is different from that of the Hamiltonian 
eigenstates, 
the expected behaviour of the local-effective-mass
 $am_{\rm eff}(t+\frac{a}{2})=\log\frac{C(t)}{C(t+a)}$ in the correlation 
function is the following: we should see an almost-flat plateau 
(corresponding 
to the excited state of the lattice IR that will evolve into a high spin
state in the continuum limit), followed by a breakdown into another, flat and
stable plateau (corresponding to the fundamental state in the given lattice
IR).

{\bf II.~} First we construct a set of lattice IR operators, 
$\{W^{(0)}_i\}_{i=1}^{N}$. Next we construct (approximate) rotated copies of these. 
We thus have a large basis of operators, $W^{(\phi)}_i$, $\phi$ 
labelling the rotation.
We  diagonalise (using the variational method~\cite{Luscher:1990ck}) 
the  correlation matrix  of  $\{W^{(0)}_i\}_{i=1}^{N}$  in order to 
extract the energy eigenstates  in this lattice IR. 
These states $\psi^{(0)}_i$
are encoded by their components in the original basis $W^{(0)}_i$:
\begin{equation}
 \psi^{(0)}_i = \sum_j v_{ij} W^{(0)}_j~. 
\end{equation} 
 Now we need to determine the angular wave function of these glueball states. 
We do so by building the linear combinations
\begin{equation}
\psi^{(\phi)}_i  = \sum_j v_{ij} W^{(\phi)}_j,\quad \forall \phi
\end{equation} 
and looking at the correlation function
\begin{equation}
G_i(t;\phi,\phi')~\equiv~
 \< ~\psi^{(\phi)}_i(0)~  \psi^{(\phi')}_i(t)~ \>~. 
\label{eqn_cwfprobe}
\end{equation} 
If the `rotated copies' are faithful, $G_i(t;\phi,\phi')$ depends only on
$|\phi-\phi'|$; this provides a useful test for the restoration of rotational 
invariance. The time-separation $t$ is chosen so that the local effective
mass of the $ \psi^{(0)}_i$ two-point function has reached a plateau.

In the case of an $A_1$ state~(\ref{eq:se}), we expect to see
\be
G(t;\phi)\equiv G(t;\phi,0) ~\sim~ \sum_{m=0,4,8,\dots} ~ c_m \cos{m\phi},
\ee
with $c_m$ related to the radial wave function of the state 
through~(\ref{eq:c_m}).
If we are reasonably close to the continuum, we should observe
a  behaviour of this correlation function corresponding to an 
approximate continuum wave function, 
i.e. $G(t;\phi)\propto \cos{m (\phi)}$ for some $m$, 
with small contributions from other modes\footnote{As
 remarked above, care has to be taken near any level crossings.}
(cf. Eqn.~\ref{eq:se}).
In Chapter~\ref{ch:regge_2d}
a generalisation of this method will be discussed where the operators
used in Eqn.~(\ref{eqn_cwfprobe}) are based on different loops.

We note that the data needed for both analyses is the same, so that 
they can easily be used in parallel. 
The second method has the advantage that there is no need to restrict
ourselves to $\frac{2\pi}{n}$-type angles in order to project out  states
corresponding to unwanted spins.  On the other hand, if high spin states
are very heavy, a large number $N$ of trial operators will be needed in order
that the variational method can resolve them. A simple case of 
this method, that does not employ the variational method and allows one
to determine the mass and quantum numbers of the lowest-lying 
state in a given lattice IR, consists in measuring the correlation matrix 
of one operator $W$ with its rotated copies $\{W^{(\phi)}\}$ at sufficiently large
 Euclidean-time separation  so that the local 
effective mass has reached a plateau.
\section{Applications of Strategy I}
%
As remarked at the beginning of this chapter, 
a robust prediction \cite{Karl:1999sz, Johnson:2000qz}
of the Isgur-Paton flux tube model \cite{Isgur:1985bm} is that the lightest
$0^-$ state should have a much larger mass than  that 
obtained in lattice calculations~\cite{Teper:1998te} 
for the lightest SU(2) glueball in the $A_2$ lattice 
representation. Moreover the latter mass is close to that 
of the spin 4 glueball as predicted by the flux tube model
\cite{Karl:1999sz, Johnson:2000qz}. Since
the $A_2$ lattice representation contains the continuum
$0^-,~4^-,\dots$ states this has led to the conjecture
 \cite{Karl:1999sz, Johnson:2000qz}   that the lightest $A_2$ state is in
fact $4^-$ rather than  $0^-$.
All this has motivated some lattice calculations 
\cite{Johnson:2002qt} which suggest that it is in fact so. 

In this section we shall use the first of the two strategies outlined 
in Section~\ref{subsec_twostrat} 
to address this question in some detail.
We shall begin with a simple approach applied directly
to states in the $A_2$ representation --- the conclusions of which
will be confirmed in a quantitatively controllable way
when we apply  `strategy II'. 
We then return to the more difficult question of how one isolates
the $4^+$ from the $0^+$ in the $A_1$ representation. We provide 
a procedure that appears to work well. Because of parity doubling
for $J\not= 0$ this provides another way to calculate
the spin 4 glueball mass. And indeed the $4^+$ and $4^-$ masses we
obtain are entirely compatible. After checking that
finite volume corrections for the higher spin states are 
under control, we perform calculations for several
larger $\beta$ values and  extrapolate our mass ratios to
the continuum limit.
%
\subsection{The $0^-~/~4^-$ puzzle\label{04m}}
To distinguish the  $0^-$ from the $4^-$ glueball we construct 
trial $0^-$ and $4^-$ wavefunctionals using suitable 
linear combinations of (approximate) rotated copies of asymmetric 
operators, as described earlier. 
The calculations in this subsection are performed at $\beta=6$ on a 
$16^3$ lattice. The links are smeared before the paths joining lattice 
sites are constructed.
The three operators on the left of Fig.~\ref{sh1} are rotated 
by $\frac{\pi}{6}$ angles.
Those on the right, which are only rotated by  $\frac{\pi}{2}$ angles,
are of a type that has been used previously~\cite{Teper:1998te}
to measure the lightest state in the $A_2$ representation.
\begin{figure}[ht]
\vspace{-1cm}
~~~~~~~~~~~~\centerline{
\begin{minipage}[c]{6.5cm}
\begin{picture}(100,160)(10,20)
\multiput(0,0)(0,10){16}{\line(1,0){150}}
\multiput(0,0)(10,0){16}{\line(0,1){150}}
{\thicklines
\put(0,0){\line(1,0){60}}
\put(60,0){\line(5,3){50}}
\put(110,30){\line(-4,1){40}}
\put(0,0){\line(5,3){70}}}
\put(70,40){\circle*{4}}
\put(0,0){\circle*{4}}
\put(60,0){\circle*{4}}
\put(110,30){\circle*{4}}
{\thicklines
\put(0,50){\line(1,0){60}}
\put(60,50){\line(1,4){10}}
\put(0,50){\line(5,3){70}}}
\put(0,50){\circle*{4}}
\put(60,50){\circle*{4}}
\put(70,90){\circle*{4}}
{\thicklines
\put(0,100){\line(1,0){120}}
\put(120,100){\line(1,4){10}}
\put(130,140){\line(-1,0){60}}
\put(0,100){\line(5,3){70}}}
\put(0,100){\circle*{4}}
\put(120,100){\circle*{4}}
\put(130,140){\circle*{4}}
\put(70,140){\circle*{4}}
\end{picture}
\end{minipage}
\begin{minipage}[c]{6.5cm}
\begin{picture}(100,160)(10,20)
\multiput(0,0)(0,10){16}{\line(1,0){150}}
\multiput(0,0)(10,0){16}{\line(0,1){150}}
\thicklines{\put(0,0){\line(1,0){60}}
\put(60,0){\line(0,1){40}}
\put(60,40){\line(-1,0){10}}
\put(50,40){\line(0,-1){20}}
\put(50,20){\line(-1,0){50}}
\put(0,20){\line(0,-1){20}}}
\put(0,0){\circle*{4}}
\put(60,0){\circle*{4}}
\put(60,40){\circle*{4}}
\put(50,40){\circle*{4}}
\put(50,20){\circle*{4}}
\put(0,20){\circle*{4}}
\thicklines{
\put(60,70){\line(1,0){60}}
\put(120,70){\line(0,1){40}}
\put(120,110){\line(-1,0){30}}
\put(90,110){\line(0,-1){20}}
\put(90,90){\line(-1,0){30}}
\put(60,90){\line(0,-1){20}}}
\put(60,70){\circle*{4}}
\put(120,70){\circle*{4}}
\put(120,110){\circle*{4}}
\put(90,110){\circle*{4}}
\put(90,90){\circle*{4}}
\put(60,90){\circle*{4}}
\thicklines{
\put(90,0){\line(1,0){50}}
\put(140,0){\line(0,1){30}}
\put(140,30){\line(-1,0){20}}
\put(120,30){\line(0,-1){20}}
\put(120,10){\line(-1,0){30}}
\put(90,10){\line(0,-1){10}}}
\put(90,0){\circle*{4}}
\put(140,0){\circle*{4}}
\put(140,30){\circle*{4}}
\put(120,30){\circle*{4}}
\put(120,10){\circle*{4}}
\put(90,10){\circle*{4}}
\end{picture}
\end{minipage}
}
\vspace*{1cm}
\caption{On the left, operators used to construct a $4^-$ wave function:
(I) trapeze  (II) triangle (III) asymmetric.
On the right, a `conventional' set of operators: (1) bottom left (2) top (3)
bottom right.}
\la{sh1}
\end{figure}
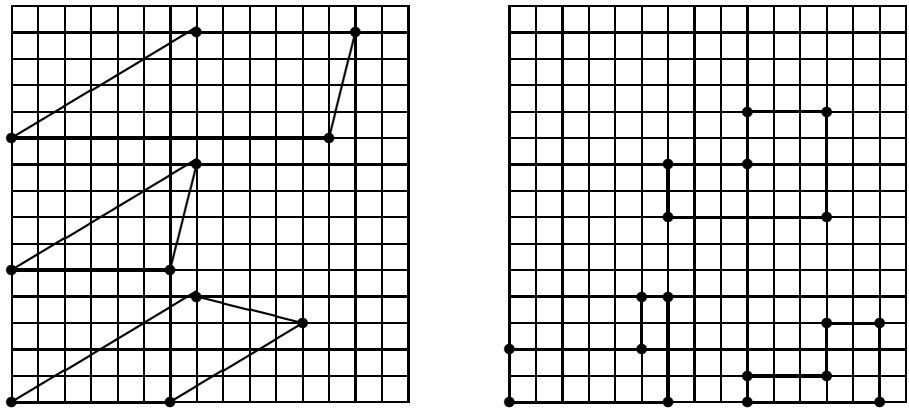
With the former three operators and their rotations, we form 
linear combinations that
correspond to trial $4^-$ and $0^-$ operators, while with the latter
three operators we construct the $A_2$ representation of the square group.
The overlaps\footnote{Errors are $\sim1\%$, 
but shape II produced a very noisy $0^-$ operator.}
between these two sets of operators are given 
in Tab.~\ref{tab:a2_overl}. 
\begin{table}[htb]
\begin{center}
\begin{tabular}{|c|c|c|c|}
\hline
overlap & 1($A_2$)  & 2($A_2$) & 3($A_2$) \\
\hline
I($4^-$) & 0.89 &  0.88 &  0.73\\
II($4^-$) & 0.95 & 0.96 &  0.97 \\
III($4^-$) &  0.38 &  0.38 & 0.24\\
\hline
I($0^-$) &  0.24 &  0.20 &  0.15 \\
II($0^-$) & - & - & -\\
III($0^-$) &  0.16 &  0.13 &  $<0.01$\\
\hline
\end{tabular}
\end{center}
\caption{Overlaps $\frac{\langle {\cal O}_1 {\cal O}_2\rangle
  }{\left(\langle{\cal O}_2
 {\cal O}_2\rangle \langle {\cal O}_1{\cal O}_1 \rangle \right)^
{\frac{1}{2}}}$ between $A_2$ and trial $4^-$ and $0^-$ operators.}
\la{tab:a2_overl}
\end{table}
The rows correspond to the three `$4^-$' and `$0^-$' operators,
while the columns refer to the  $A_2$ operators based 
on the three loops on the right of Fig.~\ref{sh1}.
Clearly, the approximate $4^-$ operators overlap much more onto
the operators of the $A_2$ representation.
Performing a variational analysis of the correlation matrix of the 
first set of operators, we obtain effective masses at one lattice
spacing of $am(4^-)=2.556(68)$ and $am(0^-)=  3.34(31)$. 
We find similar values for these masses with other sets of 
operators~\cite{Meyer:2002mk}.
Thus we have  strong indications
 that the labelling in \cite{Teper:1998te} of the lightest  
$A_2$ glueball as $0^-$ was mistaken, and that it is in fact a $4^-$.

%
\subsection{A recipe for data analysis\label{recipe}}
%
We now return to the problem of distinguishing $4^+$ and $0^+$  
states in the $A_1$ representation. Since the  $0^+$ is much
lighter than the $4^+$ there is the danger that what we will 
claim to be a  $4^+$ will in fact be an excited  $0^+$.
Indeed, if the rotated loops are only approximate copies of each other, 
it is quite possible that the cancellations induced by the oscillating
coefficients induce not only a piece of the wavefunction that
has the desired angular oscillations, but also a piece
where the cancellations, and resulting oscillations, are in
the radial rather than angular direction. The latter can
project onto an excited\footnote{Since one expects the ground
state radial wavefunction to be smooth, a significant
overlap onto the ground state  $0^+$ would be unexpected.} $0^+$.
Now, since the lightest $0^+$ is
very much lighter than the lightest $4^+$, even such an
excited $0^+$ may be lighter than the lightest $4^+$
-- as turns out to be the case here -- and may undermine a 
variational calculation. As $a\to 0$ and the rotated
loops become better copies of each other,
the radial cancellations become more extreme, the $0^+$
states being projected upon become more highly excited and
more massive, and once they become more massive than the
lightest $4^+$ states of interest the problem disappears for
all practical purposes. 
Thus one needs to perform enough checks to avoid being misled. 
This leads to a rather involved procedure which we describe in
the context of a calculation at $\beta=6$. 
To construct our trial states of spin 0 and 4 we use the four
operators in Fig. (\ref{sh4}, left) together with their twelve rotations.
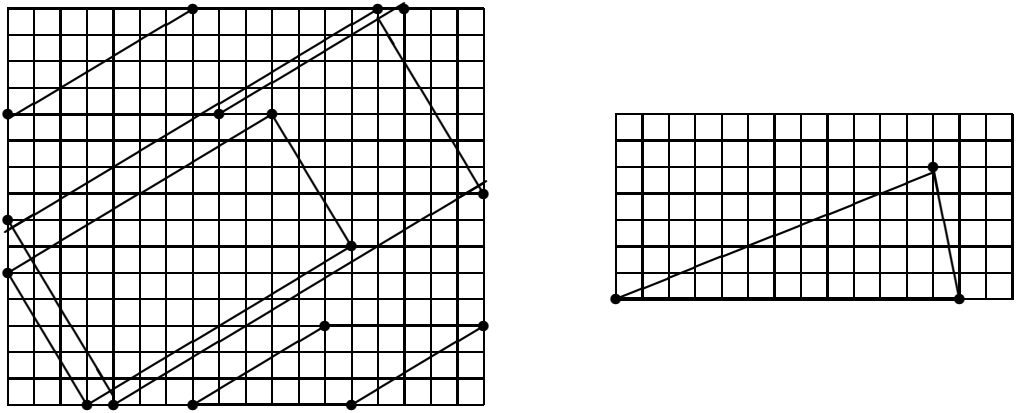
\begin{figure}[ht]
\vspace{-1.2cm}
\centerline{
\begin{minipage}[c]{14cm}~~~~
\begin{minipage}[c]{8.0cm}
\begin{picture}(100,160)(10,20)
\multiput(0,0)(0,10){16}{\line(1,0){180}}
\multiput(0,0)(10,0){19}{\line(0,1){150}}
{\thicklines
\put(30,0){\line(5,3){100}}
\put(130,60){\line(-3,5){30}}
\put(100,110){\line(-5,-3){100}}
\put(0,50){\line(3,-5){30}}}
\put(30,0){\circle*{4}}
\put(130,60){\circle*{4}}
\put(100,110){\circle*{4}}
\put(0,50){\circle*{4}}
{\thicklines
\put(70,0){\line(1,0){60}}
\put(130,0){\line(5,3){50}}
\put(180,30){\line(-1,0){60}}
\put(120,30){\line(-5,-3){50}}}
\put(70,0){\circle*{4}}
\put(130,0){\circle*{4}}
\put(180,30){\circle*{4}}
\put(120,30){\circle*{4}}
{\thicklines
\put(40,0){\line(5,3){141}}
\put(180,80){\line(-3,5){40}}
\put(140,150){\line(-5,-3){141}}
\put(0,70){\line(3,-5){40}}}
\put(40,0){\circle*{4}}
\put(140,150){\circle*{4}}
\put(180,80){\circle*{4}}
\put(0,70){\circle*{4}}
{\thicklines
\put(0,110){\line(1,0){80}}
\put(80,110){\line(5,3){70}}
\put(150,150){\line(-1,0){80}}
\put(70,150){\line(-5,-3){70}}}
\put(0,110){\circle*{4}}
\put(80,110){\circle*{4}}
\put(150,150){\circle*{4}}
\put(70,150){\circle*{4}}
\end{picture}
\end{minipage}
\begin{minipage}[c]{5.5cm}
\begin{picture}(100,80)(10,20)
\multiput(0,0)(0,10){8}{\line(1,0){150}}
\multiput(0,0)(10,0){16}{\line(0,1){70}}
\thicklines{
\put(0,0){\line(1,0){130}}
\put(130,0){\line(-1,5){10}}
\put(0,0){\line(5,2){120}}
}
\put(130,0){\circle*{4}}
\put(0,0){\circle*{4}}
\put(120,50){\circle*{4}}
\end{picture}
\end{minipage}
\end{minipage}}
\vspace*{1cm}
\caption{Left: the operators used for the determination of the even spin 
spectrum at $\beta=6$:
(1) small rectangle (2) large rectangle  (3) small parallelogram (4) large 
 parallelogram. Right: the triangular operator  used
to determine  the lightest $A_2$ state wave function at $\beta=12$.}
\la{sh4}
\end{figure}
Suppose that we have obtained the correlation of each operator
in any orientation with any operator in any orientation. 
Our experience has shown the following procedure to be reliable
at $\beta=6$. 
\begin{description}
\item[Preselection of the operators.-] 
Look at the correlation matrix of each
individual `shape' with its rotated copies. 
This $n\times n$ matrix should (ideally) be a symmetric Toeplitz 
one\footnote{The entries $M_{ij}$ should only depend on $|i-j|$.}, 
with the additional cyclic property $M_{i,j}=M_{n+2-i,j}$, $i=j+1,\dots,n$,
reflecting the fact that the correlation between angle $0$ and 
$\frac{2\pi}{n}$ is the same as between $-\frac{2\pi}{n}$ and $0$. 
 In the present case $n=12$ and the correlation matrices are given 
in Table~\ref{tab:cor_mat}.
Notice that the second is worse than the others; we therefore discard it and 
only operators 1, 3 and 4 remain in the subsequent steps.
\begin{table}[htb]
\begin{center}
\begin{tabular}{|c@{$\quad$}c@{$\quad$}c@{$\quad$}c@{$\quad$}c@{$\quad$}c|c@{$\quad$}c@{$\quad$}c@{$\quad$}c@{$\quad$}c@{$\quad$}c|}
   \hline
	 1.00 &&&&& &1.00 &&&&&\\
 
    0.52 &   1.00&&&& &  0.22&    1.00&&&&\\
 
    0.24  &  0.52  &  1.00&&& & 0.26 &   0.22  &  1.00&&&\\

    0.18 &   0.22 &   0.62  &  1.00&& &0.03 &   0.01  &  0.16 &   1.00&&\\
    0.22  &  0.13 &   0.22 &   0.52  &  1.00& &0.01 &   0.00  &  0.01  &  0.21 &   1.00&\\
 
    0.62 &   0.22  &  0.18 &   0.24 &   0.52  &  1.00 & 0.16 &   0.01 &   0.03 &   0.26  & 
 0.22 &   1.00\\
\hline
 1.00&&&&& &  1.00&&&&&\\
 
    0.84  &  1.00 &&&& &  0.65 &   1.00&&&&\\
 
    0.73  &  0.86 &   1.00&&& & 0.47  &  0.69 &   1.00&&&\\
 
    0.60  &  0.65 &   0.87  &  1.00&& & 0.37 &   0.42  &  0.69 &   1.00&&\\
 
    0.65  &  0.60 &   0.73  &  0.84  &  1.00& &0.42  &  0.37  &  0.48  &  0.65  &  1.00&\\
 
    0.87  &  0.73  &  0.76  &  0.73  &  0.86  &  1.00 & 0.69  &  0.48  &  0.49  &  0.48 &  
 0.69  &  1.00\\
\hline
 \end{tabular}
\end{center}
\caption{Correlation matrices of the operators depicted on Fig.~(\ref{sh4}, left).
Op. 1 is top left, 2 top right, 3 bottom left and 4
bottom right; for symmetry reasons only the upper $6\times6$ block is shown.}
\label{tab:cor_mat}
\end{table}
\item[Selection of the operators.-] We now concentrate on the linear 
combinations corresponding to  `spin $J_1$'
and `spin $J_2$' operators ($J_1=0$ and $J_2=4$ being the case
of interest here). Each of the spin $J_1$ operators has a certain overlap
with each of the spin $J_2$ operators. This is due partly 
to the imperfect rotations, and partly because the lattice
Hamiltonian eigenstates do not diagonalise the spin operator.
First look at the diagonal quantities, that is, the overlap  
of the $J_1$ and $J_2$ operators constructed with the same shape. 
Eliminate those which have significant  overlap.
Once bad operators are eliminated, this whole set
of overlaps should contain none larger than ${\cal O} (a)$. If this cannot be 
achieved, it either means that we are too far from the continuum ---
in the sense that the wave functions of the
physical states are very different from the continuum ones --- or that our
rotated operators are not sufficiently faithful copies of the initial
ones.  In the present case, we find the following overlaps for our three
candidate $J_2$ operators onto the corresponding $J_1$ operators: 
(0.04,0.0,0.12), (0.13,0.20,0.07) and (0.09,0.05,0.10) respectively
for 1, 3 and 4, and so we retain these operators for the  subsequent 
calculation.
\item[Diagonalisation of the $J_1$  operators.-] We now
diagonalise the remaining $n_{\rm sel}$ $J_1$ operators using the
 variational procedure. To decide how many of the orthogonal states one should
keep, the following criteria can be applied.
From the comparison of  the components
of each  linear combinations to the quantity
$\chi\equiv(\mathrm{det}O)^{1/n_{\rm sel}}$, where $O$ is the 
transformation matrix leading to the orthogonal operators with unity 
equal-time correlator, only keep those lightest states whose components 
are not significantly larger than this determinant\footnote{
If the determinant itself is large, 
try removing shapes that could be too similar to one another
and diagonalise again.}. In practice, linear combinations with large
components are found to have a very poor signal. 
 In the present case, the coordinates in units of $\chi$ read 
(0.090, 0.24, 0.37), (0.37, -1.1, 1.1) and  (-0.93, -1.19, 1.7); we keep all
three states. 
\item[An intermediate check.-]
Look at the overlaps $\langle {\cal O}_{J_2} {\cal O}^D_{J_1}\rangle$. 
Here the ${\cal O}^D_{J_1}$ are the operators obtained from the variational
procedure, and correspond to our best estimates of the  
 lightest $J_1$ glueball wavefunctionals. The operators
${\cal O}_{J_2}$ are the original un-diagonalised 
$J_2$ loop combinations. We require that the total overlaps,
which can now be calculated as  
$\left(\sum_i \langle{\cal O}_{J_2} 
{\cal O}^{(D)i}_{J_1}\rangle^2\right)^{1/2},$
should be less than ${\cal O}(a)$.  The overlaps in the present case are 
(0.07,0.21,0.26), (0.14,-0.21,-0.38) and (0.088, 0.10, 0.039).
Therefore we only keep the last operator as a candidate $J_2$
operator.

\item[Diagonalisation of the $J_2$ operators.-]
Diagonalise the $J_2$ operators. In the present case the operation is trivial
since we are left with only one operator.

\item[Final check.-] Now consider the same overlaps as above
but with both $J_1$ and $J_2$ operators being diagonalised ones. 
The total overlaps between these final $J_1$ and $J_2$ operators 
is required to be still less than ${\cal O}(a)$.
Here we obtain a total overlap of $14\%$.
\end{description}
\subsection{Results}
We give our results in terms of effective masses at 1 and 2 lattice spacings
(see Table~\ref{sIdata} at the end of this chapter). 
The `quality' of an operator is defined as
$|\langle n | \hat{\cal O} | \Omega \rangle |^2$, where $n$ is the state being 
measured and $\hat{\cal O}$ is our  operator. In calculating this quantity
we assume that the bold-faced mass values in the tables represent the
corresponding mass plateaux. The `overlap' represents the overlap
(as defined in Table~\ref{tab:a2_overl}) between the state of interest
and each of the other spin eigenstates lying in the 
same square irreducible representation (e.g. spin 4 with the
various spin 0 states obtained through the variational procedure)
and adds these in quadrature. Thus it provides a measure of the
overlap of the wavefunctional onto the basis of states of the `wrong'
spin. 

We note that our operator construction method leads to reasonably good
 overlaps onto the physical states, 
while the overlaps of operators with different quantum
numbers typically remain well under the $10\%$ level. This quality requirement
 is obviously dependent on the spectrum itself: if there is a large gap 
between the spin $0$ and the spin $4$, the spin 4 operator will have to be 
of exceptionally high purity, since the heavier state contribution to the 
correlator, even with a much larger projection onto the operator, 
 becomes negligible with respect to the lighter state at large 
time-separations.

In physical units the size of our standard spatial volume is 
$L\simeq 4/\sqrt{\sigma}$. This is the same size that was used in 
\cite{Teper:1998te}, where the choice was motivated by an explicit 
finite volume study, showing this to be about the 
smallest volume on which the lightest states in the $A_1$ and $A_3$ 
lattice irreducible representations  did not suffer significant finite 
volume corrections. Since we would expect  the size
of a glueball to grow with its spin $J$ for large enough $J$,
it is important to check that our volumes are indeed large enough to 
accommodate a glueball with $J=4$. 
We therefore carried out an explicit check of finite volume
effects at $\beta=9$ ($1/\sqrt{\sigma}\simeq 6a$), 
for $\hat L=16,~24,~32$ (see Table~\ref{vdep_tab}).  
We see  no statistically significant
 correction to the $J=4$ glueball mass~\cite{Meyer:2002mk}, 
although the spin 2 glueball becomes slightly lighter on the smallest volume.
This particular variation is presumably due to the presence of states composed
of pairs of flux tubes that wind around the spatial torus~\cite{Teper:1998te}.
Following the same procedure as we used for our $\beta=6$ calculation
in section \ref{recipe}, we perform further  mass calculations
at $\beta=9,~12~$ and 14.5 on lattices of the same physical size. 
The  results of our calculations are presented in Table \ref{sIdata}. 
We now express the glueball masses in units of the string tension,
$am/a\sqrt{\sigma} \equiv m/\sqrt{\sigma}$, using values of
the string tension,  obtained at the same values of
$\beta$ in~\cite{Teper:1998te}. 
The leading lattice correction should be ${\cal O}(a^2)$ and
so if we plot our values of $m/\sqrt{\sigma}$ against $a^2\sigma$
  we can extrapolate linearly to $a=0$ 
for sufficiently small $a$. Such extrapolations are shown in
Fig.~\ref{extrapol} and the resulting spectrum is given next to it.
%
\begin{figure}[tb]
\centerline{\begin{minipage}[c]{14cm}
\begin{minipage}[c]{5cm}
\begin{tabular}{|c|c|c|}
\hline
State & $m/\sqrt{\sigma}$ & C.L.$[\%]$ \\
\hline
\hline
$0^+$ & 4.76(11) &  99 \\
\hline
$0^{+*}$ & 6.88(23) & 14 \\
\hline
$2^+$    & 8.44(24)& 99  \\
$2^-$ &    8.81(34) & 90\\
\hline
$4^+$ &10.28(81) & 94 \\  
$4^-$ &10.70(93) & 97 \\   
\hline
\hline
\end{tabular}
\end{minipage}
\begin{minipage}[c]{9.0cm}
\psfig{file=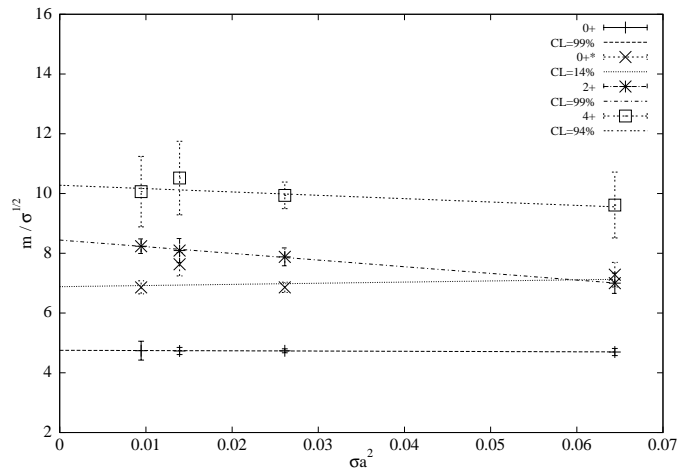,angle=0,width=9.0cm}
 \end{minipage}
 \end{minipage}}
\caption[a]{Strategy I: continuum extrapolation of the $J^P=(2n)^+$ masses.
The corresponding confidence levels (C.L.) are also given.}
\la{extrapol}
\end{figure}
We note that we obtain the  parity-degeneracies expected in the continuum 
limit for the $J=2$ and $J=4$ spins that we consider. We also observe
that the corrections to the $4^+$ mass are reasonably well
fitted by a simple $a^2$ term for $\beta \geq 9$, just as they are
for the  $0^+$ and $2^+$.
%
\section{Applications of Strategy II}
%
In this section we apply the second strategy introduced
in Section~\ref{subsec_twostrat}, in which we probe glueball
wavefunctions with  rotated operators so as to directly extract the
coefficients of the Fourier modes contributing to the angular 
variation of those wavefunctions.

As an illustration we first apply the method to a case where we 
believe we know the answer, namely the lightest states in the $A_1$ 
and $A_3$ representations. We confirm that these states are
indeed $J=0$ and $J=2$ respectively. We then return to the
$0^-~/~4^-$ puzzle. We establish that the lightest state in the 
$A_2$ representation is indeed spin 4 and that this is much lighter
than the $0^-$ ground state. The evidence is more convincing
than before not only because of the greater transparency of this
approach, but also because we repeat the calculation closer
to the continuum limit. We then go on to investigate
the angular behaviour of the lightest states falling in the 
two-dimensional $E$ representation which contains all of the 
continuum odd-spin states. Our conclusion will be that
the lightest state is a spin-3-like state, rather than spin 1. 
Finally, we reanalyse the spectrum of 
states in the $A_1$ representation, and perform a continuum 
extrapolation, obtaining results that are consistent with
those obtained with our first method.
%
\subsection{Wave functions of the lightest $A_1$ and $A_3$ states}
%
To analyse the angular content of the lightest states lying in the 
$A_1$ and $A_3$ representations we use the four operators in
Fig.~(\ref{sh4}, left) together with their twelve rotations.
 We first use the exact lattice symmetries
to form operators in each of these lattice representations, and 
then we use the variational method to determine  the linear
combinations of these operators that provide the best approximations 
to the ground state glueball wavefunctionals. We then construct
the same linear combinations of (approximately) the same operators
rotated by different angles. This provides us with rotated
versions of the ground state wavefunctionals. From the correlation 
at (typically) two lattice spacings between the original and rotated 
copies of our ground state wavefunctional, we can extract
the angular variation, as displayed in Fig.~\ref{a1a3}.
We clearly observe the characteristic features of $0^+$ and $2^+$ wave 
functions\footnote{The $A_1$ representation of the second operator varies
more with the angle $\phi$ than the others. We had already noted that 
its correlation matrix was far from being Toeplitz
and the selection criteria in Section~\ref{recipe} had led 
us to remove it from the analysis.}. This provides a simple
illustration of the method in a non-controversial context.
%
\subsection{The $0^-~/~4^-$ puzzle revisited}
%
We now proceed to analyse the angular variation of the wave function
of the ground state $A_2$ glueball on our $16^3$ lattice
at $\beta =6$. We begin with the first set of operators displayed 
in Fig.~(\ref{sh1}, left).  We obtain 
$am_{\rm eff}(a/2)=2.575(71)$,  $|c_0|=0.0525(60)$ and $|c_4|=0.9986(53)$.
The coefficients clearly show that the lightest $A_2$ state wave
function is completely dominated by the spin 4 Fourier component. 
It appears that the $4^-$ ground state is lighter than the $0^-$. 
Of course one needs to check that
this statement is robust against lattice spacing corrections. This we now
do by performing a calculation at  $\beta=12$ on a $32^3$ lattice. 

In principle we could proceed as before: constructing the 
$A_2$ square representation and looking at the two-lattice-spacing 
correlations with an operator oriented in different directions. 
However there is an interesting subtlety associated with the 
$0^-$ state and the unfamiliar parity operation which we can 
exploit. 
A linear combination of axis-symmetric operators that
corresponds to the quantum numbers of the $A_2$ representation does not 
couple to the $0^-$ component of the lattice  states. The 
reason is that the image of a symmetric operator under an axis-symmetry
can also be obtained by a rotation, so that the relative minus sign cancels
the contribution of any $0^-$ component. Thus we can 
measure the projection of the wave function onto the space orthogonal to the 
$0^-$ subspace. If the overlap onto the state whose mass we extract with
this operator is not dramatically decreasing  as we approach the continuum 
limit, we can safely conclude that the state has quantum numbers $4^-$.

The kind of operators we used are triangular, 
as drawn in Fig.~(\ref{sh4}, right).
The corresponding wave function shown in Fig.~\ref{a2e}
confirms the last paragraph's conclusions: 
the data points fall perfectly on a $\sin{4x}$ type curve. This result
was obtained with all the operators we employed.
From the fact that our overlaps onto the state at each of 
$\beta=9,12,14.5$  are  better than $90\%$, we confidently
conclude that the state carries quantum numbers $4^-$ in the continuum 
limit. As expected from parity doubling, its mass is consistent  with
the lightest $4^+$ glueball, with a mass of $am(4^-,\beta=12)= 1.365(56)$.
%
%
\begin{figure}[tb]
\centerline{~~\begin{minipage}[c]{14cm}
\begin{minipage}[c]{5cm}
\begin{tabular}{|c|c|c|}
\hline
State & $m/\sqrt{\sigma}$ & C.L.$[\%]$ \\
\hline
\hline
$0^+$ & 4.934(98) &  50 \\
$0^{+*}$ & 7.03(26) & 55 \\
$0^{+**}$ & 7.54(70)  & 36 \\
$2^+$    & 8.65(33)& 38  \\
$4^+$ & 11.6(1.3) & 84 \\
\hline
\hline
\end{tabular}
\end{minipage}
\begin{minipage}[c]{9cm}
 \psfig{file=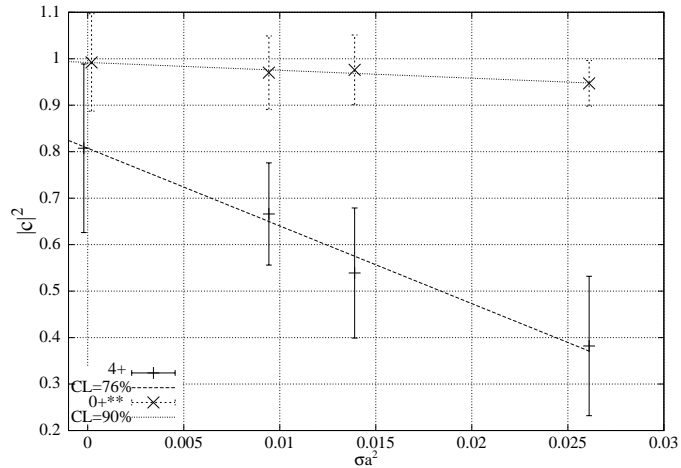,angle=0,width=9cm}
 \end{minipage}\end{minipage}}
\caption[a]{Continuum extrapolation of coefficient $c_0$
 for the $A_1^{**}$ state and coefficient $c_4$ for the  $A_1^{***}$ state. 
The continuum values are 0.996(53) and 0.81(18) respectively. Thus the former
state evolves to a $0^+$, the latter to a $4^+$ state in the continuum limit.}
\la{coeffs}
\end{figure}
\subsection{Wave functions of the lightest $E$ states}
%
Proceeding as above, we extract the  angular wave function of the 
lightest state lying in the $E$ representation at $\beta=12$ 
(Fig.~\ref{a2e}). It is  obtained from correlations 
at two lattice spacings of separation, so as to allow the excited mode
contribution to decay relatively to that of the lightest states.
This two-dimensional 
representation contains the continuum spin 1 and 3 states. 
We clearly observe the characteristic behaviour of a spin 3 wave function
in two orthogonal polarisation states.
The wave functions are well fitted by $\sim\cos{3\phi}$ and $\sim\sin{3\phi}$.
 We conclude that the spin 3 glueball is lighter than the spin 1.
%
%
\subsection{The masses and Fourier coefficients in the continuum limit}
%
Having identified at several values of $\beta$ the states
corresponding to different continuum spins, we  extrapolate
them to the continuum limit in the usual way. The result is
given next to Fig.~\ref{coeffs}. The masses agree with the
calculations performed earlier using our `strategy I'. 
In identifying the lattice states as belonging to particular
continuum spins, we assume that the appropriate Fourier
coefficient  $|c_n|^2$ will extrapolate to unity in the
continuum limit. A check that it is so is provided in
Fig.~\ref{coeffs} where we show an extrapolation of the $4^+$  
component of the $A_1^{***}$ state, as well as the $0^+$ component 
of the $A_1^{**}$ state. Although the error bars are  large, 
we are able to draw  definite conclusions about the 
quantum number of these states in the continuum\footnote{An equally
 good possibility would have been to extrapolate $|c_n|$, which are
also expected to have ${\cal O}(a^2)$ corrections.}.
\section{Conclusion}
To calculate the mass of a glueball of spin $J$ in a lattice
calculation, one must identify the lattice energy eigenstate
that tends to that state in the continuum limit. The limited
rotational invariance on a lattice introduces ambiguities
which means that at a fixed value of $a$ one cannot be
confident in one's spin assignment. Only by performing
a continuum extrapolation, while monitoring the angular
content of the glueball wavefunctional, can one be confident
in the mass one extracts for a high spin glueball.

In practice one needs to identify likely candidates for
such lattice eigenstates and we introduced two related strategies
to do so. The first tries to construct 
wavefunctionals with the required rotational symmetry,
which of course can only be approximate at finite $a$.
The second probes the angular variation of eigenstates obtained 
through a conventional lattice calculation
by examining their transformation properties under approximate rotations.
In either case one needs to be able to easily calculate smeared 
Wilson loops with arbitrary shapes, and we developed methods for
doing so.

To test our methods we applied them to the relatively simple
problem of determining whether the ground state spin $4$ glueball 
is lighter than the ground state $0^-$ glueball in  
D=2+1 gauge theories.  Our calculations
confirmed unambiguously that the former
is indeed much lighter than the latter,
so that the usual identification of the ground state of the 
$A_2$ representation as being $0^-$ is mistaken.
We showed that a degenerate spin $4$ state appears in the
$A_1$ representation among several excited scalar states.

In applying our methods we found that the second strategy
was in practice the more transparent and reliable. 
These preliminary calculations make us confident 
that computing higher spin glueball masses is a practical task. 
\clearpage
\begin{table}
\begin{center}
\begin{tabular}{|c|c|c|c|}
\hline
State & $\hat L=16$ &  $\hat L=24$ &  $\hat L=32$ \\
\hline
\hline
$0^+$ & 0.764(11) & 0.7681(93) & 0.766(27)\\
$0^{+*}$ &1.065(21) & 1.159(28)  & 1.113(48) \\
$2^+$  &  1.194(22) & 1.287(47) & 1.295(62)\\
$4^+$ & 1.620(37) & 1.623(66) & 1.57(13) \\
\hline
\hline
\end{tabular}
\end{center}
\caption{Strategy I: volume dependence of various glueball masses in lattice 
units at $\beta=9$.}
\label{vdep_tab}
\end{table}
\begin{table}
\begin{center}
\begin{tabular}{|c|c|c|c|c|}
\hline {\small
$\mathbf{\beta=6}$} & $am$ ($t=a$)& $am$ ($t=2a$)& quality$[\%]$ & overlap$[\%]$\\ 
\hline
$0^+$ &  1.2309(77)& $\mathbf{ 1.203(27)}$      &    97.2(36)  & 6.3 \\
$0^{+*}$ &1.995(23)  &$\mathbf{ 1.79(12)  }$    &  81.4(12) & 4.8 \\
$2^+$ & 1.998(12) & $\mathbf{1.777(80)}$       & 80.1(74) & 4.1 \\
$2^-$ & 1.947(11) & $\mathbf{1.70(12)}$       & 78(10) & 9.9 \\
$4^+$ &  2.509(24)& $\mathbf{  2.44(27)  }$     &   93(27) & 13.8 \\  
$4^-$ & 2.536(35) & $\mathbf{  2.42(37)}$       &   89(36) & /\\
 \hline 
\end{tabular}\\
\vspace{0.6cm}
\begin{tabular}{|c|c|c|c|c|c|}
\hline
$\mathbf{\beta=9}$ & $am$ ($t=a$)& $am$ ($t=2a$)& $am$ ($t=3a$)&
 quality$[\%]$ & overlap$[\%]$\\
\hline
\hline
$0^+$    &    0.8053(85) &$\mathbf{0.7681(93)}$& 0.739(22)  &96.3(21) & 6.9 \\
$0^{+*}$ &  1.1904(82)&$\mathbf{1.159(28)}$& 0.995(12)     & 96.9(36) & 5.9 \\
$2^+$ & 1.3311(71) &$\mathbf{ 1.287(47) }$&  1.156 (97)    &  95.6(54) & 5.5\\
$2^-$ & 1.410(13) &$\mathbf{ 1.301(49) }$&  1.16 (21)    &  89.7(58) & 9.0\\
$4^+$ &   1.721(14) &$\mathbf{ 1.623(66)}$ & 1.70(45)     & 90.7(76) & 2.0 \\
$4^-$ &   1.709(18) &$\mathbf{ 1.67(11)}$ & 1.58(52)     & 96(12) & / \\
 \hline
\end{tabular}\\
\vspace{0.6cm}
\begin{tabular}{|c|c|c|c|c|c|c|}
\hline
$\mathbf{\beta=12}$ & $am~(t=a)$  & $am~(t=2a)$ & $am~(t=3a)$ & $am~(t=4a)$ &
 quality$[\%]$ & overlap$[\%]$\\
\hline
$0^+$ & 0.6337(54) & 0.5845(66)&$\mathbf{0.567(14)}$&0.558(27)&91.8(41)&1.8 \\
$0^{+*}$ &1.054(11)&0.946(22)&$\mathbf{ 0.899(47)}$ & 0.95(14)& 81.6(96)&3.3\\
$2^+$   & 1.0991(55) & 1.030(20) &$\mathbf{0.991(54)}$& 0.86(11)&86(11)&4.0 \\
$2^-$   & 1.0928(71) & 1.009(29) &$\mathbf{0.946(98)}$& 0.70(14)&81(19)&2.6 \\
$4^+$  &1.4105(86) &1.364(45)&$\mathbf{1.24(14)}$&0.70(24) & 95.5(53)&5.7\\  
$4^-$  &1.412(10) &1.328(50)&$\mathbf{1.20(15)}$&0.95(28) &91.9(58)&/\\ 
\hline
 \hline
\end{tabular}\\
\vspace{0.6cm}
\begin{tabular}{|c|c|c|c|c|c|c|}
\hline
$\mathbf{\beta=14.5}$ & $am~(t=a)$  & $am~(t=2a)$ & $am~(t=3a)$ & $am~(t=4a)$ &
 quality$[\%]$ & overlap$[\%]$\\
\hline
\hline
$0^+$ & 0.5823(42)& 0.5107(57)&0.4921(79)&0.486(15)& 79(11)$$ & 3.3 \\
$0^{+*}$&0.7999(51)&0.732(11)&$\mathbf{0.666(20)}$&0.601(40) & 82.0(46)&8.2 \\
$2^+$ & 0.9851(52)& 0.883(12)&$\mathbf{ 0.800(22)}$&0.786(69) &76.4(47)&1.1\\
$2^-$ & 0.9413(70)& 0.867(12)&$\mathbf{ 0.826(30)}$&0.822(84) &85.4(68)&3.7\\
$4^+$    &1.413(13)&1.184(41)&$\mathbf{0.98(11)}$&1.14(38) &91.0(26)& 9.2\\  
$4^-$  &1.2698(91)&1.196(43)&$\mathbf{1.028(97)}$&/ &92.9(51)& /\\  
\hline
 \hline
\end{tabular}
\end{center}
\caption{Strategy I: The local effective masses, quality factors and overlaps 
(as defined in the text) between states of different wave functions
 at $\beta=6,~9,~12$ and 14.5. For the $0^+$ at $\beta=14.5$,
 we used the effective mass at five lattice spacings $\mathbf{0.460(30)}$.}
\la{sIdata}
\end{table}
\clearpage
\begin{table}
\begin{center}
\begin{tabular}{|c|c|c|c|}
\hline
$\mathbf{\beta=6}$ & $am~[\bar t]$ & $c$ & $c'$ \\
\hline
\hline
$A_1$     & 1.190(24)[1.5] & 1.000(10)  &  0.017(14)  \\
$A_1^*$  & 1.804(96)[1.5] & 0.990(21)   &  0.142(28)  \\
$A_3$     &1.666(81)[1.5] &   1         &   0          \\
 \hline
\hline
\end{tabular}\\
\vspace{0.6cm}
\begin{tabular}{|c|c|c|c|}
\hline
$\mathbf{\beta=9}$ & $am~[\bar t]$ & $c$ & $c'$ \\
\hline
\hline
$A_1$     & 0.7731(79)[1.5]  &   1.000(14)  &  0.028(21)  \\
$A_1^{*}$  & 1.179(37) [1.5] &   0.998(10)  & 0.063(16)  \\
$A_1^{+**}$ & 1.433(64)[1.5]  &   0.973(25)   &  0.231(32))   \\
$A_1^{+***}$     & 1.70(13) [1.5]  &   0.786(98) &    0.618(12)\\
$A_3^+$     &  1.303(47)[1.5] &   1     &        0     \\
 \hline                    
\hline
\end{tabular}\\
\vspace{0.6cm}
\begin{tabular}{|c|c|c|c|c|}
\hline
$\mathbf{\beta=12}$ & $am~[\bar t]$ & $c$ & $c'$ \\
\hline
\hline
$A_1$    & 0.572(15)~[2.5] &   1.000(18) & 0.000(25)\\
$A_1^{*}$ & 0.856(53)~[2.5] &   0.992(26) & 0.124(37)\\ 
$A_1^{**}$& 0.943(39)~[2.5] &   0.988(38) & 0.152(53)\\ 
$A_1^{***}$    &1.294(59) ~[1.5] &   0.680(68) & 0.734(97)\\ 
$A_2$    & 1.365(57)~[1.5] &  0         & 1  \\
$A_3$    & 0.990(60)~[2.5] &   1 &  0  \\
$A_4$    & 1.03(10)~~[2.5] &   0.999(32) & 0.035(21) \\
 \hline
\hline
\end{tabular}\\
\vspace{0.6cm}
\begin{tabular}{|c|c|c|c|}
\hline
$\mathbf{\beta=14.5}$ & $am~[\bar t]$  & $c$ & $c'$ \\
\hline
\hline
$A_1$    & 0.489(13)~[2.5] &  1.000(41) & 0.016(58)\\ 
$A_1^{*}$ & 0.669(24)~[2.5]  &  0.998(16) & 0.0619(22) \\
$A_1^{**}$& 0.816(56)~[2.5]  &  0.985(40) & 0.172(56)  \\  
$A_1^{***}$    &1.11(12) ~[2.5]  &  0.577(48) & 0.816(68)\\
$A_2$    & 1.04(12)~[2.5]   &  0           & 1 \\  
$A_3$    & 0.776(34)~[2.5]  &  1  &  0 \\ 
$A_4$    & 0.71(13)~~[3.5]  &  0.995(30) & 0.097(27) \\ 
 \hline
\hline 
\end{tabular}
\end{center}
\caption{Strategy II: The local effective masses, and Fourier coefficients
obtained  at $\beta=6,~9,~12$ and 14.5. 
The wave function coefficients $c$ and $c'$, 
all obtained at two lattice spacings,  correspond  respectively to the 
smallest and second-smallest spin wave function compatible with the lattice
representation (e.g., $c=c_0$ and $c'=c_4$ for the $A_1$ representation).
The number in brackets indicates at what time separation the local effective
mass was evaluated.}
\la{sIIdata}
\end{table}
\clearpage
%
\begin{figure}[tb]
\vspace{-0.75cm}
\centerline{\begin{minipage}[c]{14cm}
    \psfig{file=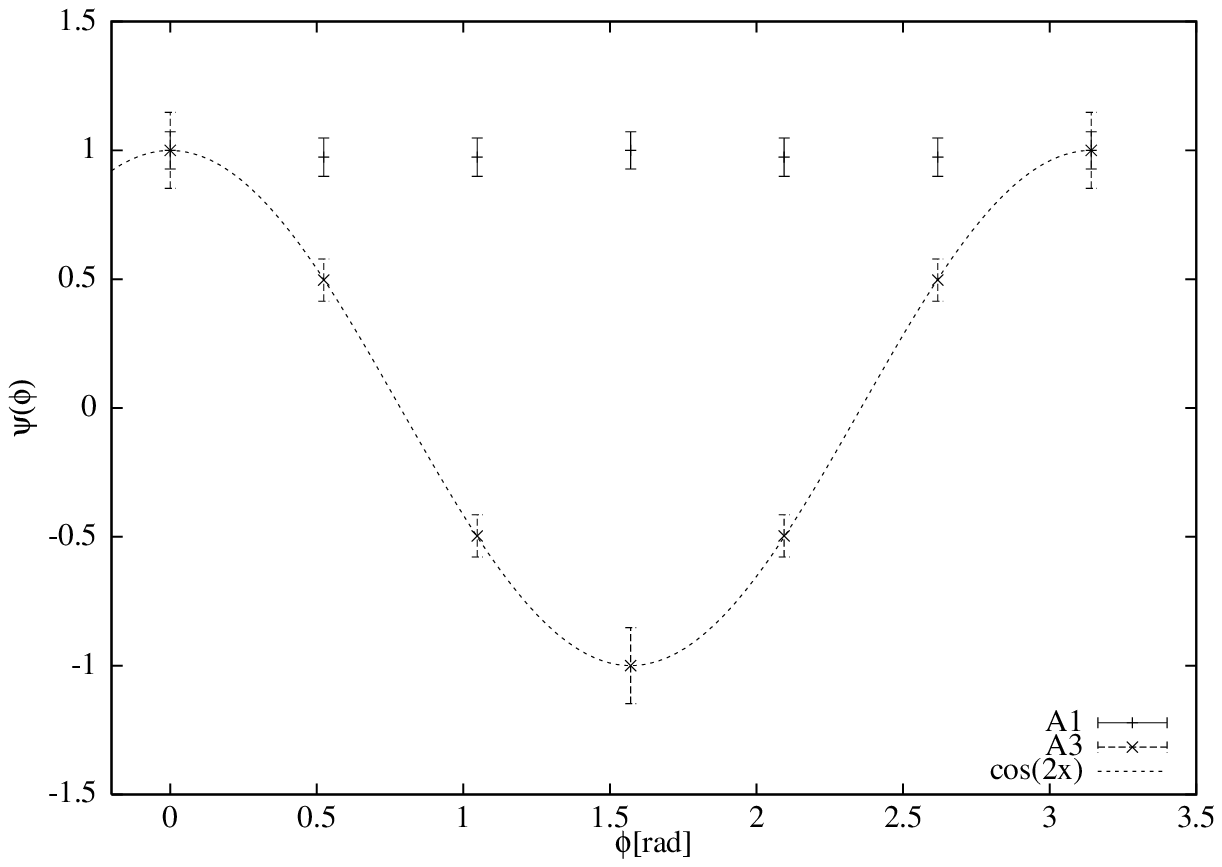,angle=0,width=7cm}
    \psfig{file=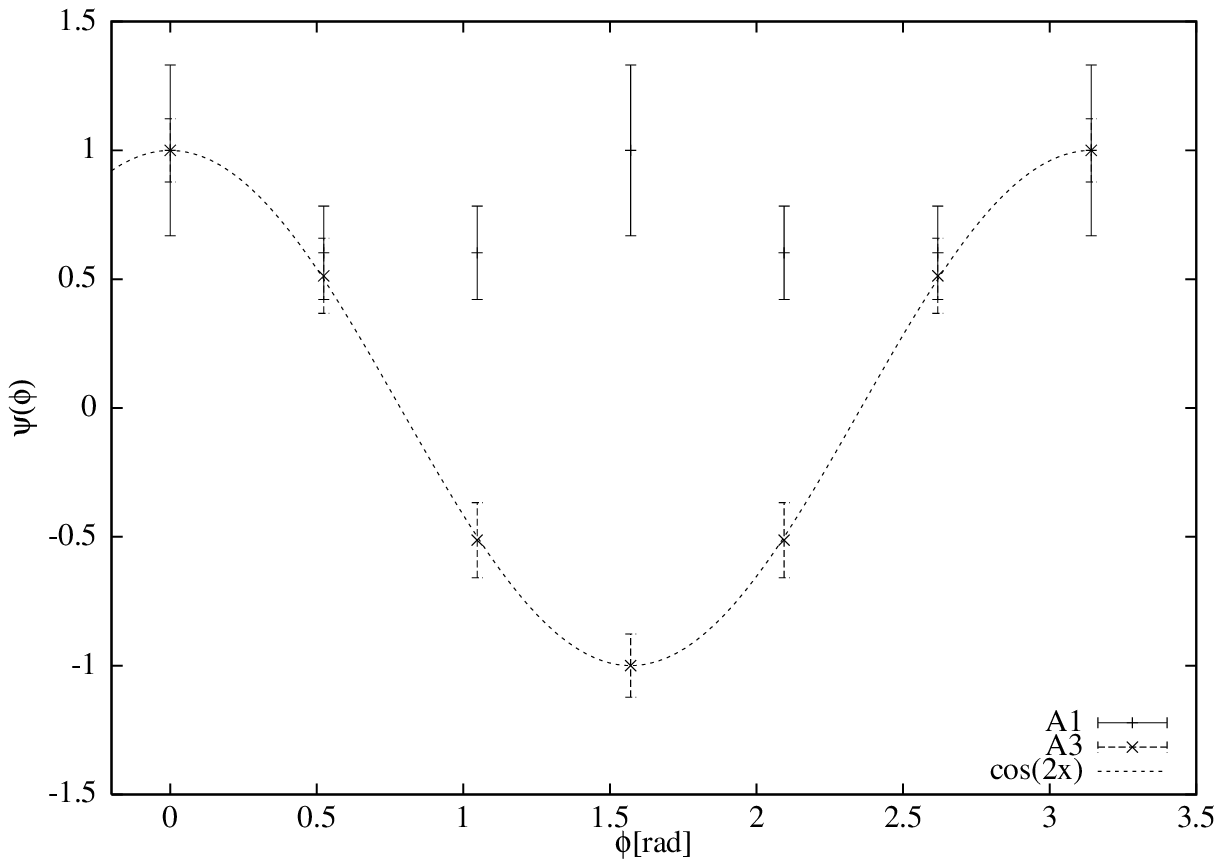,angle=0,width=7cm}
    \end{minipage}}
\centerline{\begin{minipage}[c]{14cm}
    \psfig{file=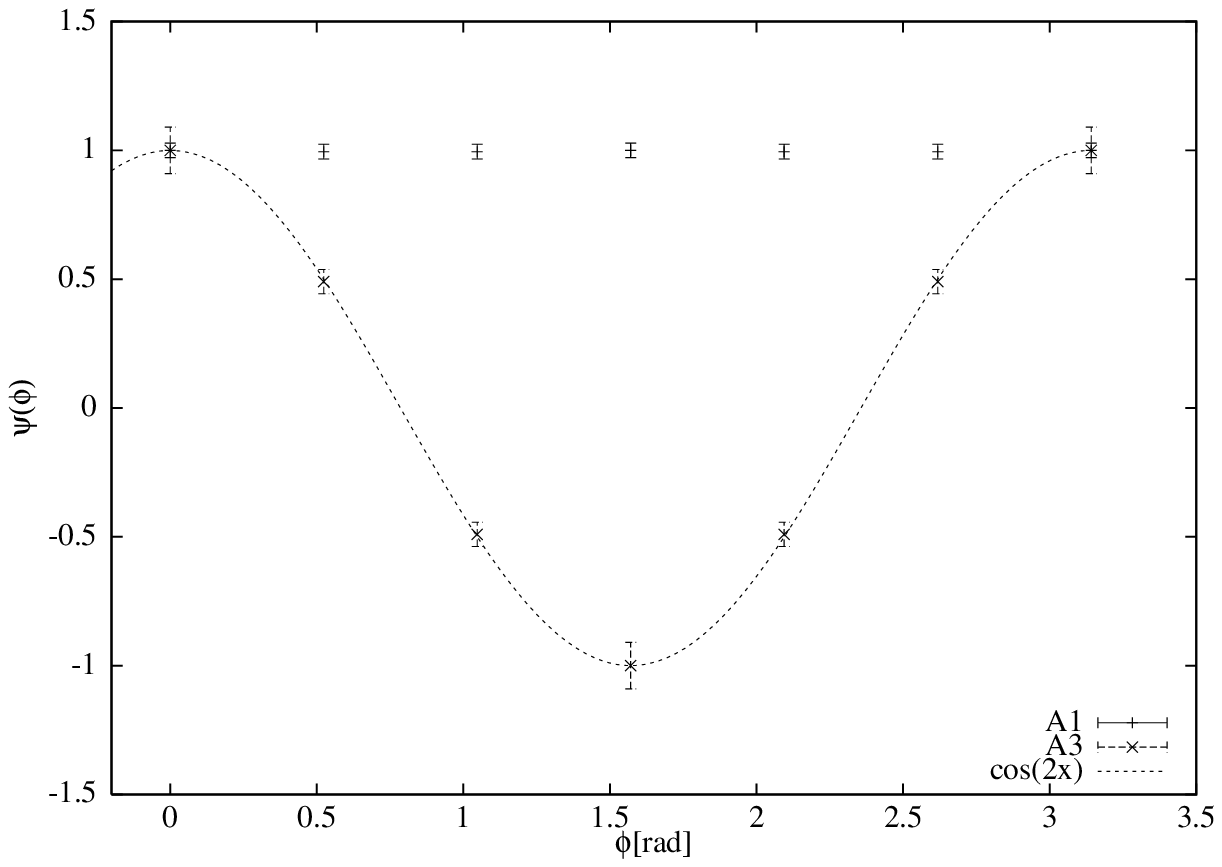,angle=0,width=7cm}
    \psfig{file=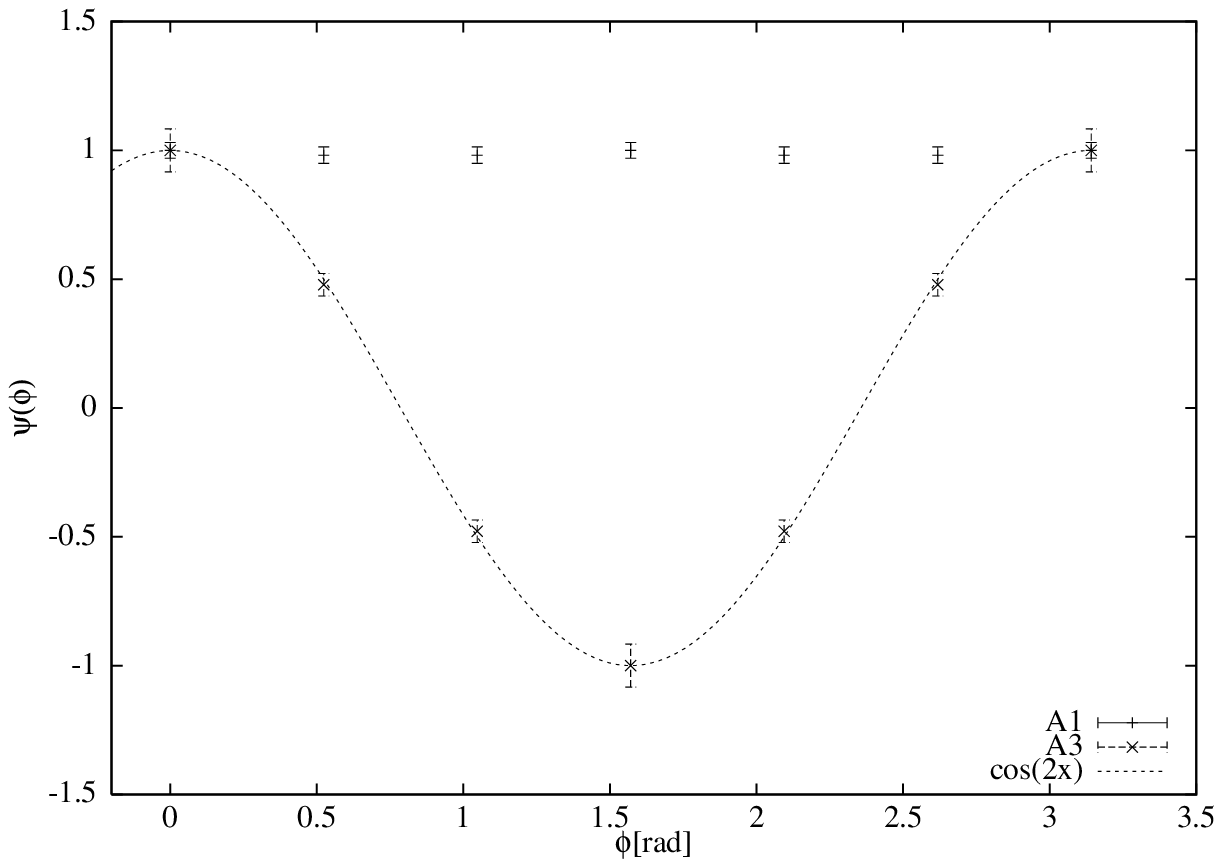,angle=0,width=7cm}
    \end{minipage}}
\caption[a]{The wave function of the lightest state in the $A_1$ and $A_3$ 
lattice representations, as measured with our four operators at $\beta=6$.
The plots can be compared to the correlation matrices given 
in section (\ref{recipe}).}
\la{a1a3}
\end{figure}
\begin{figure}[!bt]
\vspace{-0.15cm}
\centerline{\begin{minipage}[c]{14.0cm}
    \psfig{file=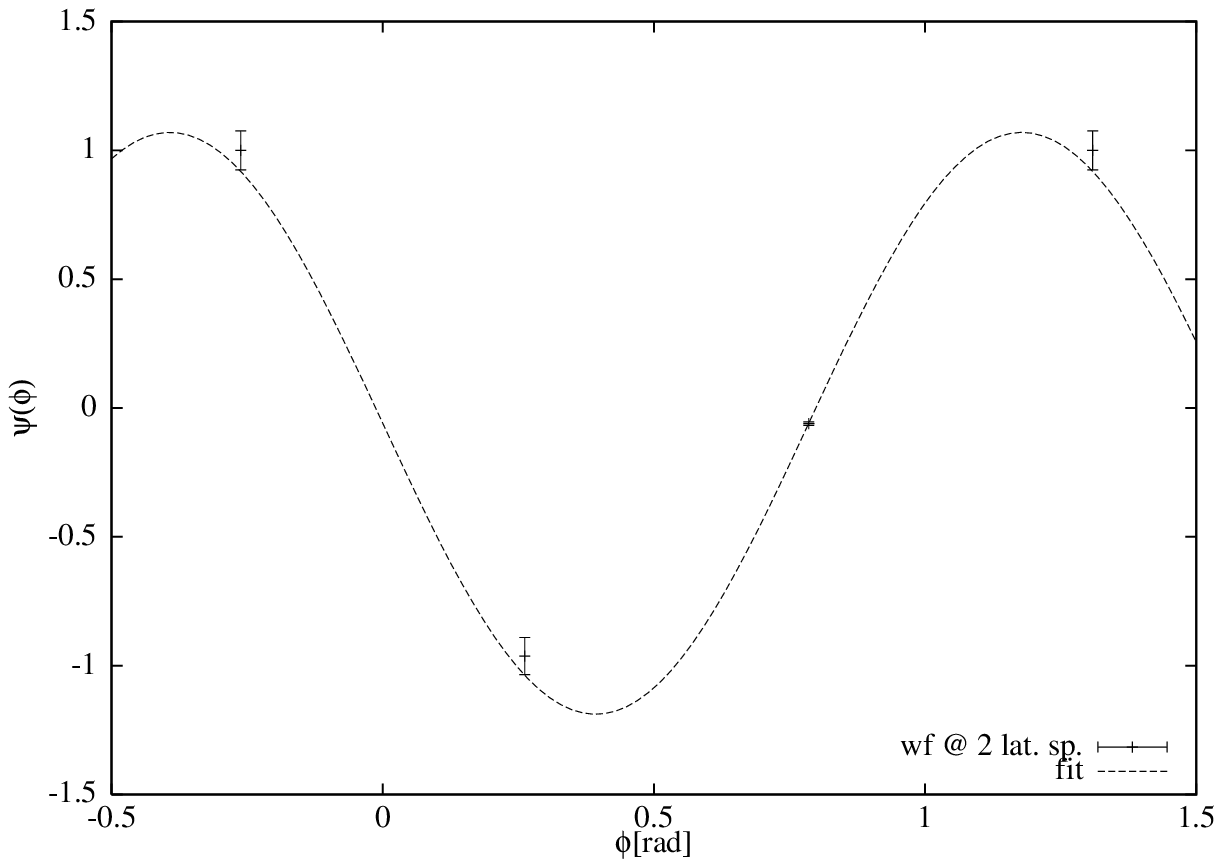,angle=0,width=7cm}
    \psfig{file=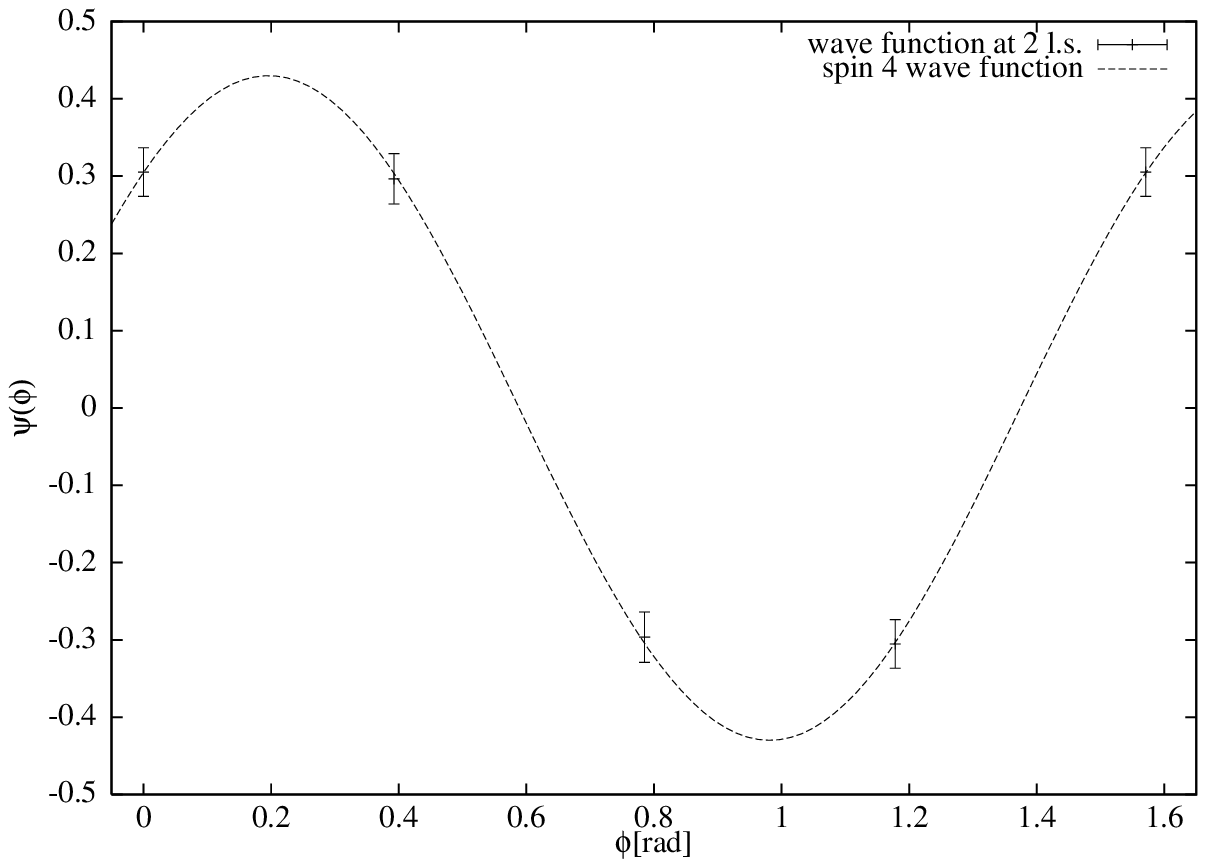,angle=0,width=7cm}
    \end{minipage}}
\centerline{\begin{minipage}[c]{14.0cm}
     \psfig{file=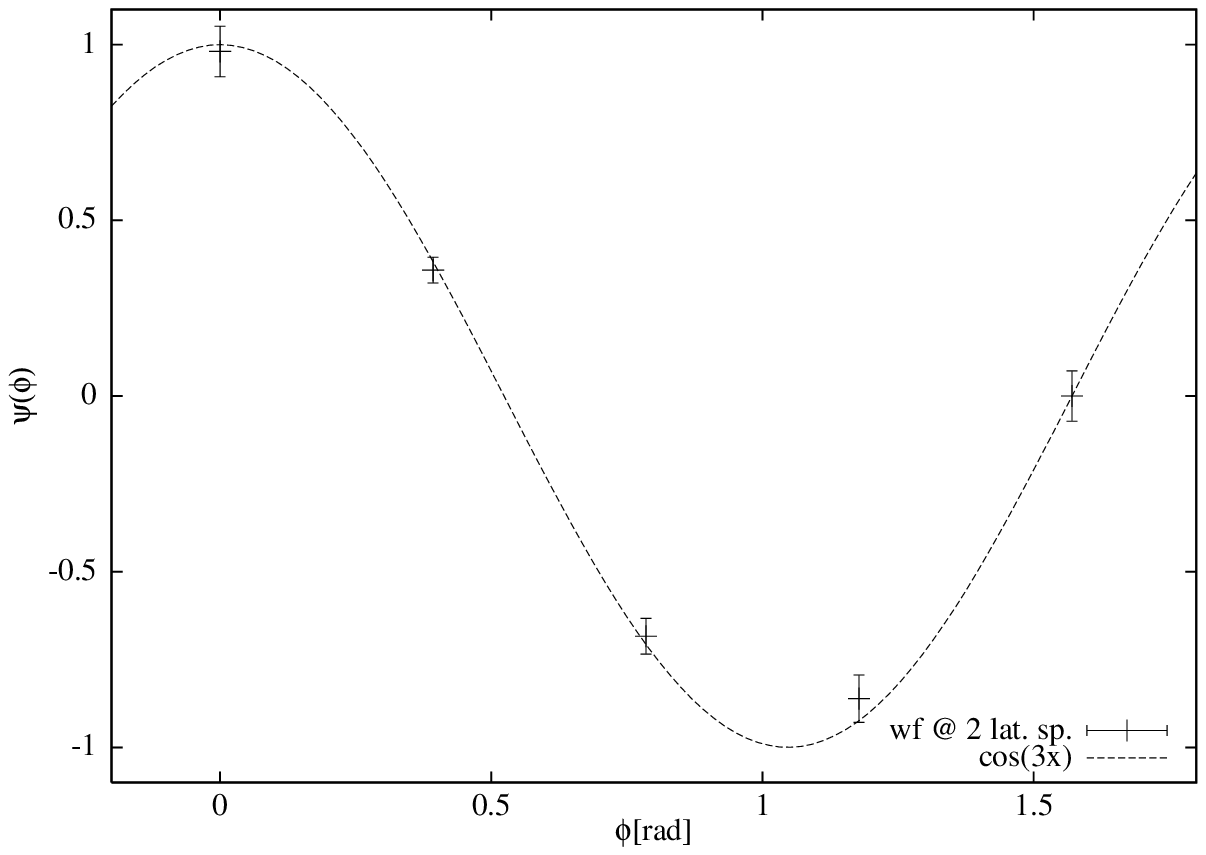,angle=0,width=7cm}
    \psfig{file=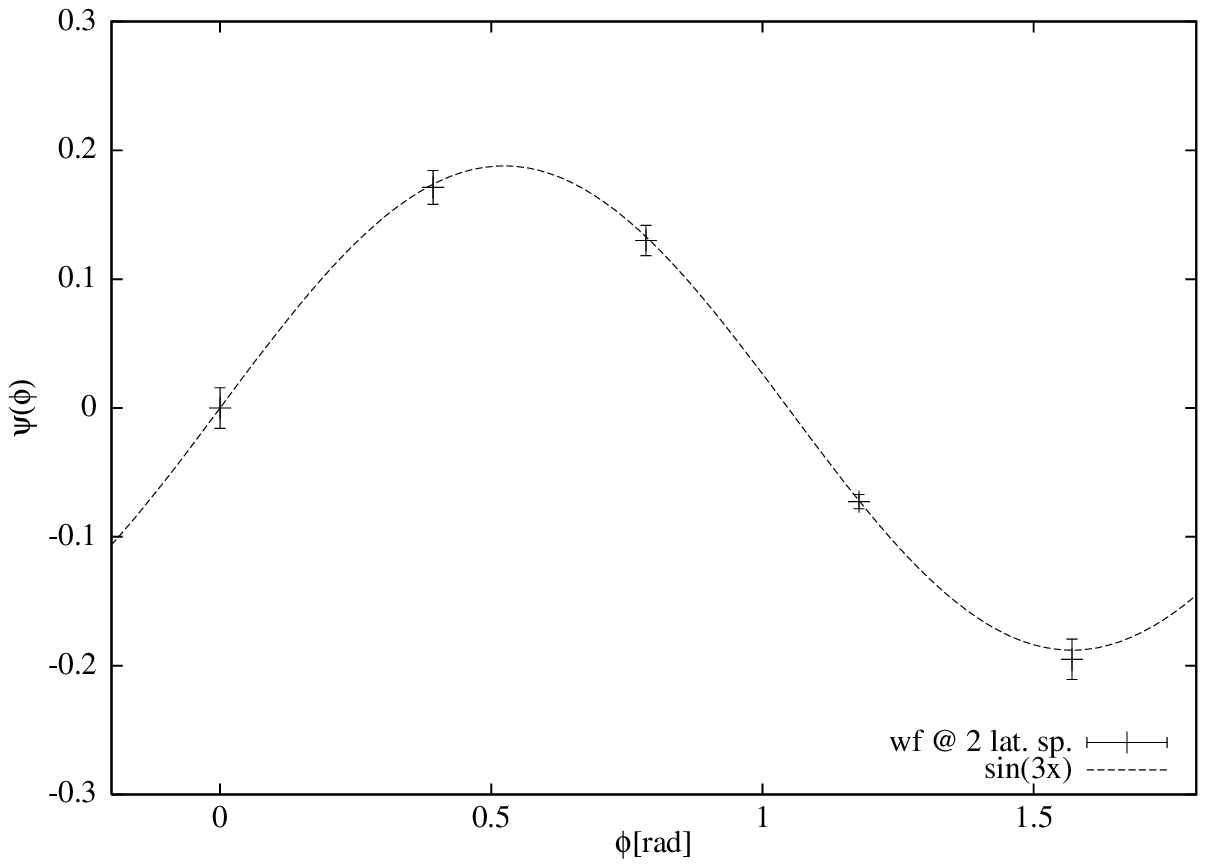,angle=0,width=7cm}
 \end{minipage}}
\caption[a]{Top: The wave function of the lightest state in the $A_2$
lattice representation; left, at $\beta=6$ and measured 
with 12-fold operators; right, at $\beta=12$ with
16-fold operators. 
Bottom: The wave functions of the lightest state in two orthogonal 
polarisations of the  lattice representation $E$, as measured at $\beta=12$.
The vertical axis has arbitrary scale.}
\la{a2e}
\end{figure}

%% file: chapter5.tex
\chapter{Multi-level algorithms}
\label{ch:mla}
\noindent 
In equilibrium statistical mechanics and quantum field theory, 
much of the physical information is encoded in $n$-point functions.
The short-range nature of interactions in the former and the causality 
requirement in the latter case lead to the property of locality of the 
Hamiltonian (resp.  action). In Monte-Carlo simulations, 
the properties of the spectrum are extracted
from numerically calculated 2-point functions 
$C(t)\equiv\langle{\cal O}(t) {\cal O}(0)\rangle$
in the Euclidean formulation.
When the theory admits a mass gap, the exponential 
decay of each term singles out the lightest state compatible
 with the symmetry of the operator, thus enabling us to extract the
 low-lying spectrum of the theory. However, it is precisely this decay that 
makes the 2-point function numerically difficult to compute at large $t$.
Indeed, standard algorithms keep its absolute variance roughly constant, 
so that the variance on the local effective mass 
\[am_{\mathrm{eff}}\left(t+\frac{a}{2}\right)=\log{\frac{C(t)}{C(t+a)}}\]
increases exponentially with the time separation. For that reason, 
it would be highly desirable to have a more efficient
method to compute correlation functions at large time separation $t$. The
task amounts to reduce uncorrelated fluctuations between the two time slices
separated by Euclidean time $t$.

In this chapter we present a `noise reduction' method that exploits the 
locality property. It has the advantage of being compatible 
with the popular link-smearing and -blocking: one can cumulate the advantages
of both types of techniques. Indeed, while the fuzzing algorithms also
help reduce short wavelength fluctuations inside a time-slice,
our method aims at averaging out the noise induced by fluctuations 
appearing in neighbouring time-slices by performing 
additional sweeps between fixed time-slices. 
As the continuum is approached, the volume
over which the average is performed ought to be kept fixed in physical units 
to maintain the efficiency of the algorithm.
Finally, we note that the idea is very general and is expected
 to be applicable in other types of theories.

We first present the idea in its full generality, 
and then formulate a multi-level scheme for the case of a 
2-point function and point out how the efficiency and parameters of the 
algorithm are determined by the low-lying spectrum of the theory.
We finally apply this algorithm to the case of $D=3+1$ $SU(3)$
gauge theory.

\section{Locality $\&$  multi-level algorithms}
The locality property of most studied actions allows us to  derive
 an interesting way of computing correlation functions. 
First we give a general, `topological' definition of locality in continuum
field theories. We use a symbolic notation; 
 if ${\cal C}$ denotes a configuration, let 
${\cal X}$, ${\cal Y}$ and ${\cal A}$ be 
mutually disjoint subsets of ${\cal C}$. If $\Omega_X$,  $\Omega_Y$ and 
$\Omega_A$ are their respective supports on the space-time
 manifold $\cal M$, suppose furthermore that any continuous path 
$\gamma: I \rightarrow {\cal M}$ joining  $\Omega_X$ and $\Omega_Y$
(i.e. $\gamma(I)\cap \Omega_X\neq \emptyset$ and  
$\gamma(I)\cap \Omega_Y\neq \emptyset$)
passes through $\Omega_A$ (i.e. $\gamma(I) \cap \Omega_A \neq \emptyset$).
 See Fig.~\ref{topol}.
\begin{figure}[htb]
\centerline{\begin{minipage}[c]{8cm}
    \psfig{file=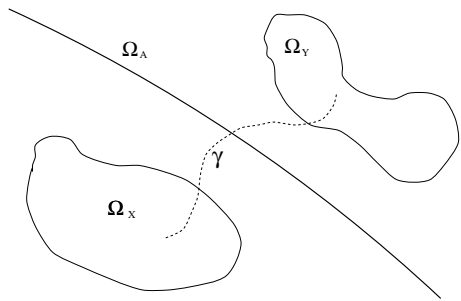,angle=270,width=6cm}
    \end{minipage}}

\vspace*{0.5cm}

\caption[a]{The sets $\Omega_X$,  $\Omega_Y$ and  $\Omega_A$ on the 
space-time manifold.}
\la{topol}
\end{figure}

  The theory with probability distribution 
$p({\cal C})$ is \emph{local} if there exists functionals $p_{A}$ and
$\tilde p_{A}$ such that, for any  setup with this topology, 
\be p({\cal X}, {\cal Y})= \sum_{\cal A} p({\cal A})~p_{A}({\cal X})~
 \tilde p_{A}({\cal Y})\la{ppp}
\ee
That is, ``${\cal X}$ and ${\cal Y}$ influence each other
only through ${\cal A}$''.
This condition is obviously satisfied by continuum Euclidean 
field theories whose 
Lagrangian density contains a finite number of derivatives. With a suitable 
notion of `continuity' of the path $\gamma$, one can extend this definition
to lattice actions. For instance, the Wilson action 
 is also local in this sense,
 but note that  $\Omega_{\cal X}$ and $\Omega_{\cal Y}$ 
must be  separated by more than one lattice spacing in order to realise
 the setup in the first place.
\paragraph{Hierarchical formula\\}
As a consequence, for two operators ${\cal O}_x$ and ${\cal O}_y$,
functionals of ${\cal X}$ and ${\cal Y}$ respectively, we have
\be \< {\cal O}_x {\cal O}_y \> 
\equiv \sum_{{\cal C}} {\cal O}_x({\cal C}) {\cal O}_y({\cal C}) p({\cal C})
=\sum_{\cal A} p({\cal A})~ \< {\cal O}_x \>_A ~ \< {\cal O}_y \>_A
\la{form}
\ee
where
\ba
\< {\cal O}_{x} \>_A &=& \sum_{\cal X} p_A({\cal X}) 
~{\cal O}_x({\cal X})\nonumber\\
\< {\cal O}_{y} \>_A &=& \sum_{\cal Y} \tilde p_A({\cal Y})
~ {\cal O}_y({\cal Y}) \label{subav}
\ea
are the average values of the operators at a fixed value of $\cal A$. Thus
the averaging process factorises into an average \emph{at} fixed `boundary 
conditions' (BCs) and an average \emph{over} these boundary conditions. There
are several ways in which this factorisation can be  iterated:
first, if the operator ${\cal O}_x\equiv{\cal O}_{x_1} ~{\cal O}_{x_2}$ 
itself factorises, the decomposition can be carried out also at this level,
where $p_A$ now plays the role of $p$. This means that the decomposition
(\ref{form}) allows us to treat the general $n$-point functions in the
 same way as the $n=2$ case that we shall investigate in more detail:
each factor can be averaged over separately.

There is another way the decomposition can be iterated: we can in turn write
  $\<{\cal O}_x\>_A$ and $\<{\cal O}_y\>_A$ as factorised averages over
yet smaller subspaces, thus obtaining a nested expression for the 
correlation function. A three-level version of (\ref{form}) would be
\ba
\< {\cal O}_x {\cal O}_y \>& =&\sum_{\cal A} p({\cal A})~\times~
\sum_{{\cal A}_1} p_A({\cal A}_1)~\sum_{{\cal A}_2} p_{A_1}({\cal A}_2)
\<{\cal O}_x\>_{{\cal A}_2}~ \times \nonumber\\
&&\sum_{\tilde{\cal A}_1} \tilde p_A(\tilde{\cal A}_1)~
\sum_{\tilde{\cal A}_2} \tilde p_{\tilde A_1}(\tilde{\cal A}_2)
\<{\cal O}_y\>_{\tilde{\cal A}_2}\la{hifo}
\ea
This type of formula is the basis of our multi-level algorithm for the 2-point 
function. 

\paragraph{Multi-Level algorithm for the 2-point function\\}
The hierarchical formula (\ref{hifo}) is completely analogous
to the expression derived in \cite{Luscher:2001up} in the case of the Polyakov loop,
where it was also proven that it can be realised in a Monte-Carlo simulation
by generating configurations in the usual way, then keeping the subset
$\cal A$ fixed and  updating the regions $\cal X$ and $\cal Y$.
Suppose we update the boundary $N_{\rm bc}$ times
and do $n$ measurements of the operators for each of these updates. 
We are thus
 performing $N_{\rm bc}\cdot n$ measurements.
But because the two sums in Eqn.~(\ref{form}) are factorised, 
this in effect achieves $N_{\rm bc} n^2$ measurements.  As long as 
\begin{itemize}
\item the latter are independent;
\item  that the
fluctuations on the boundary $\Omega_{\cal A}$ have a much smaller influence
than those occurring inside $\Omega_{\cal X}$ and $\Omega_{\cal Y}$;
\item that no phase transition occurs \cite{Luscher:2001up}
due to the small volume and the boundary conditions;
\end{itemize}
error bars reduce with Monte-Carlo time $\tau$ like $1/\tau$ rather than
 $1/\sqrt{\tau}~$: to half the variance, we double $n$. 
The fluctuations
of the boundary are only reduced in the usual $1/\sqrt{\tau}$ regime.
Thus, for a fixed overall computer time, 
one should tune parameters of the multi-level algorithm
 so that the fluctuations in $\cal X$ and $\cal Y$ 
are reduced down to the level of those coming from $\cal A$. 

The one-level setup
we shall use in practice is illustrated in Fig.~\ref{timesl}:
$\Omega_{\cal X}$, $\Omega_{\cal Y}$ and $\Omega_{\cal A}$ are time-slices.
\begin{figure}[tb]

\centerline{\begin{minipage}[c]{8cm}
    \psfig{file=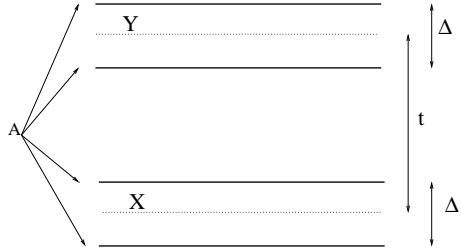,angle=270,width=6cm}
    \end{minipage}}

\vspace*{0.5cm}

\caption[a]{Our choice of $\Omega_{\cal X}$, $\Omega_{\cal Y}$ and 
$\Omega_{\cal A}$ to implement the hierarchical formula.}

\la{timesl}
\end{figure}
An indication of how many 
`submeasurements' should be chosen at each level
is given by the following consideration:
if we measure an operator located in the middle of a time-block 
 (of width $\Delta$) bounded by $\cal A$, then
any fluctuation occurring on $\cal A$ can be decomposed on the basis which
diagonalises the Hamiltonian of the theory. If  the lowest-lying state
 compatible with the symmetry of the operator being measured has  mass $m_0$
and $\Delta  > 1/m_0$, that state will act as the main `carrier' of
 the fluctuation, so that it will induce a fluctuation of relative 
magnitude $e^{-m_0\Delta /2}$ on the time-slice where the operator is 
measured (see Fig.~\ref{topol}); indeed the propagation of fluctuations is damped
 exponentially in a system that develops a mass gap.
Thus it is worth performing roughly $n\sim e^{m_0\Delta }$ 
measurements\footnote{These may be  nested, in which case it is the total
number of measurements under fixed boundary $\cal A$ that is meant here.}
  under fixed boundary $\cal A$ in order to reduce the 
fluctuations coming from  $\cal X$ down to the level of those of $\cal A$.
In fact, this estimate is an upper bound, because the vacuum state
under the fixed BCs could lie at a higher energy level
than the full-lattice vacuum.
Finally, if zero modes are present, we expect a power law 
$n\propto (\Delta )^\eta$.

At any rate, we may want to optimise
the parameters of the multi-level Monte-Carlo algorithm `experimentally', 
but we shall see in a practical example 
that our order-of-magnitude estimate is in qualitative
agreement with the outcome of the optimisation procedure.
This simple argument also shows that the  purpose of the 
multi-level scheme is to reduce fluctuations occurring
at all separations (from the time-slice where the operator is measured) 
ranging from 0 to $\Delta /2$ with an appropriate number of 
updates, in order to reduce their influence down to the level of the
 outermost boundary, which is then averaged over in the standard way. 

\paragraph{Variance reduction\label{errestim}\\}
Suppose the theory has a mass gap and that for a given $t$, 
the correlation function $C(t)\sim e^{-mt}$ 
is determined with equal amounts of computer time with the standard algorithm 
and the multi-level one. 
If we now want to compute $C(2t)$ with the same \emph{relative} precision
with the former, we need to increase the number of measurements by a factor
$e^{2mt}$. With the multi-level algorithm however, we would 
increase the number of submeasurements by a factor $e^{mt}$, 
as explained in the preceding section. Thus in this situation,
the gain in computer time  is a factor $e^{mt}$; 
turned the other way, it achieves
a variance reduction $\propto e^{-mt/2}$ compared to the
 standard algorithm\footnote{Note however that the computer
 effort is still increasing exponentially with $t$.}. 
Since the quantity determining the variance reduction
is the product $mt$, the variance reduction is in first approximation 
independent of $\beta$, as long as we measure the correlation function
at fixed $t$ in physical units. It would be inefficient to perform
sweeps over time-blocks that are much thinner than the physical length scale: 
none of the three conditions highlighted in the preceding section is likely
to be satisfied.
%
\section{A first application}
We consider pure $SU(3)$ lattice gauge theory in
 $(3+1)$ dimensions. 
A simulation is done at $\beta=5.70$ on an $8^4$ lattice;
as was noted in~\cite{Luscher:2001up}, the most elegant 
way to implement a  multi-level Monte-Carlo program is to use 
a recursive function.
We compute the $0^{++}$ and $2^{++}$ correlation functions
formed with a $4\times 2$ rectangular Wilson loop  at 4 lattice spacings. 
Two smearing steps are applied on the operator. 
A two-level scheme is implemented: 
the 8 time-slices are split into 2 time-blocks of width 4,  which are
in turn decomposed into 2 time-blocks of width 2. 
Here we restrict ourselves to
the measurement of the correlation function at even time-separations.

For the $0^{++}$ correlation function at 4 lattice spacings, 
one `measurement' comprises  
10 submeasurements at the lower level, 40 at the upper level. When
performing the latter, we are free to compute the 0 and 2 lattice spacing
 correlation in the standard way (thus the variance reduction  only 
applies to the $t=4a$ correlation).
We collected 260 of these compound measurements. 

We proceed similarly in the $2^{++}$ case with following parameters:
 one `measurement' comprises  8 submeasurements
at the lower level, 150 at the upper level, 
each of these being preceded by 5 sweeps;
our program needs about 8.3 minutes on a standard Alpha workstation to perform
one of these compound measurements. We collected 520 of them; 
because they are time-consuming, we perform $\sim 200$ sweeps
between them to reduce their statistical dependence. 

The following values for the correlation functions, 
as well as their corresponding local effective masses 
(extracted from a cosh fit),  were obtained:
\begin{center}
\begin{tabular}{|c|c|c||c|c|}
\hline
$t/a$ &  $\< {\cal O}_0(0)~ {\cal O}_0(t)\>$&$am_{\rm eff}^{(0)}(t)$ 
 &$\< {\cal O}_2(0)~ {\cal O}_2(t)\>$& $am_{\rm eff}^{(2)}(t)$ \\
\hline
\hline
0   &  1.0000(65)&       &  $  1.0000(14)$ &  \\
1   &            &  1.017(35)   &     &  2.151(75)    \\
2   &   0.1331(99)&       &   $ 1.36(20) \times 10^{-2}$ &  \\
3   &            & 0.929(49)   &     &   1.794(74)   \\
4   &   0.0406(39)&      & $ 7.49(70)\times10^{-4}$&  \\
\hline
\hline
\end{tabular}
\end{center}
\vspace{0.5cm}
The $t=4a$ correlation of the $2^{++}$ operator shows that the variance has been 
reduced by a factor 20  with respect to the $t=0$ point. It is already
at an accuracy that would be inaccessible on similar single-processor
 machines with the standard algorithm. Indeed the latter yields  error bars 
that are roughly independent of $t$ and would have given an error comparable
to our error on the $t=0$ data, where we do not achieve any error 
reduction\footnote{It would however benefit somewhat from the unbroken 
translational invariance in the time direction.}.  
 The naive error-reduction  estimate
of Section~\ref{errestim} evaluates to $\exp(1.8\times4/2)\simeq36$. Not
unexpectedly, the observed reduction factor is somewhat 
lower than this naive estimate, presumably because the configurations
 generated at fixed BCs are quite strongly correlated.
%
\paragraph{Summary $\&$ outlook\\}
So far we have proposed a general scheme in which the accuracy of 
numerically computed $n$-point functions in local field theories could be 
improved by the use of nested averages. While the procedure is known to be 
exact, the efficiency of the algorithm must ultimately be tested on a 
case-by-case basis. A simple application to $SU(3)$ Wilson loop
correlations showed that, in some cases at least, the multi-level algorithm
drastically reduces statistical errors. 
It can straightforwardly be used in conjunction with the smearing and blocking
techniques. A further nice feature of the 2-point function
 case is that previous knowledge of the low-energy spectrum provides useful
guidance in the tuning of the algorithm's many parameters. 

 While it now seems clear that for asymptotically large 
Euclidean time separation, the multi-level algorithm 
becomes more efficient than the 1-level algorithm, 
the question of real practical interest is whether
one can truly improve the efficiency of realistic calculations. Typically,
the operators have reached mass plateaux already at 0.2fm in the case 
of glueballs. In such a regime, one cannot expect a statistical error 
reduction by orders of magnitude. Only a numerical  analysis can
reliably address the question formulated above.

It is equally important to determine whether the efficiency of the algorithm
 is maintained as the lattice spacing is decreased. Indeed the correlation
length becomes larger and larger in lattice units 
and one might wonder whether the low-level measurements at fixed BCs
 are still helping to reduce the dominant fluctuations on the correlator.

\section{A 2-level version of the algorithm\la{description}}
In this section, we describe the implementation of a 2-level algorithm 
for the measurement of  2-point correlation functions in some more detail. 
The operators are smeared, blocked,
definite-momentum operators in 2+1  dimensional $SU(2)$ gauge theory. 
We shall use glueball operators as examples, 
however the conclusions are applicable to the measurement
of fuzzy spatial flux-tubes as well~\cite{Meyer:2003hy}.

The lattice size is $\hat L_x\times \hat L_y\times \hat L_t$.
After a number $N_{\rm up}$ of compound update sweeps, we freeze $\hat L_t/\Delta$ 
time-slices separated by distance $\Delta$, and measure the average values
$\langle {\cal O}(t_i)\rangle_{\rm bc}$
of the operators in all the other time-slices $t_i$ 
between the fixed time-slices by doing $n$ updates under these
fixed BCs. These average values 
in each time-slice are kept separately.
They are written to disk before updating the full lattice again.
$N_{\rm up}$ is typically chosen to be $n/10$, so as to represent
a negligible amount of computer time, and nevertheless ensure good
statistical independence of the `compound measurements' (this will
be checked in Section~\ref{performance}).

After $N_{\rm bc}$ of these compound measurements, 
the correlator for $t\geq 2a$ can easily be computed `off-line' from 
\be
\langle {\cal O}(t_i) {\cal O}(t_j)\rangle
= \frac{1}{N_{\rm bc}}\sum_{\rm bc} \langle {\cal O}(t_i)\rangle_{\rm bc} 
\langle {\cal O}(t_j) \rangle_{\rm bc},
\ee
if the time-slices $t_i$ and $t_j$ do not belong to the same `time-block'.
This equation holds because 
the BCs have been generated with the weighting
given by the full lattice action~\cite{Luscher:2001up}.

A few comments on the data storage are in order.
If $N_{op}$ are being measured, the amount of data generated is
\[ {\rm nb(data)_{II}} = N_{op} \times N_t \times N_{\rm bc}. \]
This is to be contrasted with the ordinary way of storing the data:
the correlation matrix is computed during the simulation, and stored in 
typically $N_{bin}={\cal O}(100)$:
\[ {\rm nb(data)_{I}} \simeq N_{op}^2 \times N_{t} \times N_{bin}. \]
The ratio is thus
\be
\frac{\rm nb(data)_{II}}{\rm nb(data)_{I}} = \frac{1}{N_{op}}~~
 \frac{N_{\rm bc}}{N_{bin}}
\ee
As an example, for a large production run with a total of $10^6$ measurements,
 we may do $n=10^3$ measurements under $N_{\rm bc}=10^3$ fixed BCs.
Therefore, for $N_{op}\gg 10$, which is usually the case, the data size
is smaller than with the 1-level algorithm.
The obvious advantage of version II is that 
one can use a much larger number of operators (e.g. include non-zero momenta, 
scattering states, the square of the traces of operators,~\dots). 
Further advantages  include:
\begin{itemize}
\item one can  choose the binning \emph{a posteriori}, thus making a 
 more detailed check of auto-correlations possible;
\item if e.g. $\Delta=8$ and one is computing the 2-point function
at $t=5$, there are several ways to obtain it, which of course
all have the same average, but different variances; it is very convenient 
to be able to choose which combination is optimal \emph{a posteriori}
(see Section~\ref{physics}).
\item 
in principle, one can extract 3- and 4-point function from the same data set,
as long as one correlates operators that have been averaged 
in different time-blocks.
Derivatives of the 2-point function can be computed just as easily.
\end{itemize}
A downside of our computational scheme is that one looses
 information on the short-range correlator 
(0 and 1 lattice spacing of Euclidean time separation). 
The time-zero correlator is useful because it allows one
to evaluate the overlap of the original operators onto the physical states.
In some cases it may be desirable to store the BC-averages of the short-range
 correlators  since 
\be
\langle {\cal O}(t_i) {\cal O}(t_j)\rangle = \frac{1}{N_{\rm bc}}
\sum_{\rm bc} \langle {\cal O}(t_i) {\cal O}(t_j)\rangle_{\rm bc}
\ee
if the time-slices $t_i$ and $t_j$ belong to the same time-block. 
Incidentally, for $t_i=t_j$, 
we shall see in Section~\ref{performance}
that this measurement can be useful to predict  the 
performance of the algorithm for the larger time-separations.

\section{The algorithm and its parameters\la{parameters}}
We now present  data obtained at 
$\beta=12$, $V=32^3$ in the 2+1D $SU(2)$ pure gauge theory. 
Note that $\sqrt{\sigma}a = 0.1179(5)$~\cite{Teper:1998te}, 
which means that $a=0.053$fm 
(if we use $\sqrt{\sigma}=440$MeV) and we are indeed
well in the scaling region, as far as the low-energy observables are 
concerned.

We perform a check of the auto-correlation of compound measurements done 
at fixed BCs. We then proceed to a study of the efficiency
of the algorithm as a function of its various parameters: first, the width
$\Delta$ of the time-block inside which the submeasurements are made;
secondly, the number of submeasurements. We will then look at 
the dependence of the error bars
on the mass of the state being measured.
\paragraph{Binning analysis}
On Fig.~\ref{fig:binning}, we show the error bar on the correlator and its 
local-effective mass (LEM) 
for an operator lying in the $A_2$ irreducible representation 
 (IR, containing spins $0^-$, 4, 8, 12\dots) as function of the 
number of jacknife bins. We note that as long as the number of bins is not
much smaller than 100, the error bars are stable under the change of binning.
Obviously the error bars are subject to fluctuations themselves, and in some
cases we will give estimates of the latter. However we can draw the lesson
that the number of updates $N_{\rm up}\simeq n/10$ is apparently sufficient
to decorrelate the BCs.
\paragraph{The distance between fixed boundary conditions}
Fig.~\ref{fig:a2_plain} 
show the dependence of the error bar on the $A_2$ correlator and its LEM 
as a  function of the number of submeasurements $n$, 
at roughly fixed CPU time. We consider that the comparison of error bars
 is meaningful at the $20\%$ level. The fundamental state
in that lattice IR is relatively heavy and has
rotational  properties very similar to a continuum spin 4 glueball
 (Chapter~\ref{ch:hspin}).
 The left graphs show the $\Delta=4$ data, the right ones the $\Delta=8$ data. 

In the first case, the
smallest error bar is achieved for the smallest number of measurements (here
$n=100$) for all time-separations ($ t=2,~3,~4$). On the 
right-hand side, the situation is different: for time-separations 
$ t=2,~3,~4$, a small number of submeasurements ($n=200$) is
 more favourable, while, interestingly, the error bar for the 
 $ t =5$ correlator is practically independent of $n$.
For $ t \geq 6$, the hierarchy is inverted: the runs with a 
large number of subsweeps ($n=800,~1600$) yield smaller error bars.
This is consistent with the rule of thumb proposed above, namely
that the optimal number of submeasurements should be of order $e^{mt}$.
In the present case, this evaluates to $\sim1300$, given that
 $am(A_2)\simeq1.2$.

We can already draw  the conclusion that $\Delta=4$ is too small a time 
block and a significant number of submeasurements does not
 lead to further variance reduction on the correlator. 
$\Delta=8 \simeq \frac{1}{\sqrt{\sigma} a}\simeq 
0.5{\rm fm}/a$ on the other hand 
seems well suited for that purpose. This conclusion is expected
to hold also in 3+1D pure gauge systems, since their long distance correlations
are very similar to the present 2+1D case.

It is also interesting to look at the error bars on the 
LEM (bottom graphs). 
Indeed, when a large number of submeasurements is performed, the 
2-pt function at time $t$ and $t+a$ can be expected to be numerically 
more strongly correlated, thus reducing the fluctuations on their 
ratio. On the left (concerning $\Delta =4$), the  variance on the 
LEM at 3.5 lattice spacings is practically constant. On the right, we observe
that even at the smaller time separations $t = 2,~3,~4$, the 
runs with a large number of subsweeps (800) are at least as good as the 
runs with the smaller number of subsweeps (200). At $t=5.5$ (and beyond), there
is a clear advantage at performing a large number of subsweeps. For instance, 
the error bar for the $n=800$ run is roughly 3 times smaller than that for 
the  $n=200$ run, and this at equal CPU time.
\paragraph{Time-separation dependence of the error bars}
It is also instructive to look at the same data from another point of view:
for a fixed number of submeasurements, how does the error bar on the 
correlator and on its LEM vary as a function of time separation?
On Fig.~\ref{fig:a2_plain_YX}, it is clearly seen that the error bar 
decreases exponentially as the operators are measured further away from
the fixed boundaries. For $\Delta=8$, 
 the variance drops by a factor 100 between $t=2$ and $t=7$
for the runs with $n=800$ and 1600. 

\paragraph{Mass dependence of the error bars}
We plot the LEM as well as the local decay constant 
of the error bar on the correlator together on Fig.~\ref{fig:LEM}.
 The upper figure illustrates the situation
with a large number of submeasurements ($n=1600$), 
while the lower shows what happens with only $n=200$.

We show two light states, the fundamental $A_1$ and $A_3$ states as well 
as the fundamental $A_2$ that was considered up to here.
 For the $A_1$ and $A_3$, the error bar decays along 
with the signal, since the former's decay constant matches the LEM 
of the corresponding operator.  As a consequence, long mass plateaux are
seen, with error bars increasing only very slowly. For the heavier $A_2$ 
state, the error bar decay constant keeps up only to 4.5 lattice spacings, 
resulting in a fast loss of the signal beyond that. It is 
nevertheless much more favourable a situation than with only 200 
submeasurements: while the lightest glueball plateau is obtained just as 
well, the $A_3$ data is much more shaky and the $A_2$ is essentially
 lost beyond 4.5 lattice spacings. We note that although the basis of $A_3$
 operators was the same for  $n=1600$ as for $n=200$, the variational
calculation performed slightly less well in the latter case.

In fact, the time separation where the error bar decay constant falls off 
on Fig.~\ref{fig:LEM}
gives us an idea of the time-separation for which the number of 
submeasurements is optimal.
Indeed, if the error bar continues to fall off, it means that 
 the measurements have a large degree of statistical dependence
through the common BC,
since moving further away from the fixed BC makes them less dependent.
Once, far away from the BC,  the error bar is constant (i.e.
its decay constant is now zero), the signal to noise ratio is falling 
exponentially to zero.
Thus $n=1600$ is best suited for measuring the $A_2$ mass ($am\simeq 1.2$)
at 5.5 lattice spacings.

\section{Optimisation procedure $\&$ performance\la{performance}}
We proceed to a more systematic study of the efficiency of the 
2-level algorithm. We shall consider three states, in the $A_1$, 
 $A_2$ and $A_3$ lattice IRs. 
The lightest  states in these  representations correspond to the 
$ J^P=0^+$, $J=4$ and $J=2$ continuum states (see Chapter~\ref{ch:hspin}).
The procedure we adopt is to measure these three correlators 
 at fixed physical Euclidean time separation $ t$. We do so at 
three values of $\beta=6,~9$ and 12 -- recall that in the scaling region, the 
lattice spacing simply scales as $1/\beta$. The correlator is evaluated for 
different numbers of submeasurements under fixed BCs:
\be
1\leq n \leq 200.
\ee
We then plot the \emph{inverse efficiency}
 $\xi^{-1}$ as a function of the number of  submeasurements:
\be
\xi^{-1}(n) \equiv [\Delta C_n(t)]^2 ~\times~n, \la{effi}
\ee
where $\Delta C_n(t)$ is the error bar on the correlator when measured 
$n$ times under fixed BCs. In some cases, we shall also consider the 
efficiency with respect to the LEM, in which case $\Delta C_n(t)$ is replaced
by $\Delta m^{(\rm eff)}_n(t)$.

In this study the number of BCs was 100. They are separated
by 80  sweeps. The 
 individual measurements done under fixed BC were stored separately, to 
allow us to combine them in different ways. In particular, to obtain the 
efficiency corresponding to 10 submeasurements, for each BC
we can split the 200 submeasurements into
 20 `independent' sequences of 10 submeasurements. These 20 sequences are then
used to estimate the variance on the error bars themselves. 
On Fig.~\ref{fig:a3},~\ref{fig:a2} and~\ref{fig:a1} we show 
these roughly estimated variances for $n\leq 20$, after what
the number of `independent' sequences becomes smaller than 10 and these
variance estimates become unreliable. The aim here is only to give the order
of magnitude of the uncertainty on $\xi$, so as to be able to reach 
meaningful conclusions concerning its minimum as a function of $n$.

Eventually of course it is desirable to have an easier way to optimise the 
parameters of the algorithm. When we have an operator with exactly vanishing
vacuum expectation value (VEV),  we define a quantity $\omega$ as
the zero-time-separation correlator, measured with $n$ submeasurements, 
multiplied by the number of submeasurements $n$:
\be
  \omega(n,t_i) = \frac{1}{N_{\rm bc}}
\sum_{\rm bc} ~~\sum_{{\rm meas}=1}^{n}
 \langle {\cal O}(t_i)^2 \rangle_{\rm bc}
\ee
Obviously $\omega$ is a function of the distance between the time-slice
where the operator is measured and the fixed time-slices. 
It is easy to evaluate this quantity   accurately:
one of the objectives of this analysis is to check for the validity of this
 quantity as a predictor of the  optimal number of submeasurements
 of the 2-level algorithm. 
The absolute value of $\omega$ will not interest us, rather we will
check whether its minimum is reached at the same $n$ 
as $\xi^{-1}(n)$.
 
It  is also interesting to compare the efficiency of the 2-level algorithm to 
the standard 1-level algorithm with an equal number of  measurements.
In this case the translational invariance in the time direction is not 
broken by the algorithm. The sweeps between BCs have 
no \emph{raison d'\^etre} here; on the other hand 
 the measurements are done in each time-slice, 
including those that are kept fixed in the 2-level algorithm.
Thus the comparison of algorithms is fair.

Let us first consider the lightest $A_3$ state (Fig.~\ref{fig:a3}).
The graphs correspond, from top to bottom to $\beta=6,~9$ and 12. We 
keep the physical time separation approximatively fixed at about 0.22fm
(2, 3, and 4 lattice spacings respectively), 
and similarly the separation of the fixed 
time-slices is augmented in lattice units ($4a$ at $\beta=6$, $6a$ at 
$\beta=9$ and $8a$ at $\beta=12$; 
we also show the case $\Delta=4a$ at $\beta=12$ for comparison). 
The first observation is that the 2-level algorithm
performs better at all three lattice spacings. If the number of 
submeasurements is chosen `reasonably', the inverse efficiency is 
smaller by a factor $\sim3$ at the coarsest lattice spacing, and by a factor
$\sim2$ at both of the smaller lattice spacings, provided $\Delta$ is kept 
fixed in physical units. Secondly, the curve for $\xi^{-1}$ is extremely flat
around its minimum. For instance, at $\beta=9$ 
it seems that it does not matter whether one does 10 or 40 submeasurements, 
the performance for this particular observable will be unchanged. The flatness
 becomes even more pronounced closer to the continuum. This however is not
true for the case $\Delta=4a$ at $\beta=12$. Although the curve has a narrow
minimum at a small number of submeasurements, the efficiency then decreases 
rapidly and this setup becomes less favourable than the standard algorithm.
Thirdly, we note that the quantity $\omega$ shown at $\beta=12$ 
(it has been rescaled in such a 
way that it can be plotted along with the other curves) is a very good 
predictor of the minimum of the inverse efficiency curve $\xi^{-1}$, 
and this both
when $\Delta=4a$ and $\Delta=8a$. Its qualitative aspect (including the 
flatness) is very similar to the $\xi$ curve.

The qualitative statements that have been made for the $A_3$ correlator
 also apply to the $A_2$ correlator (see Fig.~\ref{fig:a2}), whose mass
is larger by a factor $\sim4/3$. As one might expect, the higher mass 
favours the use of the 2-level algorithm even more: the gain in CPU time
for constant error bars is roughly a factor 6 at all three values of $\beta$.
Again the $\xi$ curve is extremely flat, but the optimal number of 
submeasurements has shifted to the right: in fact, 100 submeasurements
seems to be a good choice at all three lattice spacings.  Choosing a narrow
width for the time-blocks has the clear disadvantage of leading
to a smaller gain in efficiency and that this efficiency varies much more 
rapidly with the number of submeasurements. These facts are again 
well predicted  by the curve $\omega$.
\paragraph{The $0^{++}$ case}
We now move to the $A_1$ correlator, which gives the mass of the
 lightest glueball. Since this
is the trivial representation, the operator has a non-zero VEV, which has 
to be subtracted in one way or the other in order to extract information on 
the glueball spectrum. With the ordinary 1-level 
algorithm, it is customary to subtract the VEV \emph{a posteriori}:
\ba
C(t) & =& \sum_{t'}~\langle ({\cal O}(t')-\langle {\cal O}\rangle) 
({\cal O}(t+t')-\langle {\cal O}\rangle)  \rangle \nonumber\\
&\equiv& \sum_{t'=1}^{N_t}~\langle {\cal O}(t') {\cal O}(t+t') \rangle
- \frac{1}{N_t} \left(\sum_{t'=1}^{N_t} ~ \langle {\cal O}(t') \rangle\right)^2   
\label{vev}
\ea
This way of proceeding is perfectly applicable to the 2-level algorithm, 
\emph{provided that only those measurements incorporated in
 the 2-point function are included in 
the VEV evaluation}. In other words, exactly the same measurements must 
appear in the second  sum as in the first in Eqn.~\ref{vev}:
\be
C(t) = \sum_{t'\in\Theta_t}~\langle {\cal O}(t') {\cal O}(t+t') \rangle
- \frac{1}{\#(\Theta_t)} \left(\sum_{t'\in\Theta_t} ~ \langle {\cal O}(t')
 \rangle\right)^2 \la{vev2}
\ee
where $\Theta_t$ is a subset of  $\{1,\dots,\hat L_t\}$. It varies with $t$:
depending the time-separation, the measurement of the correlator uses 
different time-slices. It is recommended to store the measurements
in  double precision, since the cancellation between the two sums grows 
with the time-separation.

Our experience is that failing to do the subtraction in this way leads
to a very large variance on the correlator ($30-50\%$ in a typical
run). The explanation is that in this way, one is really measuring, on a 
large but finite set of configurations, the fluctuation of the operator
around its average value \emph{measured on these configurations}.
Naturally, in the infinite statistics limit, both schemes give the 
same answer, but the proposed one benefits from the strong correlation
between the 2-point and 1-point function when they are measured on the same
 configurations.

There are of course many alternative possibilities\footnote{I thank Urs Wenger
for discussions on this point.}. One of them 
relies on the variational method~\cite{Luscher:1990ck}, 
which is widely used to improve
 the projection onto the fundamental state and to extract information on the 
excited spectrum. It was applied for instance 
in~\cite{Meyer:2003wx} and consists 
in feeding  the \emph{unsubtracted} correlation matrices
 into the variational calculation. The generalised eigenvalue problem 
then yields the massless vacuum, followed by the fundamental glueball, the 
first excited, etc. The determination of the vacuum is very accurate in our 
experience, and the variance on the masses of the physical states did not
 seem to be higher. Naturally, one of the operators in the basis is wasted 
to project out the vacuum, but  this is not an issue when
one disposes of a large set of operators, as is usually the case.

Finally, we note that a lattice group~\cite{Majumdar:2003xm}
 has used the 2-level algorithm for compact $U(1)$ 
scalar glueball calculations, where the forward-backward 
symmetric derivative of the correlator was taken. It is clear that at
small temporal lattice spacing, the finite-difference formula can evaluate
the derivative accurately, due to 
the large correlations between time-slices. The idea is thus related to 
that expressed by Eqn.~\ref{vev2}.

These different methods are illustrated on Fig.~\ref{fig:a1}:
 the inverse efficiency of the 1- and 2-level algorithms are plotted 
as a function of $n$. The VEV has been subtracted either by use of 
Eqn.~\ref{vev2} or by applying the variational method to a set of three 
operators (the resulting operator had very large overlap onto the lightest
state in either method, and therefore a  comparison is meaningful).
We see that with either algorithm, the two VEV-subtraction methods perform
equally well. The second observation is that the 2-level algorithm is 
performing poorly here, if $n\geq10$. If we turn to the LEM,
we see that both the derivative-method and the direct VEV-subtraction method
have the same efficiency, once  $n\geq{\cal O}(50)$. 
For $n\leq50$, the VEV-subtraction looks better;
note however that, for discretisation reasons,
 the LEM on the derivative is actually at 4 lattice spacings, rather than 3.5.
%
\section{Glueball calculation in 3+1 dimensions\la{physics}}
One might wonder whether the conclusions reached in the previous section
carry over to 3+1 dimensions, since
the short-distance fluctuations scale differently. 
Here we shall simply present a comparison of efficiency
in a  realistic case of glueball calculations at $\beta=6.0$,  $\beta=6.2$
and $\beta=6.4$, where we can compare our data to that of the 10-year-old
UKQCD data~\cite{Bali:1993fb}. 
The parameters of the 2-level algorithm are $n=40$ for all three values of 
$\beta$, while $\Delta=8$ for $\beta=6.2$ and 6.4, and $\Delta=6$ for
$\beta=6.0$.
Let us focus on the lightest states in the 
$A_1^{++}$, $E^{++}$ and $T_1^{++}$ representations (see Table~\ref{g3+1}). 
We compare the efficiency in terms of the error bars on the
LEMs by scaling the 1-level error bar to the number of 
sweeps done in the run where the 2-level algorithm was implemented (see
Eqn.~\ref{effi}).
The same conclusions hold as in 2+1D: apart from the lightest glueball, 
the efficiency of the 2-level algorithm is greater than that of the 1-level
one, and increases rapidly with the mass of the state.
Admittedly, the comparison to the UKQCD data is less robust, because the 
operators used are not the same and the statistics are quite different. 
The difference in the extent of the time
direction was compensated by scaling up the statistics of the 2-level run.
Still, the same trend is observed as in the comparisons at  coarser
lattice spacings.

Consider the correlator at four lattice spacings. It can be obtained 
by correlating the time slices situated symmetrically around the fixed 
time-slice, or asymmetrically. Naturally, the first way is more favourable.
However, for a very massive state, the measurements are expected to
be very weakly correlated to the fixed BC; therefore the asymmetric
correlator can increase the statistics and reduce the final error bar.
In fact, one can make any mixture of both measurements. If $\bar t$ is the 
time-coordinate of the fixed BC:
\ba
C(t=4)& \propto& \frac{\alpha}{N_{\rm bc}}\sum_{\rm bc}
   \left[{\cal O}(\bar t+1){\cal O}(\bar t-3)
+ {\cal O}(\bar t-1){\cal O}(\bar t+3)\right]\nonumber  \\
&+& \frac{1}{N_{\rm bc}}\sum_{\rm bc}{\cal O}(\bar t+2){\cal O}(\bar t-2)
\ea
The parameter $\alpha$ can be optimised \emph{a posteriori}. We find 
 that the optimal value of $\alpha$ increases with the mass of the state, 
but the dependence on $\alpha$ is weak for $\alpha\geq0.2$.
%
\section{Conclusion}
It is time to summarise what we have learnt about the 2-level algorithm.
We have emphasised the linear dependence of the data size on the number of 
operators; auto-correlations can be checked for easily, and the precise
way in which the correlator is computed can be optimised \emph{a posteriori}.

The optimisation study of the parameters led to the conclusion that
$\Delta\simeq\frac{1}{\sqrt{\sigma}a}$ is a good choice for the separation
of the fixed time-slices. In that case,  the variance of the correlator
decreases exponentially $\sim e^{-mt}$ as long as the number of measurements
at fixed boundary conditions $n>e^{mt}$. As a consequence, longer mass 
plateaux are seen, even for the more massive states. This feature should
help in reducing the systematic bias to overestimate the masses 
being calculated.

Suppose we want to compute the correlator at time-separation $t$
from measurements in time-slices $\bar t+t/2$ and $\bar t-t/2$ with respect
to the fixed time-slice position $\bar t$.
The optimisation of $n$ can be achieved by minimising 
  [the $t=0$ correlator measured $n$ times at distance $t/2$
from fixed time-slices] $\times~n$. This  is an easy quantity to compute
as function of $n$; it is sufficient to  store
the individual measurements  separately. For a fixed physical separation 
$t$, the optimal number of measurements $n$ is only weakly
dependent on the lattice spacing. A possibility that we have not explored
is to let the number of measurements depend on the boundary conditions, 
with a termination condition determined by the desired accuracy (it
would presumably be chosen to be proportional to $1/\sqrt{N_{\rm bc}}$).

The efficiency of the 2-level algorithm was compared to that of the 1-level
algorithm in 2+1 and 3+1 dimensions for different gauge groups. 
We found that the 2-level algorithm performs better
 for all glueball states but the lightest. The kind of gain in computing-time
 one can expect in realistic glueball spectrum calculations
varies between 1.5 and 7 for the lightest states
in the lattice irreducible representations of 2+1D $SU(2)$. 
The gain then increases exponentially with the mass of the state.
If high accuracy is required for the lightest glueball, it might make 
sense to do a separate run using the 1-level algorithm: at any rate, it will
use far less computing time than is required for the heavy states.
The same qualitative statements apply in computations of flux-tube 
masses~\cite{Meyer:2003hy}.
The 2-level algorithm starts to become more favourable at a string length
of $\sim2.5$fm; and it is always more performant for the excited states and 
the  strings of higher representations.

We would like to conclude by mentioning two further applications 
of the 2-level algorithm. As was suggested in~\cite{Meyer:2002cd}, 
the method should be well suited to compute 3-point functions
of glueballs~\cite{Tickle:1990gw} and flux-tubes~\cite{Pennanen:1997qm}, 
since these observables involve 3 factors, each subject to UV fluctuations. 

An alternative to variational calculations
in conjunction with a large number of fuzzy operators is the  spectral
function method~\cite{Asakawa:2000tr} in conjunction with the 
maximal entropy method to perform the inverse Laplace transform.
It would be  interesting to investigate the possibility of 
using  UV operators (e.g. a bare plaquette, which couples equally 
to many states) to extract the glueball spectrum. The
correlator would need to be measured very accurately -- and here we
expect  the 2-level algorithm to be of great help --
on a lattice with a very fine temporal resolution.
We leave this line of research open for the future.
\clearpage
%
\begin{figure}[tb]
\centerline{\begin{minipage}[c]{15cm}
    \psfig{file=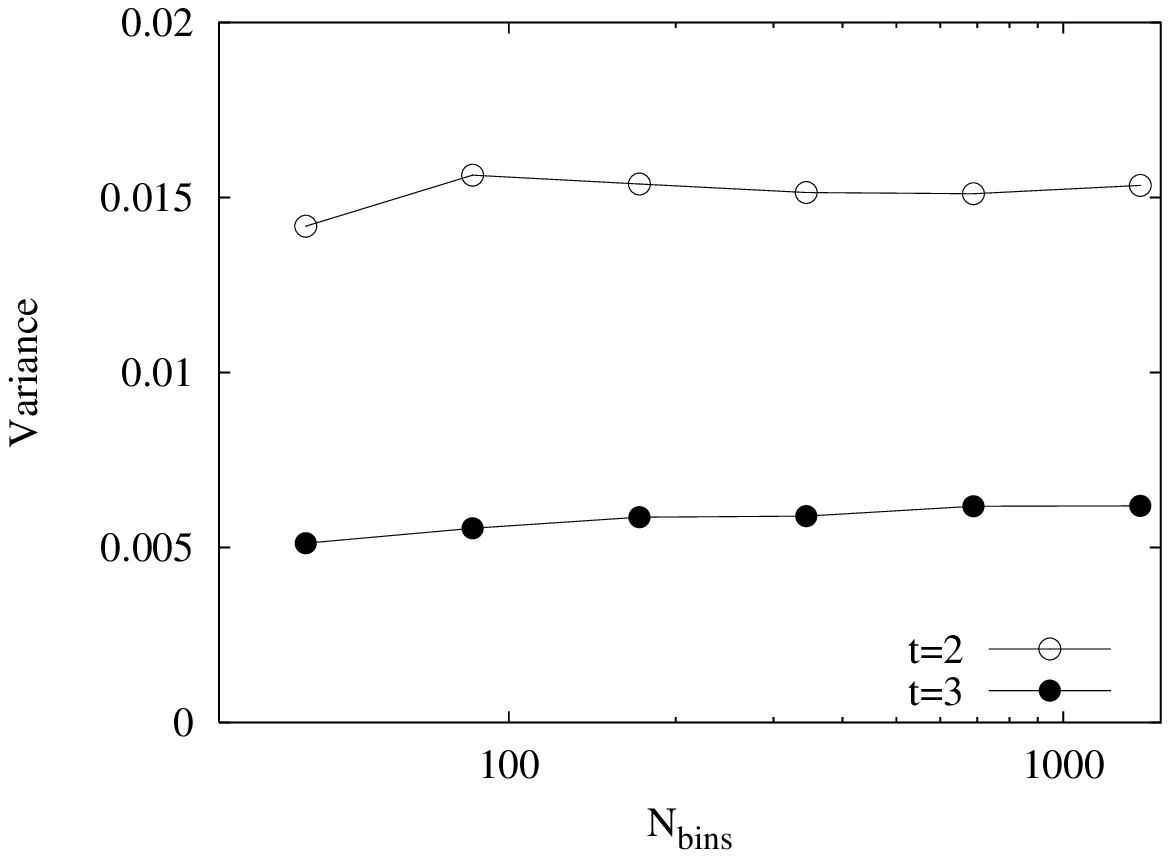,angle=0, width=7.5cm}
	\psfig{file=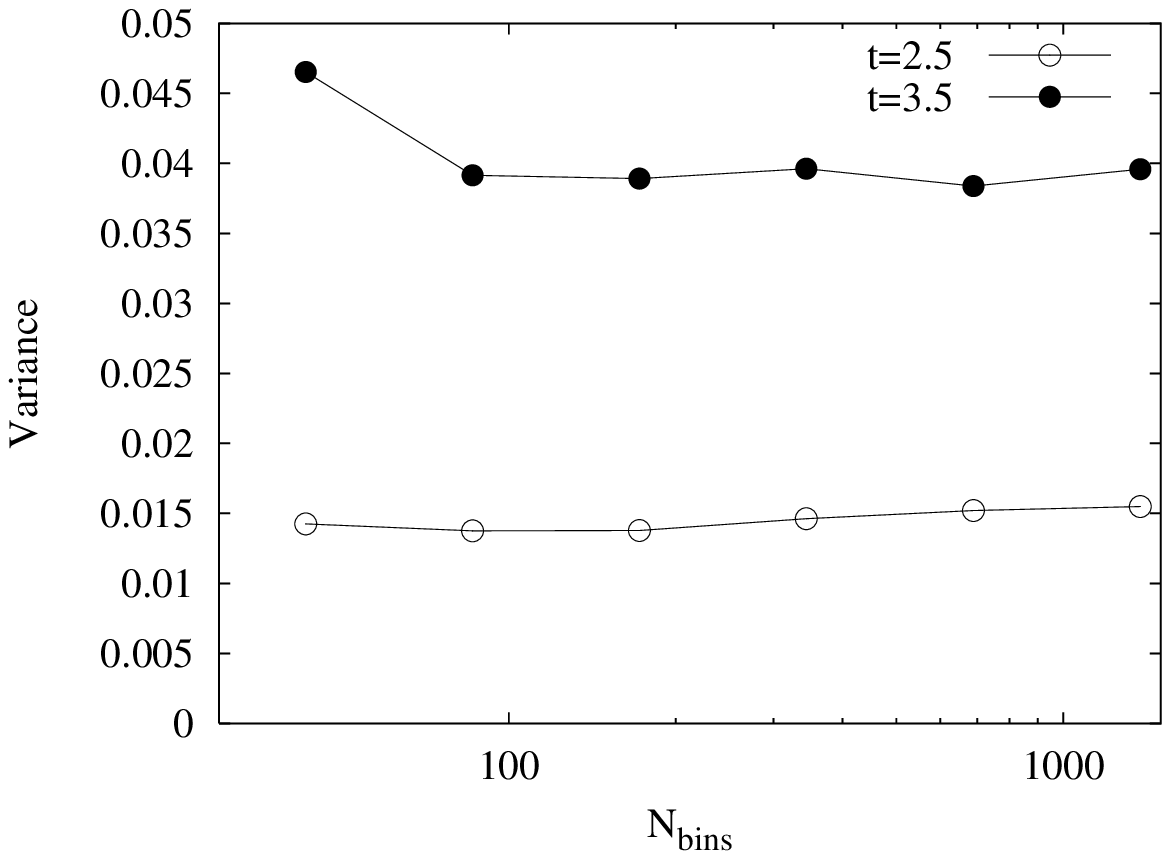,angle=0, width=7.5cm}	
    \end{minipage}}
\vspace*{0.3cm}
\caption[a]{Jacknife-bin-size dependence of the statistical error on 
the $A_2$ correlator (left) and its local-effective-mass (right). The 
separation of the fixed time-slices is  $ \Delta = 4$. For 2+1D $SU(2)$, 
at $\beta=12,~V=32^3$ and $n=100,~N_{\rm bc}=1400$.}
\la{fig:binning}
\end{figure}
\begin{figure}[bt]
\centerline{\begin{minipage}[c]{15cm}
    \psfig{file=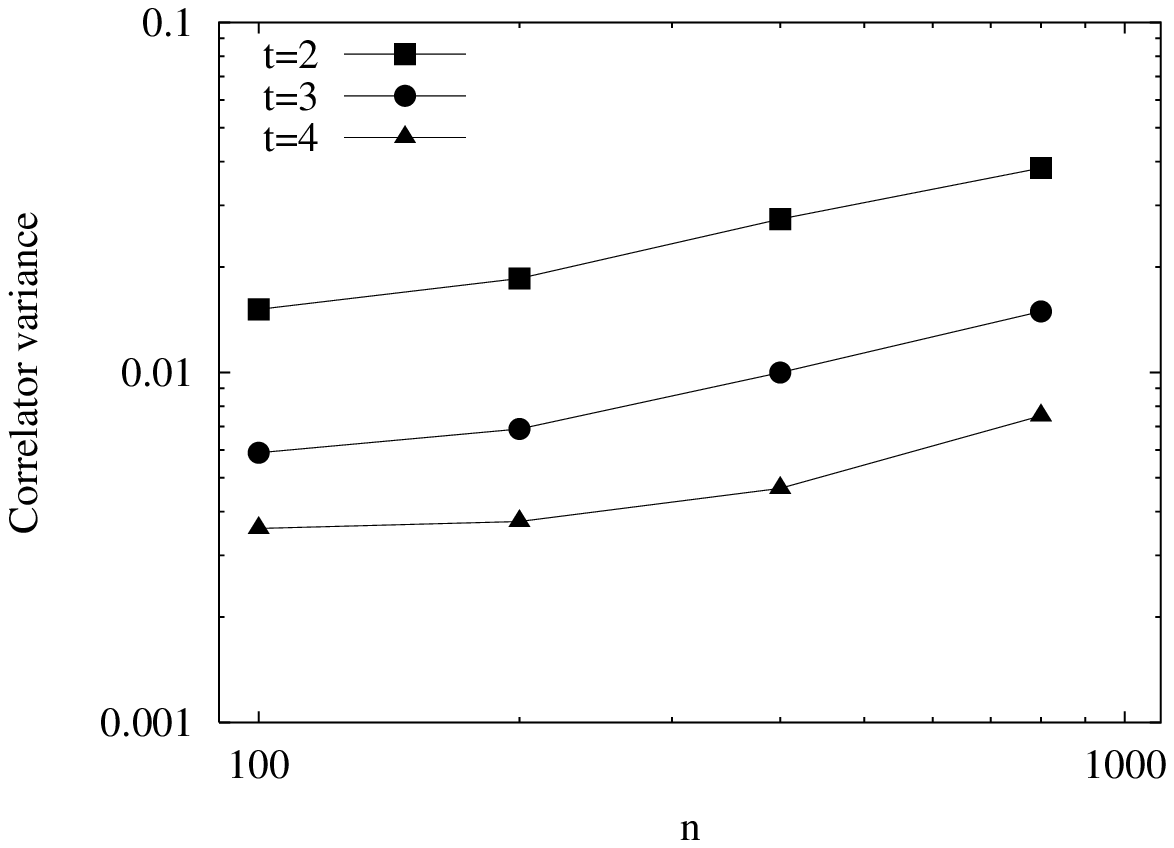,angle=0,width=7.5cm}
	\psfig{file=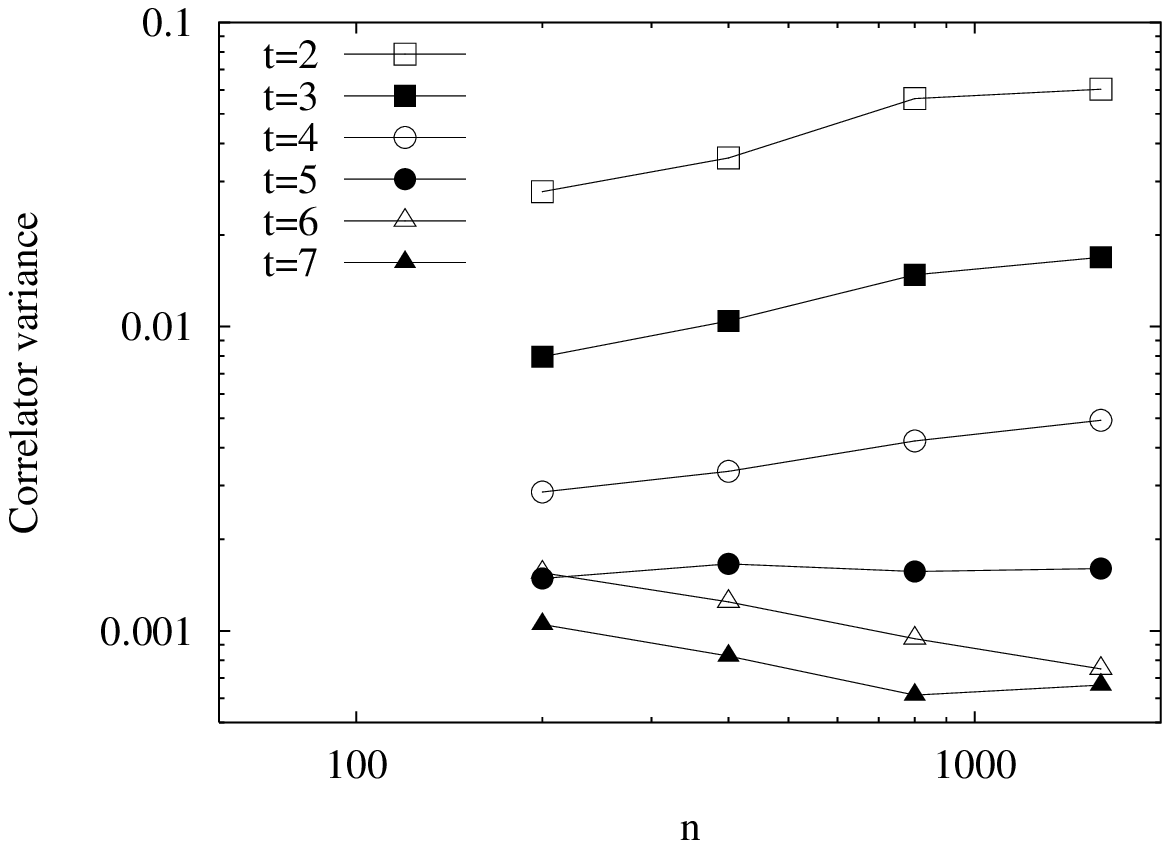,angle=0, width=7.5cm}
      \psfig{file=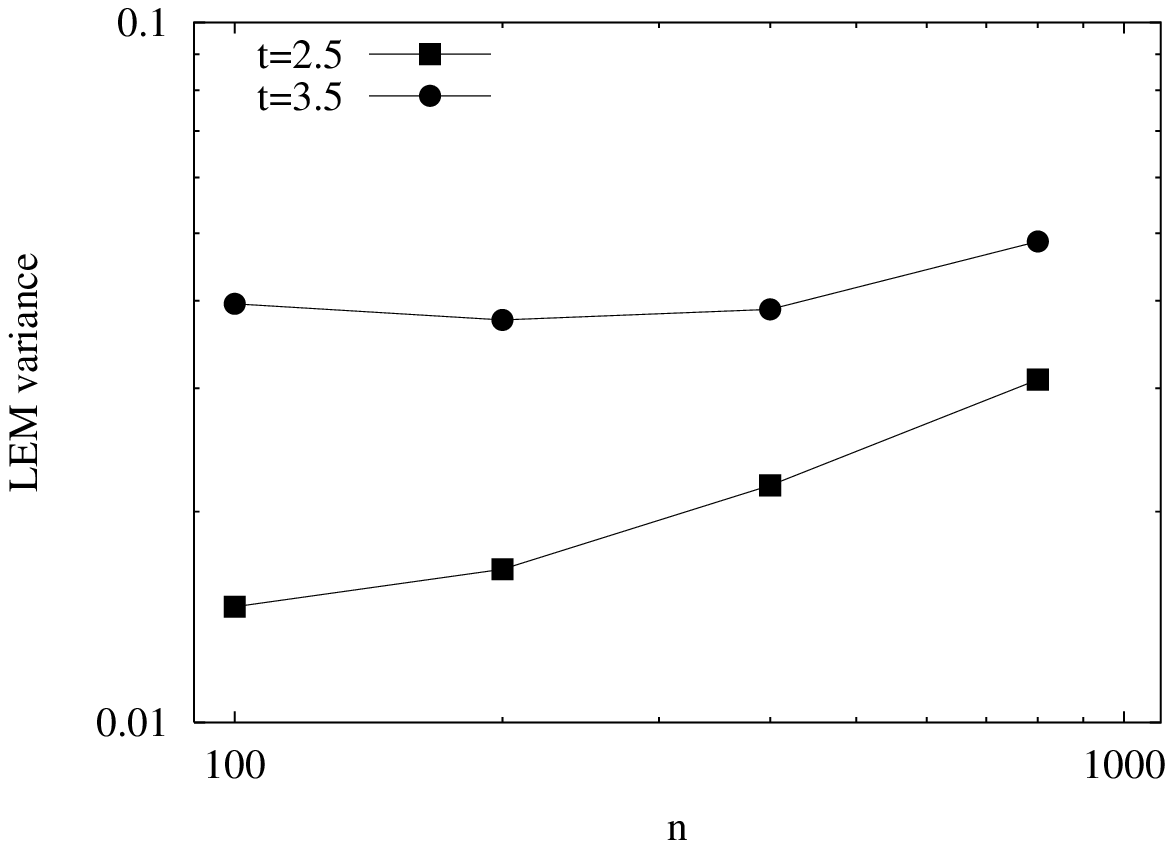,angle=0,width=7.5cm}
	\psfig{file=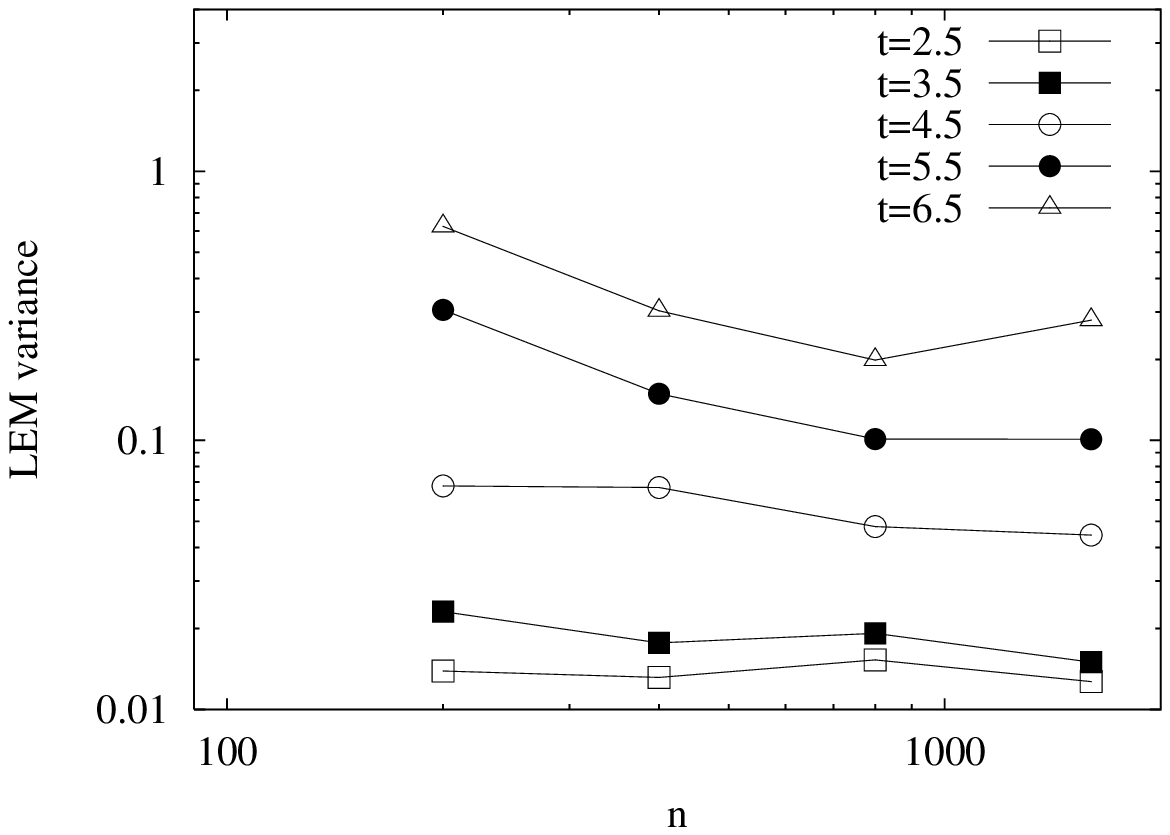,angle=0,width=7.5cm}
    \end{minipage}}
\vspace*{0.3cm}
\caption[a]{The variance of the correlator (top) and the local effective mass
 (bottom), as function of  the number of 
measurements under fixed boundary conditions $n$, for fixed computing
time. The separation of the fixed time-slices is $\Delta  =4 $ on the left
 and  $\Delta  = 8 $ on the right. The operator is a linear combination of
fuzzy magnetic Wilson loops lying in the $A_2$ square lattice irreducible
representation. For 2+1D $SU(2)$, at $\beta=12,~V=32^3$.}
\la{fig:a2_plain}
\end{figure}
\begin{figure}[tb]

\centerline{\begin{minipage}[c]{14cm}
    \psfig{file=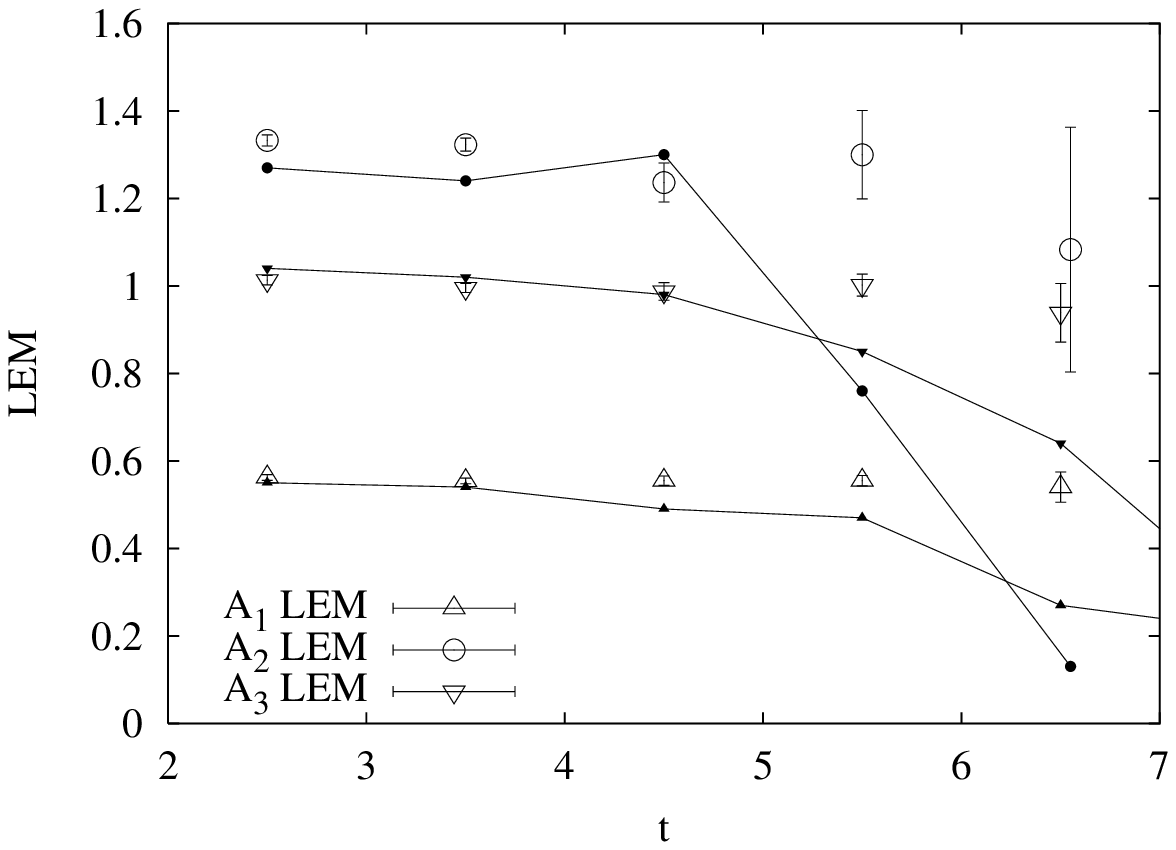,angle=0, height=9cm}\\
	\psfig{file=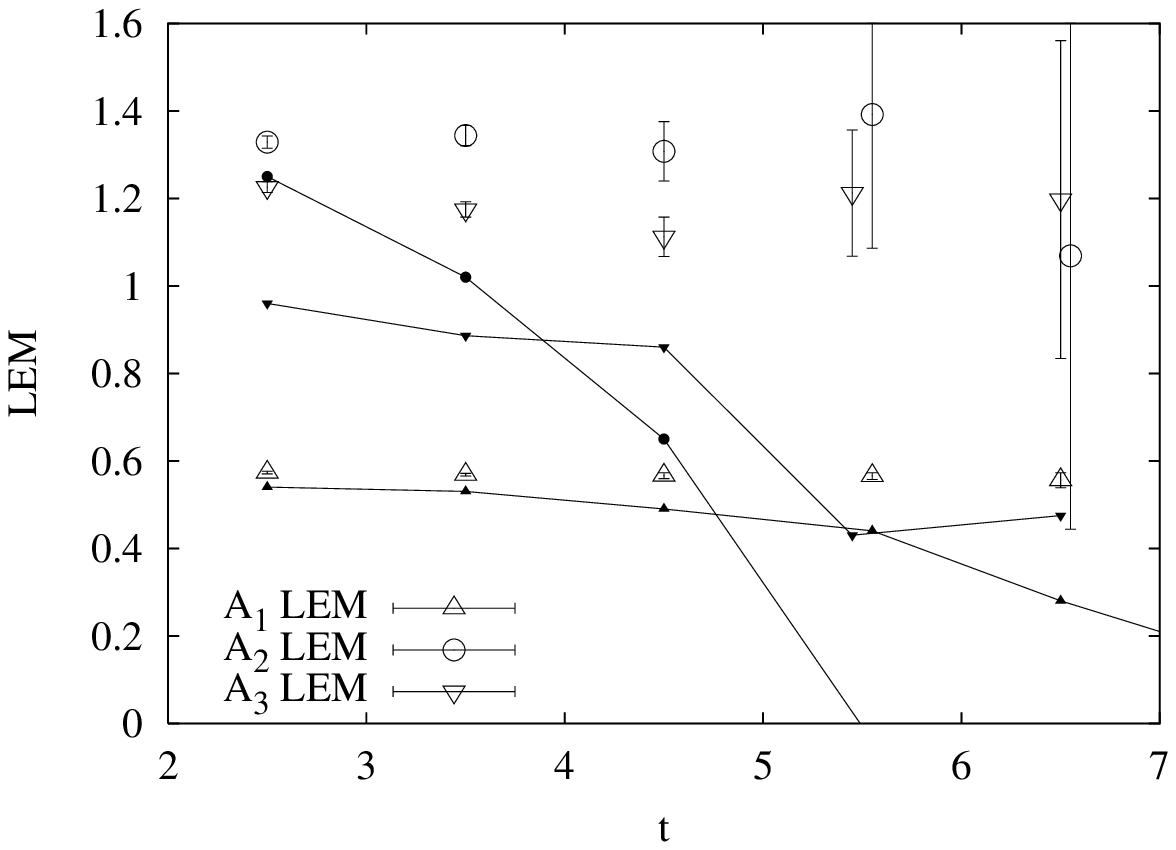,angle=0, height=9cm}	
    \end{minipage}}
\vspace*{0.5cm}

\caption[a]{The local-effective-mass of various correlators (white), and of the 
variance  on the latter (black), as function of the Euclidean-time separation $t$;
the geometric shapes of the data points match.
The distance between fixed time-slices is $\Delta = 8$ and 
the number of measurements under fixed boundary
conditions is $n=1600$ for the top plot and $n=200$ for the bottom plot.
For 2+1D $SU(2)$, at $\beta=12,~V=32^3$.}

\la{fig:LEM}
\end{figure}

\begin{figure}[tb]
\vspace{-2cm}
\centerline{\begin{minipage}[c]{11cm}
	\psfig{file=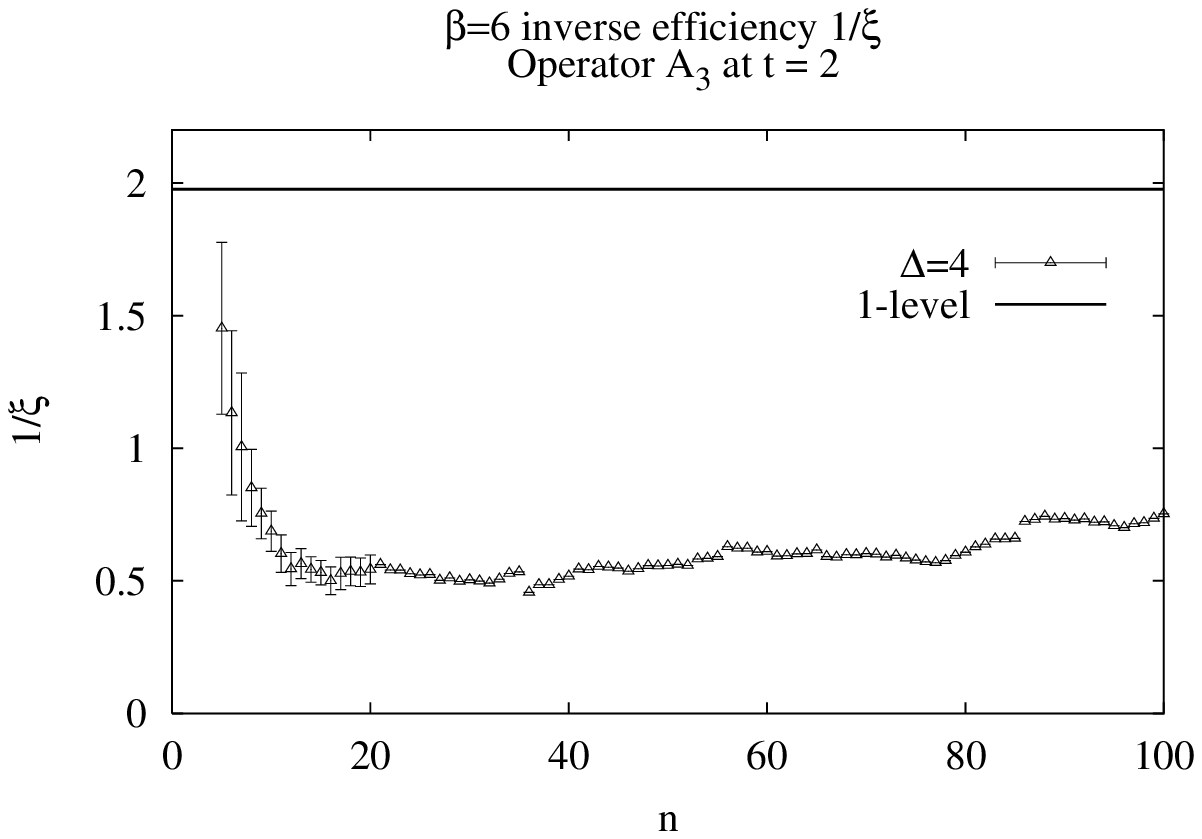,angle=0,width=10cm}\\
\psfig{file=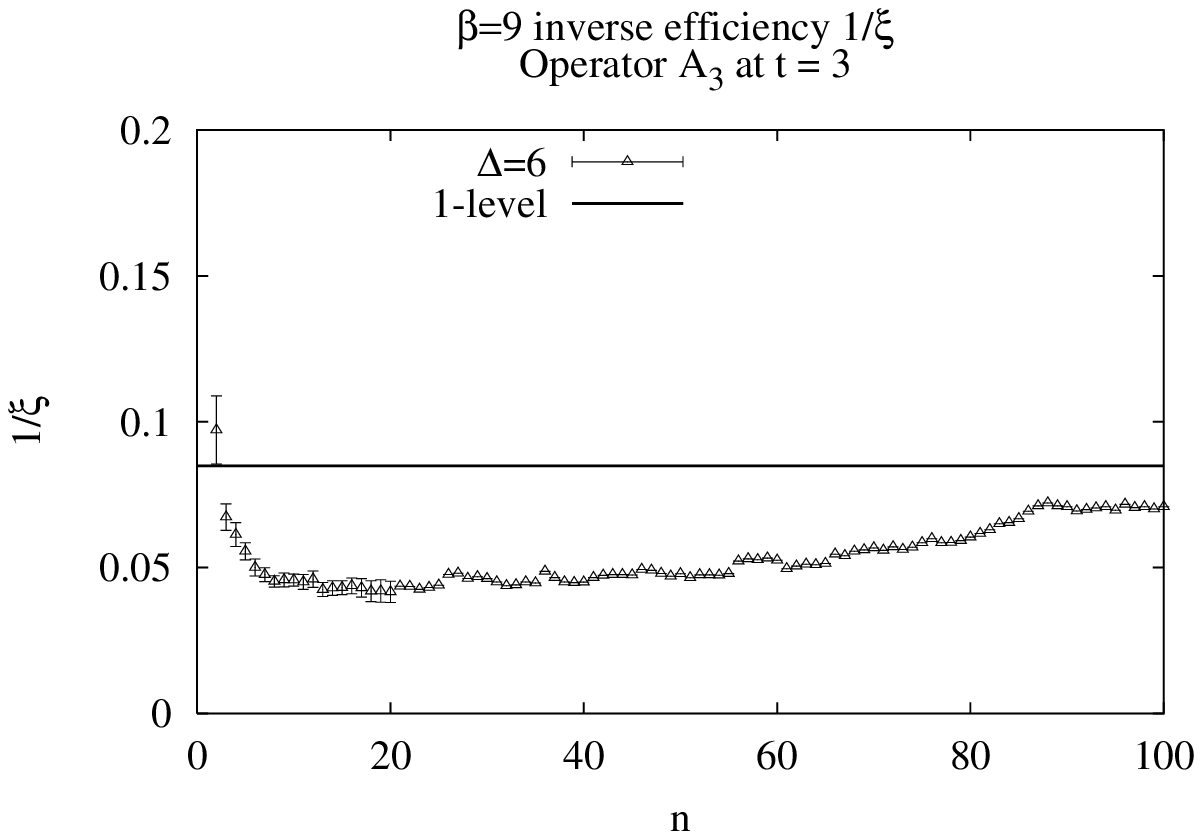,angle=0,width=10cm}\\
\psfig{file=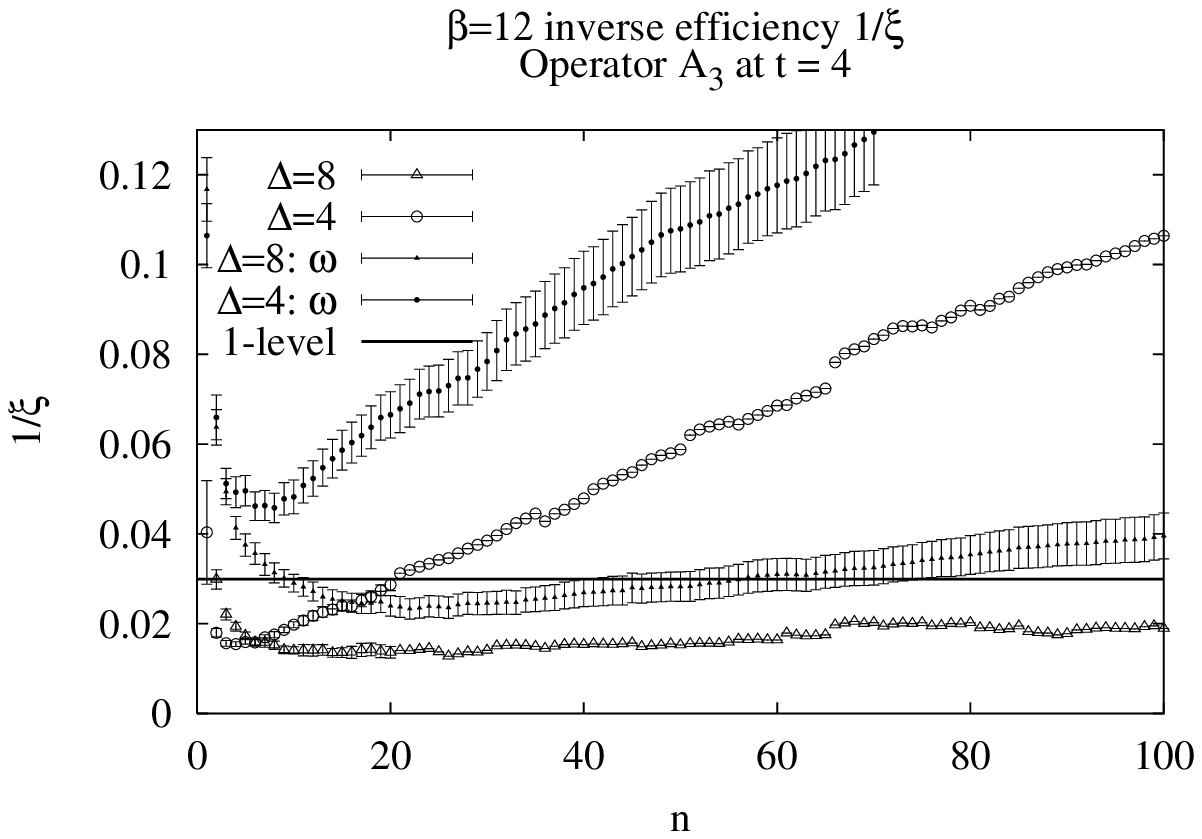,angle=0,width=10cm}
    \end{minipage}}
\vspace*{0.2cm}
\caption[a]{$A_3$ inverse efficiency and its predictor $\omega$ 
in 2+1D $SU(2)$.}
\la{fig:a3}
\end{figure}

\begin{figure}[tb]
\vspace{-2cm}
\centerline{\begin{minipage}[c]{11cm}
\psfig{file=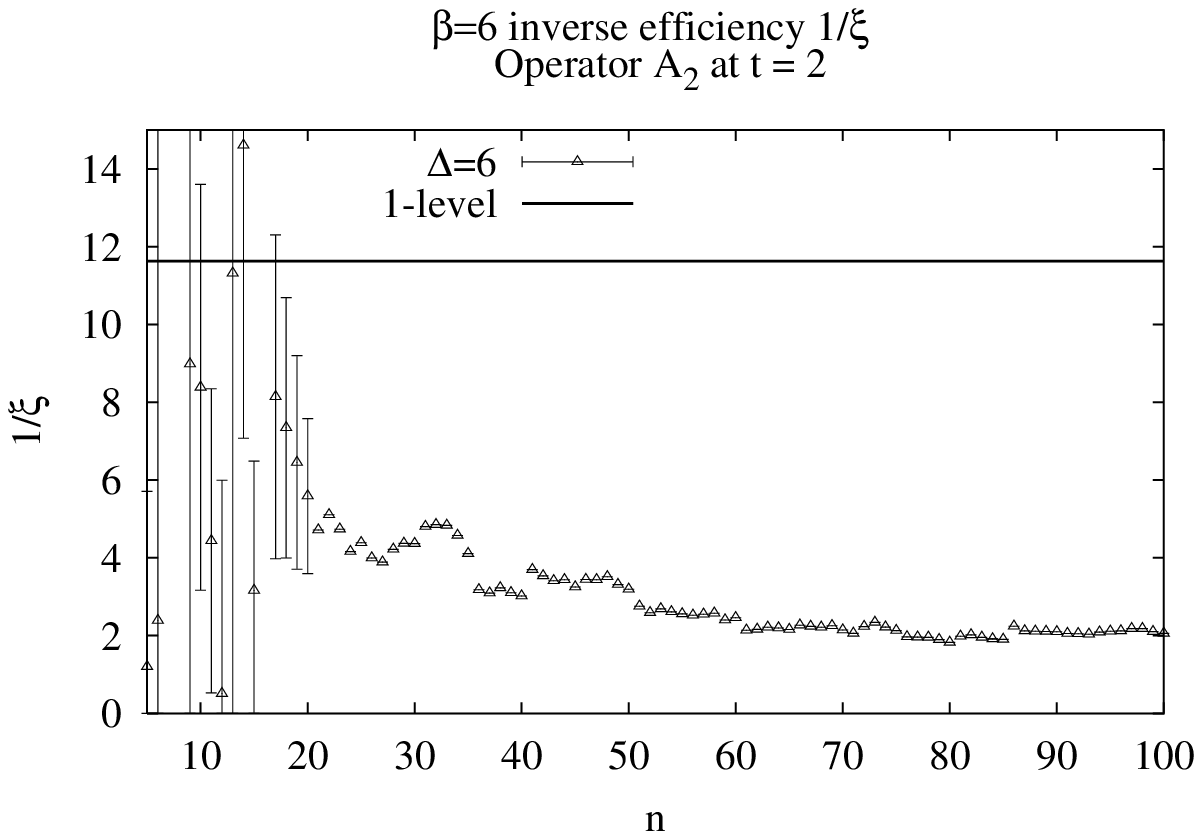,angle=0,width=10cm}\\
\psfig{file=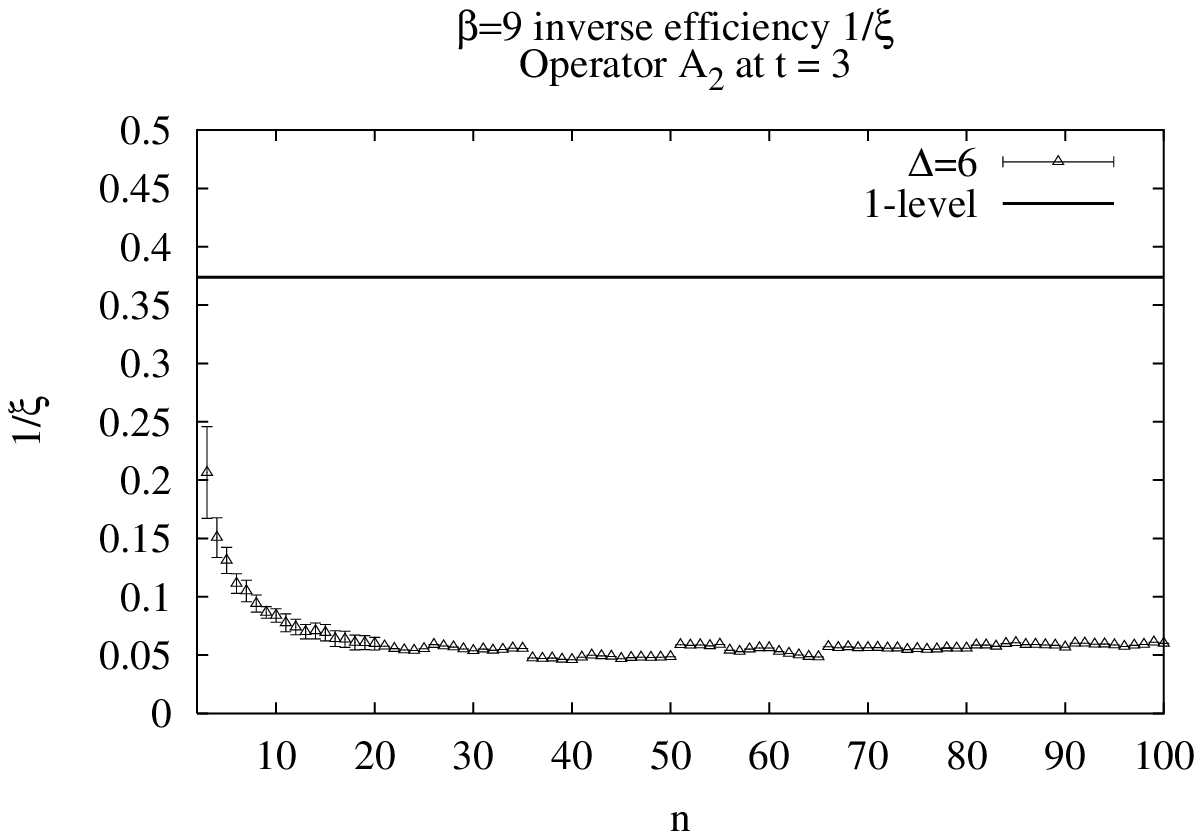,angle=0,width=10cm}\\
\psfig{file=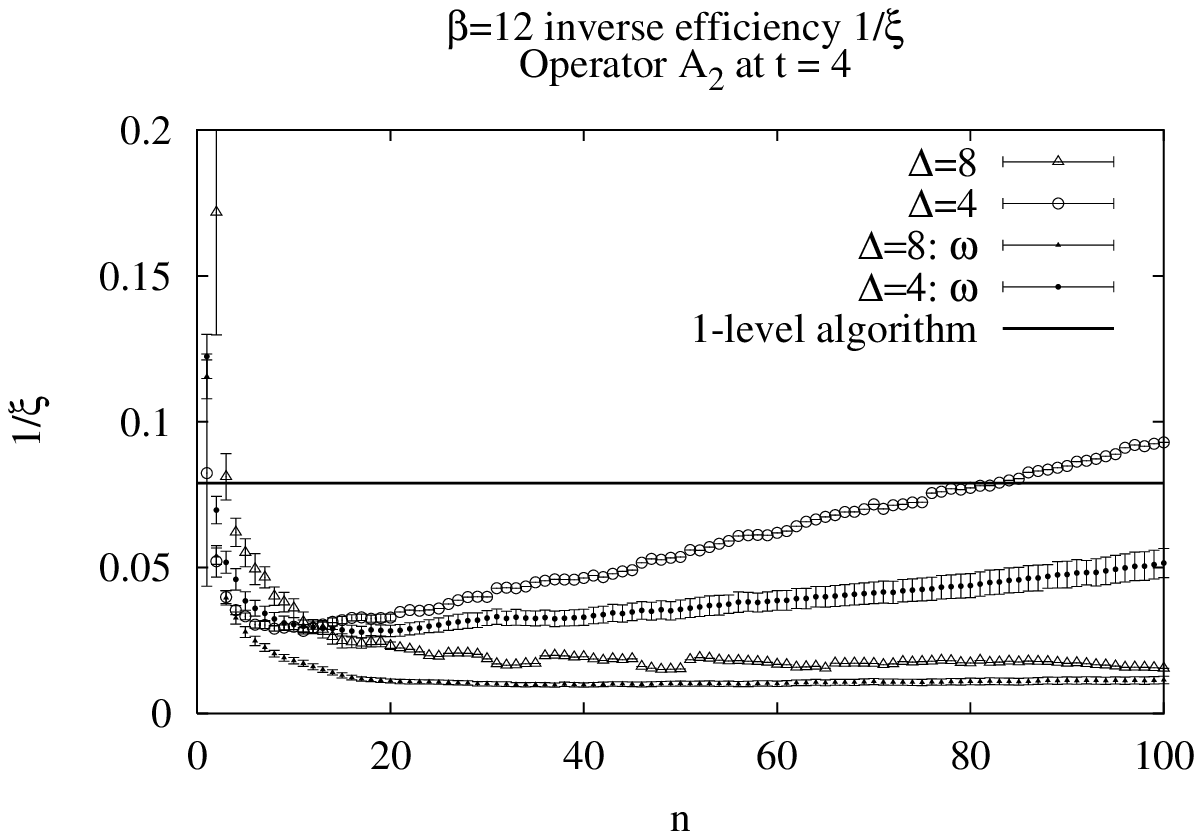,angle=0,width=10cm}
    \end{minipage}}
\vspace*{0.2cm}
\caption[a]{$A_2$ inverse efficiency and its predictor $\omega$ 
in 2+1D $SU(2)$.}
\la{fig:a2}
\end{figure}

\begin{figure}[tb]

\centerline{\begin{minipage}[c]{13cm}
        \psfig{file=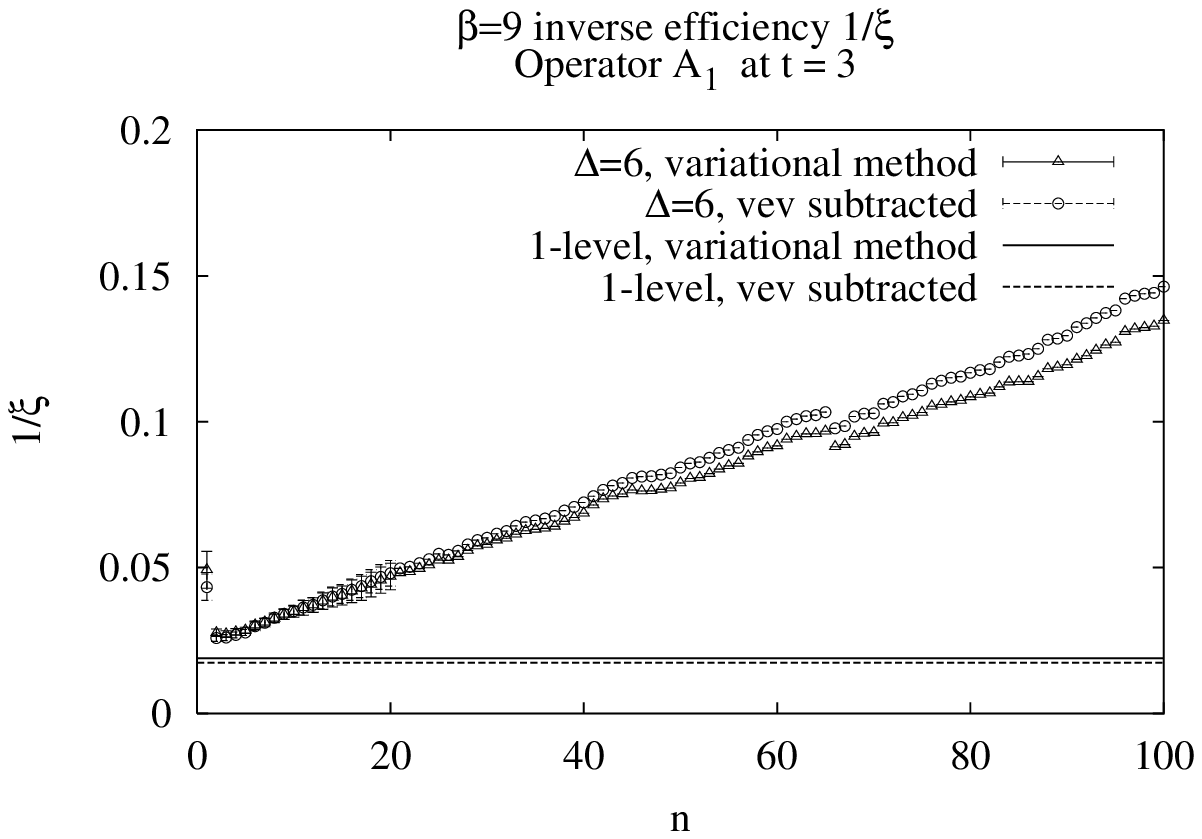,angle=0,width=13cm}\\
	\psfig{file=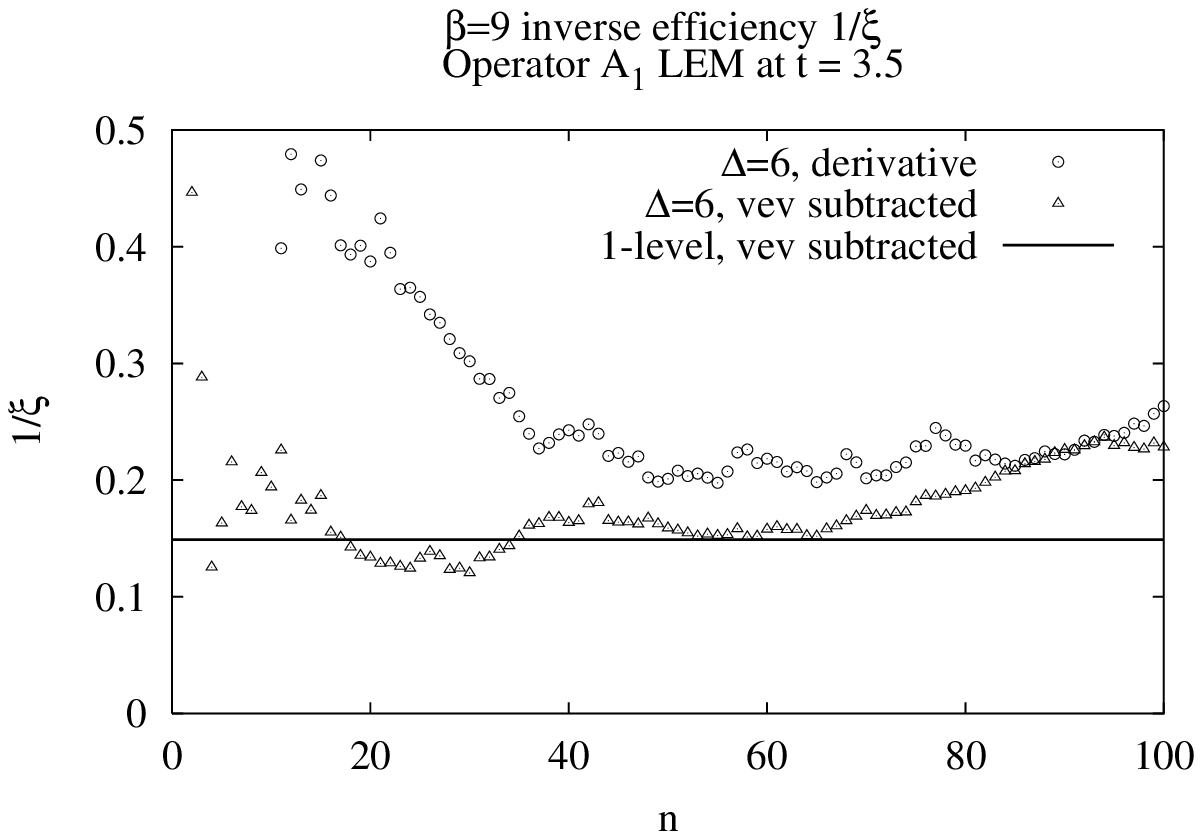,angle=0,width=13cm}
    \end{minipage}}
\vspace*{0.5cm}

\caption[a]{$A_1$ correlator (top) and LEM (bottom) efficiency curves  using
various methods of VEV subtraction. In 2+1D $SU(2)$.}

\la{fig:a1}
\end{figure}

\begin{figure}[tb]
\vspace{-1.0cm}
\centerline{\begin{minipage}[c]{12cm}
\psfig{file=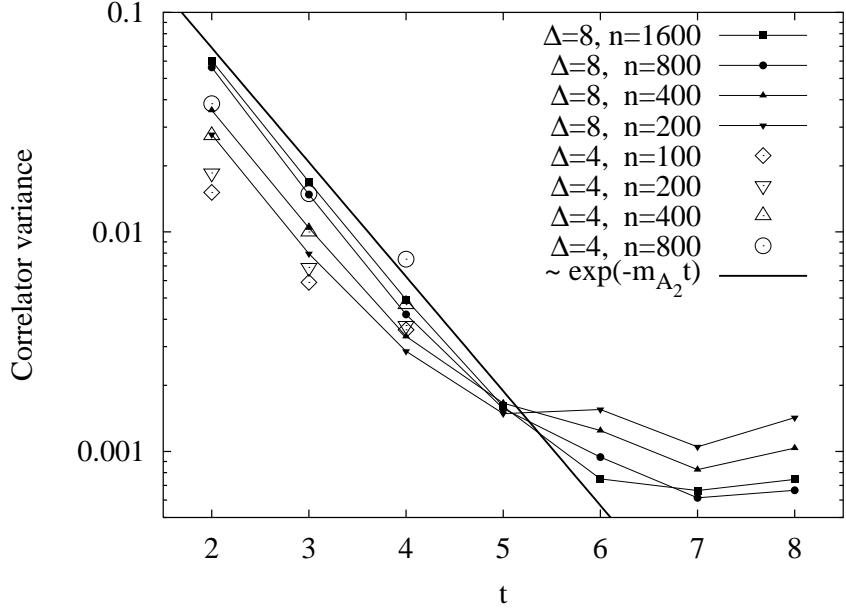,angle=0,width=12cm}
    \end{minipage}}
\caption[a]{$A_2$-correlator variance as function of  the Euclidean-time
 separation $t$, for  different numbers of measurements under fixed
boundary conditions $n$ and separation of the fixed time-slices $\Delta$. 
The computing time is the same for all points.
For 2+1D $SU(2)$, at $\beta=12,~V=32^3$.}
\la{fig:a2_plain_YX}
\end{figure}
\vspace{0.3cm}
\begin{table}[b]
\begin{center}
\begin{tabular}{|c|c|c|c|}
\hline
$\beta=6.0$& $m^{\rm 1-lev}_{\rm eff}(2.5a)$ & $m^{\rm 2-lev}_{\rm eff}(2.5a)$ &
\large{$\frac{\xi_{\rm 2-level}}{\xi_{\rm 1-level}}$ }\\
 $16^3\times36$   & $4.16\cdot10^5$ sweeps&  $15.04\cdot10^5$ sweeps & \\
\hline
$A_1^{++}$& 0.7106(87) & 0.7248(55) &  0.69\\
$E^{++}$  & 1.078(16)   & 1.0776(63) & 1.80 \\
$T_1^{++}$&  1.605(90)  &  1.612(18) & 6.55 \\
\hline
\end{tabular}
\vspace{0.3cm}\\
\begin{tabular}{|c||c|c||c|}
\hline
$\beta=6.2$& $m^{\rm 1-lev}_{\rm eff}(3.5a)$ & $m^{\rm 2-lev}_{\rm eff}(3.5a)$ &
\large{$\frac{\xi_{\rm 2-level}}{\xi_{\rm 1-level}}$} \\
 $24^3\times32$       & $2\cdot10^5$ sweeps&  $9.28\cdot10^5$ sweeps & \\
\hline
$A_1^{++}$ & 0.531(12) & 0.5273(62) &  0.83 \\
$E^{++}$   & 0.768(22)  & 0.7819(64) &  2.46 \\
$T_1^{++}$  & 0.99(15) & 1.250(27)  &  6.60\\
\hline
\end{tabular}
\vspace{0.3cm}
\begin{tabular}{|c||c|c||c|}
\hline
$\beta=6.2$& $m^{\rm 1-lev}_{\rm eff}(2.5a)$ & $m^{\rm 2-lev}_{\rm eff}(2.5a)$ &
\large{$\frac{\xi_{\rm 2-level}}{\xi_{\rm 1-level}}$} \\
  $24^3\times32$       & $2\cdot10^5$ sweeps&  $9.28\cdot10^5$ sweeps & \\
\hline
$A_1^{++}$ & 0.5269(77) &  0.5369(52) &  0.48 \\
$E^{++}$  & 0.8079(99)   & 0.8026(43) & 1.12  \\
$T_1^{++}$  & 1.260(39) & 1.294(11)  & 2.31 \\
\hline
\end{tabular}
\vspace{0.3cm}
\begin{tabular}{|c|c|c|c|}
\hline
$\beta=6.4$ & $m^{\rm UKQCD}_{\rm eff}(2.5a)$~\cite{Bali:1993fb}
&$m^{\rm 2-lev}_{\rm eff}(2.5a)$ &   
\large{$\frac{\xi_{\rm 2-level}}{\xi_{\rm 1-level}}$}   \\
    & $V=32^4$:  $0.322\cdot10^5$ sw 
&$V=32^3\times48$: $1.11\cdot10^5$ sw &  \\
\hline
$A_1^{++}$&0.415(14) &  0.4000(73) &  0.64 \\
$E^{++}$  & 0.620(17)&   0.5894(72)& 1.08\\
$T_1^{++}$& 1.06(8)  &   0.946(10) & 12.4\\
\hline
\end{tabular}
\end{center}
\vspace{-0.5cm}
\caption{Comparison of local effective masses 
 using the ordinary 1-level and the 2-level algorithms in 3+1D $SU(3)$. 
The ratios of efficiencies $\xi$,
representing the inverse ratio of CPU time 
required for fixed accuracy,  is given in the last column. In the last
case, the statistics of the 2-level run were scaled up by 1.5 in the 
efficiency computation to take the different volume into account.\label{g3+1}}
\end{table}

%% file: chapter6.tex
\chapter{Glueball Regge trajectories in 2+1 dimensions}
\label{ch:regge_2d}
We are now in a position  to determine if glueballs fall
on linear Regge trajectories and if so whether the leading
trajectory has the characteristics of the pomeron.  
In this chapter we address the question in the context of
the D=2+1 $SU(2)$ gauge theory. At first glance this may seem
far removed from the case that is of immediate 
physical interest, $SU(3)$ in D=3+1.
Apart from the reduced computational cost, 
 one finds that  D=2+1 non-Abelian gauge theories resemble those in 
D=3+1 in a number of relevant respects. They become free
at short distances, the coupling sets the dynamical
length scale, and the (dimensionless)  coupling becomes 
strong at large distances. They are linearly confining, and
the confining flux tube appears to behave like a simple
bosonic string at large distances~\cite{Lucini:2001nv}. 
We also note that the link
to string theories (at least at large $N_c$) can be made
in D=2+1 just as in D=3+1~\cite{Witten:1998zw}. 
For all these reasons we believe that our 
exercise is of significant theoretical interest.  

At a more heuristic level, one is motivated to search
for a pomeron trajectory where one has high-energy
cross-sections that are roughly constant in energy.
Although the scattering of glueballs has not been observed 
experimentally,
one's intuition is that they will behave as `black disks',
just like the usual mesons and hadrons, and so it makes
sense to speculate that the pomeron might be the leading
(glueball) Regge trajectory in the D=3+1 SU(3) gauge theory.
We do not expect this to depend strongly on the number of
colours, so it should be a property of all $SU(N_c)$ gauge theories.
Finally, since we can think of no obvious reason why going
from 3 to 2 spatial dimensions should prevent
colliding glueballs from having roughly constant cross-sections 
at high energies -- although as `black segments' rather than as
`black disks' -- we believe it makes sense to search for
something like the pomeron in D=2+1 $SU(N_c)$ gauge theories.
Our results will also be used  to test models 
(see Chapter~\ref{ch:string}).

We start by  discussing high-energy scattering in two space 
dimensions (with the details relegated to Appendix~\ref{ap:regge2d}).
In particular we  review perturbative pomeron calculations that
 investigate what happens  when one moves from
3 to 2 space dimensions. 
We then turn to our lattice calculation to obtain quite accurate
continuum extrapolations of the  glueball masses.
We find that the lightest glueballs of even $J$ lie
on a linear trajectory in a Chew-Frautschi plot of
$J$ versus $m^2$, and that the slope is small, just as one
would expect for a pomeron pole. However the intercept
is much too low to provide a constant high-energy cross-section,
and we discuss the physical implications of this result.
Finally we present some results for the leading glueball 
trajectory in $SU(N_c>2)$ gauge theories, showing that there
is no qualitative change as $N_c$ varies from 2 to $\infty$.
%
\section{High-energy scattering in 2+1 dimensions}
\label{section_scatt}
%
\subsection{Regge theory predictions}
\label{subsection_regge}
%
The optical theorem relates the total cross-section to the
 scattering amplitude $A(s,t)$ through Eqn.~(\ref{eq:opt_thm}).
In two space dimensions the elastic scattering amplitude has dimension of energy
and the `cross-section' has dimension of length.
As is shown in Appendix~\ref{ap:regge2d}, 
it receives the following contributions:
\be
A(s,t)=a_0(s)~~+~~{\rm background~integral}~~+~~\sum ~\left[{\rm
Regge~pole~terms}~\propto~\left(s^{\alpha(t)}\right)\right],
\ee
where $\alpha(t)$ describes the Regge trajectory in the Chew-Frautschi plot.
This equality is based on the analytic continuation  of the partial
waves in $\lam$, the angular momentum, and on crossing symmetry. 
There are two differences with 
respect to the 3+1 dimensional case: the background integral gives a constant
contribution to the amplitude, rather than decreasing as $\frac{1}{\sqrt{s}}$;
and the s-wave exchange is not included in the Sommerfeld-Watson transform. 
In potential scattering, and even more general situations, 
$\lambda=0$ can be shown to be a branch point in the complex
$\lam$ plane at threshold (see~\cite{Chadan:1998qm} and Appendix~\ref{ap:regge2d}). 
%
\subsection{QCD$_2$ at high energies}
\label{subsection_QCDhighE}
%
We first give the simplest estimates of the colour-singlet exchange
for high-energy scattering. We then comment on the failure of gluon 
reggeisation and review the results of Li and Tan~\cite{Li:1994et}
for colour-singlet exchange  obtained in the leading logarithmic approximation.
In order to develop some intuition for 2+1 dimensional physics,
we finish with a discussion of the $Q^2$ dependence of hadronic 
structure functions.
%
\paragraph{Colour-singlet exchange in leading order\label{subsubsection_singlet}\\}
If we compute the colour-singlet part of a two-gluon exchange diagram
between two `quarks' in 2+1 dimensions (the Low-Nussinov pomeron, 
see Section~\ref{sec:low-nussinov}), we find
\be 
A_1^{(1)}=i\alpha_s^2s  \frac{N_c^2-1}{N_c^2}\int\frac{dk}{k^2(k-q)^2},
\ee
 implying, by use of the optical theorem in 2+1 dimensions, 
\be 
\sigma_{\mathrm{tot}}(qq\rightarrow qq)=\alpha_s^2 \frac{N_c^2-1}{N_c^2}
 \int \frac{dk}{k^4} \la{low}
\ee
The result is entirely analogous to the D=3+1 case, except that the IR 
divergence is worse by one power -- $\sigma_{\rm tot}$ has units of length.
The pomeron exchange amplitude is finite
 once impact factors are introduced for the hadrons. 

In the dipole formalism~\cite{Mueller:1999yb}, the leading order (large $N_c$) 
dipole-dipole cross-section reads
\be
\sigma_{\rm dd}(d,d')=4\alpha_s^2\int_{-\infty}^\infty \frac{dk_T}{k_T^4}
(1-\cos{k_T d})(1-\cos{k_T d'})=\pi \alpha^2 d_<^3
(3\frac{d_>}{d_<}-1)\qquad(2+1)
\ee
where $d_>$ ($d_<$) is the greater (lesser) of the two dipole sizes $d$
and $d'$. This is to be  compared to Eqn.~(\ref{eq:sig_dipole_4d}).
In both cases, we find a constant cross-section. 

To go beyond the leading contribution, several calculational schemes
 are available.
In particular, the BFKL pomeron is obtained by keeping, order by order
 in $g^2$, only the leading logarithmic contribution in the perturbative
 expansion. The first step in calculating the amplitude for pomeron exchange
is to establish gluon reggeisation.
%
\paragraph{The issue of gluon reggeisation\\}
\label{subsubsection_gluon}
In the Regge limit $s\gg t \gg g^4$, where $s,t$ are the Mandelstam variables,
it is natural to compute the amplitude for colour-octet exchange
in the leading logarithmic approximation.
At least formally, the gluon reggeises~\cite{Ivanov:1998we}:
\be
A^{(8)}=A_0^{(8)}~\left(\frac{s}{k^2}\right)^{\epsilon_G(t)}
\ee
where $A_0^{(8)}$ is the one-gluon exchange amplitude and 
\be
\epsilon_G(t)=N_c\alpha_s\int^{+\infty}_{-\infty}
\frac{dk}{2\pi} \frac{t}{k^2(k-q)^2}\le 0,\qquad (t=-q^2).
\ee
The infrared divergence in the quantity $\epsilon_G(t)$ is linear
 (as opposed to logarithmic in 3+1 dimensions), 
and it must be so since $\alpha_s$ carries dimension of mass. 
A `gluon mass' $M$ has to be introduced, in which case
$\epsilon_G(t)=\frac{N_c\alpha_s}{M}$. Physically $M$ can be interpreted 
as a non-perturbative mass that the gluon acquires at the confining scale; 
therefore we expect $g^2/M=\mathcal{O}(1)$. This however shows that, due to
the infrared divergence, the result of the perturbative calculation has 
a linear sensitivity  to physics at the confinement scale $g^2$, 
where the perturbative expansion breaks down. 

In the Verlinde approach~\cite{Verlinde:1993te} 
to high-energy scattering adopted by Li and Tan~\cite{Li:1994et}, 
gluon reggeisation fails. However, as the authors remark, 
this is not necessarily in contradiction with conventional perturbative
calculations, since the truly physical quantity is the 
colour-singlet exchange.
\paragraph{The 2+1 perturbative pomeron\label{subsubsection_pertpomeron}\\}
The BFKL equation was solved exactly in the presence of a gluon mass in
\cite{Ivanov:1998we}. However when this mass is taken to zero, 
the IR divergence shows up in the fact that the BFKL exponent $\omega_0$ 
runs as $\sim\alpha_s/M$; this fact could be guessed on dimensional grounds.
Within perturbation theory, such a mass $M$ can only appear as an IR 
regulator. The situation is in radical contrast to  the 3+1 dimensional case, 
where the cancellation of  IR divergences in the BFKL equation makes
 it self-consistent.
In the detailed calculation, the simple structure of the infinite series
is spoilt in the $M\rightarrow 0$ limit 
by the re-emergence of a power dependence on $s$ at each 
order due to the IR divergences. Thus, in this framework, 
 a power-like dependence of the cross-section on $s$ in the limit of
 zero gluon mass is not possible in 2+1 dimensions.

A thorough investigation of $QCD_2$ high-energy scattering 
 was undertaken by Li and Tan~\cite{Li:1994et}. In their first 
paper, they used the Verlinde approach~\cite{Verlinde:1993te} to obtain a 
one-dimensional action, where they are able to compute the (finite) 
colour-singlet exchange exactly. They predict a 
$\sigma\propto 1/\log{s}$ dependence
of the total cross-section on the centre-of-mass energy. In a second 
publication, they rederive this result using the dipole picture 
\cite{Mueller:1994rr} of high-energy scattering. 
In this case all quantities are naturally IR-safe.
\subsubsection{Deep inelastic scattering in 2+1 dimensions}
\label{subsubsection_deepinel}
A standard prediction of the BFKL pomeron in 3+1 dimensions 
is the strong rise of the deep inelastic structure functions
as  $x\to 0$ when $Q^2$ is large but fixed (for an introduction, see
\cite{Forshaw:1997dc}): 
\be
F(x,Q^2)\sim \frac{x^{-\omega_0}}{\sqrt{\log{1/x}}}
\ee
where $\omega_0=4\frac{N_c}{\pi}\alpha_s \log{2}$ is the BFKL exponent.

On the other hand,
the DGLAP~\cite{Gribov:ri} equation for the evolution in  $Q^2$ of the
 moments $M(n,Q^2)$ of the parton distributions leads to the behaviour
\be 
M(n,Q^2)=C_n~\left(\log{\frac{Q^2}{\Lambda^2}}\right)^{-A_n}
\ee
where the pure number $A_n$ is an `anomalous dimension' computed in 
perturbative QCD. The $Q^2$ dependence comes from the running of the coupling 
$\alpha_s$; in 2+1 dimensions, the equation is therefore replaced by
\be
\frac{\partial M(n,Q^2)}{\partial \log{Q^2}}=A_n \frac{\alpha_s}{Q} M(n,Q^2)
\ee
yielding the following large $Q^2$ behaviour:
\be
M(n,Q^2)=M(n,\infty)~\exp{\left(-\frac{2A_n\alpha_s}{Q} \right)}\simeq 
M(n,\infty)-2A_n\frac{\alpha_s}{Q}.
\ee
That is to say, the structure functions tend to finite constants at large $Q^2$.
The physical reason for this is that at high energy, the theory 
becomes free very rapidly (the effective coupling scales as  $1/E$), 
and this does not allow for an evolution of the structure functions. Above
the confinement scale, we qualitatively expect a rapid evolution in $Q^2$ 
of the structure function toward its asymptotic value; in other words,
Bjorken scaling becomes exact. Once a high $Q^2$ has been reached, 
the virtual photon $\gamma^*$ does not `see' more partons when
its resolution is increased, because the amplitude that they be emitted is 
suppressed by $\alpha_s/Q$. 
%
\section{The $SU(2)$  spectrum}
\label{section_glueballs}
\subsection{Operators and spin identification}
The operators we use lie in definite lattice irreducible 
representations (IRs), and we use the variational method
\cite{Luscher:1990ck}
to extract estimates for the eigenstates (in our operator basis)
and their masses. In this way we calculate the mass of the lightest
state and of several excited states in the given lattice IR ---
typically the number is one third of the number of operators we 
are using. To identify which $J$ each of these states tends to,
we do a  Fourier analysis of the wave function of the 
corresponding diagonalised operator.
We consider a generalisation of `strategy II' described in 
Chapter~\ref{ch:hspin}: we 
measure the correlations between the glueball operator
 and a set of `probe' operators that 
we are able to rotate to a good approximation by angles smaller 
than $\frac{\pi}{2}$:  $\{P^{(n)}\}_{n=0}^N$, $\< P^{(n)}~P^{(n)} \>=1$.
If the glueball operator $\Psi$ and $P^{(0)}$ belong to  $A_1$,
\ba
|\Psi\> &= &|\psi_0\>~|0\>~+~ |\psi_{4}\>~|\{4\}\>   ~+~\dots \nonumber \\
|P\> &= &|p_0\>~|0\>~+~ |p_{4}\>~|\{4\}\>        ~+~\dots \nonumber
\ea
we get for the probe -- glueball operator correlation
\be
g_P(\phi) = \<p_0|\psi_0\> ~+~  \<p_{4}|\psi_4\>~\cos4\phi.\la{eq:pg_cor}
\ee
Thus the $\cos4\phi$ Fourier coefficient is proportional to $||\psi_4||$.
The precise ratio of  the Fourier coefficients, however, depends 
on the details of the overlap of  $P$ and $\Psi$: a  continuum
extrapolation of these Fourier coefficients is not meaningful.
Nevertheless, if we assume that the breaking of rotational symmetry is
 small, then either $||\psi_0||$ or  $||\psi_4||$ is small. So
 as long as  the overlap between  $P$ and $\Psi$
 is substantial, the dominant Fourier coefficient in (\ref{eq:pg_cor})
still identifies the dominant spin component of the state $\Psi$.
Usually there is 
already a very  dominant coefficient at finite lattice spacing -- 
except when a crossing of states occurs -- as we shall see in an 
example later on.

Our data consists of two sets of simulations. The first set provides the 
states in the $A_1$ and the $A_3$ representations and was published in
\cite{Meyer:2003wx}, 
whereas the more recent data on the remaining lattice representations
($A_2$, $A_4$ and $E$) was partially published in~\cite{Meyer:2003hy}.
The multi-level algorithm was used on all but the smallest 
lattice spacings (i.e. on $\beta=6,~7.2,~9,~12$). 
In the first data set, for $6\le \beta \le 9$, we 
used $n=\mathcal{O}(500)$ submeasurements, while we decreased their number to 50 
at $\beta=12$. These submeasurements are done on sub-lattices 
which represent `time-blocks' of width 4 at the three coarsest $\beta$.  
The second set has higher statistics. The number of sub-measurements is 
5000, 1000 and 800 and the width of the fixed time-blocks is 
4, 6 and 8 at $\beta=6, 9, 12$ respectively.
%
\subsection{Results}
\label{subsection_results}
%
In Table~\ref{tab:g2+1} we list the values of the masses we
calculate on $L^3$ lattices at various values of $\beta=4/ag^2$.
The masses are in lattice units and are labelled both by
the lattice IR to which they belong, and by the spin $J$
of the state to which they tend in the continuum limit.
The latter assignment is achieved as described above,
and an explicit example will be given below.
We  have also calculated the confining string tensions as indicated.
%
\subsubsection{Finite volume effects}
\label{subsubsection_volume}
%
As one can see from Table~\ref{tab:g2+1}, the spatial size that we
use for most of our calculations satisfies $L\surd\sigma \sim 4$. 
This choice was based on earlier finite-volume studies~\cite{Teper:1998te}
where it appeared to be  large enough for the lightest glueball states.
In particular, on such a volume the lightest state of two periodic flux 
loops (which can couple to local glueball operators) will be heavier 
than the lightest few  $A_1$ states and the lightest $A_3$ state. 
In this paper, however, we are interested in higher spin states that 
may be significantly more extended than these lightest states, 
so it is important to check for finite-volume corrections by performing 
at some $\beta$ the same calculations on much larger volumes.
We do this at $\beta=7.2$, where the spatial extent of our comparison 
volume is twice as large.

We see from Table~\ref{tab:g2+1} that there is in fact no significant 
change in any of the masses listed when we double the lattice size
from  $L\surd\sigma \sim 4$ to $L\surd\sigma \sim 8$
at $\beta = 7.2$. In particular this is true for the $J=4$ and $J=6$
states where our concern is greatest. We also note  that on the
$L=40$ lattice a state composed of two periodic flux loops will
have a mass $am_T \sim 2 La\sigma \simeq 3.45$ which is much
heavier than any of the masses listed and so it will not be a
source of finite-volume corrections there. From the comparison
it would appear that these `torelon' states cause no problem on
the $L=20$ lattice even though their mass $am_T \sim 1.7$ is small
enough for it to mix with the states of interest. 

The degeneracy seen at $\beta=9$ between
 the $A_3$ and $A_4$ states (which have $J=2$ mod 4) is a powerful cross-check, 
because torelon pairs do not couple to $A_4$. 
We note that the degeneracy is broken at $\beta=12$. At this lattice spacing
it seems that the excited
$A_3$ states are displaced by the presence of torelonic states, an
effect of the mixing. In particular, the first excited state was not seen 
at $\beta=12$  and it appeared to be very light at $\beta=18$. For these reasons, we
will use  the  $A_4$ data to estimate the first-excited spin 2 state
in the continuum\footnote{This state was also problematic in~\cite{Teper:1998te}.}.
%
\paragraph{The Fourier coefficients \label{subsubsection_fourier}\\}
In Table~\ref{coefdata} we give the Fourier coefficients calculated
at the lattice spacings $\beta=7.2,~9,~12,~18$.
The table shows the normalised $|c|^2$ coefficients corresponding to the 
spin that the state is assigned in the continuum limit. We see that the 
states that become $0^+$  have very isotropic wavefunctions even 
at the finite lattice spacings considered. The spin 4 coefficients of the 
spin-4-to-be states vary a lot more. Let us look at the fundamental spin 4 
glueball in more detail.

The coefficient is very close to one at $\beta=7.2,~9$ and $18$, but shows
a big dip at $\beta=12$. While this could simply 
be due to the choice of the probe, 
we attribute this to the crossing of the lightest
spin 4 state and the $0^{+***}$. Indeed looking at the masses in 
Table~\ref{tab:g2+1}, we observe that these two states are always 
nearly degenerate, the spin $4$ being slightly heavier on the coarse lattices
and slightly lighter on the finer lattices, while they are closest precisely at $\beta=12$. As was pointed out in Chapter~\ref{ch:hspin},
an `accidental' 
degeneracy like this automatically  leads to maximal quantum mechanical 
mixing between the states, since there is no lattice symmetry to
prevent that. Taking this into account, the observed evolution of the
Fourier coefficient is not implausible.
%
\paragraph{Continuum extrapolation\label{subsubsection_continuum}\\}
\begin{table}
\begin{center}
\begin{tabular}{|c|c|c|c|c|c|}
\hline
spin& IR  & $m/\sqrt{\sigma}$ &  $\chi^2/(\nu-2)$ & $\nu$ &$\bar m/\sqrt{\sigma}$\\
\hline
\hline
$0^+$ & $A_1$ &    4.80(10)     & 0.36  & 5 &  4.80(10)  \\
\hline
$0^{+}$ &$A_1$ &  7.22(24)      & 0.46  & 4 &   7.22(24)      \\
\hline
$2$ &$A_3$ &    7.85(15)        & 0.80  &5 & 7.875(76)  \\
 & $A_4$ & 	7.881(76) & 	0.85 & 3  &\\
\hline
2	& $A_4$ & 9.54(12)  & 0.29 & 3 & 9.54(12) \\
\hline
3	& $E$ & 10.84(11) & 1.23 & 3  & 10.84(11) \\
\hline
$4$  & $A_1$ &    9.75(45)      & 0.27  & 5 & 9.96(12)   \\
     &	$A_2$	 & 9.98(12) & 0.99 & 3 &\\
\hline
$4$ &$A_1$ &    12.06(88)      & 1.1  & 3 & 11.76(39)   \\
	& $A_2$ & 11.70(39)& 1.09 & 3 &\\
\hline
$6$&$A_3$&  12.09(40)        & 1.0  &5 &  12.60(20) \\
    & $A_4$ & 12.73(20) & 0.01 & 3  &\\
\hline
\hline
\end{tabular}
\end{center}
\caption{The lightest 2+1 $SU(2)$ glueball states in the continuum limit.}
\la{contdata}
\end{table}
 We extrapolate the masses in units of the string tension according 
to \ref{eq:cont_extrapol} and require that at least three points are used 
in the extrapolation.
We observe as in~\cite{Meyer:2002mk} that the evolution in $a$ is weak.
Table~\ref{contdata} gives the continuum spectrum in units of the 
string tension, as well as the $\chi^2$ and the number of
different lattice spacings included in the fit. For the fundamental
states of spin 0, 2, 4 and 6, the confidence levels are good and 
include all five lattice spacings. 
Not surprisingly, the second and third excited states 
have less reliable extrapolations: here we conservatively only keep the 
best determined ones.
We note that most energy levels in units of the string tension appear to be 
slightly lower at $\beta=18$ than at the other lattice spacings;
this is most likely due to an over-estimation of the string tension.
%
\section{SU($N_c>2$)\label{section_allN}}
As we remarked earlier, it is only in the
$N_c\to\infty$ limit, where all glueballs become stable, that
one can hope to identify the ideal linear Regge trajectory.
In principle all one needs to do is to repeat the
above SU(2) calculation for $N_c=3,~4,~5,..$. We know from
\cite{Teper:1998te}
that the approach to $N_c=\infty$ is rapid so that the first
few values of $N_c$ should suffice for a good extrapolation
to all values of $N_c$. We can use the fact that in $SU(2)$, 
 the lightest state in the lattice $A_2$ IR, which contains
$J^P=0^{-+},~4^{-+},~8^{-+}$, is  the 
$4^{-+}$ rather than the $0^{-+}$ and assume that the ordering
will be the same for $N_c>2$.
The  flux tube model predicts~\cite{Johnson:2000qz}
that the lightest $0^{-+}$ should be much more massive
than the lightest observed $A_2$ state, while the latter is
consistent with the model prediction for the lightest
$4^{-+}$ state. 
Due to parity doubling in D=2+1
this mass is the same as that of the $4^{++}$
(in the infinite-volume continuum limit). Thus we 
can use the lightest states in the $A_1$, $A_3$ and $A_2$
lattice representations, as calculated for various
$SU(N_c)$ groups in 
\cite{Teper:1998te},
to provide us with the lightest $J=0,~2,~4$ glueball masses.

The assumption that  for $SU(N_c>2)$ the lightest  $A_2$ state is 
the $4^{-+}$ is very reasonable but it should be checked.
We have therefore performed such a check in the $SU(5)$ case, 
at $\beta=64$, $L=24$, where $\sigma^{-1/2}\simeq 6a$. 
Using a 16-fold rotated triangular probe operator reveals
that the wave function of our best $A_2^{C=+}$ operator, measured at 
a Euclidean time separation of one lattice spacing,
behaves like $\sin{4x}$.
This confirms the correctness of our assumption.

A similar, but more extensive analysis of the $SU(3)$ spectrum
at $\beta=21$, $L=24$ reveals that the first two states in the 
$A_2^{C=+}$ representation have spin 4. Also, the first two 
states in the $E^{C=+}$ representation have spin 3 (rather than 1).
On the other hand, the sequence in the $E^{C=-}$ representation 
is: $1^{\pm-},~3^{\pm-}$. In the latter representation, a twisted, 
`8' shaped  operator turned out to be the best probe.
This completes the relabelling of the glueball states published in
\cite{Teper:1998te}.
%
\section{Physical discussion}
\label{section_physics}
We begin by asking what our glueball spectrum tells us about
the nature of the leading glueball Regge trajectories, both 
for $SU(2)$ and for larger $N_c$. We then compare what we find
to the predictions of the simple glueball models presented in 
Chapter~\ref{ch:string}.
Finally we discuss what role these trajectories will play
in high-energy scattering. 
%
\subsection{The glueball spectrum in a Chew-Frautschi plot}
\label{subsection_trajectory}
%
In Fig.~\ref{fig:cf_su2_3d} we plot our continuum $SU(2)$ glueball spectrum 
in a Chew-Frautschi plot of ${m^2}/{2\pi\sigma}$ against the spin $J$.
We see that the lightest $J=0,~2,~4$ masses appear to lie on
a straight line. If we fit them with a linear function
$J=\alpha(t)$, where $\alpha(t)=\alpha_0+\alpha't$ and $t=m^2$,
then we obtain  
\be
2\pi\sigma\alpha_{(m)}'=0.327(9)\qquad \alpha^{(m)}_0=-1.21(9)
\ee
with a confidence level of $61\%$. If we drop the $J=0$ state
from the fit, the errors become somewhat larger, but the
trajectory is essentially the same. Thus we reach the remarkable 
conclusion that the lightest glueballs of spin $J$ fall 
on a linear Regge trajectory. This is the leading trajectory, 
hence the index $(m)$ standing for `mother trajectory':
it has the striking feature that only even spins appear on it. 
The position of the spin 6 state hints at a bending of the trajectory,
as one must expect for unstable states. 

We also fit the $0^{+*}$, 
$2^{*}$, $3$ and $4^*$ states to a straight line and find
\be
2\pi\sigma\alpha_{(d)}'=0.288(18)\qquad \alpha^{(d)}_0=-2.3(3)
\ee
with a confidence level of $30\%$. This `daughter' trajectory
is approximately parallel to the leading one, and its
intercept is down by about one unit. It seems to contain all spins
except spin 1.

As we explained in Section~\ref{section_allN}, we can also say 
something about the leading Regge trajectory in $SU(N_c>2)$ gauge 
theories, if we use the masses calculated in
\cite{Teper:1998te}
in conjunction with our relabelling of the spin quantum number.
For instance, the Chew-Frautschi plot for the continuum $SU(3)$ 
spectrum is shown on Fig.~\ref{fig:cf_su3_3d}.
The parity doublets have been averaged and are represented by a
single point on the graph.
A linear fit through the  $0^{+}$, $2^{\pm+}$ and $4^{\pm+}$ states
 works for all $N_c$ and yields:
\begin{center}
\begin{tabular}{|c|c|c|c|}
\hline
\cite{Teper:1998te} data & $2\pi\sigma\alpha_{(m)}'$ & $\alpha^{(m)}_0$ & conf. lev.          \\
\hline
$N_c=2$ &  0.324(15) & -1.150(75) & $89\%$ \\
$N_c=3$ &  0.384(16) & -1.144(71) & $54\%$ \\
$N_c=4$ &  0.374(18) & -1.068(75) & $71\%$ \\
$N_c=5$ &  0.372(22) & -1.036(88) & $86\%$ \\
\hline 
\end{tabular}
\end{center}
We have also included the result for $SU(2)$ and we note that the parameters 
of the trajectory are in very good agreement with the present data.
It is clear that for all the number of colours available, the linear fit 
has a very good confidence level.

We conclude that all $SU(N_c)$ gauge theories possess an approximately linear,
even signature and $C=+$ leading Regge trajectory, 
with slope 0.3---0.4 in units of $2\pi\sigma$, and intercept close to -1, 
which appears to be approaching that value as $N_c\to\infty$.

Equally striking in the $SU(3)$ spectrum are the (near) 
$C=\pm$ degeneracies for states not lying on the leading trajectory:
$0^{++*}$ and $0^{--}$, $0^{++**}$ and $0^{--*}$, $2^{\pm+*}$ and $2^{\pm-}$.
An exception to that is the presence of a relatively light $1^{\pm-}$ state
with no $1^{\pm+}$ counter-part.
The $3^{\pm-}$ (the next state in the $E^{-}$ representation) was not 
extrapolated to the continuum in~\cite{Teper:1998te}, but it appears to be
only slightly heavier than the $1^{\pm-}$ on the smallest lattice spacing;
it could end up being near-degenerate with the $3^{\pm+}$ state. It is tempting 
to group the ($0^{++*}$,  $0^{--}$), ($2^{\pm+*}$, $2^{\pm-}$)
and $3^{\pm+*}$ states into a trajectory, as we did for our $SU(2)$ data.
It would appear that this subleading trajectory now carries $C=\pm$ doublets, 
but is otherwise very similar to the subleading trajectory in the $SU(2)$ spectrum.
%
\subsection{Comparison to glueball models}
\label{subsection_datamodel}
%
As we saw in Chapter~\ref{ch:string}, the flux-tube model of glueballs
predicts a leading Regge trajectory that is linear, with a slope
that is independent of $N_c$:
$2\pi\sigma~\alpha_{FT}'=\frac{1}{4},~ \forall N_c$. 
All spins are present on the trajectory, except spin 1, 
and the states come in $C=\pm$ doublets (except of course for $SU(2)$).
The adjoint string model also predicts a linear Regge trajectory
but with a slope  $2\pi\sigma_a~\alpha_{AS}'=1$ that in general 
depends on $N_c$ through the $N_c$ dependence of $\sigma_a/\sigma$. 
Lattice calculations in D=3+1~\cite{Deldar:1999vi}
and D=2+1~\cite{Kratochvila:2003zj}
support a dependence that is close to Casimir scaling, 
$\frac{\sigma_a}{\sigma} = \frac{C_A}{C_F} =2\frac{N_c^2}{N_c^2-1}$.
Assuming this, the slope predicted by the adjoint string model becomes
3/8 for $SU(2)$, 4/9 for $SU(3)$, and tends toward 2 for $N_c\to\infty$.
It should be remembered that the numerical values for the slopes 
correspond to the asymptotic $J\to\infty$ limit.
Because the adjoint string is unoriented and the gluonic sources are 
bosonic, only even spins are generated by the spinning adjoint string.

What we observe is a leading trajectory with only even spin glueballs and
a sub-leading trajectory which has precisely the features of the flux-tube
phononic trajectory. The picture that emerges naturally is that of a 
mixed spectrum of a spinning open adjoint string and a vibrating
closed fundamental string. The leading trajectory is associated with the 
spinning adjoint string. Perhaps 
 the large mass-offset of the states on the closed-string
trajectory calls for a curvature term (see Chapter~\ref{ch:string}). 
A feature that the flux-tube model does not predict
is the presence of the relatively light  $1^{\pm-}$ state with no $C=+$ partner. 
Such a state is natural (see Chapter~\ref{ch:string})
 if the oriented flux-tube can adopt a twisted, `8' type configuration, which 
is not considered in the flux-tube model.

For all the $N_c$ considered, the lattice result 
for $\alpha'_m$ is almost exactly midway between the
two model predictions. We illustrate this fact in Fig.~\ref{a1}. 
We might speculate that even if both models are valid, 
thus producing two glueball trajectories with different slopes,
at finite $N_c$ mixing will deform these trajectories
from exact linearity and that such a deformation will be
greatest at some lower $J$ where the states of the two
trajectories are closest and also where we perform
our calculations. 
We observe that the intercept of the leading Regge trajectory that
we have obtained is close to -1, and becomes  even closer 
at larger $N_c$, as we see on Fig.~\ref{a0}. 

At finite $N_c$, Regge trajectories are not expected to rise linearly at 
arbitrarily large $t=m^2$. In particular
we should  expect that due to mixing between high spin glueballs
and multi-glueball scattering states, for which
$\alpha(t)\propto \sqrt{t}$,
the local slope of the trajectory decreases as $J$ increases. 
We might be seeing the beginning of such an effect with the spin 6
state in $SU(2)$.
This effect is, however, suppressed by $\frac{1}{N_c}$ in the 
large $N_c$ limit.
%
\subsection{Implications for high-energy reactions}
\label{subsection_highE}
%
The contribution of the leading glueball trajectory to the total cross-section
behaves as $\Delta \sigma\propto s^{\alpha_0-1}$,
which means, given our calculated value $\alpha_0\simeq -1$ , that it is
suppressed as $\sim s^{-2}$. Thus the high-energy
scattering of glueballs is not dominated by Regge pole exchange in 2+1
dimensions; at least if we believe that cross-sections should
be constant at high energies up to powers of $\log{s}$. 

Going back to 
Section~\ref{subsection_regge}, we note that the other terms contributing to
the scattering amplitude are the `fixed-pole' amplitude $a_0$ and the 
`background integral'. Because there is a unitarity bound on 
each partial wave, 
the contribution of any partial wave amplitude to the total cross-section
is bounded by $\sim s^{-1}$. Thus the $s$-wave amplitude will
not dominate either at high energies. That, then, only leaves the 
background integral. If the partial wave amplitude $a(\lam,t)$ were
meromorphic in the region $0<{\rm Re}~\lam < \frac{1}{2}$, 
we would simply get additional Regge pole contributions, which should show
up as physical states by analytic continuation. Therefore there must be 
a more complicated  singularity structure in that region. For instance 
it is well known that $\lam=0$ is a logarithmic 
branch point of the partial wave amplitude
$a(\lam,t)$ at low energies (see \cite{Chadan:1998qm} and 
Appendix~\ref{ap:regge2d}). Also, Li and Tan~\cite{Li:1994et}
remark  that the dipole-dipole forward scattering amplitude
can be written as a contour integral in the complex $\lam$ plane around 
$\lam=0$:
\be
A(d,d',s)=\frac{2\pi g^2 d~ d'}{ N_c}\frac{1}{\log{s}}=
-\frac{2\pi g^2 d~ d'}{ N_c}\int\frac{d\lam}{2\pi i}s^\lam \log{\lam}
\ee
where $d,~d'$ are the sizes of the scattering dipoles; again, the 
logarithmic branch point seems to dominate the scattering process.
This intriguing similarity suggests a universal contribution from the point
$\lambda=0$.
%
\section{Conclusion}
\label{section_conc}
We computed part
of the higher-spin mass spectrum of $SU(2)$ gluodynamics 
in 2+1 dimensions.  We have also
revisited the data published in~\cite{Teper:1998te} and
reassigned the spin quantum number.

Such calculations can tell us what 
the leading glueball Regge trajectory looks like and, in
particular, whether it resembles the pomeron.
Apart from the predictions of string models of glueballs, 
 our motivation for a study in D=2+1 is an intuition 
that in high-energy scattering the colliding 
glueballs should behave like `black 
segments' (analogous to the `black disks' of D=3+1) so
that the cross-section is approximately constant at high $s$.
Of course in D=2+1 we have no experimental support for such an
intuition and we therefore investigated how various field theoretic 
approaches to high-energy scattering can be translated from
D=3+1 to D=2+1. The generic change is that infrared divergences
become much more severe so that one can no
longer predict a power-like dependence
of the cross-section in $s$ directly from the BFKL equation 
\cite{Ivanov:1998we}. However there exist alternative 
calculational schemes~\cite{Li:1994et} which lead to constant
  cross-sections, up to logarithms.

The framework for Regge poles is Regge theory and we saw
that there are qualitative changes when we go from 3 to 2 
spatial dimensions. In particular the $\lambda=0$ partial wave
is not included in the Sommerfeld-Watson transform and the Froissart
bound is $\sigma_{\rm tot}\leq {\rm const.} \log s$.
 We speculate that the singularity structure at $\lambda=0$,
which is not associated with particles of the theory, is
 promoted to a dominant contribution to the high-energy  cross-section.

With this theoretical background in mind, we presented the results
of our lattice calculation of the higher spin glueball spectrum.
Extrapolating our masses to the continuum limit shows that 
the leading Regge trajectory in the (mass)$^2$ versus spin plane
is in fact linear (to a good approximation) and contains only even-spin
states.  Moreover it
has a small slope that lies roughly midway between the large-$J$ predictions
of the flux tube and adjoint string models. The intercept at $t=m^2=0$ is
$\alpha_0\simeq -1$. We identified a parallel daughter trajectory, lying about
one unit of $J$ lower, containing all spins but spin 1.  
After reassigning the spin quantum number in data~\cite{Teper:1998te},
we were also able to determine the two leading
Regge trajectories for other $SU(N_c)$ groups. We
found that the result depends very little on $N_c$ --
except that the subleading trajectory now  contains approximate $C=\pm$ doublets.
A very natural interpretation is that we are seeing
a combined spectrum of a spinning open adjoint string and a vibrating
closed fundamental string, the leading trajectory being associated with the former.
Presumably states produced by either dynamics undergo a certain amount of mixing 
with those generated by the other degrees of freedom. The presence of the 
isolated $1^{\pm-}$ state is a hint at more complicated topologies of the string. 
Finally, we note that
our results are relevant to the theoretically interesting $SU(\infty)$ limit.

The very low intercept of the leading glueball trajectory
($\alpha_0\simeq -1$) indicates that the moving Regge pole corresponding 
to these glueball states gives a negligible contribution to high-energy 
scattering in 2+1 dimensions. We concluded that there must be a 
more complicated singularity
structure of the partial wave amplitude in the complex angular momentum plane
$\lam$. Evidence for a possibly universal branch point at $\lam=0$ comes
mainly from low-energy potential scattering 
(where the result is independent of the potential~\cite{Chadan:1998qm}) 
and is suggested by the $\frac{1}{\log{s}}$ scattering amplitude 
found by Li and Tan in $QCD_2$ high-energy scattering.
These statements are quite different to what one expects in 3+1 dimensions. 
\newpage
\begin{table}
\vspace{-0.5cm}
\begin{center}
\begin{tabular}{|c|c|c|c|c|c|c|c|}
\hline
IR& spin  & $\beta=6$ &$\beta=7.2$ &$\beta=7.2$ &  $\beta=9$  & $\beta=12$ & $\beta=18$ \\
  &      & $ L=16 $  &$ L=20 $  & $ L=40 $  & $ L=24 $ &  $ L=32 $ &$ L=50 $ \\
$\sqrt{\sigma}$ & &$0.2538(10)^*$&0.2044(5)&  0.2072(46)   
 &$0.1616(6)^*$&$0.1179(5)^*$   & 0.0853(14)\\
\hline
\hline
$A_1$& $0^+$ &  1.198(25) &0.981(14)& 0.951(14)  &0.7652(78)  &  0.570(11)& 0.3970(78) \\
&$0^{+}$&  1.665(43) &1.396(21) & 1.394(18) &1.108(23) &0.847(18) & 0.584(32)   \\
&$0^{+}$& 2.198(76) &1.859(25)&1.778(34)  &1.426(37)   &0.980(28)& 0.717(76)   \\
&$0^{+}$&2.27(10)  &2.084(41) & 2.067(54) &1.522(36)   &1.226(17) & 0.845(37)   \\
&$4$ &    2.44(27)  &2.07(33)& 2.146(64)   &1.570(39)   &1.195(47) &  0.798(32)   \\
&$4$ &          &2.53(13) &   2.50(14) &1.700(52)   &   1.419(90) & 0.963(45)  \\
\hline
$A_3$&$2$ & 1.957(48) &1.584(18)&1.567(18)  & 1.232(38)   &0.933(11)&  0.634(18)   \\
&$2$&  2.08(18)  &1.870(37)  & 1.891(39)   & 1.421(44)   & /   &  0.667(20)\\ 
&$2$& 2.34(25)  &2.219(90)  &  2.242(77) & 1.660(54)   &1.152(42)  &  0.862(14)   \\
&$2$&2.65(29)  &2.451(71) &  2.47(12) & 1.746(56)   &1.459(29)  &   1.019(92)   \\
&$6$&   2.93(21)  &2.51(19) & 2.64(15) & 1.878(86)   &1.438(28)  & 0.906(69)   \\
\hline
\hline
\end{tabular}
\vspace{0.5cm}

\begin{tabular}{|c|c|c|c|c|}
\hline
IR    & spin & $\beta=6$ & $\beta=9$ & $\beta=12$  \\
\hline 
\hline
$A_2$ & 4    & 2.423(34)   & 1.5766(96)  & 1.180(17)    \\
      & 4    & 2.638(85)   & 1.772(44)   & 1.384(50)    \\
\hline 
$A_4$ & 2    & 1.908(48)   & 1.2372(73)  & 0.9191(33)  \\
      & 2    & 2.04(13)    & 1.4619(73)  & 1.0920(51)  \\
      & 2    & 2.310(36)   & 1.685(22)   & 1.2568(73)  \\
      & 2    & 2.521(62)   & 1.777(39)   & 1.321(10)  \\
      & 6    & 2.780(65)   & 1.932(84)   & 1.456(16)  \\
\hline 
$E$   & 3    & 2.520(60)   & 1.6732(97)   & 1.2547(54)   \\
\hline
\hline 
\end{tabular}
\end{center}
\vspace{-0.3cm}
\caption{The lightest states in the 2+1D $SU(2)$ gauge theory. 
The string tension values with an asterisk are taken from~\cite{Teper:1998te}.
The data for the $A_1$ and $A_3$ representations was published in~\cite{Meyer:2003wx},
the  data  for the $A_2$, $A_4$  and $E$ representations is more recent and unpublished.}
\label{tab:g2+1}
\end{table}
\vspace{0.7cm}

\begin{table}
\begin{center}
\begin{tabular}{|c|c|c|c|c|c|}
\hline
IR & spin    & $\beta=7.2$    & $\beta=9$ & $\beta=12$ &  $\beta=18$  \\
\hline
\hline
$A_1$   & $0^+$     &  1         &  1 & 1        & 1      \\  
	& $0^+$  &  1         &  1 & 1        & 1      \\
	& $0^+$ &  1         &  1 & 1        & /      \\
	& $0^+$&  0.59(12)  & 0.68(62)  & 0.97(13) &  1       \\
	& $4$     &  0.94(9)   & 0.98(4) & 0.38(2)&   0.95(2)     \\
	& $4$  &    /        & 0.67(16) & 0.59(4)&   0.98(3)  \\
\hline
$A_2$	& 4  &  	&		&  1  & \\
	& 4  &		&		& 0.98(2)  & \\
\hline
$A_3$ 	& $2$ &      1   &  1         &        1 &  \\ 
	& $2$&    1   &  1         &        1 &  \\ 
	& $2$&   1   &  1         & 0.97(11) &  \\ 
	& $2$&  1   &  1         &        1 &  \\ 
	& $6$&     /   &  0.87(8)  & 0.88(4)&  \\
\hline
$A_4$ 	& 2 &   	&		&  1  & \\
	& 2 &   	&		&  1  & \\
	& 2 &   	&		&  /  & \\
	& 2 &   	&		&  /  & \\
	& 6 &   	&		&  0.19(2)  & \\
\hline
$E$ 	& 3 &		&		&   0.96(2) &\\
\hline
\hline
\end{tabular}
\end{center}
\vspace{-0.3cm}
\caption{The Fourier coefficients of the spin $J$ states given in 
Table~\ref{tab:g2+1}: $|c_J|^2$  at $\beta=~7.2,~9,~12$ and 18. 
When the coefficient is larger than 0.99, we round it off to 1.}
\la{coefdata}
\end{table}
\begin{figure}[bt]
\vspace{-1cm}
\centerline{\begin{minipage}[c]{14.5cm}
    \psfig{file=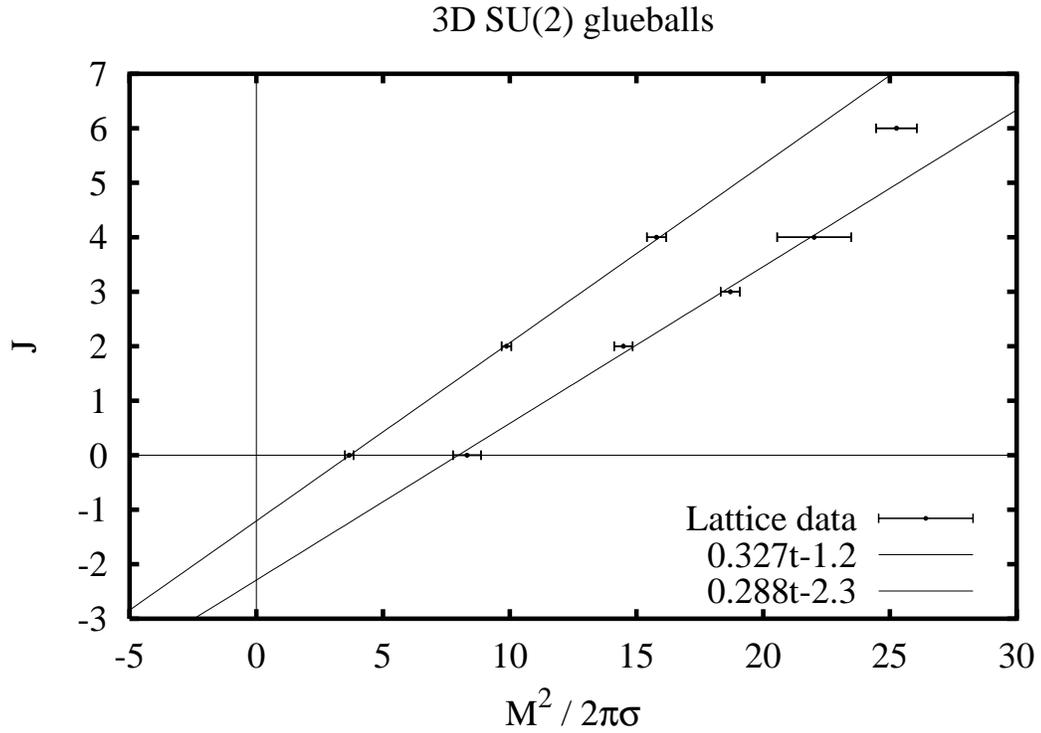,angle=0,width=14.5cm}
    \end{minipage}}
\vspace*{0.2cm}
\caption[a]{The Chew-Frautschi plot of the continuum D=2+1 $SU(2)$ 
glueball spectrum.}
\la{fig:cf_su2_3d}
\end{figure}
\begin{figure}[bt]
\vspace{-0.8cm}
\centerline{\begin{minipage}[c]{14.5cm}
    \psfig{file=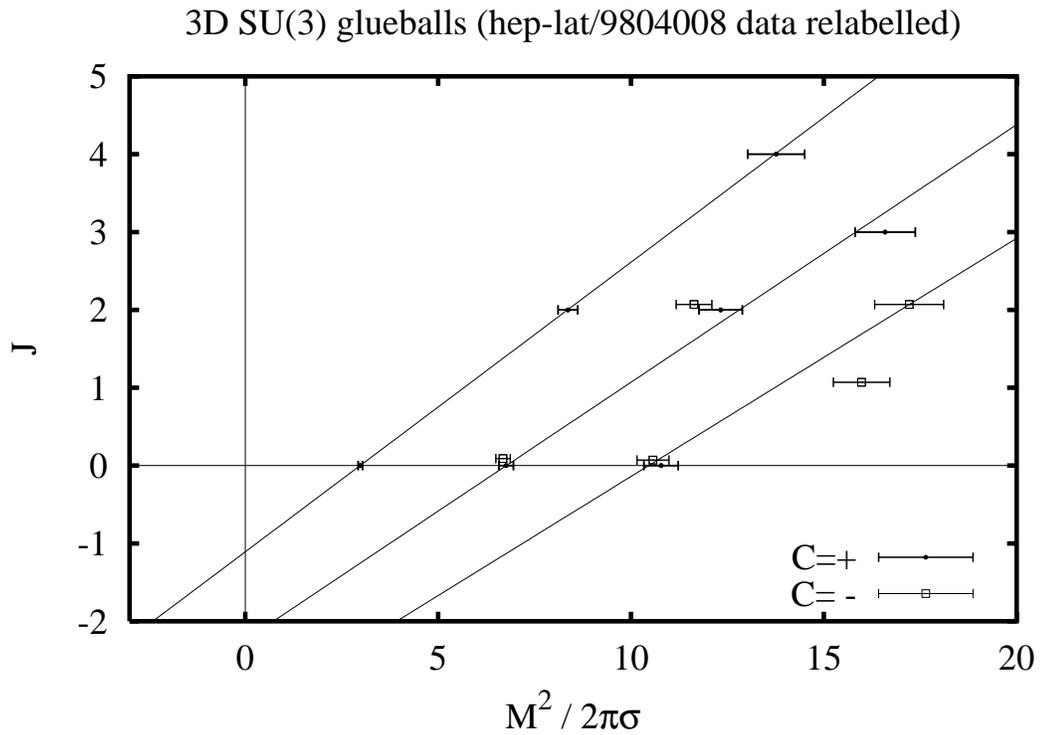,angle=0,width=14.5cm}
    \end{minipage}}
\vspace*{0.2cm}
\caption[a]{The Chew-Frautschi plot of the relabelled continuum D=2+1 $SU(3)$ 
glueball spectrum obtained in~\cite{Teper:1998te}.}
\la{fig:cf_su3_3d}
\end{figure}
\begin{figure}[bt]

\centerline{\begin{minipage}[c]{13cm}
    \psfig{file=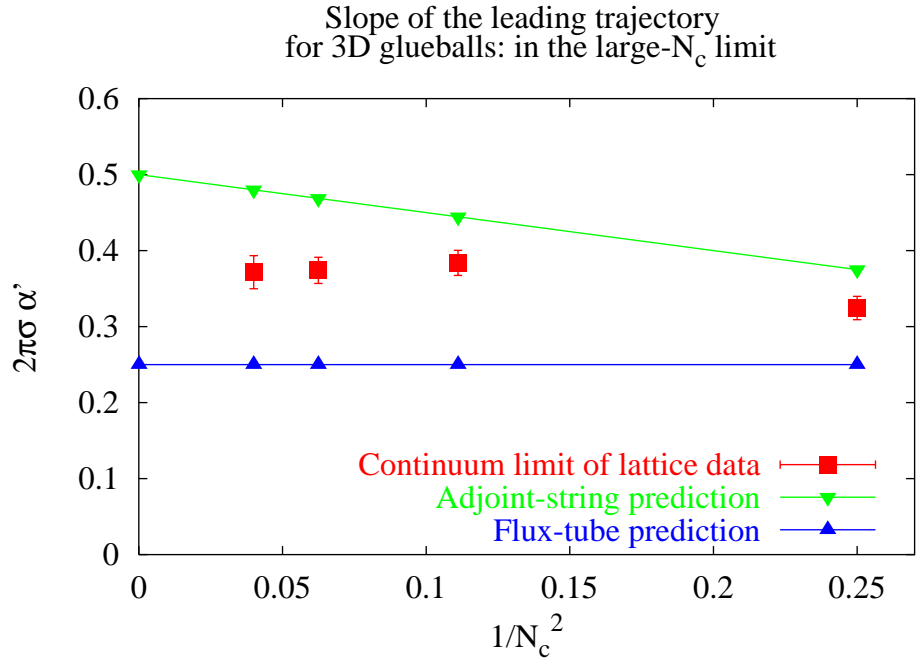,angle=0,width=13cm}
    \end{minipage}}
\vspace*{0.5cm}
\caption[a]{The slope $\alpha'$ of the leading Regge trajectory in 2+1 $SU(N_c)$
gauge theory, in units of
$\frac{1}{2\pi\sigma}$, as a function of $\frac{1}{N_c^2}$}
\la{a1}
\end{figure}
\begin{figure}[bt]

\centerline{\begin{minipage}[c]{13cm}
    \psfig{file=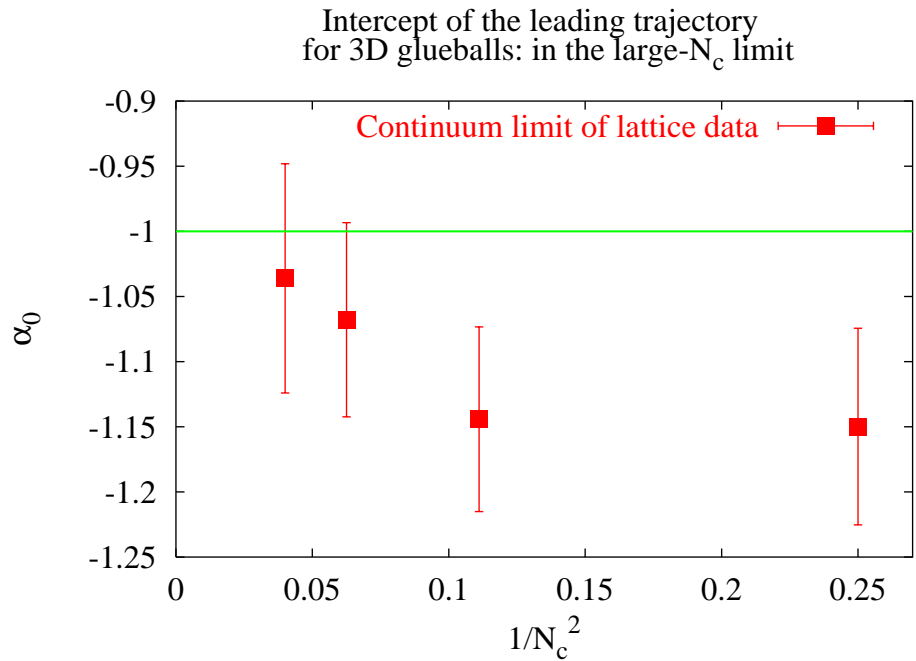,angle=0,width=13cm}
    \end{minipage}}
\vspace*{0.5cm}
\caption[a]{The  intercept of the leading Regge trajectory, 
as a function of $\frac{1}{N_c^2}$ in 2+1 $SU(N_c)$ gauge theory.}
\la{a0}
\end{figure}

%% file: chapter7.tex
\chapter{Glueball Regge trajectories in 3+1 dimensions\label{ch:regge_3d}}
We  investigate the spectrum of $SU(3)$ gauge theory in 3+1 dimensions 
 and its relation to the pomeron. We shall also present
a comparison with the spectrum of $SU(8)$ gauge theory, and argue that the 
latter is representative of the planar limit.

We first explain some lattice technicalities, and then show how our spin
identification technique generalises to three space dimensions. 
We present our lattice data, 
along with a finite-volume study and a first attempt to understand
the influence of double-torelon and scattering states on the extracted spectrum.
An interpretation of the latter in terms of Regge trajectories is given.
In particular, a pomeron-like trajectory is identified and its physical
consequences at negative $t$ are discussed. 
The general conclusion is postponed to the next chapter.
\section{Lattice technology\la{techni}}
In this section we present the details of our operator construction 
procedure, as well as the strategy to label the glueball states with
the correct spin quantum number.
\subsection{Operator construction\la{sec:probe}}
It is well-known that the overlap of operators onto the lightest
physical states of the theory can be greatly enhanced by methods
such as smearing~\cite{Albanese:1987ds} and blocking~\cite{Teper:1987wt}.
We used several levels of the  iterative procedure described below.
Each step produces out of the links of the previous step a link
that is `fatter' and doubled in length. Thus a Wilson line with
an index $(B)$ represents a line of length
$2^B$ times the initial length. The operations at level $B=0,\dots N_B-1$
take the following form (initially, $U^{(0)}_\mu(x)=U_\mu(x)$):
\begin{enumerate}
\item 
If $B\neq 0$, 
double the length of the links produced at the previous level:
\be
U^{(B)}_\mu(x)~ =~ U^{(B-1)}_\mu(x)~~U^{(B-1)}_\mu(x+2^{B-1}a\hat\mu)
\ee
\item 
Smear the links $U^{(B)}$ 
 $n_B$ times: omitting the site and direction indices
of the links, 
\be
 U^{(B)}_{S} = \left[~U^{(B)}_{S-1}~+~w^{(B)}_S~ 
\Sigma^{(B)}_{S-1}~\right]_{\cal U},\qquad S=1,\dots,n_B
\ee
where $\Sigma^{(B)}_{S-1}$ is the sum of staples made of
the links $U^{(B)}_{S-1}$ and the lower index represents
the smearing level. The operation $\cal U$ is a unitarisation
procedure. We choose to maximise $\re\tr( U V^\dagger)$
where $V\in GL_{N_c}(C)$ is the matrix to be reunitarised and 
$U\in SU(N_c)$ is the result of this operation (see App.~\ref{ap:NRLGT}).
 After the last smearing step, we call $U^{(B)}_{n_B}\equiv U^{(B)}$.
\end{enumerate}
The number of smearing parameters $\{n_B\}^{N_B-1}_{B=0}$ and 
$\{w^{(B)}_S\}$
is large and it is hardly possible to undertake an optimisation program. 
Fortunately, it is well-known that the smoothness of 
 the links is rather insensitive to the precise choice of parameters.
We typically choose $0.35\leq w \leq 0.40$.
We can now construct gauge-invariant operators out 
of $N_B$ sets of links, covering a range of physical sizes.

One simple observation is computationally useful. Any product
of links along a closed path uses an even number of them. 
Therefore, if a large number of operators is to be measured, 
one can spare half of the $SU(N_c)$ matrix multiplications if
 all paths of length 2 (we call them `wedges' and `double-links')
in a time-slice are initially computed and stored. 
In 3 spatial dimensions there are 3 double-links going in the `up'
direction and (3 planes) $\times$ (4 wedges per plane) = 12 wedges
to be computed and stored per site. This represents thus 15 
matrix multiplications per site. In 2 spatial dimensions, there are 
6 of them (2 double-links and 4 wedges per site). Any closed loop can
then be computed by multiplying these Wilson lines end-to-end.
Since the double-links need to be computed for the next level of the 
link-fattening procedure $B+1$,
 the actual computational overhead to obtain these building blocks
is 12 (4) matrix multiplications in 3 (2) spatial dimensions.
For instance, suppose the simplest 
operator without any intrinsic symmetry is being measured (sometimes
called the `hand' in $D_s=3$, while in $D_s=2$  it is 
the `knight's-move'). These  operators come in a large number of copies
(48 and 8 respectively), each of which requires  6 (resp. 8) 
$SU(N_c)$ multiplications. If the wedges and double-links have been stored, 
each of the copies only requires half as many multiplications.
Thus the initial overhead is seen to be very small indeed, and the 
gain is practically a factor 2 in speed. Note that we did not have to make
any restricting assumptions about the operators that 
are going to be measured.
Naturally, further increase in speed is possible if it is decided
at compiling time which operators are going to be measured, or if 
the program is `intelligent' enough to find by itself, before
the very first measurement of the simulation, the optimal
order in which to do all the necessary matrix multiplications for 
the measurements.

The memory requirements of this operator-construction program
are relatively large but not prohibitive: 
only one level $B$ of the fattened links needs
to be stored at a time, if measurements and link-fattening steps
are alternated. The use of the double-links and wedges requires 15 (6) 
of them  to be stored per site in $D_s=3~(2)$, rather than 3 (2) links.
\paragraph{}Our spin-identification method requires a set of `probe' 
operators (Chapter~\ref{ch:hspin}). Therefore
we want to construct a set of spatial Wilson loops roughly of the  size
of glueballs $d$
which are obtained from each other through rotations by an
angle $\varphi_n=\frac{2\pi}{n}$. Because the lattice breaks the rotation
group $SO(3)$ down to the cubic group $O_h$, this can only be done 
(and is only meaningful) up to order $\left(a/d\right)^2$ corrections.

In ref.~\cite{Johnson:2002qt} and~\cite{Meyer:2002mk}, an elegant definition of 
the propagation of flux was formulated (see Chapter~\ref{ch:hspin}). 
In brief, it involves
computing the propagator of a fundamental scalar field, of bare mass
$m_0$ and  confined to a time-slice, on the gauge field background. 
The propagator taken between two points is a sum of Wilson lines along
all possible paths, whose weighting is determined by their length 
$\ell$ and is given by $\alpha^\ell$ where
\be
\alpha = \frac{1}{(am_0)^2+2d}
\ee
($d$ is the number of space dimensions).
In~\cite{Meyer:2002mk} it was shown in the free case 
(all links are unit matrices) that in the regime $a\ll d \ll m_0^{-1}$,
the  propagator is dominated by paths of length $\ell\gg d$ and that 
the rotation symmetry is restored up to $\left(a/d\right)^2$.
This means that at least in principle, for any desired accuracy and fixed
physical length scale, there exists a small enough lattice spacing such that 
the propagation of a scalar particle 
on that physical length scale looks spherically symmetric.
If we use this propagator to construct gauge-invariant operators, the 
same statement will apply to them.

We did not pursue that particular method because of its 
computational cost. Instead we use smeared links to construct our operators, 
that is, the level $B=0$ set of operators described above. 
In Chapter~\ref{ch:hspin} it was shown in 2+1D simulations that such an approach
worked just as well at the lattice spacings typically used.
%
%
\subsection{Spin identification on a cubic lattice\la{sec:spin}}
The decomposition of 
$SO(3)$ irreducible representations in cubic group representations leads
to a set of degeneracies in the continuum limit. These degenerate lattice 
states  merge into a single continuum state of definite spin. They
 can thus be considered as different polarisations of the same
state. The set of degenerate states depends on the spin.
The expected degeneracies (see Appendix~\ref{ap:group})
lead to some simple signature rules: for instance, the spin 4 is the 
smallest that yields a degeneracy between an $A_1$ and an $E$ state, while
the spin 6 is the smallest that yields a degeneracy between an $A_2$
 and an $E$ state. Naturally higher spins can lead to the same degeneracies:
just relying on these rules, one could mistake a spin 8
for a spin 4, or a spin 7 for a spin 6. They gradually become less
useful: for instance, a spin 12 state simply appears 12 times in the 
$A_1$ representation!

But in practice, there are more severe limitations to this method. 
`Accidental' degeneracies can appear for dynamical reasons. If we were
to measure the hydrogen atom spectrum on the lattice, the method would be
completely inapplicable, due to the independence of the spectrum on the 
angular momentum quantum number; a specificity of the Coulomb potential. 
While we do not expect such an extreme situation to occur
in the case of glueballs, the density of states increases very rapidly
with energy, and by the time the splitting between states becomes comparable
to statistical error bars, the method becomes insufficient.

We therefore need additional information, and we rely on a direct 
measurement~\cite{Meyer:2002mk}
of the transformation properties of the lattice states under `approximate
rotations', as defined in Section~\ref{sec:probe}, by angles smaller than
$\pi/2$. Before we describe the procedure in detail, 
we work out what wavefunctions we expect to find in the continuum limit.
Let us consider for instance a spin 2 state. In the 
continuum, its five polarisations, labelled by the projection $m$ of the 
spin on a chosen $z$ axis, are described by the spherical harmonics $Y_2^m$,
$-2\leq m\leq 2$. On the lattice, these polarisations get rearranged into
two lattice irreducible representations, $E$ (2-dimensional) and $T_2$
(3-dimensional). Since we are always measuring states belonging to
 definite lattice  representations, the question is then, which 
linear combinations of the $\{Y_2^m\}$ 
are bases of $E$ and $T_2$. It is easy to check that
 $\{Y_2^2+Y_2^{-2},Y_2^0\}$ form an orthogonal  basis  of $E$, 
while $\{Y_2^2-Y_2^{-2},Y_2^1,Y_2^{-1}\}$ span $T_2$.

A more complicated case arises for the spin 4 state. We know from  group
theoretical arguments that a unique linear  combination of $\{Y_4^m\}$ 
belongs to the trivial lattice representation $A_1$. Given that an $A_1$ 
state is invariant under $\frac{\pi}{2}$-rotations around the $z$ axis, 
only $Y_4^0$ and $Y_4^{\pm4}$ can contribute. Symmetry about the $xz$ and 
$yz$ planes imposes the couple $Y_4^{\pm4}$ to come in the combination
$Y_4^4+Y_4^{-4}$. The relative weight of the latter and $Y_4^0$ is 
finally determined by requiring that the combination be invariant under
a $\frac{\pi}{2}$ rotation around the $y$-axis. This linear combination
can be obtained by diagonalising the $D$-matrix $D(0,\frac{\pi}{2},0)$
which describes the rotation parametrised by Euler angles 
$(\alpha,\beta,\gamma)$ in the basis of spherical harmonics:
the correct linear combination 
\[ \lambda Y_0^0 + \frac{\mu}{\sqrt{2}} (Y_4^4 + Y_4^{-4}) \]
is the eigenvector with eigenvalue $+1$. $\lambda$ and $\mu$ are thus known
numerical coefficients.
A complete decomposition of the $J\leq6$ representations obtained 
in this way is given in Table~(\ref{tab:decomp}). Let us repeat what 
these linear combinations represent in the case of the 
polarisation of the spin 4 state lying in the $A_1$ representation:
it is the wave function of 
 a true, continuum spin-4 particle prepared in a state that is invariant
 under all the symmetry operations of the cubic group. 

The linear combinations thus merely reflect the fact that we  measure the 
states in specific polarisations, which we choose to correspond to
lattice irreducible representations. Naturally, the lattice states will
in general have wave functions which are a mixture of these specific 
polarisations. A discretised wave function will be defined below, 
with a discretisation angle $\Delta \varphi={\cal O}(a/d)$, 
where $d$ is the size of the state. As $a\to 0$, we expect such a wave
function to approach one of the allowed linear combinations appearing
in Table~(\ref{tab:decomp}) that corresponds to a continuum state.
There is however  a circumstance where even this expectation can
be in default: that is the mixing of `accidentally' near-degenerate 
states, when $\Delta E/\sqrt{\sigma} = {\cal O}(\sigma a^2)$.
Because the glueball density of states increases strongly beyond the 
first $\sim 10$ states, at any finite lattice spacing 
 the near-degeneracies eventually overwhelm the strict separations 
between states belonging to different continuum multiplets and the 
lattice state wave functions bear no resemblance with their continuum
counterparts. Such mixings would also affect the  energy levels themselves.
Thus a small lattice spacing is mandatory to compute the excited spectrum.

We now consider the actual wave function measurement. 
We shall make use of the notation
\ba
|J~\{m\}~\> &=& \frac{1}{\sqrt{2}}\left(~|J,~m\> ~+~ |J,~-m\>~\right) 
\nonumber \\
|J~[m]~\> &=& \frac{1}{\sqrt{2}}\left(~|J,~m\> ~-~ |J,~-m\>~\right).
\ea
The phases are chosen such that $\<\theta~\phi | J,~m\> = 
Y_J^m(\theta,\phi)$, where 
$\theta,\phi$ are the spherical coordinates with respect to the lattice axes.
Let  $\{\Psi^{(n)}\}_{n=0}^N$ be a set of  operators, such that
$\Psi^{(n)}=U^\dagger(\phi_n)\Psi^{(0)} U(\phi_n)$, 
where $\phi_n\simeq\frac{2\pi}{n}$ and 
$U$ is a representation of the $SO(2)$ rotation group approximately realised
around the (0 0 1) direction of the lattice on length scales $d\gg a$.
Suppose furthermore that $\Psi\equiv\Psi^{(0)}$ 
acting on the vacuum creates an eigenstate
of the transfer matrix (we slightly abuse the notation by using the
 same symbol to denote the state and the operator). 
The angular function
\be
g(\phi_n) ~\equiv ~\< \Psi^{(n)} | \Psi^{(0)} \>
\ee
characterises the state created by $\Psi$. Let us consider as an example 
$\Psi\in A_1$. Then 
\be
| \Psi\> = |\psi_0\>~|0,~0\> ~+~ |\psi_4\>~\left(~ \lambda ~|4,~0\>~ +~ 
\mu ~| 4~\{4\}\> ~\right) ~+~\dots
\ee
where $\lambda, \mu$ are known ($\lambda^2+\mu^2=1$)
and the dots refer to contributions of higher spin states. We have
\be
g(\phi) = ||\psi_0||^2 ~+~ \lambda^2 ||\psi_4||^2 ~+~\mu^2||\psi_4||^2\cos4\phi
\la{eq:g_phi}
\ee
Therefore the measurement of the Fourier coefficients of $g(\phi)$ 
allows us to determine $||\psi_0||^2$ and $||\psi_4||^2$. These numbers, whose
sum we normalise to 1, represent the `probability' for the lattice state
to be a spin 0 or spin 4 state. Indeed, if we consider an effective
 continuum theory that is equivalent to the lattice theory~\cite{Symanzik:1983dc}, 
then the lowest-dimensional 
(irrelevant) operator that breaks rotational symmetry is of dimension
six (we are using the Wilson action). We know that such an operator has the 
effect of introducing ${\cal O}(a^2)$ corrections in the mass ratios of the 
theory. Suppose that we start with a basis of states lying in irreducible 
representations of the $SO(3)$ rotation group. 
First-order perturbation theory then tells us that the energy 
eigenstates acquire new components
of ${\cal O}(a)$ in presence of the irrelevant operator; 
and therefore, in the example  considered above, 
either $||\psi_0||^2$ or  $||\psi_4||^2$ is ${\cal O}(a^2)$.

\begin{figure}[ht]
\centerline{
\begin{minipage}[c]{10cm}
\begin{picture}(100,130)(10,20)
\put(0,0){\line(1,0){180}}
\put(10,6){\line(1,0){180}}
\put(20,12){\line(1,0){180}}
\put(30,18){\line(1,0){180}}
\put(40,24){\line(1,0){180}}
\put(50,30){\line(1,0){180}}
\put(60,36){\line(1,0){180}}
\put(70,42){\line(1,0){180}}
\put(80,48){\line(1,0){180}}
\put(90,54){\line(1,0){180}}
\put(100,60){\line(1,0){180}}
\put(110,66){\line(1,0){180}}
\put(120,72){\line(1,0){180}}
\put(130,78){\line(1,0){180}}
\put(140,84){\line(1,0){180}}
\put(150,90){\line(1,0){180}}
\put(0,0){\line(5,3){150}}
\put(12,0){\line(5,3){150}}
\put(24,0){\line(5,3){150}}
\put(36,0){\line(5,3){150}}
\put(48,0){\line(5,3){150}}
\put(60,0){\line(5,3){150}}
\put(72,0){\line(5,3){150}}
\put(84,0){\line(5,3){150}}
\put(96,0){\line(5,3){150}}
\put(108,0){\line(5,3){150}}
\put(120,0){\line(5,3){150}}
\put(132,0){\line(5,3){150}}
\put(144,0){\line(5,3){150}}
\put(156,0){\line(5,3){150}}
\put(168,0){\line(5,3){150}}
\put(180,0){\line(5,3){150}}
{\thicklines
\put(150,90){\line(1,0){156}}
\put(306,90){\line(0,1){40}}
\put(20,12){\line(5,3){130}}
\put(20,12){\line(0,1){40}}
\put(20,52){\line(5,3){130}}
\put(168,36){\line(-1,3){18}}
\put(168,36){\line(0,1){40}}
\put(168,76){\line(-1,3){18}}
\put(90,18){\line(5,6){60}}
\put(90,18){\line(0,1){40}}
\put(90,58){\line(5,6){60}}
\put(244,60){\line(-3,1){92}}
\put(244,60){\line(0,1){40}}
\put(244,100){\line(-3,1){92}}
\put(150,90){\circle*{3}}
\put(150,90){\line(0,1){40}}
\put(150,130){\line(1,0){156}}
}
\end{picture}
\end{minipage}}
\vspace{1cm}
\caption{A typical set of Wilson-loop probes  $P_n$ 
used for the wave-function measurement 
in the $PC=++$ representations.}
\label{fig:probe}
\end{figure}
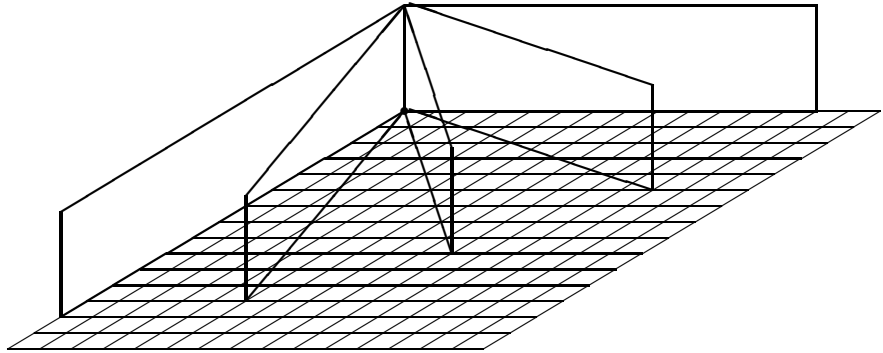

In general, to obtain the wave functional $\Psi$ 
of a glueball, the variational 
method~\cite{Luscher:1990ck} has to be applied on a large basis of operators. This
 method yields a matrix (ideally) giving the components of energy eigenstates
in the original basis of operators. The same linear combinations of the rotated
copies $|\Psi^{(n)}\>$ can be taken to extract the angular 
wave function as described above.
Although the method is general and has been applied in practice
 (see \cite{Meyer:2002mk} and Chapter~\ref{ch:hspin}),
the condition that we must be able to rotate the operators by small angles
 is quite a restrictive one to impose on the trial operators.
Therefore a generalisation was considered in~\cite{Meyer:2003wx}
 (see Chapter~\ref{ch:regge_2d}).
We give ourselves a set of `probe' operators $\{P^{(n)}\}_{n=0}^N$, 
$\< P^{(n)}~P^{(n)} \>=1$. The method now consists in measuring
\be
g_P(\phi_n)\equiv\langle P^{(-n)} \Psi \rangle = \langle P_0 \Psi^{(n)}\rangle 
 \qquad \forall n, \la{eq:wf}
\ee
 The equality relies on the invariance of the vacuum under the rotation.
 An example of probe operators used in the $PC=++$ sector 
is illustrated on Fig.~(\ref{fig:probe}).
In that sector the identification requires the most care, given the 
relatively large density of states. 
%
%
In the case of $\Psi\in A_1$, we could use for $P$
the sum of two orthogonal 
rectangular Wilson loops illustrated on Fig.~(\ref{fig:probe}). If $P$ is 
the combination lying in planes (1 0 0) and (0 1 0), then
 it belongs to a reducible representation $A_1\oplus E$. Therefore
\be
|P\> = |p_0\>~|0,0\>~+~|p_2\>~|2,0\>~+~|p_{40}\>~|4,0\>~+~|p_{44}\>~|4\{4\}\>
	~+~\dots
\ee
and 
\be
g_P(\phi) = \<p_0|\psi_0\>~+~\lambda\<p_{40}|\psi_4\>
 ~+~ \mu \<p_{44}|\psi_4\>~\cos4\phi.
\ee
The Fourier coefficient of $\cos4\phi$ thus informs us about
\be
c_4 = \mu \<p_{44}|\psi_4\>.
\ee
 By the Schwarz inequality, 
\be
||\psi_4||^2 \geq \frac{|\<p_{44}|\psi_4\>|^2}{||p_{44}||^2}= 
\frac{|c_4|^2}{\mu^2||p_{44}||^2}.
\ee
$||p_{44}||^2$ may be determined by `probing the probe' itself, as 
described above, or the probe can be made to have a spin 4 wave function 
by construction.  We can thus determine lower bounds for the amount of 
spin 4 admixture in the $A_1$ states. Now 
 the Fourier coefficients can no longer be extrapolated to the continuum, 
but in practice it turns out that this is not absolutely
necessary on the relatively  small lattice spacings that we are using.
Although we have through this generalisation decoupled the task of constructing
good glueball operators from the spin assignment problem, it is crucial
that the probes should have a substantial overlap onto the glueball operators $\Psi$.
If that is not the case, the lower bound set by the Schwarz inequality will
not allow us to discriminate between spin 4 and spin 0. Given that close to the 
continuum, $||\psi_4||^2$  is of order ${\cal O}(a^2)$ for a quasi-spin 0 
state, the overlap $\<P|\Psi\>$ must be significantly larger than 
${\cal O}(a)$ for the bound to be useful. Clearly,
 as smaller lattice spacings are used, this condition on the probe  weakens, 
but it is still non-trivial at the lattice spacings at which we can afford
to do simulations.

\section{The quenched QCD spectrum from the lattice \la{sec:su3}}
We performed six Monte-Carlo runs at four different lattice spacings (see
Table~\ref{tab:su3runs}). The two extra runs were used to estimate 
finite-volume effects, as well as the importance of mixing of the 
single-glueball states  with 2-torelon and 2-glueball states.  
We were using a 2-level algorithm~\cite{Meyer:2002cd} as described 
in Chapter~\ref{ch:mla}.
The number of measurements performed at fixed time-slices was 40.
The choice  resulted  as a compromise between improving the accuracy of the 
heavy-glueball correlators and maintaining a reasonable efficiency for the 
lighter ones. Ten compound sweeps separated two sets of `fixed' time-slices. 
While the boundary time-slices are kept fixed, 
we perform two compound sweeps between the measurements, since the 
 number of measured operators was large. 

In the following we describe the spin identification procedure in
 practice, the study
of finite-volume effects and the evaluation of mixing between our operators
and 2-glueball and 2-torelon states.
%
%
%
%
\subsection{Spin identification in action}
Let us consider for instance the data at $\beta=6.4$ ($a\simeq 0.05$fm)
and start with the $A_1^{++}$ representation. We will first try and see
how far we can go with the degeneracy arguments. The smallest
spins appearing in this representation are 0, 4, 6 and 8.
The first two states clearly have no degenerate partners in the 
other representations, therefore they must be spin-0 like.
The first two states in the $E$ and $T_2$ obviously match,
and match no other states; therefore they must be spin-2 partners.

In the region $0.85<am<1.0$, there are altogether eight lattice states
(see Table~\ref{tab:su3p}).
Since one of them is $A_2$, a state with spin 3, 6 or higher must
 be involved; and we ignore the $A_2^*$ for the moment.
One possibility is that this collection of  states corresponds 
to a spin 6, a spin 0 and a spin 1 state. 
Another possibility is that we have a spin 3, a 
spin 4 and a spin 0 state. A third possibility is a spin 3, a spin 2,
a spin 1 and two spin 0. We got stuck!

Now we will make use of the information coming from the wave function
measurements, Table~\ref{tab:fourier}. The state in the $A_2$ 
representation appears to have no $\cos{6x}$ component in its wave 
function, within the systematic uncertainty of the wave function 
measurement. Thus  the first-mentioned 
possibility is practically excluded. We can pair the ($A_2$, $T_1$, $T_2$) 
together to form the seven polarisations of a spin-3 state. The question
is now whether a spin 4 state is present or not.
The wave function measurement provides no evidence for the $A_1^{**}$ being
 a spin 4. But the 
$A_1^{***}$, the $E^{**}$ and the $T_1^*$ all have spin-4 components that 
are at least five times larger than the expected  scale of the rotational
 symmetry breaking. Therefore with all likelihood  these lattice 
states correspond to different polarisations of a spin 4 state. 
The $T_2^{***}$ completes the nine polarisations.
The absence of additional states in the $T_1$ and $T_2$ representations that are
near-degenerate with the $A_1^{**}$ state
confirms the scalar-like nature of the latter. 
It should be noted that there is an extra state in the $E$ representation, 
but that state was seen at all lattice spacings without a partner state in
the $T_2$ representation and is found to have a strong volume dependence.
 The only way this can happen systematically is 
if it corresponds to a two-torelon state. It is an infrared
breaking of rotational invariance,  which we shall come back to in the next
section.

The wave function analysis of 
$A_2^*$ exhibits no $\cos6\phi$ component, and therefore it almost
certainly corresponds to a spin 3 state. 
The next state, $A_2^{**}$ has a large component  of this type and hence it must
be one of $J=6,7,9,\dots$ A near-degenerate state is seen in the $E $
representation, although its overlap onto the probe is too small to allow us to
determine the angular dependence of its wave function with confidence. Also, 
a degenerate state in the $A_1$ representation is seen, and it has
 a strong $\cos4\phi$ component; if the degeneracy 
 is not accidental, this rules out the possibility $J=7$.
While we cannot rule out accidental degeneracies, the ($A_1$, $A_2$,
$E$) triplet thus most probably corresponds to a spin-6 state.

The combination of the degeneracy arguments and the wave function 
measurements thus allows us to disentangle the states. In the other $PC$ sectors, 
the density of states below 4GeV is not as high and  the degeneracies 
are usually sufficient to determine the spin-multiplets. An interesting case
arises in the $PC=-+$ sector. The first two $T_1$ states are almost degenerate
at all lattice spacings; and the $A_2$ state is very high-lying, and so is 
the $A_1^{**}$ above the two low-lying pseudo-scalars. This excludes the 
possibility that one of the $T_1$ states can belong to a spin 3 or spin 4 
multiplet. Furthermore, we see that the $T_2^{**}$  state is quasi-degenerate.
So unless we are seeing an accidental  degeneracy between three (!) states 
(two spin 1 and a spin 2), the $T_1$, $T_1^*$, $E^{**}$ and $T_2^{**}$
must correspond to a spin 5. Furthermore, an unpublished
 set of data, where the glueball operators allowed us to 
measure the spin components directly (see Eqn.~\ref{eq:g_phi}), tells us 
that at least one of the $T_1$ states has large variations in its $\phi$ 
dependence, which excludes the $J=1$ interpretation.

In the $PC=+-$ sector, a striking degeneracy at all lattice spacings 
of the $E$, $T_1^{***}$ and $T_2^{***}$ states  suggests a  spin 4 or 
spin 5 multiplet.  We are unable to say whether another $T_1$ state is 
near-lying (which would complete the spin 5 multiplet), or whether instead
the $A_1$ state is degenerate with these lattice states. If we combine our data
with that given in~\cite{Morningstar:1999rf}, 
where the $A_1$ state was found to be definitely 
heavier than the $E$ state, it would appear that we are actually seeing a 
spin 5 multiplet (a gap above the $T_2^{***}$ excludes the possibility of a spin
 3  -- spin 2 degeneracy). 
In this case we feel however that the labelling requires confirmation, 
in particular the missing  $T_1$ state should be seen to complete the multiplet.
%
\subsection{Bi-torelon states and other finite-volume effects}
Tables~(\ref{tab:fvp}) and~(\ref{tab:fvm}) show the relative variation of 
the energy levels as the volume is varied between 1.4 and 2.0fm
 (runs IV, V, VI at $\beta=6.2$). Almost all levels undergo variations that 
are consistent with zero, within 1-1.5$\sigma$. 
Exceptions are highlighted in the table: they concern 
the $A_1^{++*}$ and the $E^{++}$  states, as well as a state appearing in the 
$E^{++}$ representation which has no analog in the $T_2^{++}$, and which we labelled
as `2T'. It is the only state for which we see a definite trend of increasing
mass with the volume. We therefore interpret it as a finite-volume effect.
Apart from this special case, 
the largest  effect is observed for the $0^{++*}$ state
on the 1.4fm lattice (2.7  standard deviations). 
A similar volume dependence for this state was also observed 
in~\cite{Morningstar:1999rf}. 

 Naturally these variations (apart from the 2T case) 
could simply be statistical fluctuations.
We noticed empirically that the variational calculation can in some 
cases amplify  the statistical fluctuations, since it finds  by definition 
the linear combination of operators that minimises the local effective mass,
which has a non-zero variance; the systematic error due to this enhancement is
 typically not included in quoted statistical errors.
If it is a  physical effect, it is somewhat peculiar that these states
are lighter both on the smaller and the larger volumes. Also, in the case
of the lightest $E^{++}$ state, the partner state in the $T_2^{++}$ 
representation shows no sign of volume dependence. 
There is a well-known finite-volume
effect that can explain a mass difference between two  states
in the $E^{++}$ and $T_2^{++}$ representations corresponding to the same
continuum state, as well as the appearance of an extra state in $E^{++}$.

The torelon, a state created by a Polyakov loop in the fundamental 
representation, transforms non-trivially under the centre symmetry ${\bf Z}_{N_c}$
of the gauge group $SU(N_c)$. The product of two such operators winding in 
opposite direction, however, has trivial transformation properties under the 
global ${\bf Z}_{N_c}$ symmetry and can therefore create states that mix with glueballs.
In physical terms, fusion of flux-tubes can occur, after which
the newly formed loop is contractible.
The general consequence  is that extra states are seen 
in the finite-volume `glueball' spectrum with respect to the infinite volume
limit, where the bi-torelon states become infinitely massive and decouple.
The two cubic representations that can be affected by the extra states are 
the $A_1^{++}$ and $E^{++}$.

In the two  simulations (run IV and VI, Table~\ref{tab:su3runs}) designed 
to check for finite-volume effects, we explicitly included the measurement
of products of (fuzzy) Polyakov loops. We use operators which have zero 
total momentum, and zero relative momentum. A plot of their local effective 
masses as function of Euclidean time is shown in Fig.~(\ref{fig:bitorelon}).
We firstly note that although there is no symmetry preventing the decay of the
observed plateaux toward the lightest glueball energy level, this does not 
happen in the range where we are able to measure the correlator accurately.
We conclude that the bitorelon has very little overlap onto the lightest 
glueball. Secondly, the mass plateaux are very much lower than the value
for two non-interacting flux-tubes, and the lowering increases with the length
of the torelons. This behaviour is qualitatively 
very different from the case of a two-glueball scattering state (see below).

To understand their effect on the 
$A_1^{++}$ and $E^{++}$ spectrum, we give the energy levels resulting from
the variational calculation with, and without the bi-torelon operator included
in the starting basis (Table~\ref{tab:bitorelon}). 
In the $A_1$ case, we use the superposition of 
bi-torelon states winding around all three cycles of the torus, while for 
the $E$ we use only the one winding in the $z$-direction (whilst the other $E$
operators are of the type $Y_2^0$). The 
overlap $|\langle{\cal O}_G {\cal O}_{2T}\rangle|$  between the bi-torelon
operator and the ordinary operators prior to diagonalisation
 is less than 0.1 in the $E$ case and $0.2-0.3$ in the $A_1$ case; they
only decrease slightly as the spatial extent of the lattice is increase from
1.4 to 2fm.
The generic effect we observe is that a new, relatively light state
appears - which  we did not `see' on the original volume. 

The $A_1$ and $E$ spectra without the
 inclusion of the bi-torelon operator are very similar
on all three volumes, with one exception: on the small volume (run IV), 
there is an  extra light state in the  $A_1$.  
It may correspond to the lightest
 scattering state or a bi-torelon state. In both the 1.4 and 2.0fm boxes,
 the inclusion has the general effect of producing a relatively 
`noisy' extra state. We note again that the new state is significantly lighter
than the naive sum of two torelon masses (Table~\ref{tab:su3runs}), especially
in the $E$ case.
We infer that the force between two torelons is strongly attractive.
Interestingly, the $A_1^{++*}$ in the 2.0fm box, which appeared to be lighter by
$~2\sigma$, is now back to the value obtained in the 1.7fm box 
(Table~\ref{tab:su3p}), although it is now much noisier.  
\begin{figure}[t]
\vspace{-0.5cm}
\centerline{\begin{minipage}[c]{14cm}
   \psfig{file=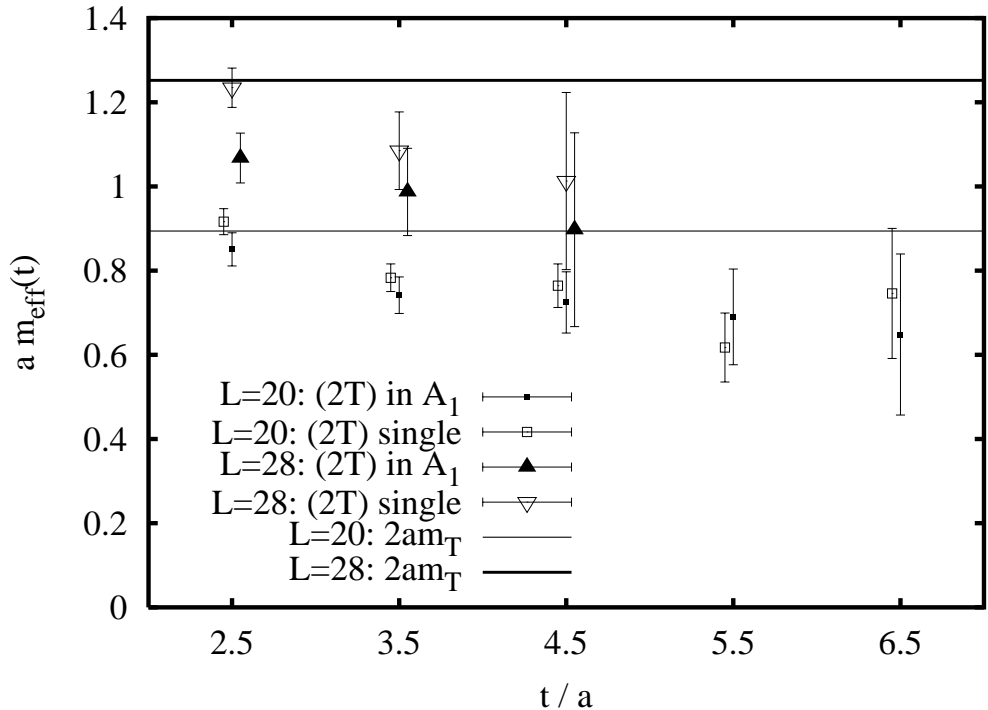,angle=0,width=14cm} 
    \end{minipage}}
\caption[a]{The local effective mass of bi-torelon operators, at $\beta=6.2$
(runs IV and VI). The $A_1$ representation is obtained by superposition 
of the bi-torelon operators winding around each cycle of the hypertorus, 
while the `single' corresponds to just one of them. The expectation for 
two non-interacting flux-tubes is indicated by the horizontal lines.
A fit to the $A_1$ mass plateaux yields 0.787(39) and  1.02(6) for $L=20$
and 28 respectively. }
\la{fig:bitorelon}
\end{figure}
In conclusion, the bitorelon states mix with the $A_1,E^{++}$ glueball states,
but the extra states introduced as a result of the mixing become heavier
as the extent of the spatial lattice is increased. The anomalous behaviour of
 the $A_1^{++*}$ energy level as a function of the volume may be due to 
the proximity of a two-torelon energy level, which only appears explicitly
in the variational spectrum if two-torelon operators are included in the basis. 
The case of the fundamental state
in the $E^{++}$ representation is harder to understand: it comes out lighter
on either smaller or larger volume than the standard volume ($L\simeq1.7$fm), 
whether or not the bi-torelon operator is included. We are therefore 
inclined to consider this particular variation as a 
 statistical fluctuation.

\subsection{Scattering states $\&$ decays}
When we apply the variational method to extract the fundamental and 
several excited states in a given symmetry channel, we do not know 
\emph{a priori} the nature of these states. We need additional
information to label the state with continuum quantum numbers, as discussed
above. Another procedure needs to be followed in order to conclude
that the linear combination obtained corresponds to a single glueball state, 
and not to a scattering state of several glueballs.
In~\cite{Morningstar:1999rf}, the issue was sidestepped by only considering states
that are below the two-particle threshold, which was estimated using the 
free relativistic dispersion relation. This approach proved sufficient to
 obtain a large number of states in the low-lying spectrum.

It is well-established~\cite{Luscher:1991cf} that decay widths of infinite-volume 
states and energy levels
of scattering states in a finite volume are intimately connected.
As a first attempt to study the effects 
of the two-particle threshold, we include direct products of traced Wilson
loops with zero relative momentum into our basis of operators. 
Let us suppose that each of these 
operators, having very high overlaps onto physical states, 
creates one particular glueball. First consider for simplicity 
the planar limit $N_c\rightarrow\infty$.
Due to the factorisation property we expect that the direct product creates
two non-interacting glueballs. In this free-particle limit, each of the Wilson
loops can be assigned a definite momentum ${\bf p_{1,2}}$. 
If the two glueball states have equal mass, in the centre-of-mass frame we 
simply have  ${\bf p_1}=-{\bf p_2}$. In what follows we only consider
the simplest case: the product of two of the operators creating the lightest
 glueballs ($0^{++}$) with ${\bf p_1}={\bf p_2}=0$. Now at finite
$N_c$, this is not an eigenstate of the Hamiltonian, and therefore it can 
\emph{a priori} have a finite overlap with any of these eigenstates, provided
it has quantum numbers $0^{++}$ and ${\bf p}=0$. If we normalise
the eigenstates and the direct product operator ${\cal O}$ such that
 $\langle n | n\rangle=1$ and $
\langle  {\cal O}~{\cal O}^\dagger\rangle= 1$,
unitarity implies that the sum of all the overlaps must add in quadrature
 to 1. In Table~(\ref{tab:2G-1G}), we show the overlaps 
$|\langle \Omega |  {\cal O} | n \rangle|$
of such `2-lightest-glueball' (2LG) 
operators onto the orthogonal operators
 whose  mass plateaux yielded the spectrum in Table~(\ref{tab:su3p}). 
The overlaps are under the $10\%$ level.
 The dependence on which of the two 2LG operators is being used
gives us a measure of the systematic uncertainty on the values 
obtained\footnote{Statistical errors are negligible in this case, 
since it is a correlation in a single time-slice.}. 
The fact that the overlaps are tiny tells us 
that the 2LG operators are creating states that are practically orthogonal
to our original basis of single-glueball operators.

The same conclusion can be reached in a different way.
If we include either of the  2LG operators in the variational calculation, 
we obtain an extra state with mass $am=1.047(33)\simeq 2am_{0^{++}}$, 
which gets most of its contribution from the 2LG operator.
The other 0++ states are unaffected within statistical errors by the inclusion
of the extra operator. This provides some evidence that the states we 
extract above 2-particle threshold (starting with the $0^{++**}$) 
are narrow glueball resonances, since our variational basis does not
resolve the 2-particle threshold. Two of the lightest
glueballs seem to fit easily in the 1.75fm box without interacting (much): 
the extra mass plateau we obtain by including 2LG operators 
is within statistical errors twice the mass of the lightest glueball.

 In general, the mass of resonances must be inferred 
from energy shifts~\cite{Luscher:1991cf} in the two-particle spectrum as the 
size of the periodic box is varied. However, for very narrow resonances,
the energy levels become volume-independent (up to exponentially small corrections).
 And the correlator of the interpolating operator creating the quasi-state from
 the vacuum will have a long mass plateau before `decaying' toward the
 energy of the scattering state. 
 If operators are used that overlap very little onto the decay
product of a quasi-stable state, it can be very hard to see the transition
occur; a typical example is the measurement of the adjoint string tension 
with Wilson loops~\cite{Kratochvila:2003zj}. 
Given our observation that in general 
single-trace and double-trace operators have little overlap, we  expect our 
unstable-glueball operators to produce long `meta-stable' mass plateaux
 before the effects of the decay show up -- and this is largely what is seen in
the data. Finally, the observation of the multiplets corresponding to the $SO(3)$
representations are again useful cross-checks in this context, since 
the representations of $O_h$ contain different sets of scattering states.
 
It should also be remembered that the finiteness of the  variational basis
implies that the linear combinations of operators obtained cannot be 
strictly orthogonal in the infinite-dimensional space of operators.
This means that we can never strictly obtain stable mass plateaux above
the fundamental one -- although in practice it is sufficient that they
be long enough that a fit over several points may be done. It remains
a small source of systematic error nonetheless. 
To illustrate these points we 
show a large number of mass plateaux in the $PC=++$ sector
on Fig.~(\ref{fig:meff_64a1a2et2pp})  at our 
smallest lattice spacing ($\beta=6.4$). Recall that we usually do not store the 
very-short-distance correlator, so that the first local-effective-mass value
occurs at 2.5 lattice spacings ($\sim$0.13fm).
The representation where we could
resolve the most states, and where the 2-particle threshold is lowest,
is naturally the trivial one ($A_1$). According to Table~(\ref{tab:su3p}),
the threshold is around $aE=0.80$. Therefore only the fundamental 
$A_1^{(0)}$ and the first radial excitation $A_1^{(1)}$ 
are actually stable states. While the $A_1^{(2)}$ and $A_1^{(3)}$
(corresponding to a mixture of $0^{++**}$ and $4^{++}$) show reasonably
long plateaux, $A_1^{(4)}$ is  seen to decay into a lighter state 
quite `rapidly' in Euclidean time, which could  be the lightest
scattering state of two of the lightest glueballs with zero relative
momentum. In the $A_2$ representation (bottom plot), the first
two states (corresponding to $3^{++}$) are  stable, and 
the third mass-plateau (corresponding to $6^{++}$), 
which is above threshold, shows reasonable stability as well, although
the error bars are larger. 

Similar qualitative observations can be made about the mass plateaux in
 the $E$ and $T_2$. Although we know that they must be stable, because
they are below threshold, the first excited state in each of these
representations seems to be decaying into the lightest state. This is 
therefore most probably the above-mentioned effect of the finiteness of
 the variational basis, which implies that we cannot project out exactly
 the wave function of the fundamental state. Whilst the $E^{(2)}$ and 
$T_2^{(3)}$ states (corresponding to different polarisations of the spin
4 glueball) have very good plateaux, the effective-mass of the $T_2^{(4)}$
exhibits no plateau at all, and thus shows that the method has
its limitations; in this particular case,  the basis used in the 
variational calculation contained ten operators.
%
\subsection{Continuum extrapolation and polarisation-averaged spectrum}
Our continuum extrapolation is entirely conventional.
Having obtained the spectrum at lattice spacings 0.05-0.10fm, we extrapolate
the glueball spectrum in units of the string tension to the continuum, using
a linear fit in $\sigma a^2$.
This is the most natural scheme to set the scale in view of our forthcoming
discussion of Regge trajectories. The resulting continuum spectrum is given
in Tables~(\ref{tab:extrapol_p}) and~(\ref{tab:extrapol_m}). We are usually
able to include all four points in the fit and have an acceptable value of 
 $\chi^2$. In a few cases we drop the data at $\beta=6.0$, our coarsest 
lattice spacing. The extrapolations are illustrated on 
figs.~(\ref{fig:extrapol_pp_a}--\ref{fig:extrapol_m}).

The extrapolations of a few states produce bad $\chi^2$, for instance the 
lightest $0^{-+}$ state and the $3^{++}$ state in the $T_1$ representation.
Upon inspection of Fig.~(\ref{fig:extrapol_mp}), it seems very unlikely
that this should be due to a violation of the scaling behaviour: the data points are 
`oscillating' around the best fit line. We rather take it as an indication
 that the  error bars are somewhat underestimated -- they are purely statistical.
 Another general
comment is that the data points at the finest lattice spacing ($\beta=6.4$) 
are often above the best-fit line, or have a significantly larger error bar.
This is due to the fact that it is harder to achieve good overlaps onto the 
physical states at smaller lattice spacings; it is precisely this difficulty
 that prompted the development of techniques such as smearing~\cite{Albanese:1987ds}
and blocking~\cite{Teper:1987wt}.

When a continuum state appears in several cubic  representations,
we note a remarkable agreement of the corresponding energy levels 
in the continuum limit. Given this fact, it is reasonable to average the 
energy levels to produce a final estimate of the energy level in units of the 
string tension. To do so, we weight the representations inversely 
proportionally to the square of their error bars. The final error bar 
assigned is conservatively taken to be  the smallest error bar of the 
individual representations (rather than the expression
 $(\sum_i \Delta_i^{-2})^{-1/2}$ valid for a large number of data points).
The result is given in the last column of Tables~(\ref{tab:extrapol_p}) and 
(\ref{tab:extrapol_m}). To obtain the spectrum in physical units, we use
$\sqrt{\sigma}=(440\pm20)$MeV, which leads to the values given in 
Table~(\ref{tab:continuum}).
\subsubsection{Previous work}
Globally, our spectrum is in good agreement with that of Morningstar and 
Peardon~\cite{Morningstar:1997ff} (see Fig.~\ref{fig:spec_all}). 
There is a tendency for us to get the 
energy levels lower, in particular the mass gap is smaller in our data; also,
 our $1^{--}$ state is quite a bit lighter than in~\cite{Morningstar:1997ff}, 
although the accuracy of our continuum $PC=--$ energy levels suffered from the 
poor data at our  smallest lattice spacing. 
Nevertheless, the $1^{--}$ level 
shows very little dependence on the lattice spacing, so that any of our 
data points would lie well below the value of~\cite{Morningstar:1997ff}.

In Ref.~\cite{Liu:2001wq}, Liu and Wu obtain the value 3.650(60)(180)GeV
 for the lightest $4^{++}$ state, 
using very different methods from ours, and this value is 
in good agreement with our estimate. It is however 
intriguing that their estimate
of the lightest state in the $E^{++}$ representation should be systematically
heavier than the one of Morningstar and Peardon for simulations done 
at exactly the same parameters - in several cases by ten standard deviations.
\subsubsection{Sources of systematic error on the spectrum\la{sec:systematics}}
Several sources of error have already been mentioned and discussed
 in the text. We simply enumerate them once again to give a comprehensive picture: 
\begin{enumerate} 
\item the dependence on the variational basis of operators; in particular, 
      its finiteness
\item finite-volume effects
\item the difficulty to maintain good overlaps as the continuum is approached
\item Euclidean correlators give upper bounds on the energy levels
\item even with the 2-level algorithm, the variance increases rapidly
      with the time-separation
\end{enumerate}
It is of course impossible to estimate the uncertainty of the spectrum
due to these effects. Effects (1) and (2) can either raise or lower the 
energy levels, while the other items lead to a systematic 
over-estimations of the energy  levels.
The most accurately determined states deserve the most care with 
 respect to systematic errors. On the positive side, the consistency
of continuum energy levels obtained in different lattice representations
demonstrates that the systematic effects are not much larger than 
 the statistical error bars. In particular, the averages over
polarisation should reduce the dependence of the final spectrum on the 
variational basis used.

\section{The lattice glueball spectrum at large $N_c$\la{sec:sun}}
Given the theoretical prejudice that the large $N_c$ gauge theory should 
be in many ways simpler than the generic case~\cite{Witten:1979kh}, 
it is useful to estimate
the similarity between (quenched) QCD and the $SU(\infty)$ theory. 
According to standard $1/N_c$ counting rules, the spectrum should have only
$1/N_c^2$ corrections. This has been verified for the states $0^{++},~
0^{++*}$ and $2^{++}$ in~\cite{Lucini:2001ej}. Here we largely 
extend the survey of the glueball spectrum for $N_c=8$, at which point the 
$1/N_c^2$ corrections are certainly smaller than our statistical error bars.
Our methodology is entirely similar to the $SU(3)$ case. The 
wave function measurements were not needed to do the spin assignments for the 
states where we have reasonably small statistical errors. It should be noted
that the computing cost grows roughly as $N_c^3$, and that it is dominated 
by matrix multiplications at large $N_c$.

The parameters of the various runs are given in Table~(\ref{tab:su8runs}).
The lattice spectrum can be found in Tables~(\ref{tab:su8p}) and 
~(\ref{tab:su8m}) and its continuum extrapolation in~(\ref{tab:su8cont}).
Fig.~(\ref{fig:spec_all}) gives a direct comparison of the two spectra in
units of $(2\pi\sigma)^{1/2}\simeq 1$GeV. The similarity is striking. In fact
the only statistically significant difference is the mass of the $0^{++*}$.
It is quite a bit lighter than its $SU(3)$ cousin, and also well below the 
$N_c=\infty$ estimate given in~\cite{Lucini:2001ej}. A possible explanation 
is that we are presently seeing a `new' state which has no analog in 
$SU(3)$. New states can naturally arise at larger $N_c$ 
in the flux-tube model, if the fundamental string is replaced by 
$k$-strings~\cite{Johnson:2000qz}. 
In the present case, the position of the $0^{++*}$ and  $0^{++**}$
energy levels with respect to the 
fundamental state is compatible with the known ratios of the 
$k=2,3$ string tensions to the fundamental one~\cite{Korthals2004}, 
$m^*/m\simeq\sqrt{\sigma_2/\sigma_1}$ and 
$m^{**}/m\simeq\sqrt{\sigma_3/\sigma_1}$. 
Naturally, further investigation is required to prove or disprove
 this speculation.
\section{Physical discussion\la{sec:discussion}}
We intend to interpret the spectrum in terms of Regge trajectories.
It is thus natural to present the spectrum in a Chew-Frautschi plot (see
Fig.~\ref{fig:cf}). The data is most accurate and 
complete in the $PC=++$ sector. This is also  the most interesting
part of the spectrum in this context, given the possible connexion with 
the pomeron. 
Having in mind the string models of glueballs (Section~\ref{sec:models}), 
we propose an interpretation
of the spectrum based on the existence of `orbital' and  `phononic' 
trajectories. 
Drawing a straight line going through the lightest $2^{++}$, $4^{++}$ states
in the Chew-Frautschi plot we obtain
\be
2\pi\sigma~\alpha'=0.281(22)\qquad\alpha_0=0.93(24).
\qquad(2^{++}, 4^{++})\la{eq:dat_orb_traj}
\ee
Also, we can draw a straight line through the lightest $0^{++}$, the 
first excited $2^{++}$ and the $3^{++}$ glueballs:
\be
2\pi\sigma~\alpha'=0.395(21)\qquad \alpha_0=-0.704(66)\qquad \chi^2=0.01
\qquad(0^{++},~2^{++*},~3^{++})\la{eq:dat_phon_traj}
\ee
Given the selection rules associated with the spinning adjoint-string and phononic
 trajectories (Section~\ref{sec:isgur}), we associate the leading trajectory
with  the former, and the subleading with the latter.
We further observe a $6^{++}$ state which plausibly belongs to 
the leading trajectory. 
If we adopted the orbital motion of the closed oriented string as interpretation
for the leading trajectory, then we would predict
 additional  states to lie on the trajectory with $C=-$ and odd spin.
The  $3^{+-}$ state would have to be associated with the leading trajectory, 
while the data favours its belonging to the `phononic' trajectory.
On the other hand, the intercept of the trajectory
is so high that the  crossing with the horizontal axis corresponds to negative $t$
and no scalar glueball appears on the trajectory. 

It is interesting that the observed slopes are close to those corresponding
to the spinning adjoint string (4/9 assuming Casimir scaling) 
and the closed fundamental string phononic trajectory (1/4); 
except that they are inverted!
 This can easily be accounted for by mixing, 
since the straight lines cross around $J=5$; 
the classical values for the slopes are  valid at large $J$. The 
hyperbolae corresponding to the resulting curved trajectories are
  sketched on the Chew-Frautschi plot as a suggestion 
(the splitting at the crossing point was chosen by hand).
The intercept is thus raised by `repulsion' with the phononic trajectory:
there cannot be a double pole in a unitary quantum field theory.
%

For the phononic trajectory, states with all combinations of $PC$ are 
expected on the leading trajectory, with the exception of spin 0, where 
only the $++$ combination should appear, and the spin 1, where there should
be no state at all, due to the absence of an $m=1$ phonon. We observe an
 almost perfect parity doublet at $J=2$: there is a
near degeneracy of the  $2^{++*}$ and $2^{-+}$ states. Also, there
is a  $3^{+-}$ lying close to the $3^{++}$ state. 
This is a non-trivial observation from the point of view of simple 
operator-dimension counting rules, since the  $3^{++}$ is created by a 
dimension 5 operator and the $3^{+-}$ by a dimension 6 operator.
On the other hand, 
the $2^{\pm-}$ and $3^{-\pm}$ states are either missing from the spectrum
or much heavier. In general it seems that  light states with 
quantum numbers $J~{\rm even}$, $C=-$ or $J~{\rm odd }$, $P=-$
are missing from the spectrum. 
It would be interesting to see whether string corrections to the flux-tube
model can provide a natural explanation for these large mass 
splittings (Chapter~\ref{ch:string}). The 
$m=2$ phonons play a special role in this context, because they lead to 
spin-orbit coupling terms linear in the deformation of the circular 
configuration. The lightest of these heavy
states in our spectrum is the $1^{--}$, followed 
by the $2^{--}$ and $3^{--}$. These states  are of particular
interest because they could be related to the odderon trajectory. 

This brings us to the discussion of subleading trajectories, which is 
necessarily more speculative. Nonetheless,
within the flux-tube model it is hard to understand that the $1^{--}$ 
is lighter
than the $3^{--}$, because the only way to obtain the $1^{--}$ is by 
subtracting an $m=2$ $\hat \rho$-type phonon from an $m=3$  
$\hat \rho$-type phonon. This would locate the $1^{--}$ on a subleading 
trajectory with respect to the $3^{--}$ state, which is easily obtained by
exciting an $m=3$  $\hat \rho$-type phonon. 
 The twisted,  `8' type configuration (Chapter~\ref{ch:string})
of the oriented closed string provides
a natural explanation for this discrepancy.  The orbital trajectory 
 built  on such a configuration leads to the sequence of states~(\ref{eq:tw_traj}).
As remarked
earlier, the $1^{--}$ and $3^{--}$ states apparently have very  small lattice
spacing corrections, and therefore we will use the data at $\beta=6.2$ 
($a\simeq0.07$fm), which is much more accurate than the continuum values,
to test this idea. 
A straight line can be drawn through the $0^{++*}$, the $1^{--}$ and the 
$3^{--}$ states\footnote{This statement also holds for the continuum spectrum.}:
\be
2\pi\sigma~\alpha'=0.351(40)\qquad \alpha_0=-1.93(16)\qquad\chi^2=0.04
\qquad(0^{++*},~1^{--},~3^{--})
\ee
The slope is similar to that of  the leading  trajectory; it would be interesting
to see if an excited $2^{++}$ state lies on the trajectory.
Also, it would be worth testing  if `planar twisted plaquette' lattice
 operators have large overlaps on the lightest $1^{--}$ (such an operator
was not included in our variational basis).
%
%
\subsection{Implications for high-energy reactions\la{he_reactions}}
The energy dependence of high-energy cross-sections is related to the 
Regge trajectory with largest intercept through~Eqn.~(\ref{eq:sigma_tot}),
$
\sigma_{\rm tot} ~\propto ~ s^{\alpha_0-1}.
$
\paragraph{The pure gauge case\\}
From our data, we infer that in the $SU(3)$ gauge theory (without quarks), 
the largest intercept corresponds to quantum numbers $PC=++$ and positive
signature (i.e. even spins). The intercept was found to be 0.93(24).
Thus our data is compatible with the idea that high-energy 
cross-sections would also be roughly constant in the pure gauge theory.

Our interpretation of the data implies that the interplay of several 
trajectories with different slopes is essential to obtain an intercept
of order 1 for the leading trajectory. Also it explains why the slope 
at the origin of the pomeron trajectory is closer to the phononic value 
of 1/4 (in units of $2\pi\sigma$) than to that of the spinning string 
configurations. 

One might speculate than an `odderon', a $C=-$ trajectory of odd signature, 
goes through the $1^{--}$ and $3^{--}$ states. This would however imply that 
its intercept is very low ($\sim -2$), and
it would appear to have little to do with the phenomenological odderon.
\paragraph{QCD\\}
The importance of mixing of trajectories  was  emphasised by 
Kaidalov and Simonov  in~\cite{Kaidalov:1999yd}, 
but in their article the mixing referred to mixing of a
single gluonic trajectory with the flavour-singlet meson 
trajectories ($f$ and $f'$). In their model, there is only one type 
of trajectory, which has 
an intercept around 0.5 in the pure gauge case. It is the mixing with 
the mesonic trajectories which raises the intercept up (and above) 1.

Our data suggests that the glueball spectrum is quite complex. Mixing 
effects are already essential in the pure gauge case and lead to a large 
 intercept of a pomeron-like trajectory. 
It is a physically appealing picture that the approximate constancy
of total high-energy cross-sections is essentially governed by gluodynamics
and has only a subleading dependence on the 
 number of light quarks and their masses. The large-$N_c$
counting rules tell us that such a picture should become exact in the 
planar limit, since glueball-meson mixing amplitudes are suppressed by 
$1/\sqrt{N_c}$. Naturally, for phenomenological applications this is hardly
a suppression and the issue of mixing of the leading glueball trajectory 
with the $f$ and $f'$ trajectories remains essential.
We indicate the  position of some $f$ and $f'$ (flavour-singlet) 
mesons~\cite{Hagiwara:fs} on the Chew-Frautschi plot (Fig.~\ref{fig:cf}), 
using $\sqrt{\sigma}=440$MeV. Mixing in the small $t$ region is
inevitable, and would generically lead to an enhancement of the leading 
intercept.
\clearpage
\input{tabs}
\input{figs}

%% file: tabs.tex
\begin{table}
\vspace{-1cm}
\begin{center}
\begin{tabular}{|c|c|c|c|c|c|}
\hline
{\Large $J^{++}$} & IR  & $\beta=6.0$ & $\beta=6.1$ & $\beta=6.2$ &$\beta=6.4$ \\
\hline 
\hline
$0$      &$A_1^{(0)}$&0.7005(47)&0.6021(85)&0.5197(51)&0.3960(93)\\
\hline
$0^{*}$  &$A_1^{(1)}$&1.167(25) &1.038(15) &0.929(10) &0.690(18)\\
\hline
$0^{**}$ &$A_1^{(2)}$&1.515(28) &1.298(30) &1.151(14) &0.918(40)\\
\hline
$0^{***}$&$A_1^{(5,4)}$&1.854(57) &1.584(58) &1.378(31) &1.160(98)\\
\hline 
2       & $E^{(0)}$  &1.0596(64)&0.916(11) &0.7784(79)  & 0.5758(82)  \\
        & $T_2^{(0)}$&1.0674(49)&0.8990(50)&0.7764(42)  & 0.5837(51)  \\
\hline
$2^*$   & $E^{(1)}$  &1.433(14) &1.180(34) &1.032(20)  & 0.795(28)  \\
	& $T_2^{(1)}$& $\leq1.502(12)$ &1.203(11) &1.047(17)  & 0.777(48)  \\
\hline
2T 	& $E^{(2,3)}$ & 1.515(21)& 1.219(69) & 1.138(32) & 0.967(26) \\
\hline 
3	& $A_2^{(0)}$&1.557(36)  & 1.298(33) &1.173(18)   &  0.869(15)   \\
	& $T_1^{(0)}$&1.604(17)  & 1.367(26) &1.173(25)   &  0.943(34)   \\
	& $T_2^{(2)}$&1.580(21)  & 1.330(22) &1.220(20)   &  0.897(18)   \\
\hline
$3^{*}$ & $A_2^{(1)}$& /  & $\leq$1.551(94) & 1.367(24) &  0.990(41) \\
\hline
4	& $A_1^{(3)}$  & 1.648(52) & 1.399(38)  & 1.244(23) & 1.006(35)  \\
	& $E^{(3,2)}$    & 1.601(24) & 1.366(27)  & 1.195(14) & 0.920(13)  \\
	& $T_1^{(1)}$  & 1.54(13) & 1.400(84) & 
	$\leq 1.360(64)$ & $\leq1.015(70)$\\
	& $T_2^{(3)}$  & 1.613(18) & 1.395(16)  & 1.227(36) & 0.962(17) \\
\hline
$4^{*}$	& $A_1^{(4,5)}$  & 1.686(40) & 1.482(37)  &1.446(33)  & 1.150(46)  \\
\hline 
$6$	& $A_1^{(6)}$  & / & /  & 1.562(50) & /  \\
        & $A_2^{(2)}$  & / & 1.867(73)  & 1.609(62) & 1.179(32)  \\
	& $E^{(6+)}$    & / & 1.766(50)  & 1.526(37) & 1.17(10)  \\
\hline
\hline 
\end{tabular}
\vspace{0.6cm}\\
\begin{tabular}{|c|c|c|c|c|c|}
\hline
{\Large$J^{-+}$} & IR  & $\beta=6.0$ & $\beta=6.1$ & $\beta=6.2$ &$\beta=6.4$ \\
\hline 
\hline
$0$      &$A_1^{(0)}$&1.151(10)&0.982(10) & 0.815(15)&0.615(14)\\
\hline
$0^{*}$  &$A_1^{(1)}$&1.47(10) & 1.312(36)& 1.125(22)&0.885(25)\\
\hline
?	&$A_1^{(2)}$& $\leq2.10(13)$&1.563(81)& 1.482(31) & $\leq1.389(78)$\\
\hline 
2       & $E^{(0)}$  &1.363(11)&1.139(21)&0.9986(86)&0.729(12)\\
        & $T_2^{(0)}$&1.370(10)&1.159(10)&1.0067(74)&0.760(11)\\
\hline
$2^*$   & $E^{(1)}$  &1.758(47)&1.468(25)&1.250(15)&0.964(52)\\
	& $T_2^{(1)}$&1.68(14)&1.497(24)&1.278(15)&1.009(47)\\
\hline 
?	& $A_2^{(0)}$&2.06(12)&$\leq1.830(97)$&1.529(63)& /       \\
\hline 
5 	& $E^{(2)}  $&$\leq1.954(74)$&$\leq1.733(58)$&  1.43(14)  &  /    \\
	& $T_1^{(0)}$&1.772(33)&1.569(41)&1.319(38)&1.024(43)\\
	& $T_1^{(1)}$&1.830(56)&1.608(28)&1.356(42)& 1.061(44) \\ 
	& $T_2^{(2)}$& 1.75(18)& 1.583(26) & 1.410(50) & 1.10(11)  \\
\hline
\hline
\end{tabular}
\end{center}
\caption{The lightest $SU(3)$ glueball states in the $C=+$ sector. 
We put together the lattice states corresponding to different polarisations
 of the continuum states. The index on the lattice IR indicates the energy level
in that representation.}
\label{tab:su3p}
\end{table}
\begin{table}
\begin{center}
\begin{tabular}{|c|c|c|c|}
\hline
Run     &   I  &  II  & III \\
\hline
$\beta$ & 6.0 & 6.1 & 6.4 \\
$V$     & $16^3\times36$ & $20^3\times24$ &  $32^3\times48$\\
sweeps  & $1.5\times10^6$   & $1.1\times10^6$    & $0.25\times10^6$ \\
$a\sqrt{\sigma}$  & 0.2150(10) & 0.1835(16)   & 0.1176(15)\\ 
$a$[fm] & 0.10 & 0.083 & 0.053 \\
$L$[fm]& 1.58 & 1.66 &  1.69 \\
$M_{2T}$& 1.346(13) & 1.242(22) &   0.8196(20) \\
\hline
\end{tabular}
\vspace{0.5cm}\\
\begin{tabular}{|c|c|c|c|}
\hline
Run     &   IV  &  V  & VI \\
\hline
$\beta$ & 6.2 & 6.2 & 6.2  \\
$V$     & $20^3\times32$ & $24^3\times32$ & $28^3\times32$\\
sweeps  &  $0.14\times10^6$ & $0.93\times10^6$ & $0.16\times10^6$ \\
$a\sqrt{\sigma}$  & 0.1580(52) &  0.15812(54)  & 0.1539(23)\\ 
$a$[fm] &0.0708   & 0.0708 & 0.0708 \\ 
$L$[fm]& 1.42 & 1.70  & 1.98\\  
$M_{2T}$& 0.894(58)& 1.1128(76)& 1.252(34) \\
\hline
\end{tabular}
\caption{The parameters of the $SU(3)$ simulation runs.  The L\"uscher correction
is used to extract string tension from the measured torelon masses. 
 For the three runs at  $\beta=6.2$ we use the string tension from run 
V to set the scale.}
\label{tab:su3runs}
\end{center}
\end{table}

\begin{table}
\begin{center}
\begin{tabular}{|c|c|c|c|c|c|}
\hline
{\Large$J^{+-}$} & IR  & $\beta=6.0$ & $\beta=6.1$ & $\beta=6.2$ &$\beta=6.4$ \\
\hline
\hline 
?	& $A_1^{(0)}$&$1.91(18)$ & $1.62(33)$&	1.39(14) & / \\
\hline 
1	& $T_1^{(0)}$&1.305(17)&1.138(15)&0.955(10)&0.718(16)\\
\hline 
3	& $T_1^{(1)}$&1.504(53)&1.324(15)&1.161(12)&0.827(20)\\
	& $T_2^{(0)}$&1.488(50)&1.306(22)&1.149(10)&0.857(14)\\
\hline
$3^*$   & $T_1^{(2)}$&1.659(44)   & 1.446(19)  & 1.260(15) & 1.005(22)     \\
	& $T_2^{(1)}$& $\leq1.909(41)$& 1.33(15)&$\leq1.344(57)$ & 1.090(93)\\
\hline 
5 (?)	& $E^{(0)}$  &1.729(47)&1.551(42)&1.371(31)&1.052(46)\\
	& $T_1^{(3)}$& 1.70(14)  & 1.541(53)  &  1.373(16)  & 1.028(38)   \\
        & $T_2^{(2)}$&1.69(15)&1.581(58)& 1.386(18)&1.058(75)\\
\hline  
\hline
\end{tabular}
\vspace{1cm} \\
\begin{tabular}{|c|c|c|c|c|c|}
\hline
{\Large$J^{--}$} & IR  & $\beta=6.0$ & $\beta=6.1$ & $\beta=6.2$ &$\beta=6.4$ \\
\hline 
\hline
? & $A_1^{(0)}$  &2.13(13) &1.79(13) &	1.584(73) & 1.230(74) \\
\hline
1	& $T_1^{(0)}$&1.624(84)&1.422(54)&1.154(50)&0.98(14)\\
\hline 
$1^*$	& $T_1^{(1)}$&1.829(69)&1.618(30)&1.461(23)&1.29(14)\\
\hline 
2       & $E^{(0)}$  &1.668(49)&1.445(22)&1.269(17)&0.968(33)\\
        & $T_2^{(0)}$&1.679(49)&1.463(24)&1.241(30)&0.987(51)\\
\hline	
$2^*$	& $E^{(1)}$  &1.902(60)&1.571(67)&1.391(35)&0.991(50)\\
	& $T_2^{(1)}$&1.794(38)&1.617(40)&1.462(59)&1.32(17)\\
\hline 
3	& $A_2^{(0)}$&1.941(76) &$\leq1.746(66)$&1.45(11) & / \\
	& $T_1^{(2)}$&1.993(89) &  1.710(51)    &1.483(25) & /  \\
	& $T_2^{(2)}$&1.97(12)  &  1.730(50)    &1.530(27)& $\leq1.485(54)$ \\
\hline
\hline
\end{tabular}
\end{center}
\caption{As Table~(\ref{tab:su3p}), for the $C=-$ sector.}
\label{tab:su3m}
\end{table}

\begin{table}
\begin{center}
\begin{tabular}{|c|c|c|c|}
\hline
{\Large$ J^{++}$} & IR  & $1-\frac{m(L_s=20)}{m(L_s=24)}$ 
&$1-\frac{m(L_s=28)}{m(L_s=24)}$  \\
\hline 
\hline
$0$      &$A_1^{(0)}$ & 0.008(27)& -0.024(27) \\  
\hline
$0^{*}$  &$A_1^{(1)}$ & {\bf 0.078(29)}&  {\bf0.076(33)} \\  
\hline
$0^{**}$ &$A_1^{(2)}$ & -0.009(34) & 0.001(48)  \\ 
\hline
$0^{***}$&$A_1^{(5)}$ & 0.014(48)& 0.057(81) \\  
\hline 
2       & $E^{(0)}$   & {\bf 0.040(18)} & {\bf 0.053(24)} \\ 
        & $T_2^{(0)}$ & 0.007(15) & 0.009(16) \\ 
\hline
$2^*$   & $E^{(1)}$   & 0.003(29) &0.011(29)\\  
	& $T_2^{(1)}$ & -0.027(23) & 0.006(25)\\  
\hline
2T	& $E^{(2,3)}$   & {\bf0.077(51)}  &  {\bf-0.073(36)} \\  
\hline 
3	& $A_2^{(0)}$& 0.013(40) & -0.019(55)  \\ 
	& $T_1^{(0)}$& 0.009(32) & -0.073(63) \\ 
	& $T_2^{(2)}$& 0.001(32) & 0.065(35)\\ 
\hline
$3^{*}$ & $A_2^{(1)}$&  0.049(45) &  -0.015(52)\\  
\hline
4	& $A_1^{(3)}$  & -0.032(54)& 0.039(51)\\ 
	& $E^{(3,2)}$    &0.001(25)& 0.014(37)  \\ 
	& $T_1^{(1)}$  & / & / \\
	& $T_2^{(3)}$  & 0.037(39)&  0.010(43) \\ 
\hline
$4^{*}$	& $A_1^{(4)}$  &/ & -0.004(69)\\  
\hline 
$6$	& $A_2^{(2)}$ &   0.070(79)  & -0.019(94)\\ 
	& $E^{(6+)}$  &  /  & /\\
\hline
\hline
\end{tabular}
\vspace{0.5cm}\\
\begin{tabular}{|c|c|c|c|}
\hline
{\Large $J^{-+}$} & IR  & $1-\frac{m(L_s=20)}{m(L_s=24)}$ 
& $1-\frac{m(L_s=28)}{m(L_s=24)}$ \\
\hline 
\hline
$0$      &$A_1^{(0)}$&  -0.039(37)  & -0.012(41)\\
\hline
$0^{*}$  &$A_1^{(1)}$& 0.068(43) &  0.070(38) \\
\hline 
2       & $E^{(0)}$  & 0.008(19) & 0.002(25) \\
        & $T_2^{(0)}$& 0.007(15) &  -0.007(22)\\
\hline
$2^*$   & $E^{(1)}$  & 0.028(34)  & 0.026(33)\\
	& $T_2^{(1)}$&  0.030(33) &  0.010(30)\\
\hline 
3	& $A_2^{(0)}$&  0.045(74) & -0.05(10)\\
\hline 
5	& $T_1^{(0)}$& -0.021(52) &  0.038(48)\\
	& $T_1^{(1)}$& -0.052(44) &  0.001(49)\\
	& $T_2^{(2)}$& 0.021(47)  &  0.050(48)  \\
\hline
\hline
\end{tabular}
\end{center}
\caption{Finite volume study at $\beta=6.2$, comparing $V=20^3\times32$
and $V=28^3\times32$ with $V=24^3\times32$ (runs IV, V, VI); 
$C=+$ sector. Statistically significant variations are highlighted. }
\label{tab:fvp}
\end{table}

\begin{table}
\begin{center}
\begin{tabular}{|c|c|c|c|}
\hline
\Large{$J^{+-}$} & IR  & $1-\frac{m(L_s=20)}{m(L_s=24)}$ 
& $1-\frac{m(L_s=28)}{m(L_s=24)}$\\
\hline 
\hline
1	& $T_1^{(0)}$& -0.023(16)&  0.023(20)\\
\hline 
3	& $T_1^{(1)}$& 0.001(18)&  0.044(31)\\
	& $T_2^{(0)}$& 0.003(21)&  0.001(27)\\
\hline 
$3^*$	& $T_1^{(2)}$&  -0.005(26)  & 0.030(31) \\
\hline 
5 (?)   & $E^{(0)}$  & -0.001(42)& -0.016(53)\\
	& $T_1^{(3)}$  & 0.029(30) & 0.088(46)\\
        & $T_2^{(2)}$&  0.033(26)& 0.016(43)\\
\hline
\hline
\end{tabular}
\vspace{0.5cm}\\
\begin{tabular}{|c|c|c|c|}
\hline
\Large{$J^{--}$} & IR  & $1-\frac{m(L_s=20)}{m(L_s=24)}$ 
& $1-\frac{m(L_s=28)}{m(L_s=24)}$\\
\hline 
\hline
1	& $T_1^{(0)}$& -0.024(54)&  0.046(61)\\
\hline 
1	& $T_1^{(1)}$& -0.013(60)   & / \\
\hline 
2       & $E^{(0)}$  & 0.007(26) & -0.028(33)\\
        & $T_2^{(0)}$& -0.003(39)&  0.017(44)\\
\hline	
$2^*$	& $E^{(1)}$  & 0.014(50) & 0.022(55)\\
	& $T_2^{(1)}$ & -0.010(50)  & -0.021(41)   \\
\hline 
3	& $T_1^{(1)}$&  -0.014(60)&  0.08(11)\\
	& $T_2^{(1)}$&   0.034(61)&  0.003(58)\\
\hline
\hline
\end{tabular}
\end{center}
\caption{As Table~(\ref{tab:fvp}), for the $C=-$ sector.}
\label{tab:fvm}
\end{table}

\begin{table}
\begin{center}
\begin{tabular}{|r|c|c||c|c|}
\hline
State  & $A_1^{++}$  &$A_1^{++}~\&~$2T &  $E^{++}$ & $E^{++}~\&~$2T \\
\hline
\hline
0         & 0.515(13)  & 0.515(13) &  0.747(13)     &    0.756(13)   \\
      1   & 0.856(26) & 0.842(22)  &  1.030(21)     &   {\bf0.786(31) }    \\
$L_s=20$~~~2&{\bf1.064(42)}&{\bf0.874(36)}&1.111(31)&  1.058(23) \\
 3        & 1.161(36) & {\bf1.100(34)} & 1.185(27)&  1.169(50)     \\
 4        & 1.284(52) & 1.159(36) &                &   1.191(54)    \\
  5       &          & 1.261(57) &                &       \\
\hline 
0           & 0.534(17) & 0.536(18) &  0.737(17)    &    0.738(17)   \\
  1         & 0.858(29) & 0.923(57)  &  1.021(23) &   1.013(25)  \\
$L_s=28$~~~2& 1.159(53) & 1.158(65)  &  1.178(44)& {\bf 1.134(49)}  \\
 3          & 1.213(56) & {\bf1.21(13)} &             &    1.186(53)   \\
 4          &           & 1.212(74) &             &       \\
\hline
\hline
\end{tabular}
\end{center}
\caption{Effect on the spectrum 
of including a bi-torelon operator into the variational basis
 at $\beta=6.2$, at $V=20^3\times32$ and  $V=28^3\times32$ (runs IV and VI).
Bold-face values correspond to states that do not appear in 
Table~(\ref{tab:su3p}).}
\label{tab:bitorelon}
\end{table}
\begin{table}
\begin{center}
\begin{tabular}{|c|c|c|c|}
\hline
$SO(3)$ IR & $O_h$ IR & Projected basis \\
\hline
\hline
$\mathbf{D_0}$ & $A_1$ & $Y_0^0$ \\
\hline
\hline
$\mathbf{D_1}$ & $T_1$ & $Y_1^{\pm1},Y_1^0$ \\
\hline
\hline
$\mathbf{D_2}$ & $E$   & $Y_2^0,~\frac{1}{\sqrt{2}}(Y_2^2+Y_2^{-2})$ \\
\hline
      & $T_2$ & $Y_2^{\pm1},~\frac{1}{\sqrt{2}}(Y_2^2-Y_2^{-2})$  \\
\hline
\hline
$\mathbf{D_3}$ & $A_2$ & $\frac{1}{\sqrt{2}}(Y_3^2-Y_3^{-2})$ \\
\hline
      & $T_1$ & $Y_3^0$\\
\hline
      & $T_2$ & $\frac{1}{\sqrt{2}}(Y_3^2+Y_3^{-2})$ \\
\hline
\hline
$\mathbf{D_4}$ & $A_1$ & $0.7637Y_4^0+0.6456(Y_4^4+Y_4^{-4})/\sqrt{2}$  \\
\hline
             & $E$   & $\frac{1}{\sqrt{2}}(Y_4^2+Y_4^{-2})$ \\
      &       & $ 0.6456Y_4^0-0.7637(Y_4^4+Y_4^{-4})/\sqrt{2}$ \\
\hline
       & $T_1$ &  $\frac{1}{\sqrt{2}}(Y_4^4-Y_4^{-4})$\\
\hline
     & $T_2$ & $\frac{1}{\sqrt{2}}(Y_4^2-Y_4^{-2})$\\
\hline
\hline
$\mathbf{D_5}$ & $E$ & $\frac{1}{\sqrt{2}}(Y_5^2-Y_5^{-2})$ \\
		& &  $\frac{1}{\sqrt{2}}(Y_5^4-Y_5^{-4})$\\
\hline
 & $T_1$ & $\frac{\cos{\mu}}{\sqrt{2}}(Y_5^4+Y_5^{-4}) + \sin{\mu}~Y_5^0$\\
\hline
 & $T_1$ & $\cos{\mu}~Y_5^0-\frac{\sin{\mu}}{\sqrt{2}}(Y_5^4+Y_5^{-4})$\\
\hline
 & $T_2$ & $\frac{1}{\sqrt{2}}(Y_5^2+Y_5^{-2})$\\
\hline
\hline
$\mathbf{D_6}$ & $A_1$ & $0.354Y_6^0-0.661(Y_6^4+Y_6^{-4})/\sqrt{2}$\\
\hline
    &$A_2$ & $0.830(Y_6^2+Y_6^{-2})/\sqrt{2}-0.559(Y_6^6+Y_6^{-6})/\sqrt{2}$\\
\hline
     & $E$   & $0.354(Y_6^4+Y_6^{-4})/\sqrt{2}+0.935Y_6^0$ \\
   & &  $0.559(Y_6^2+Y_6^{-2})/\sqrt{2} + 0.830(Y_6^6+Y_6^{-6})/\sqrt{2}$ \\
\hline
     & $T_1$ &  $\frac{1}{\sqrt{2}}(Y_6^4-Y_6^{-4})$\\
\hline
    & $T_2$ & $ \frac{\cos{\mu}}{\sqrt{2}}(Y_6^2-Y_6^{-2})
                + \frac{\sin{\mu}}{\sqrt{2}}(Y_6^6-Y_6^{-6})$\\
\hline
     & $T_2'$ & $ \frac{\cos{\mu}}{\sqrt{2}}(Y_6^6-Y_6^{-6})
                 - \frac{\sin{\mu}}{\sqrt{2}}(Y_6^2-Y_6^{-2}) $ \\
\hline
\hline
\end{tabular}
\end{center}
\caption{Decomposition of the spherical-harmonics basis 
into the cubic lattice irreducible representations.
In the $T_1$, $T_2$ cases, the two other linear combinations of $Y_\ell^m$
can be obtained  by applying the $D(0,\frac{\pi}{2},0)$ and 
$D(0,\frac{\pi}{2},\frac{\pi}{2})$ matrices on the given state.
When a state $J$ occurs twice in a lattice representation, an arbitrary
mixing angle $\mu$ appears between different polarisations $m$.}
\label{tab:decomp}
\end{table}
\begin{table}
\begin{center}
\begin{tabular}{|c|c|c|}
\hline
$J$ & IR$^{++}$  & lower bound on $||\psi_J||^2$ \\
\hline
\hline
$4$  & $A_1^{(3)}$&  0.071(1)\\
$4$  & $E^{(2)}$  & 0.094(1)\\
$4$  & $T_1^{(1)}$ & 0.11(2)\\
\hline
$4^*$ & $A_1^{(5)}$ & 0.205(3) \\
\hline
$6$ & $A_2^{(2)}$  & 0.213(5) \\
\hline
\hline
\end{tabular}
\end{center}
\caption{Lower bounds obtained 
on the higher-spin components $||\psi_J||^2$, where
for each lattice state the normalisation is such 
that $\sum_J~||\psi_J||^2=1$. At $\beta=6.4$ (run III), where the 
systematic uncertainty  on  $||\psi_J||^2$ is of order
 $\sigma a^2\simeq0.014$. We only give a value for the lower 
bound when it exceeds twice that value.}
\label{tab:fourier}
\end{table}

\begin{table}
\begin{center}
\begin{tabular}{|c|c@{\qquad}c|}
\hline
Run IV &\small{$A_1^{(I)}\times A_1^{(I)}$}& \small{$A_1^{(II)}\times A_1^{(II)}$}\\
\hline
$A_1^{(0)}$ &  0.011   & 0.004 \\ 
$A_1^{(1)}$ &   0.094  & 0.084\\
$A_1^{(2)}$ &   0.005  & 0.011 \\
$A_1^{(3)}$ &   0.062 & 0.057\\
$A_1^{(4)}$ &  0.065   & 0.055\\
$A_1^{(5)}$ &   0.015  & 0.012\\
\hline
\end{tabular}
\end{center}
\caption{The overlap between our orthonormal basis of operators, which are 
approximations to the 1-glueball-state wavefunctions, 
 and the direct product of two zero-momentum operators each 
 having $95\%$ overlap onto the lightest glueball. 
The direct products are expected to couple to 2-glueball scattering
states; two of them are used to estimate the systematic uncertainty on the 
overlap.}
\label{tab:2G-1G}
\end{table}





\begin{table}
\begin{center}
\begin{tabular}{|c|c|c|c|c|c|}
\hline
\Large{$J^{++}$ }   & IR & $m/\sqrt{\sigma}$ &$\nu$ &  $\chi^2/(\nu-2)$&Average $m/\sqrt{\sigma}$\\
\hline
\hline
$0$      &$A_1$&  3.347(68)      & 4   &  0.19 & 3.347(68) \\
\hline
$0^{*}$  &$A_1$&  6.26(16)       & 4   &  0.71 &  6.26(16)    \\
\hline
$0^{**}$ &$A_1$&  7.65(23)       & 4   &  0.71 &  7.65(23)   \\
\hline
$0^{***}$&$A_1$&  9.06(49)       & 4   &  0.74 &  9.06(49) \\
\hline
2       & $E$  &  4.916(91)  & 4 & 0.37 &  4.891(65) \\
        & $T_2$&  4.878(65)  & 4 & 0.65 &\\
\hline
$2^*$   & $E$  &  6.48(22)   & 4 & 0.90 & 6.54(22)\\
	& $T_2$&  6.76(41)   & 3 & 0.04 & \\
\hline 
3	& $A_2$&  7.52(20)  & 4 & 0.83 & 7.69(20)\\
	& $T_1$&  7.70(27)  & 4 & 1.3& \\
	& $T_2$&  7.85(20)  & 4 & 2.00& \\
\hline
4	& $A_1$  & 8.49(33) & 4 & 1.2 & 8.28(21)\\
	& $E$    & 8.06(28) & 3 & 0.14& \\
	& $T_2$  & 8.31(21) & 4 & 0.76 & \\
\hline
$4^{*}$	& $A_1$  & 10.48(38) & 4 &2.2  & 10.48(38)\\
\hline 
$6$	& $A_2$  & 9.92(58) & 3 & 0.02 & 9.91(58)\\
	& $E$   & 9.90(99) & 3 & 0.06 &\\
\hline
\hline 

\end{tabular}
\\
\vspace{0.5cm}
\begin{tabular}{|c|c|c|c|c|c|}
\hline
\Large{$J^{-+}$ }   & IR & $m/\sqrt{\sigma}$ &$\nu$ &  $\chi^2/(\nu-2)$&Average $m/\sqrt{\sigma}$\\
\hline
\hline
$0$      &$A_1$&5.11(14) & 4 & 0.86& 5.11(14) \\
\hline
$0^{*}$  &$A_1$& 7.66(35) & 4 & 0.45& 7.66(35) \\
\hline 
2       & $E$  &6.23(12) & 4 & 0.52& 6.32(11)\\
        & $T_2$&6.40(11) & 4 & 0.45& \\
\hline
$2^*$   & $E$  &7.70(31) & 4  & 0.39& 7.91(31)\\
	& $T_2$&8.32(43) & 4   & 0.73& \\
\hline 
5 	& $T_1$&8.76(37) & 4 & 0.45& 8.96(37)\\ 
	& $T_1$& 9.03(44) & 4 & 0.45& \\ 
	& $T_2$& 9.82(91) & 4 & 0.00& \\ 
\hline
\hline 

\end{tabular}
\end{center}
\caption{The continuum extrapolation for the $SU(3)$ $C=+$ sector.}
\label{tab:extrapol_p}
\end{table}

\begin{table}     
\begin{center}
\begin{tabular}{|c|c|c|c|c|c|}
\hline
\Large{$J^{+-}$}    & IR & $m/\sqrt{\sigma}$ &$\nu$ &  $\chi^2/(\nu-2)$&Average $m/\sqrt{\sigma}$\\
\hline
\hline
1	& $T_1$&6.06(15) & 4 & 0.97 &  6.06(15)\\
\hline 
3	& $T_1$&  7.32(23) & 4 & 1.8 &  7.43(20)\\
	& $T_2$& 7.52(20)  & 4 & 0.13 &\\
\hline
$3^*$   & $T_1$& 8.26(32)  & 3 & 0.01 & 8.26(32) \\
\hline
5 (?)   & $E$  &9.39(38) & 4 & 0.02  &  9.35(38)\\ 
	& $T_1$&9.25(48) & 4 & 0.26 & \\ 
        & $T_2$&9.43(63) & 4 & 0.27 & \\ 
\hline 
\hline 
\end{tabular}
\vspace{0.5cm}\\

\begin{tabular}{|c|c|c|c|c|c|}
\hline
\Large{$J^{--}$}    & IR & $m/\sqrt{\sigma}$ &$\nu$ &  $\chi^2/(\nu-2)$&Average $m/\sqrt{\sigma}$\\
\hline
\hline
?	& $A_1$ & 10.51(81)    & 4 &  0.13 & 10.51(81) \\
\hline 
1	& $T_1$&7.36(76)	& 4 & 0.68 &7.36(76)\\
\hline 
$1^*$	& $T_1$&10.26(48)	& 4 & 0.68 &10.30(48)\\
\hline 
2       & $E$  &8.39(29)	& 4 & 0.04 &8.32(29)\\
        & $T_2$&8.19(40)	& 4 & 0.57 &\\
\hline	
$2^*$	& $E$  &8.50(45)	& 4 & 0.34 &8.50(45)\\
\hline 
3	& $T_1$& 9.51(59)	& 3 & 0.01 & 9.84(59)   \\
	& $T_2$& 10.31(71)	& 3 & 0.01 &    \\
\hline
\hline 
\end{tabular}
\end{center}
\caption{The continuum extrapolation for the $SU(3)$ $C=-$ sector.}
\label{tab:extrapol_m}
\end{table}

\begin{table}
\begin{center}
\begin{tabular}{|c|c@{}l@{ }l|}
\hline
$J^{PC}$   &  $m[GeV]$  && \\
\hline
\hline
$0^{++}$   &1.475&(30) &(65)\\
$0^{++*}$  &2.755&(70) &(120)\\
$0^{++**}$ &3.370&(100)&(150)\\
$0^{++***}$&3.990&(210) &(180)\\
$2^{++}$   &2.150&(30) &(100) \\
$2^{++*}$  &2.880&(100) &(130)\\
$3^{++}$   &3.385&(90) &(150)\\
$4^{++}$   &3.640&(90) &(160)\\
$6^{++}$   &4.360&(260) &(200)\\
\hline
$0^{-+}$   &2.250&(60) &(100)\\
$0^{-+*}$  &3.370&(150) &(150)\\
$2^{-+}$   &2.780&(50) &(130)\\
$2^{-+*}$  &3.480&(140) &(160)\\
$5^{-+}$   &3.942&(160) &(180)\\
\hline 
$1^{+-}$   &2.670&(65) &(120)\\
$3^{+-}$   &3.270&(90) &(150)\\
$3^{+-*}$   &3.630&(140) &(160)\\
$5^{+-}$ (?)  &4.110&(170) &(190)\\
\hline 
$1^{--}$   &3.240&(330) &(150)\\
$2^{--}$   &3.660&(130)  &(170)\\
$2^{--*}$  &3.740&(200)  &(170)\\
$3^{--}$   & 4.330&(260)  &(200)\\
\hline
\hline 
\end{tabular}
\end{center}
\caption{The final 4D $SU(3)$ spectrum in physical units setting
$\sqrt{\sigma}=440(20)$MeV.
The first error is the statistical error from the continuum-limit 
extrapolation and the second is the uncertainty on the string tension
 $\sigma$. Sources of systematic error are discussed in 
section~(\ref{sec:systematics}).}
\label{tab:continuum}
\end{table}

\clearpage

\begin{table}
\begin{center}
\begin{tabular}{|c|c|c|c|c|}
\hline
Run  & A   &  B   &   C  &  D   \\
\hline
$\beta$ & 44.0 & 44.35 & 44.85 & 45.50 \\
$V$     & $10^3\times16$ & 
$12^4$ &$12\times16\times20\times16$ &   $16^3\times24$\\
sweeps  & $4.0\times10^5$& $2.6\times10^5$   & 
$2.4\times10^5$    & $1.1\times10^5$ \\
$a\sqrt{\sigma}$  &0.3406(20)$^*$  &0.2991(20)$^*$ &  0.2579(24)& 0.2153(24) \\
$a$[fm] & 0.153 & 0.134& 0.115& 0.096\\
$L$[fm]&1.53 &1.61 &1.38, 1.84, 2.30 &1.54 \\
$M_{2T}$&2.11(2) & 1.97(3)& 1.42(3)& 1.35(3)\\
\hline
\end{tabular}
\end{center}
\caption{The parameters of the 4D $SU(8)$ runs. Values with a $^*$ are taken
from~\cite{Lucini:2003zr}.}
\label{tab:su8runs}
\end{table}
\begin{table}
\begin{center}
\begin{tabular}{|c|c|c|c|c|c|}
\hline
\Large{$J^{PC}$ }   & IR & $m/\sqrt{\sigma}$ &$\nu$ &  $\chi^2/(\nu-2)$&Average $m/\sqrt{\sigma}$\\
\hline
\hline
$0^{++}$      &$A_1$&  3.32(15)      & 4   &  0.41 & 3.32(15)\\
\hline
$0^{++*}$  &$A_1$&  4.71(29)      & 4   &  0.39 &  4.71(29)      \\
\hline
$2^{++}$  &$E$&  4.74(21)       & 4   &  0.20 &  4.65(19)   \\
  &$T_2$&  4.57(19)       & 4   &  0.45 &     \\
\hline
$2^{++*}$  &$E$&  6.47(50)      & 4   &  1.0 &    6.47(50)    \\
\hline
$3^{++}$  &$A_2$&  7.2(1.3)       & 3   &  0.08 &   7.2(1.3)  \\
\hline
\hline
$0^{-+}$      &$A_1$&  4.72(32)      & 4   &  1.1 &  4.72(32)   \\
\hline
 $?^{-+}$    &$T_1$&  7.87(77)      & 4   &  0.70 &  7.87(77) \\
\hline
$2^{-+}$      &$E$&  6.21(53)      & 4   &  0.28 & 5.67(40)\\
      &$T_2$&  5.36(40)      & 4   &  0.22 &  \\
\hline
\hline
$1^{+-}$   &$T_1$&  5.70(29)      & 4   &  0.85 & 5.70(29)  \\
\hline
$3^{+-}$   &$A_2$& 7.2(1.5)      & 3   &  0.09 & 7.74(79)  \\
           &$T_2$& 7.89(79)      & 3   &  0.18 &   \\
\hline
\hline
$1^{--}$   &$T_1$&  7.45(60)      & 4   &  0.07 & 7.45(60) \\
\hline
$2^{--}$      &$E$&  7.4(1.4)     & 3   &  0.87 &  7.3(1.4)\\
	      &$T_2$&  7.2(1.5)      & 3   &  0.01 &  \\
\hline
$3^{--}$   &$T_1$&  7.1(1.2)      & 3   &  0.001 &7.5(1.1)  \\
	   &$T_2$&  7.9(1.1)      & 3   &  0.004 &  \\
\hline
\hline
%
%
%
%
\end{tabular}
\end{center}
\caption{The 4D $SU(8)$ continuum spectrum.}
\label{tab:su8cont}
\end{table}

\begin{table}
\begin{center}
\begin{tabular}{|c|c|c|c|c|c|}
\hline
\Large{$J^{++}$ }& IR& $\beta=44.00$  & $\beta=44.35$ & $\beta=44.85$ & $\beta=45.50$  \\
\hline 
\hline
$0$       &$A_1$&0.838(21) &  0.822(30)  & 0.721(25)&  0.642(22) \\
\hline
$0^{*}$   &$A_1$&1.560(50)  &1.376(68)  & 1.262(78)&  0.906(60)  \\
\hline
$0^{**}$  &$A_1$& 1.72(17) & 1.64(17) & 1.58(16) &  1.029(86) \\
\hline
$2$       &$E$& 1.483(43) & 1.362(44)  & 1.168(36) &  0.987(26) \\
         &$T_2$&1.546(29)  &1.324(33) & 1.175(21) & 0.984(33)\\
\hline
$2^*$       &$E$& 1.79(19) & 1.44(18)  & 1.42(12) & 1.329(72)  \\
            &$T_2$&$\leq 1.92(15)$  & 1.58(12)  & / &  1.380(57)  \\
\hline
$3$       &$A_2$& / &1.80(21)   & 1.63(27) &  $\leq1.59(11) $\\
          &$T_1$&$\leq$2.68(23)  &$ 1.78(30)$ & 1.669(87) & 1.420(95)\\
          &$T_2$& / &1.79(16) & /            & 1.466(59) \\
\hline
\hline 
\end{tabular}
\vspace{0.5cm}\\
\begin{tabular}{|c|c|c|c|c|c|}
\hline
\Large{$J^{-+}$ }& IR& $\beta=44.00$  & $\beta=44.35$ & 
            $\beta=44.85$ & $\beta=45.50$  \\
\hline 
\hline
$0$       &$A_1$ &1.543(74)  & 1.413(54)  & 1.131(51)& 1.025(40)  \\
\hline
$2$       &$E$ &  2.00(24)  &   1.61(16) & 1.500(73) & 1.283(43)  \\
          &$T_2$& 2.027(80)   &  1.63(18)    & 1.450(66) & 1.214(53) \\
\hline
?       &$T_1$&2.07(28)  & 1.63(21)   & 1.749(14) & 1.503(75)\\
\hline
\hline 
\end{tabular}
\end{center}
\caption{The 4D $SU(8)$ $C=+$ spectrum.}
\label{tab:su8p}
\end{table}

\begin{table}
\begin{center}
\begin{tabular}{|c|c|c|c|c|c|}
\hline
\Large{$J^{+-}$} & IR & $\beta=44.00$  & $\beta=44.35$ & $\beta=44.85$ & $\beta=45.50$  \\
\hline 
\hline
$1$ & $T_1$ &1.863(70)  & 1.61(16)  & 1.360(58) & 1.221(31) \\
\hline 
$3$ & $A_2$ &/  & 1.98(32)  &   1.84(27) &  1.49(10) \\
    & $T_2$ & 2.35(17) &1.67(20)  &  1.652(91)  & 1.450(46)   \\
\hline
?  & $A_1$ &/  & 2.15(52)  &   2.21(36) & 1.66(18)  \\
 &$T_2$ &/ & 2.33(33) &1.92(14)  & 1.73(13) \\
\hline
\hline 
\end{tabular}
\vspace{0.5cm}\\
\begin{tabular}{|c|c|c|c|c|c|}
\hline
\Large{$J^{--}$} & IR& $\beta=44.00$  & $\beta=44.35$ & $\beta=44.85$ & $\beta=45.50$  \\
\hline 
\hline
? & $A_1$  &/ & 1.76(46)  &  1.81(42) &  1.47(23) \\
\hline
$3$ & $A_2$  &/  & /  &  1.96(26) &  $\leq 1.73(13)$ \\
    & $T_1$ &/ &  2.25(37) & 1.92(15)  &   1.576(63)\\
    & $T_2$ & / & 2.32(25)  & 2.00(14) & 1.688(75)  \\
\hline
$2$ & $E$  &/ & 1.93(24)  &   1.89(13) &  1.48(12) \\
    &$T_2$ &/ & 1.99(39)  &   1.72(23) &   1.485(87)\\
\hline
$1$ & $T_1$ & 2.29(29) &1.99(15)  & 1.75(15) &  1.528(50)\\
\hline
\hline 
\end{tabular}
\end{center}
\caption{The 4D $SU(8)$ $C=-$ spectrum.}
\label{tab:su8m}
\end{table}

%% file: figs.tex
\begin{figure}[t]
\vspace{-1cm}
\centerline{\begin{minipage}[c]{15.2cm}
   \psfig{file=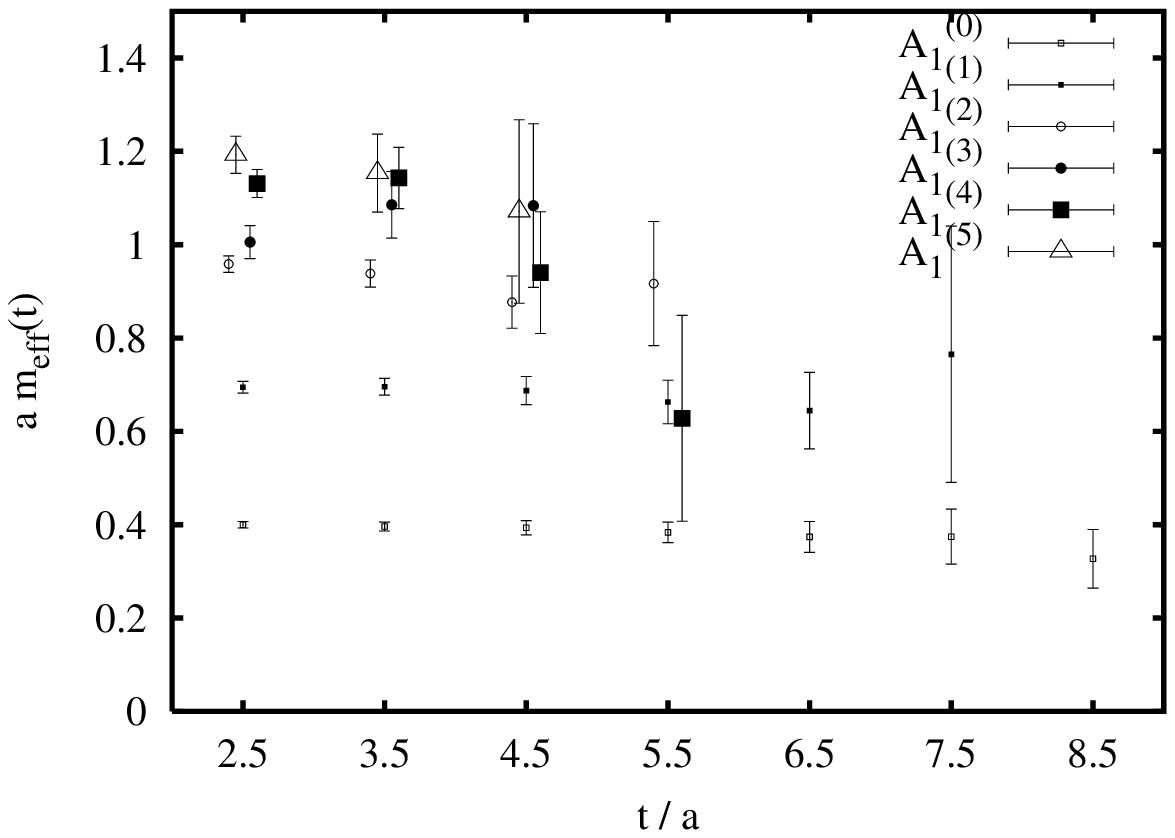,angle=90,width=7.5cm} 
 \psfig{file=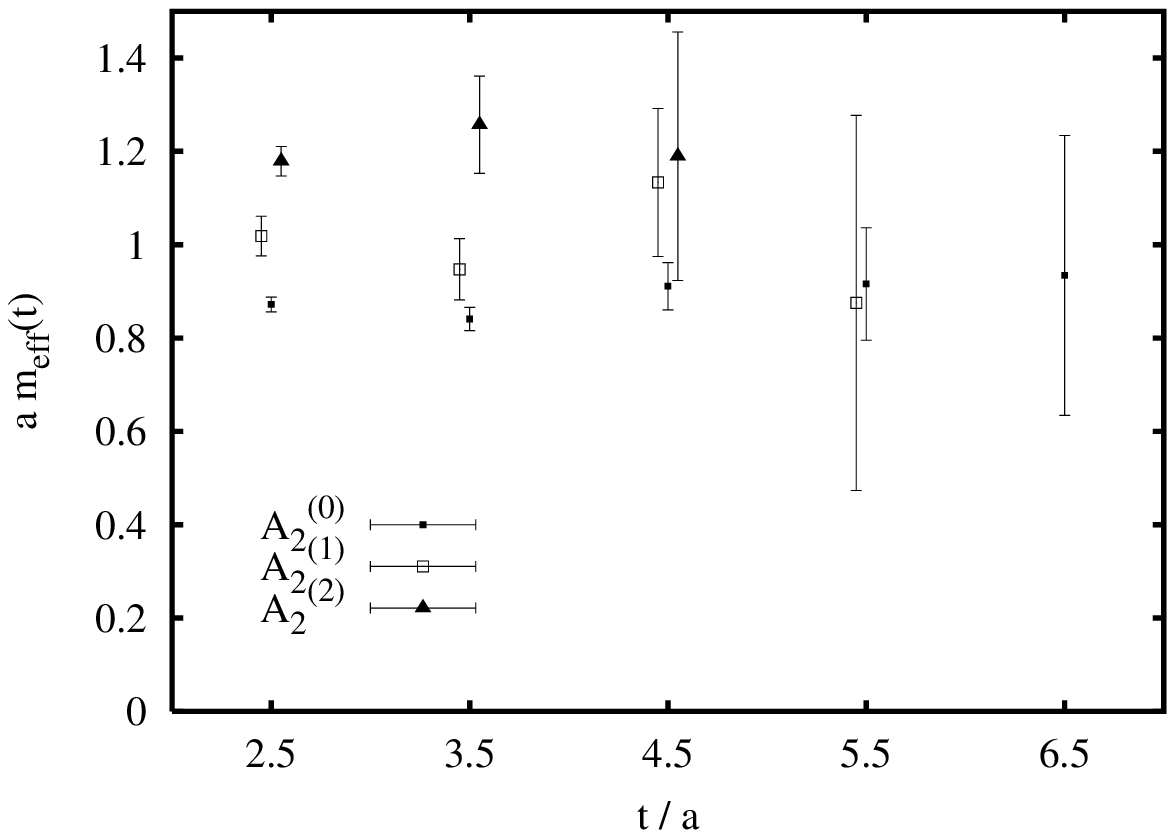,angle=90,width=7.5cm} 
\vspace{-0.5cm}\\
 \psfig{file=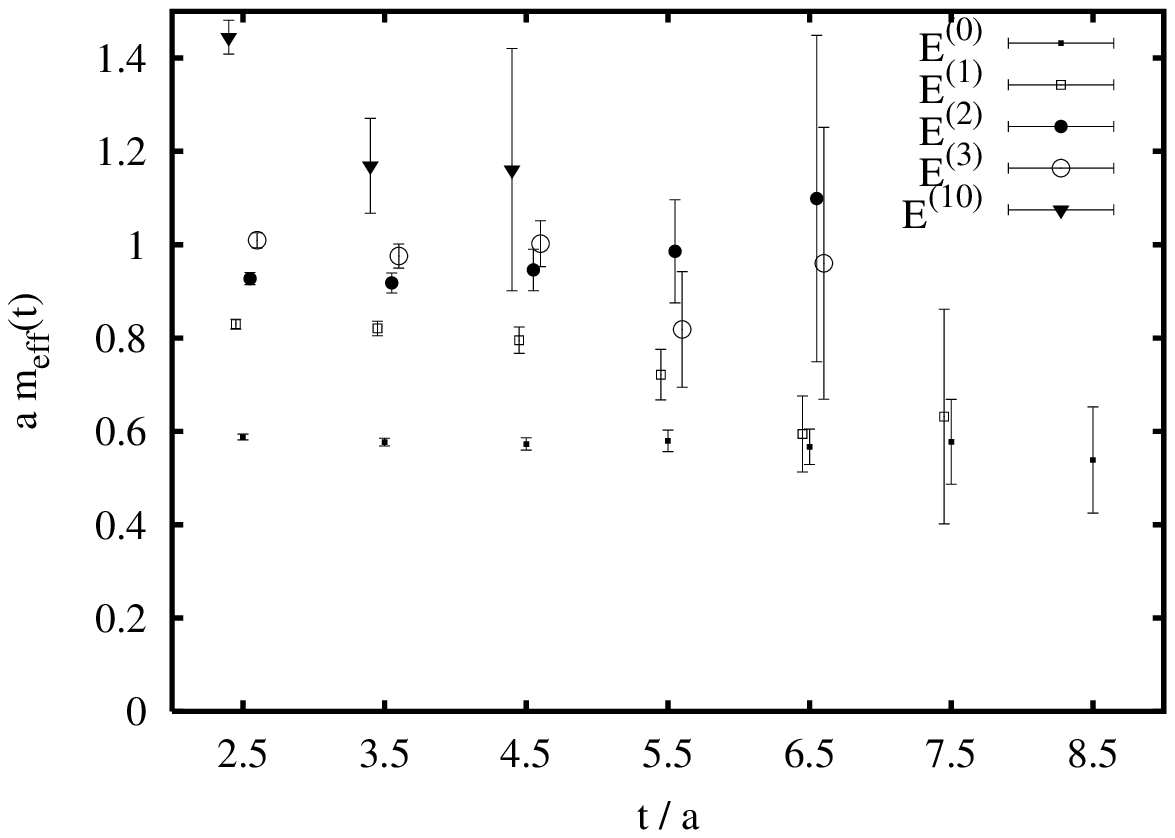,angle=90,width=7.5cm} 
 \psfig{file=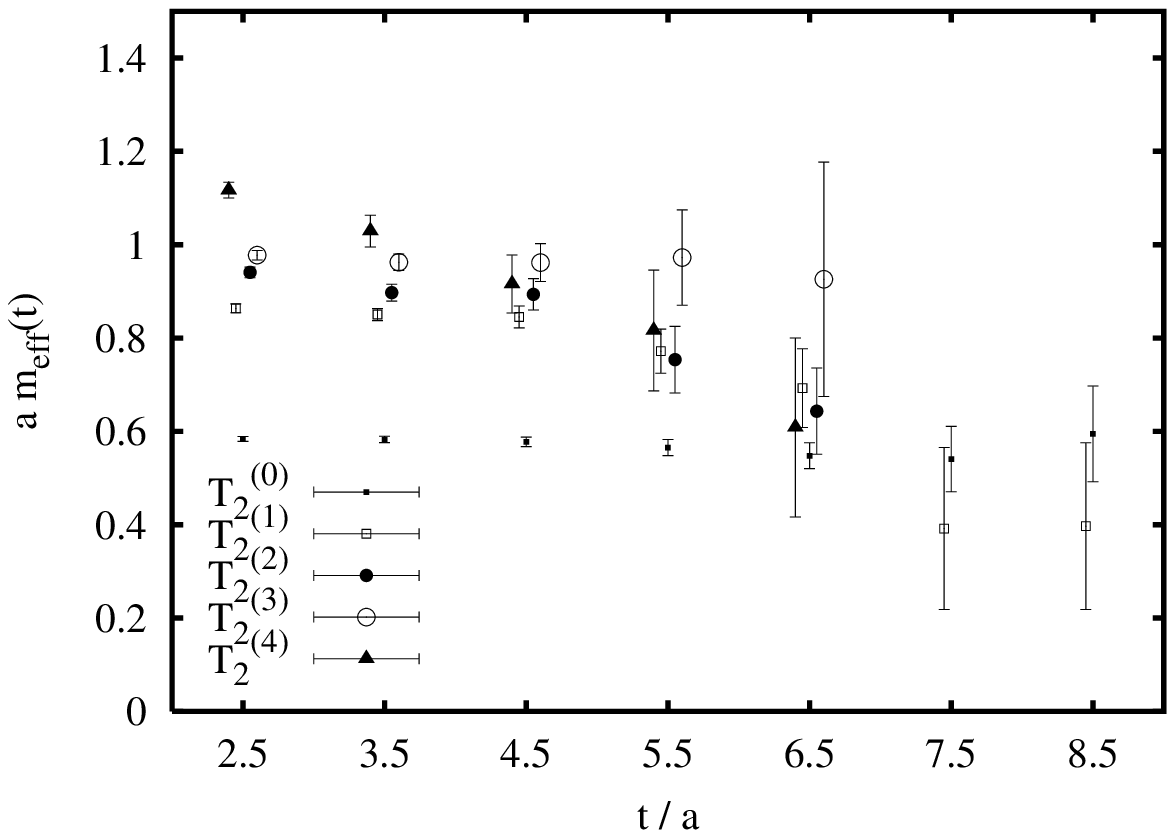,angle=90,width=7.5cm} 
    \end{minipage}}
\vspace{0.5cm}
\caption[a]{The local effective mass as a function 
of Euclidean time, for the lightest states in the 
$A_1^{++}$, $A_2^{++}$, $E^{++}$  and $T_2^{++}$
 representations at $\beta=6.4$ (run III).}
\la{fig:meff_64a1a2et2pp}
\end{figure}
\clearpage
\begin{figure}[t]
\vspace{-1.5cm}
\centerline{\begin{minipage}[c]{14cm}
   \psfig{file=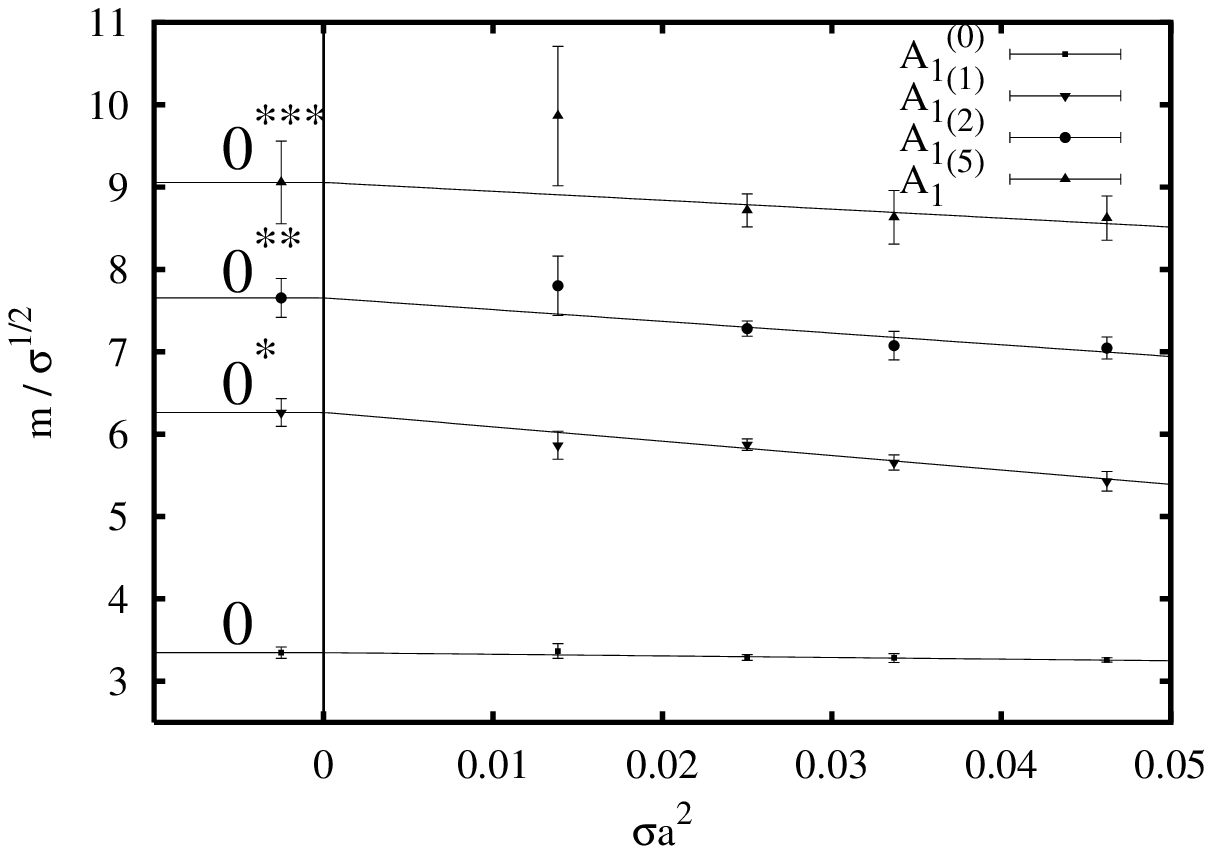,angle=0,width=14cm}
\vspace{0.2cm}\\
\psfig{file=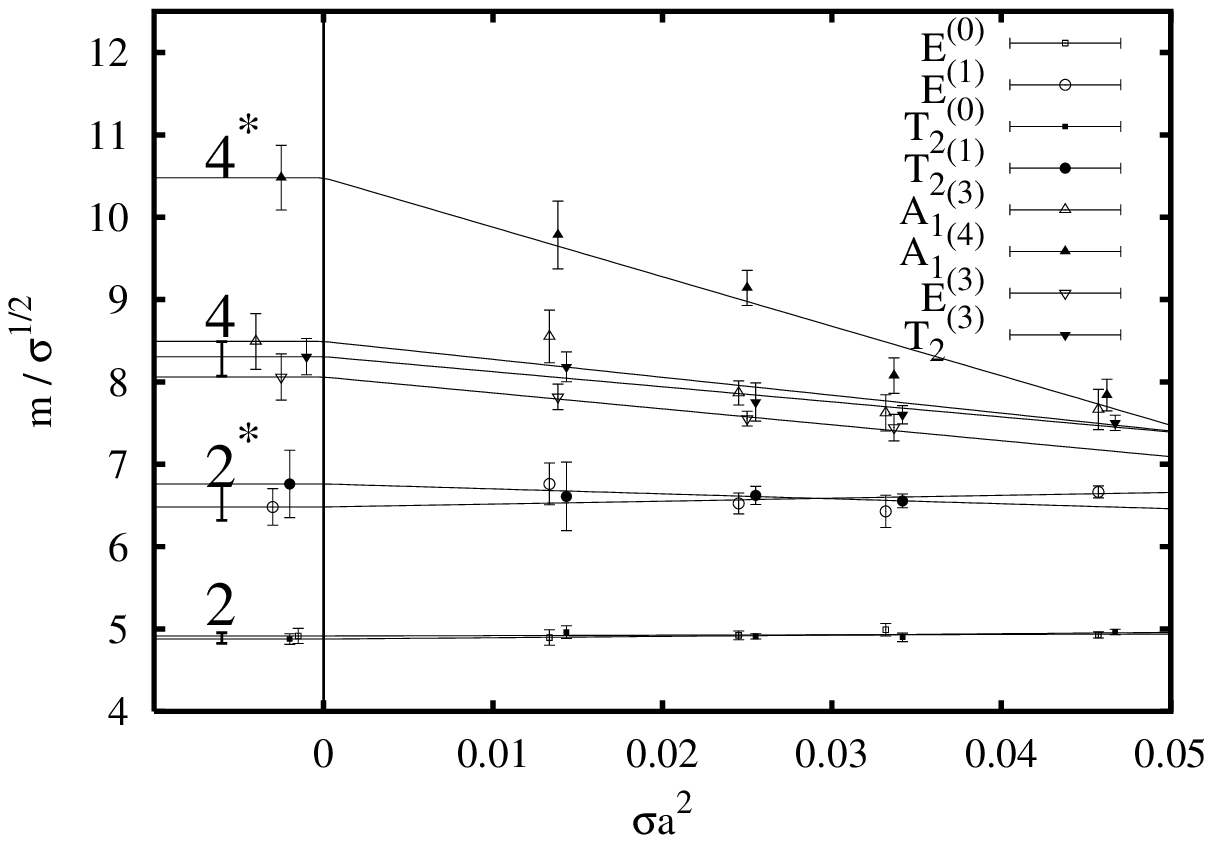,angle=0,width=14cm}
    \end{minipage}}
\vspace{0.5cm}
\caption[a]{The $SU(3)$ continuum extrapolation in the $PC=++$ sector.}
\la{fig:extrapol_pp_a}
\end{figure}
\clearpage
\begin{figure}[t]
\vspace{-1.5cm}
\centerline{\begin{minipage}[c]{14cm}
   \psfig{file=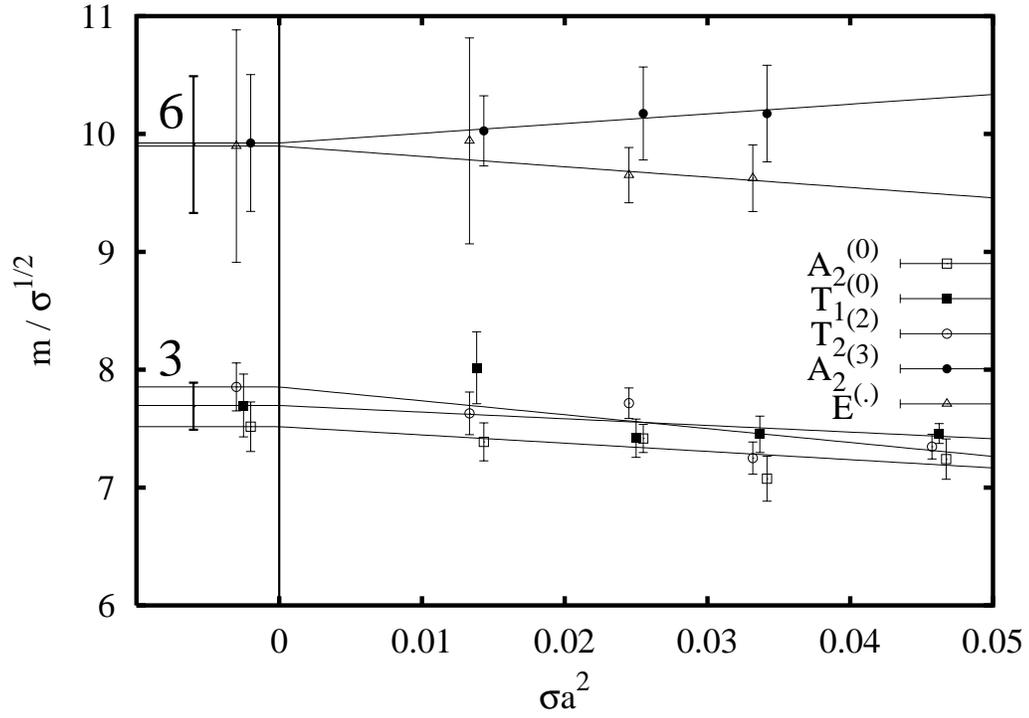,angle=0,width=14cm}
\vspace{0.2cm}\\
	\psfig{file=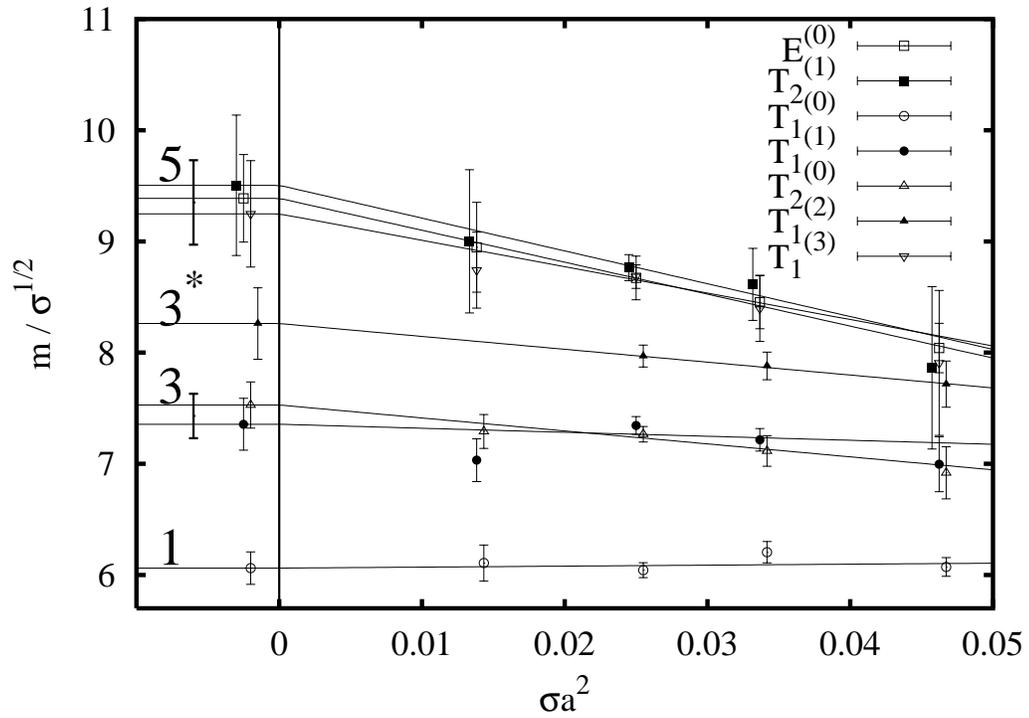,angle=0,width=14cm}
	    \end{minipage}}
\vspace{0.5cm}
\caption[a]{The $SU(3)$ continuum extrapolation in the $PC=++$ sector, 
continued (top); bottom: the $PC=+-$ sector.}
\la{fig:extrapol_pp_b}
\end{figure}
\clearpage
\begin{figure}[t]
\vspace{-1.5cm}
\centerline{\begin{minipage}[c]{14cm}
   \psfig{file=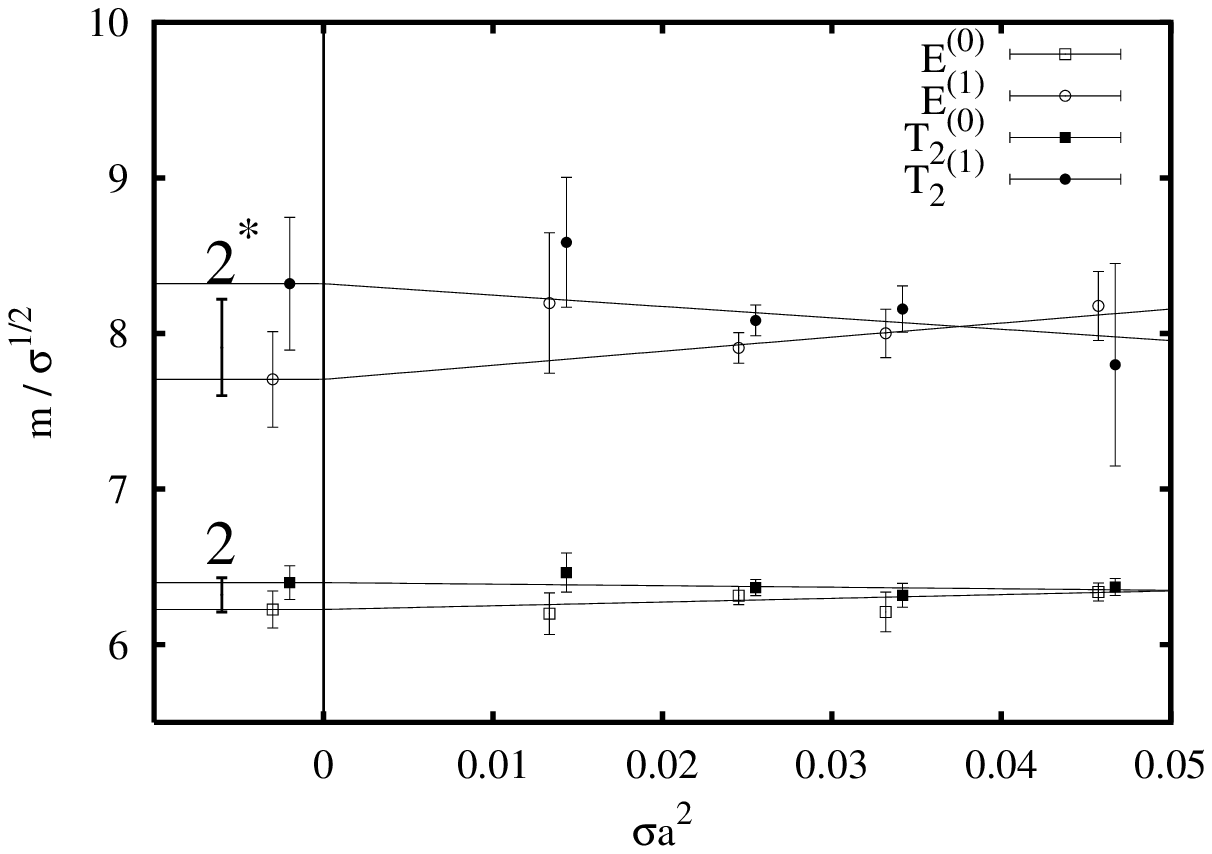,angle=0,width=14cm}
\vspace{0.2cm}\\
   \psfig{file=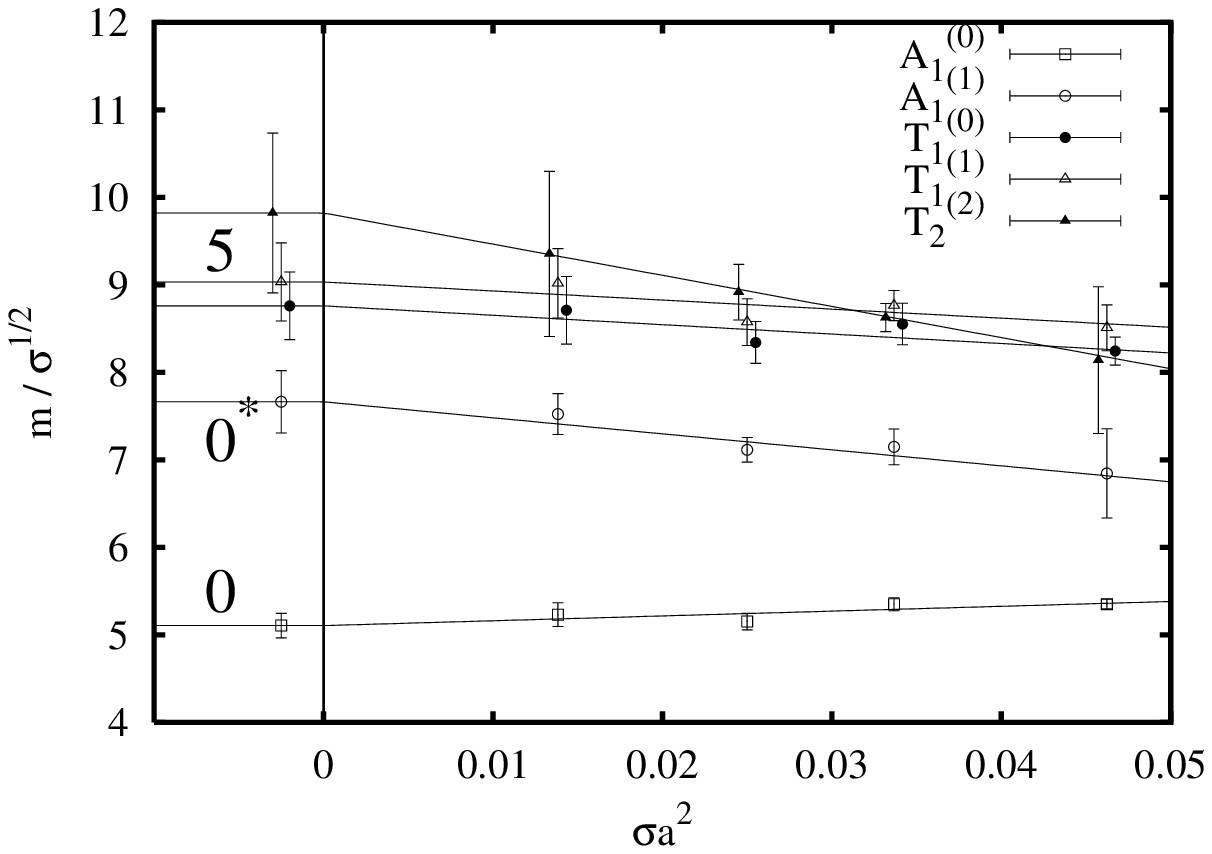,angle=0,width=14cm}
    \end{minipage}}
\vspace{0.5cm}
\caption[a]{The $SU(3)$ continuum extrapolation in the $PC=-+$ sector.}
\la{fig:extrapol_mp}
\end{figure}
\clearpage
\begin{figure}[t]
\vspace{-1.5cm}
\centerline{
\begin{minipage}[c]{14cm}
\psfig{file=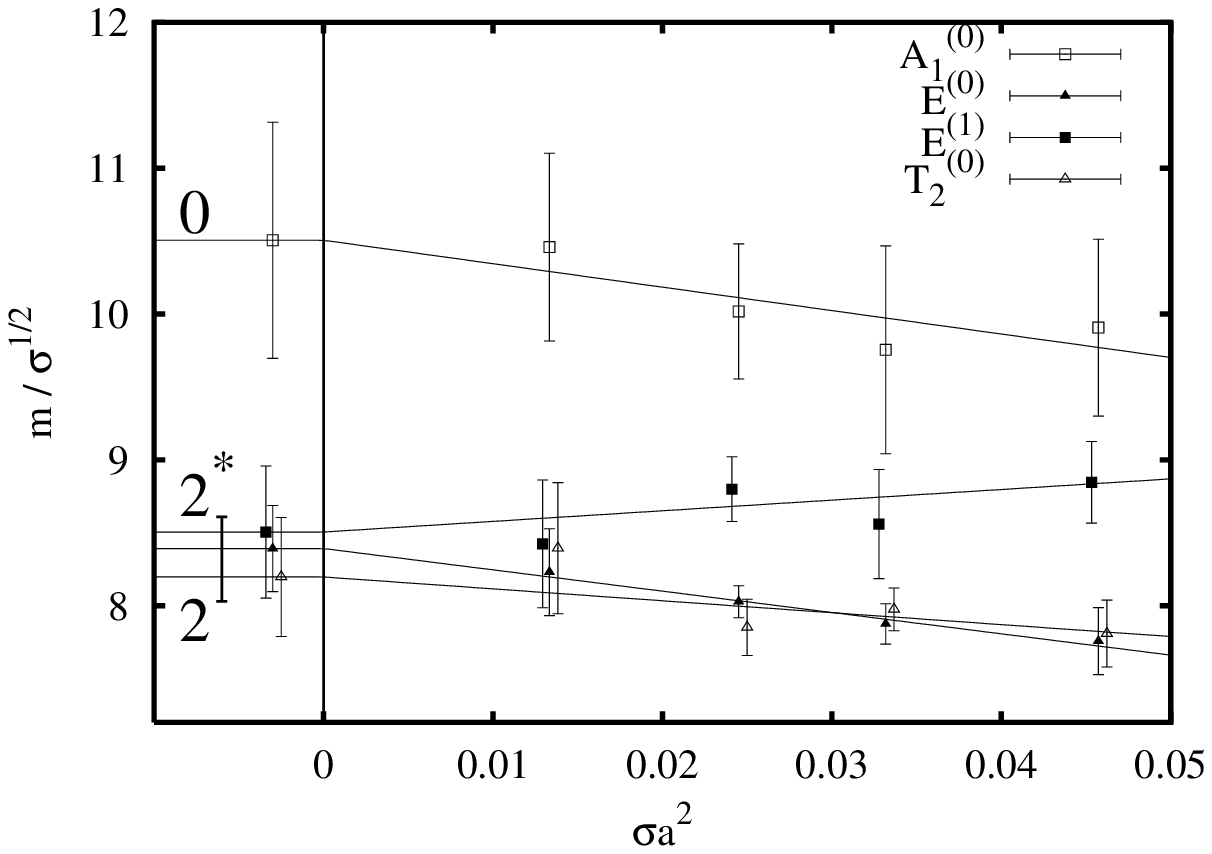,angle=0,width=14cm}\\
\vspace{0.2cm}\\
   \psfig{file=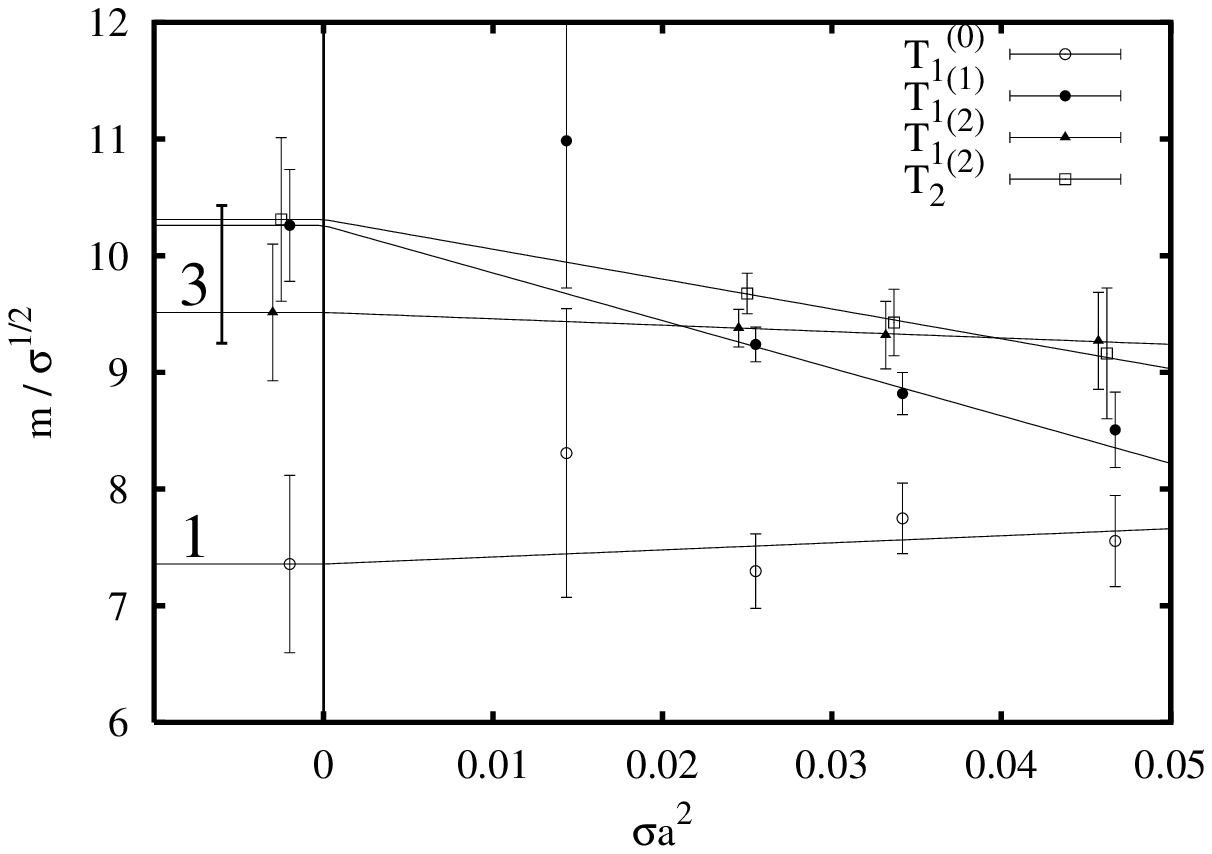,angle=0,width=14cm}
	    \end{minipage}
}
\vspace{0.5cm}
\caption[a]{The $SU(3)$ continuum extrapolation in the $PC=--$ sector.}
\la{fig:extrapol_m}
\end{figure}
\clearpage
\begin{figure}[tb]
\vspace{-1.5cm}
\centerline{\begin{minipage}[c]{14cm}
   \psfig{file=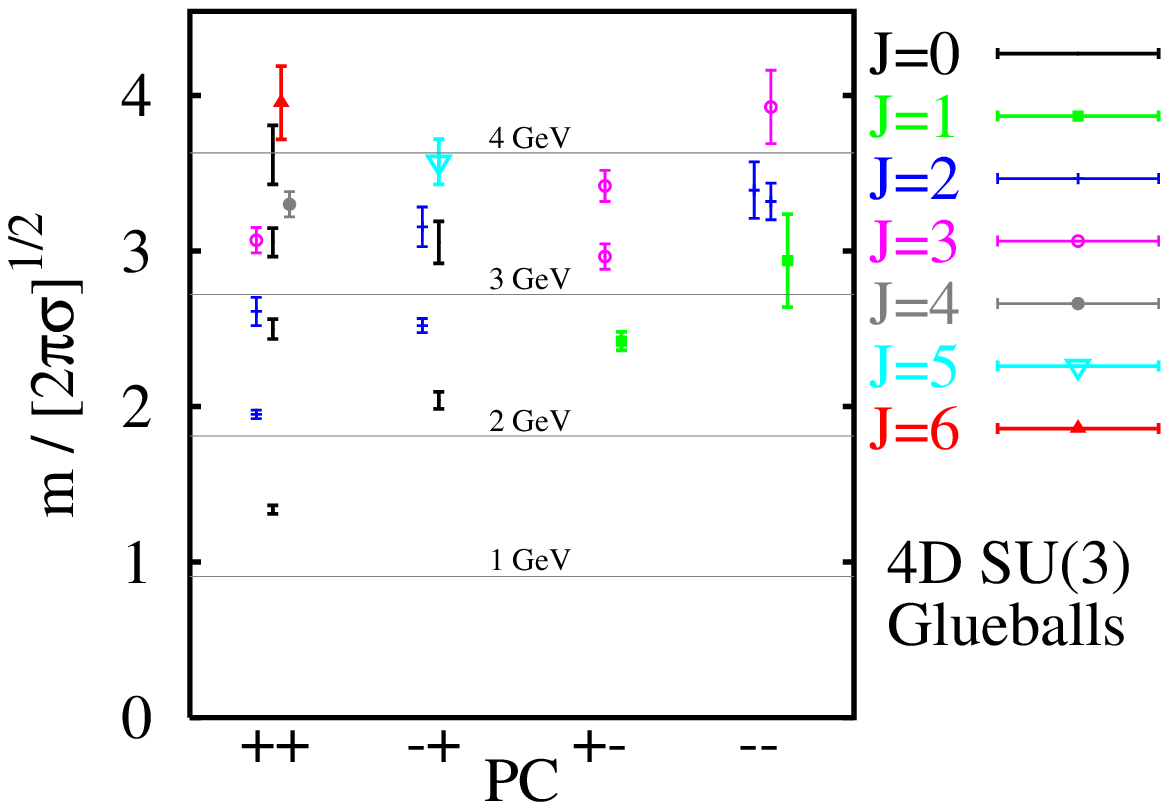,angle=0,width=14cm}
\vspace{0.2cm}\\
   \psfig{file=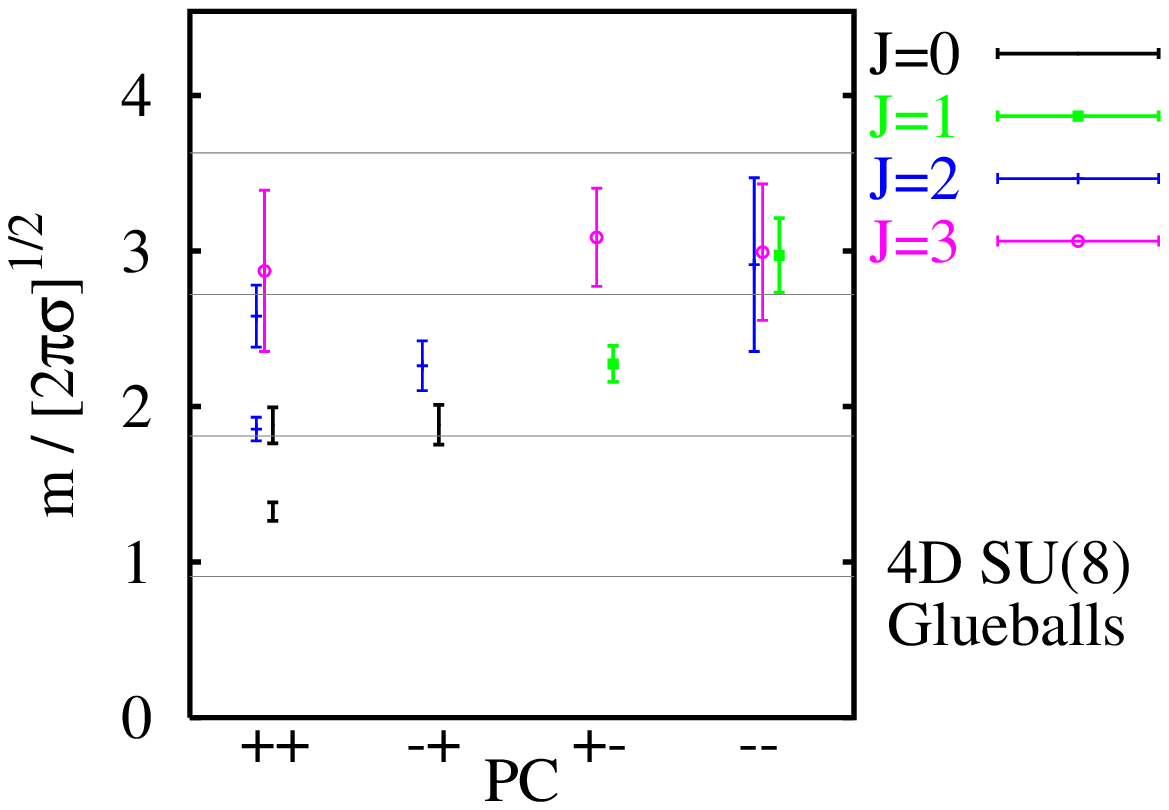,angle=0,width=14cm}
	    \end{minipage}}
\vspace{0.5cm}
\caption[a]{The continuum spectrum of glueballs in the 4D pure $SU(3)$ and 
$SU(8)$ gauge theories. The physical scale was set using 
$\sqrt{\sigma}=440$MeV.}
\la{fig:spec_all}
\end{figure}
\begin{figure}[t]
\vspace{-2cm}
\hspace{-0.5cm}
\centerline{\begin{minipage}[c]{15cm}
\psfig{file=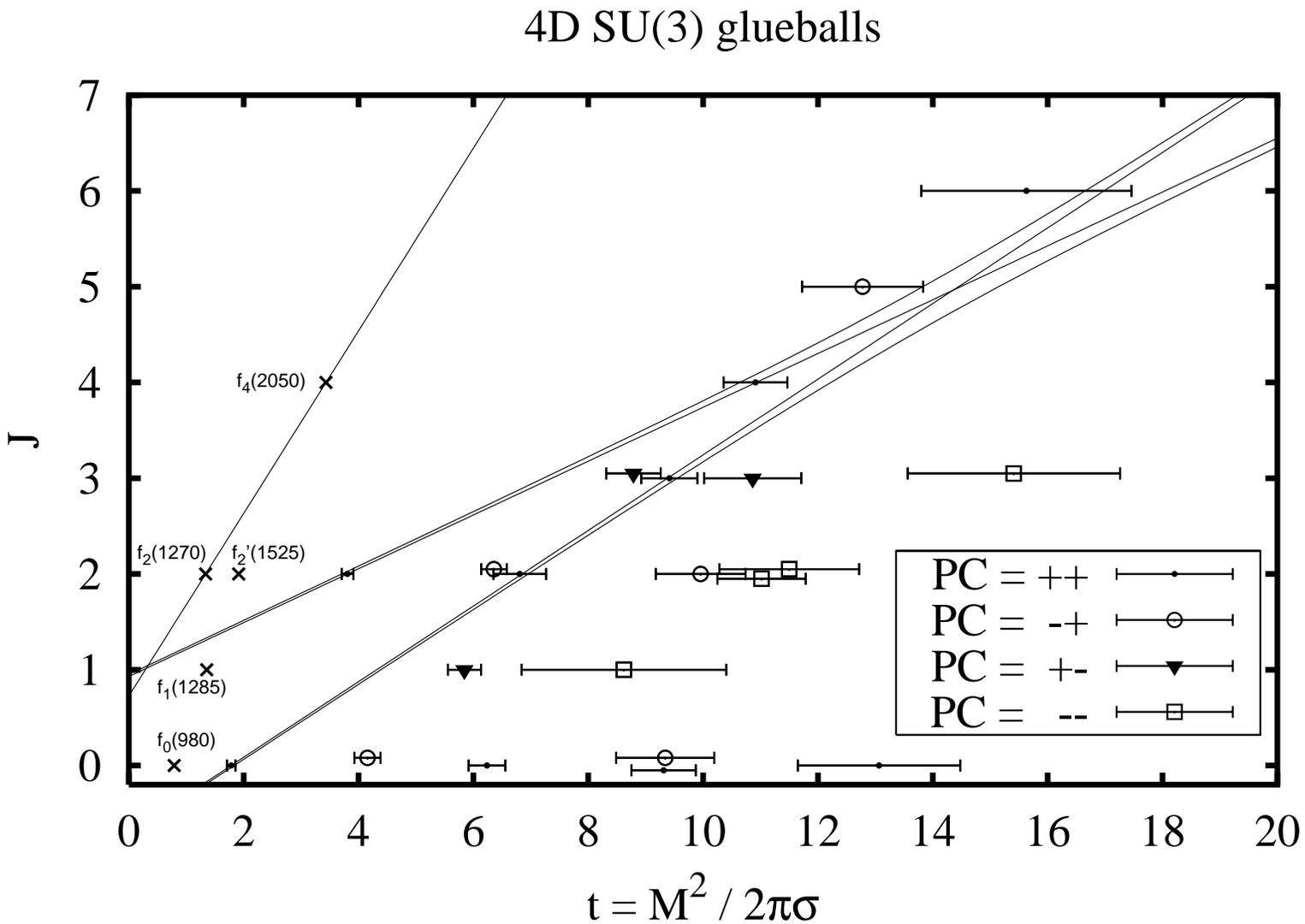,angle=90,width=15cm}
	    \end{minipage}}

\caption[a]{Chew-Frautschi plot of the continuum 4D $SU(3)$ gauge theory.
The hyperbolae are drawn to suggest the mixing of the two leading
trajectories. The position of some flavour-singlet mesons is 
indicated~\cite{Hagiwara:fs}.}
\la{fig:cf}
\end{figure}

%% file: chapter8.tex
\chapter{Conclusion\label{conclusion}} 
Total hadronic cross-sections slowly increase at large centre-of-mass 
energies and are well described
by the exchange of a simple Regge pole with intercept just above 1: 
the pomeron. It is known to have positive signature and 
vacuum quantum numbers $PC=++$.
In recent years, deep inelastic scattering experiments 
have shown that  the rise of the 
$\gamma^* p$ cross-section is more rapid at large virtualities
and this phenomenon is called
the 'hard' pomeron. Whether it is a separate object from the 'soft'
pomeron or rather a perturbatively evolved version of the latter is 
an open question among phenomenologists~\cite{Landshoff:2000mu}.
Deep inelastic scattering at intermediate $Q^2$ and large
$\log1/x$ is particularly 
interesting because it can provide insight in the
transition from a perturbative description (such as the Colour Glass Condensate 
formalism) to the Regge description in terms of hadrons.

In QCD the pomeron  corresponds
to the exchange of excitations of the gluon field. Since it has a simple
pole structure, by crossing symmetry
one should find physical states lying on the pomeron trajectory
at positive $t$. These states are bound states of gluons, or 'glueballs'.

Glueballs have been studied for over twenty years, both by numerical means and 
by modelling. We started by  investigating
 the predictions of models based on an effective-string description. 
In the pure gauge theory,
 the 'flux-tube' that forms between distant static fundamental charges
leads to a potential which agrees with the predictions of 
bosonic string theory~\cite{Lucini:2001nv, Luscher:2002qv, Juge:2003ge}.
This fact suggests that the long-distance degrees of freedom of gauge theories
are those of a string. Classical configurations of 
spinning or vibrating strings lead to straight Regge trajectories $J=\alpha' m^2$
at large angular momentum. Semi-classical corrections give rise to a positive
 'offset'. Because even very massive glueballs are stable~\cite{Witten:1979kh}
in the planar limit $N_c\to\infty$,  we reached the conclusion 
that the large $N_c$, large $J$ glueball spectrum provides another test for the
string nature of the long-distance degrees of freedom of gauge theories.
In the region of small angular momentum, however, quantum mechanical effects
such as mixing become essential and the intercept $\alpha_0$ is not a universally 
calculable quantity. Nevertheless, even without making precise numerical predictions
 for the energy levels,
the quantum numbers of Regge trajectories generated by spinning and
 vibrating strings carry very distinctive signatures of the geometry of the string. 
A few cases attracted our attention:
a spinning open adjoint string (binding two constituent gluons together)
yields a leading trajectory with positive signature and $PC=++$, just like 
the pomeron; the vibrations of a closed fundamental string produce all
spins but spin 1 on their leading trajectory. 
\paragraph{}
The fact that the lattice regularisation of gauge theories break the continuous
Lorentz symmetries down to those of the lattice has been an obstacle to
the numerical resolution of the higher angular momentum glueball spectrum.
Given the considerable theoretical interest of those states, we investigated various
techniques to overcome the problem of labelling them with the correct spin
quantum number in the continuum limit. 

Highly improved operators and the variational method allow one 
 to extract the excited states appearing in the handful
of irreducible representations of the lattice symmetry group.
Character tables tell us what degeneracies of states to expect across 
these representations in the continuum limit. This knowledge
alone is in general insufficient to uniquely determine the spin, 
and we therefore developed independent ways to do this.

An elegant way to approach the problem is to consider an effective continuum
theory which is equivalent to Wilson's lattice theory~\cite{Symanzik:1983dc}. 
The effective Lagrangian contains irrelevant operators, 
suppressed by powers of the lattice spacing $a$, which break rotational invariance.
States lying in irreducible representations of the rotation group $SO(3)$
do not diagonalise the Hamiltonian. 
However, sufficiently close to the continuum, the lattice states are in an obvious 
one-to-one correspondence with the continuum states, since rotational symmetry 
gets restored dynamically. The eigenstates of the Hamiltonian
have a dominant component corresponding to a definite spin $J$, with an
  ${\cal O}(a)$ admixture of different representations.
In principle, by measuring these components at different lattice spacings, 
one can uniquely determine the spin of any lattice glueball.
The measurement is done by determining the phase that the state acquires under
(approximate) rotations. 
The smaller the resolution on the rotation angle, the more candidate spins
can be excluded; naturally one has to make the assumption that \emph{very} 
high spin states are very massive. For a given physical length scale $\lambda$,
the angular resolution that can be achieved is ${\cal O}(a/\lambda)$, and the
spin assignment can be made unambiguous up to $J={\cal O}(\lambda/a)$.

Since the glueball masses are extracted from the exponential fall-off of 
Euclidean two-point functions, it becomes increasingly difficult to extract the
highly excited spectrum by Monte-Carlo methods. 
Multi-level algorithms exploit the locality of the action
to average out the uncorrelated fluctuations of the operators separately. This
leads to an improvement in efficiency  over the ordinary 1-level algorithm which 
is exponential in the product of the glueball
mass with the time separation, when the latter product is large.
\paragraph{}
We used these new numerical techniques to compute the spectrum of pure gauge 
theories in 2+1 and 3+1 dimensions. In the former case, we investigated the
$SU(2)$ spectrum with high accuracy and applied the spin identification methods
to the $N_c=3$ data published in~\cite{Teper:1998te}.
We found several cases where the energy level ordering did not follow the 
spin ordering. The final continuum spectra are presented in the form of 
Chew-Frautschi plots, $J$ vs. $M^2$ (Fig.~\ref{fig:cf_su2_3d} 
and~\ref{fig:cf_su3_3d}). They show striking evidence for quasi-linear
Regge trajectories: a leading trajectory with even spins $PC=++$ and intercept
approximately -1, followed
by a trajectory containing all spins but spin 1. The main effect 
of moving from $SU(2)$ to a larger number of colours is the appearance of $C=\pm$
doublets for the low-lying states, except for those lying on the leading trajectory.
Such degeneracies are most naturally explained in the flux-tube model, where
symmetric and anti-symmetric superpositions of the closed oriented string 
winding clockwise and anti-clockwise can be taken. On the other hand, 
the leading trajectory with only even spins finds no simple 
explanation in the flux-tube model~\cite{Johnson:2000qz}
and must correspond to different dynamics. The most natural
picture is that of the spinning adjoint string binding two constituent gluons 
together. In the large $N_c$ limit, 
the slope of the leading trajectory was found to be $0.37(3)/2\pi\sigma$,
slightly less than the value $1/4\pi\sigma$ expected in this model at large $J$.


The contribution to the cross-section of the leading Regge pole,
with its intercept near -1,  
decreases as $s^{-2}$ at large centre-of-mass energies $\sqrt{s}$.
We argued that a contribution to the scattering amplitude 
not associated with physical states will give the dominant contribution
to high-energy scattering.
\paragraph{}
In 3+1 dimensions, we mainly investigated the $SU(3)$ case relevant to the 'real
world'. The low-lying $SU(8)$ spectrum was found to be extremely similar, which
we take as a confirmation that the low-energy sector of 
$SU(3)$ gauge theory is close to that of $SU(\infty)$~\cite{Lucini:2001ej}.
Not surprisingly, the spectrum is much more complicated than in 2+1 
dimensions~(Fig. \ref{fig:cf}).
A major qualitative difference is that the lightest 
$0^{++}$, $2^{++}$ and $4^{++}$ states are no longer aligned on the Chew-Frautschi
plot. Instead, the straight line passing through
the latter two points has an intercept of 0.93(24). 
The value is thus compatible with the idea that high-energy 
cross-sections would also be roughly constant in the pure gauge theory, and 
it is larger than that of the mesonic trajectories $\alpha_0\simeq 0.5$.
The slope is $0.28(2)/2\pi\sigma$, in surprisingly
good agreement with the phenomenological value of $0.25$GeV$^{-2}$. 
A straight trajectory going through the lightest glueball, 
the $2^{++*}$ and the $3^{++}$ states shares some of the features 
expected for a flux-tube model trajectory: the spin 1 is absent, as it should be,
and  some of the expected degeneracies with states of other $PC$ are seen. 
On the other hand, some
states are missing or significantly heavier. Such large splittings can perhaps
be accounted for by the 'spin-orbit' interactions described in 
chapter~\ref{ch:string}. At any rate, if the two leading trajectories are taken 
seriously, they will mix strongly in the region $J\simeq 5$, and this can account
both for the interchange of their slopes and the raise of the leading intercept.
On the other hand, we see no evidence for a  'hard' pomeron trajectory 
with even larger intercept in the glueball spectrum.

We considered the possibility of an odd signature, $C=-$ trajectory of glueballs
that would correspond to the phenomenological 'odderon'. The latter 
is the object  responsible for the presence of a dip in the $pp$ elastic
scattering amplitude, which is absent in the $p\bar p$ case. 
In leading logarithmic perturbation theory, where it corresponds to the exchange 
of at least three gluons, 
it is predicted to have an intercept of 1~\cite{Bartels:1999yt}.
Our data contains no evidence for such a trajectory. We speculated that the lightest
$1^{--}$, $3^{--}$ states could lie on a  trajectory together with the $0^{++*}$ state
and that it corresponded to the orbital trajectory of a twisted closed string. 
At any rate the intercept of such a  trajectory would be negative.

Our approach, to interpret the glueball spectrum in terms of Regge trajectories, 
sheds light from a new direction on the subject. It tries to establish a 
 relation between states rather than on the absolute positions of the energy levels
and this largely reduces the dependence of the conclusions on model details.
In some cases it leads to predictions for the masses of 
states that have not been measured yet.
But now that the spectrum of gauge theories is numerically 
quite well established both in 2+1 and 3+1 dimensions, 
the most important step forward is to determine the size and structure of glueballs, 
so as to provide more detailed insight into their dynamics and help model building.
This can be achieved through three- and four-point function measurements
on the lattice~\cite{Tickle:1990gw}.
We believe that the computing power is now sufficient 
to undertake this task systematically, at least for the lightest states.
Multi-level algorithms could play an important role in this program.
Finite-volume techniques allow one to measure the widths of unstable 
glueballs~\cite{Luscher:1991cf}; in particular, 
their dependence on $N_c$ is of theoretical interest.

Naturally it is important to experimentally 
establish the existence of gluonic degrees of freedom
in the hadron spectrum. Unfortunately, the glueballs with exotic quantum numbers
seem to be very massive, $M\geq 4$GeV. However the interpretation~\cite{Close:2002zu} 
of the scalar mesons in the 1---1.7GeV range as a mixture
of a $q\bar q$ nonet with the scalar glueball is widely accepted.
Lattice simulations with light dynamical
quarks can contribute significantly to the clarification of the experimental
 situation~\cite{McNeile:2000xx}.
\paragraph{}
Total cross-sections are dominated by peripheral collisions, and the transverse
size of the virtual 'cloud' surrounding hadrons is determined by the mass gap.
In the presence of light quarks, the latter corresponds to the pion mass, while
in the pure gauge case, it is given by the lightest glueball mass. 
The pion mass is driven by the presence of 
light quarks and chiral symmetry breaking, whereas the glueball mass is 
related to the confinement phenomenon. In a world with only bottom quarks, 
the Froissart bound $\sigma_{\rm tot}\leq \frac{\pi}{m^2}
\log^2\left(\frac{s}{m^2}\right)$ is stronger by two orders of magnitude!
Our glueball data strongly suggests that high-energy cross-sections are
approximately constant  in the quenched world and that its 'pomeron'
trajectory has properties very similar to the real-world pomeron.
It provides a (partial) justification for perturbative analyses such as the 
BFKL calculation that are based on the gluon field only and are meant to 
describe the real world. But it is clear that in such frameworks, 
unitarisation should be enforced with respect to the gluonic Froissart 
bound~\cite{Dosch:2002pg}.
%
%
%
%
%
%

We can also turn the argument around.  Experimentally, the 
high-energy $pp$ cross-section only lies about a factor 1.4 
under the gluonic Froissart bound.
If the $pp$ cross-section is found to exceed it at the Large Hadron Collider, 
then it will definitely 
be necessary to include the effects of light quarks in the description
of the hadronic wave-functions at that energy. Asymptotically,  
the boost-enhanced pion cloud could  be  responsible 
for the largest part of the cross-section~\cite{Bjorken:1997zp}.
We would then no longer expect the additive quark rule to hold.

%% file: appendix.tex
\appendix
\addcontentsline{toc}{chapter}{Appendices}
%
\chapter{The symmetry group of the lattice}\label{ap:group}
\section{Irreducible representations of the square group}
The character table of the symmetry group of a 2-dimensional time-slice is 
given below. $C_4$ are the rotations by $\frac{\pi}{2}$, $C_2$ is the rotation
by $\pi$, $\sigma$ is the reflexion around the $x$ axis, $\sigma'$ is the 
reflexion around the $y=x$ axis.
\begin{table}[!h]
\begin{center}
\begin{tabular}{|c|c||c|c|c|c|c|}
\hline
function & IR    & $E$  & $2C_4$  & $C_2$  &  $2\sigma$ & $2\sigma'$ \\
\hline
1 &$(0^+,~4^+,\dots) \in A_1$ &  1   &   1    &   1     &   1    &   1 \\
$xy(x^2-y^2)$&$(0^-,~4^-,\dots)\in A_2$ &  1   &   1    &   1     &   -1   &   -1 \\
$x^2-y^2$& $(2^+,~6^+,\dots)\in A_3$ &1   &   -1    &   1     &   1    &   -1 \\
$xy$ &$(2^-,~6^-,\dots)\in A_4$ &1   &   -1    &   1     &   -1    &   1 \\
$(x,y)$&$(1^\pm,~3^\pm,\dots)\in E$   &2   &   0    &   -2   &   0   &   0 \\
\hline
\hline
$(x+iy)^j$& $D_j$      &   2 &$2\cos{\frac{j\pi}{2}}$ &  $2\cos{j\pi}$ &  0  & 0 \\
\hline
\end{tabular}
\end{center}
\la{tab:sq_group}
\caption{Irreducible representations of the square group, and decomposition
of the general continuum representation.}
\end{table}
It is interesting that $1^+$ and $1^-$ are exactly degenerate on the lattice 
---  they belong to the same representation on the lattice --- 
while the $2^-$ and $2^+$ are not.
Applying the projection rules for characters, we can immediately find 
how  the spin $J$ representation $D_J$ decomposes onto the irreducible 
representations of the square group. For instance:
\[  D_4 = A_1 \oplus A_2. \]
\section{Parity and rotations in continuous (2+1) dimensions}
The rotation group $SO(2)$ being abelian, its irreducible representations are 
one-dimensional: 
\[ \langle \phi | j \rangle = e^{ij\phi}\]
Here $j$ takes all positive an negative integer values. 

The parity transformation, i.e. a flip around an axis, 
takes a clockwise-winding state into an
anticlockwise-winding state, so that
\[ P |j \rangle = e^{i\theta} |-j\rangle, \]
which implies that $P$ and $J$ do not commute and therefore
cannot be diagonalised simultaneously. 
For a particular choice of axis, $\theta$ can be chosen to be zero.
The fact that the Hamiltonian
is parity-invariant implies that the $|j\>$ and  $|-j\>$ are degenerate:
\[ E_j=\< j | H | j\> = \< Pj | PHP | Pj\>= \< -j | H | -j\>= E_{-j}~.\]
This fact is called `parity doubling'.

It is also straightforward to show that $J$ and $P$ anticommute:
\[ \{J,P\}=0. \]
As a consequence, `parity' as defined with an axis rotated
 by an angle $\phi$ with respect to the 
reference axis will be related to $P$ according to
\[P_\phi\equiv e^{iJ\phi} P  e^{-iJ\phi}= e^{2iJ\phi} P=P  e^{-2iJ\phi} \]
In particular, acting on a spin $j\neq 0$, this relation implies that
\[ P_\phi | j \rangle = - P| j \rangle,\quad \phi=\frac{\pi}{2j}\]
Thus an elegant way to understand parity doubling is that
the $P=\pm 1$ labelling can be reversed by the use of  another 
convention; except for the spin 0, where all choices of parity axis
 will label the states in the same way.
\section{Irreducible representations of the cubic group}
The character table of the rotation symmetry group ($O_h$ or (432))
of a 3-dimensional time-slice is given below. The insertion of parity in the
group ((m3m): $O_h=O\times i$) does not introduce any complications as it does 
in two dimensions, because parity commutes with rotations and is realised
exactly on the lattice.
$C_4$ are the rotations by $\frac{\pi}{2}$, $C_2$
by $\pi$ (3 along the axes and 6 along face diagonals) and $C_3$ are the
ternary axes along the volume diagonal.
\begin{table}
\begin{center}
\begin{tabular}{|c|c||c|c|c|c|c|}
\hline
function& IR  & $E$  & 8$C_3$ & 3$C_2$  & $6C_2$   &  $6C_4$   \\
\hline
1 & $A_1$ &  1   &   1    &   1     &   1    &   1 \\
$xyz\propto Y_3^2-Y_3^{-2}$ &$A_2$ &  1   &   1    &   1     &   -1   &   -1 \\
$(Y_2^0,Y_2^2+Y_2^{-2})$& $ E$ &2   &   -1    &   2     &   0    &   0 \\
$(x,y,z)$,$(Y_1^1,Y_1^{-1},Y_1^0)$ & $T_1$ &3   &   0    &   -1     &   -1    &
  1 \\
$(Y_2^1,Y_2^{-1},Y_2^2-Y_2^{-2})$&$ T_2$   &3   &   0    &   -1   &   1   &   -1
 \\
\hline
\hline
$Y_j^m$& $D_j$ & $2j+1$ & $(1,0,-1) $ &  $(1,-1)$   &  (1,-1) & (1,1,-1,-1) \\
\hline
\end{tabular}
\end{center}
\la{tab:cu_group}
\caption{Irreducible representations of the cubic group, and transformation
properties of a spherical harmonic under the lattice symmetry operations.}
\end{table}
In the last line, the different values of $\chi_j(C_n)$ correspond to
$j\equiv 0,\dots,n-1~(\mathrm{mod}~n)$.
These values are easily obtained from the general formula
\[\chi_j(\alpha)=\frac{\sin{(j+\frac{1}{2})\alpha}}{\sin{\frac{\alpha}{2}}}.\]
Thus the smallest spins coupling to the various lattice representations are
\ba
A_1 \quad&\rightarrow &\quad {\rm spin}~0 \nonumber\\
T_1 \quad&\rightarrow &\quad {\rm spin}~1 \nonumber\\
E \quad&\rightarrow &\quad {\rm spin}~2 \nonumber\\
T_2 \quad&\rightarrow &\quad {\rm spin}~2 \nonumber\\
A_2 \quad&\rightarrow &\quad{\rm spin}~3~. \nonumber
\ea
Conversely, a few useful decompositions of the continuum representations read
\ba
D_0&=& A_1 \qquad ({\rm scalar})\nonumber\\
D_1&=& T_1 \qquad ({\rm vector})\nonumber\\
D_2&=& E\oplus T_2 \qquad ({\rm tensor})\nonumber\\
D_3&=& A_2\oplus T_1\oplus T_2 \nonumber\\
D_4&=&A_1\oplus E \oplus T_1\oplus T_2\nonumber\\
D_5&=&E \oplus2 T_1\oplus T_2\nonumber\\
D_6&=&A_1 \oplus A_2\oplus E \oplus T_1  \oplus2 T_2 ~.
\ea

When studying 2-glueball states, it is useful to know what the analog 
of spin composition is on a cubic lattice. Table~\ref{tab:cu_group_dir_prod}
gives the decomposition of all possible direct products in terms of  the  
original representations.
\begin{table}
\begin{center}
\begin{tabular}{|c||c|c|c|c|c|c|}
\hline
 $\times$     & $A_1$ & $A_2$ & $E$ & $T_1$  & $T_2$ \\
\hline
\hline
$A_1$ & $A_1$ & $A_2$ & $E$ & $T_1$  & $T_2$ \\
\hline
$A_2$ &  & $A_1$ & $E$ & $T_2$  &  $T_1$ \\
\hline
$E$   &    &  &$A_1\oplus A_2 \oplus E$ & $T_1\oplus T_2$ &  $T_1\oplus T_2$\\
\hline
$T_1$ &  &  &  & $A_1\oplus E\oplus T_1 \oplus T_2$  & $A_2\oplus E\oplus T_1 \oplus T_2$  \\
\hline
$T_2$ &  &  &  &   & $A_1\oplus E\oplus T_1 \oplus T_2$ \\
\hline
\end{tabular}
\end{center}
\la{tab:cu_group_dir_prod}
\caption{Direct products of cubic group IRs, and decomposition into the 
IRs themselves. The table is symmetric about the diagonal.}
\end{table}
\chapter{Numerical Recipes for Lattice Gauge Theory\la{ap:NRLGT}}
\section{$SU(N)$ update algorithms}\label{ap:SUN}
\subsection{The Cabibbo-Marinari algorithm}
%

The relevant part of the action, on which the link depends, is 
\[ -S[U] = \frac{\beta}{N}\re\tr[US^\dagger], \]
with $S$ the sum of staples linking the same points as $U$. 
A first obervation is that we can
update $U$ by multiplying it successively by  $SU(N)$ matrices $A$; there is
no loss of ergodicity in doing so, since the matrices are invertible, although 
the propagation through phase-space will in general be 
slower in Monte-Carlo time. A second obervation is that 
it is sufficient to select a collection of matrices of the type
$A=A^{[kl]},~k<l$, which 
is unity except for an $SU(2)$ subgroup $a$ (precisely: $A_{ij}=\delta_{ij}$, 
$\{i,j\}\neq\{k,l\}$, and 
$A_{kk}=a_{11}$, $A_{kl}=a_{12}$, $A_{lk}=a_{21}$ and $A_{ll}=a_{22}$, $a\in SU(2)$). The set of $[kl]$ must be such that no $SU(2)$ subgroup is left invariant.
A conventional (and non-minimal) choice in practice are the $N(N-1)/2$ subgroups
 $k=1,\dots,N-1;~l=k+1,\dots,N$.

The update of the link is now reduced to the problem of thermalising one of the 
$A$'s described above. The induced action for $A$ is 
\[ -S_{SU(2)}[A]\equiv -S[AU] = \frac{\beta}{N}\re\tr[AUS^\dagger] = \frac{\beta}{N}\re\tr[AP],\]
where $P$ is the current plaquette. 
Thus the gauge `force' acting on $A$ is the plaquette.
We now focus on the dependence of the action on $A$'s $SU(2)$ subgroup $a$:
it only involves the corresponding subset $p$ embedded in
 the plaquette $P$:
\be
p_{11}=P_{kk},\quad p_{12}=P_{kl},\quad p_{21}=P_{lk},\quad  p_{22}=P_{ll}.
\la{eq:sub_su2}
\ee

 Note however that $p$ is in general not an $SU(2)$ matrix.
It is straightforward to check that the real part of the trace of the product 
an $SU(2)$  matrix by a general complex $2\times2$ matrix
is unaffected  by replacing the latter by a matrix $\tilde p$ proportional 
(with a real coefficient) to an  $SU(2)$  matrix:
\ba 
d_{\rm eff}=\half(\re p_{11}+ \re p_{22})&\quad&
a_{\rm eff}=\half(\im p_{11}- \im p_{22}) \nonumber\\
c_{\rm eff}=\half(\re p_{12}- \re p_{21})&\quad&
b_{\rm eff}=\half(\re p_{12}+ \re p_{21}) \la{eq:su2_from_sun}
\ea
(we parametrise an $SU(2)$-proportional matrix with Pauli matrices:
$\tilde p=d + ib \sigma_1 + ic \sigma_2 + ia \sigma_3$). We now have
\[ -S_{SU(2)}[A] = \frac{\beta}{N}\re\tr(AP) =
 \frac{\beta}{N} \re\tr(a\tilde p) ~+~\dots \]
This means that we can use a heat-bath algorithm (described below) 
to find a thermalised $a$, with an 
effective `staple' $\tilde p$ and an effective $SU(2)$ coupling 
\be
\beta_{\rm eff}=\frac{2\beta}{N}.
\ee
Once this is found, the link and the plaquette are updated,
\[U \rightarrow AU \qquad P \rightarrow AP, \]
and we can move to the next $SU(2)$ subgroup.
\subsection{The Kennedy-Pendleton algorithm}
The problem is to generate an $SU(2)$ matrix according to the distribution
\be
Q(a) da ~=~ \exp\left(\frac{\beta}{2} \tr \{a \tilde p\} \right)~da, 
\ee
where $\tilde p$ is a sum of $SU(2)$ matrices and $da$ is the invariant 
Haar measure for an $SU(2)$ matrix.  
We can write $\tilde p=\xi~p$, $p\in SU(2)$ and 
$\xi\equiv\sqrt{\det \tilde p}\in {\rm R}$.
Let us introduce $u=ap$; we now have to generate $u$ according to
\be
P(u) du ~=~ \exp\left(\frac{\beta\xi}{2} \tr \{u\} \right)~du
\ee
and obtain $a$ at the end from $a=up^\dagger$.
The Haar measure for an $SU(2)$ matrix $u$ parametrised by 
$u = u_0 +i{\bf u }\cdot\sigma$, $u_0^2+{\bf u}^2=1$, 
is 
\ba
du &=& du_0~d^3u_i~\delta\left(1-u_0^2-{\bf u}\cdot{\bf u}\right) \nonumber \\
  &=& \frac{\sqrt{1-u_0^2}}{2} du_0 ~dr~d\theta~d\phi~\sin{\theta}~
\delta(r-\sqrt{1-u_0^2}).
\ea
$\bf u$ has now been parametrised in spherical coordinates.
The problem is thus reduced to generate $u_0$ according to
\be
P(u_0)du_0 ~=~\sqrt{1-u_0^2}~\exp\left(\frac{ u_0}{b}\right)du_0,\qquad
b= (\beta\xi)^{-1}
\ee
and it is then easy to 
generate the $u_i$ uniformly on a sphere of radius $\sqrt{1-u_0^2}$.

An efficient algorithm for the update of $SU(2)$ matrices was found by
Kennedy and Pendleton~\cite{Kennedy:1985nu}. The recipe is the following
($\eta$ random variables are uniformly distributed over [0,1]). Do
\begin{enumerate}
\item $x_1=\log\eta_1,\quad x_2=\log\eta_2$
\item $x_3=\cos^2(2\pi\eta_3)$
\item $s = 1+b(x_1+x_2x_3)$
\item $t = 1+s-2\eta_4^2$
\end{enumerate}
as long as $t$ is found negative. When the loop is exited, $u_0$ is set to $s$.
%
\subsection{$SU(2)$ over-relaxation}
We use the simplest possible form of over-relaxation. Instead of doing the heat-bath
update described above, the link is `flipped' with respect to $p$ 
acting on it: $a\rightarrow (p~ a~ p)^\dagger$. This leaves $\tr \{ap\}$ invariant.
\subsection{The unitarisation to $SU(N)$ matrices}
If $GL_N(C)$ is the set of invertible complex $N\times N$ matrices, the 
unitarisation problem amounts to finding a mapping $F:GL_N(C)\rightarrow SU(N)$
satisfying the following conditions:
\begin{enumerate}
\item Let $U_o\in GL_N(C)$  and $U=F(U_o)$. For any choice $g_1,~g_2\in SU(N)$, 
\be
 F(g_1 U_o g_2) = g_1  F(U_o) g_2.
\ee
\item if $U_o\in SU(N)$, $F(U_o)=U_o$.
\item F is continuous almost everywhere 
\end{enumerate}
The problem arises, for instance, in the context of 
 `smearing'~\cite{Albanese:1987ds}:
a link variable is to be  replaced by a sum over many Wilson lines 
covering many  different paths joining the same points $x$, $x+\mu a$. 
This sum must therefore be reunitarised. In this context, $g_1$ corresponds
to $g^{-1}(x)$ and $g_2$ to $g(x+\mu a)$ for a gauge transformation $g=g(x)$;
also, $U_o$ is a sum of $SU(N)$ matrices.
For $N=2$, the problem is then trivial, because a sum of $SU(2)$ matrices is 
proportional to an $SU(2)$ matrix: 
\be
U=\frac{1}{\sqrt{\det U_o}} U_o\qquad(N=2).
\ee
This simple property no longer holds for $N>2$; therefore we have to consider
the general problem as stated above. Note that the probability for the sum
to be a singular matrix is zero.

One possibility was proposed in~\cite{Liang:1993cz}. A linear algebra theorem states
that any $U_o\in GL_N(C)$ can be uniquely decomposed as $U_o=UH$, with
$U\in SU(N)$ and $H$ as positive definite Hermitian matrix. Choosing 
$F(U_o)=U$ clearly satisfies all three requirements (note that the uniqueness
is crucial).

Computationnally the decomposition is performed as follows. First note that
$U_o^\dagger U_o=H^2$. $H^2$ can be unitarily diagonalised (for instance 
by using the Jacobi method for complex matrices), $H^2=Y \Lambda^2 Y^\dagger$, 
with $\Lambda$ real and diagonal. The inverse of $H$ is then easily found:
 $H^{-1}= Y \Lambda^{-1} Y^\dagger$. Finally, $U=U_o H^{-1}$.

This procedure is quite computing-intense. Another widely used approach
is to solve the following, simpler problem. We want to find the matrix 
$T\in SU(N)$ which maximises the expression
\be
\re\tr(T^\dagger U_o).
\ee
This simpler problem appears in the context of topological charge measurements.
In that context, one reduces ultra-violet fluctuations
by  `cooling'~\cite{Teper:1988ed} the link $T$ with respect to the `gauge force'
acting on it, namely the sum of staples $U_o$, so as to drive the system to a
minimum of the action. This means maximising the trace
of the plaquette, which is precisely the problem at hand.

Let us suppose for the moment that this problem has a unique solution;
 this is obviously true for the special case  $U_o\in SU(N)$.
Then it is easy to see that the solution $T$ is also a solution of the 
unitarisation problem as stated at the beginning of this section. 

Let us now investigate the relation between the two procedures.
We shall make use of the decomposition $U_o=UH$ and 
use the same notation as previously to write
\be
{\rm max}_{T\in SU(N)} \re\tr(T^\dagger U_o) = {\rm max}_{W\in SU(N)}
\re\tr(W\Lambda) = {\rm max}_{w_i} \sum_{i=1}^{N} \re w_i \lambda_i,
\ee
with  $W=Y^\dagger (T^\dagger U) Y$ and $\{w_i\},~\{\lambda_i\}$ 
the diagonal elements of $W$ and $\Lambda$  respectively. 
The unitarity constraint on $W$ implies that $-1\leq\re w_i\leq1$.
In general the solution is then obviously
 $w_i={\rm sign}\lambda_i$ (provided the number of
negative $\lambda_i$ is even);
the matrix $W$ is then diagonal and real, with $\pm1$'s in the diagonal.
Now since $H$ is positive definite, the $\lambda_i$'s are all positive and 
therefore $W$ is simply the unit matrix.

The solution to the maximisation problem is unique, completing the proof
that this approach satisfies our unitarisation requirements and  moreover
$T=U~(Y W Y^\dagger)= U$. Thus the two methods produce the same solution.

How can we efficiently implement the maximisation procedure in practice?
A clue is given by the context of cooling. The key observation is that
 \emph{a cooling step is equivalent to a heat-bath step with $\beta=\infty$}.
Since the cooled link is the solution of the trace-maximisation problem, 
we can re-use the heat-bath algorithm with $\beta=\infty$ to find the cooled, 
unitarised link. In this particular case the procedure amounts to the following
algorithm. 
\begin{enumerate}
\item Start from an arbitrary $SU(N)$ matrix $T=T^{(\rm init)}$. 
\item Obtain the effective `plaquette' $P\equiv TU_o^\dagger$. 
\item For a pair $[k~l]$ extract  
an $SU(2)$ subgroup $\tilde p$ from $P$ according to Eqn.~\ref{eq:sub_su2}
and Eqn.~\ref{eq:su2_from_sun}.
\item Update $T\rightarrow AT$ and $P\rightarrow AP$ with $A\in SU(N)$ obtained 
from the unit matrix by embedding
$a=\tilde p^\dagger/\sqrt{{\rm det}~\tilde p}$ into the $[k~l]$ $SU(2)$ subgroup.
\item Iterate the $SU(2)$ `hits' for a covering set of subgroups $[k~l]$.
\item Repeat the whole operation until the link is completely `cooled'. 
\end{enumerate}

Two remarks are in order, concerning the starting and stopping points. 
In the context of smearing, the starting matrix $T^{(\rm init)}$ is
 usually obtained from $U_o$ by a Gram-Schmidt (GS) procedure. This still leaves
a choice whether to apply the GS procedure on the rows or columns. In the 
context of cooling, the current value of the link is used.

The  algorithm is  made to exit when a certain numerical accuracy 
has been reached, which is detected by the fact that the trace  no longer 
changes. Working in single precision, we typically exit when $\tr P/N$ has 
increased  by less than $10^{-5}$ as a result of the last sequence of $SU(2)$ 
hits.
\section{The generalised eigenvalue problem}
The variational method described in section~\ref{sec:glueball_intro} involves 
solving a generalised eigenvalue problem:
\be
{\bf A x} ~=~\lambda~ {\bf B x}
\ee
$\bf A$ and $\bf B$ are $N_o\times N_o$ real, symmetric matrices. 
The trick~\cite{Num_Rec:1999ab} is to use the Cholesky decomposition of
   ${\bf B}= {\bf L L}^t$, where $\bf L$ is a lower-triangular matrix. 
The problem is equivalent to the ordinary eigenvalue problem
${\bf C y} =\lambda {\bf y}$ with 
${\bf C}= {\bf L}^{-1} {\bf A} ({\bf L}^{-1})^{t}$
   and ${\bf y}={\bf L}^t {\bf x}$. 
${\bf C}$ is by construction a symmetric matrix, ensuring
that its eigenvalues exist and are real. Once ${\bf L} $ is known, 
to obtain ${\bf C}$, we first solve
${\bf LY}^t={\bf A}^t$ and then ${\bf LC}={\bf Y}$ by back-substitution.
\section{Jacknife error analysis}
The jacknife method allows us to estimate 
the variance of observables without having to rely on a particular
statistical law.
The (blocked) measurements $\{A_i\}_i^{N_m}$ are first redistributed 
 in as many `jacknife bins' (typically, ${\cal O}(100)$):
\be
A^{[J]}_i = \frac{1}{N_m-1}\sum_{j\neq i}~A_j
\ee
These bins have much smaller fluctuations than the original measurements, 
but of course they are extremely correlated. They can be used to 
compute variables which depend non-linearly on the measured $A_i$.
The average and error bar on these variables is then simply given by
\ba
\bar f(A)&=& \frac{1}{N_m}\sum_{i=1}^{N_m}~f(A^{[J]}_i)\la{eq:jk_av}\\
\Delta f(A) &=& \sqrt{ \frac{ N_m-1}{N_m} \sum_{i=1}^{N_m} \left(f(A^{[J]}_i) - 
\bar f(A)\right)^2 } \la{eq:jk_er}
\ea
\section{Correlated cosh fits}
Suppose we have the measurements of the correlator $\{C_i(t)\}_{i=1}^{N_m}$ stored 
in jacknife bins. The first step is to compute the jacknife error bars
on each measurement, according to Eqn.~(\ref{eq:jk_er}); they are needed
to weight the data points inversely to the square of their
 variance in the forthcoming fit.  The simplest method
 then consists in using standard numerical recipes~\cite{Num_Rec:1999ab} to fit each
jacknife bin by a hyperbolic cosine function centered at half the 
temporal extent of the lattice; the average value and error bar
 on the fitted parameters are obtained in the usual way from~(\ref{eq:jk_av}) 
and~(\ref{eq:jk_er}). This procedure
however does not take into account the correlations
between data points at different $t$. To improve on this, we compute the 
covariance matrix of the data points:
\be
{\rm Cov}(t_1,t_2)=\frac{N_m-1}{N_m}\sum_{i=1}^{N_m} ~
[~C_i(t1)-\bar C(t_1)~] \cdot [~C_i(t2)-\bar C(t_2)~]
\ee
This (symmetric) covariance matrix can be diagonalised. In this way, 
statistically independent linear combinations of  data points are found.
These linear combinations are formed for each jacknife bin;
the $N_m$ jacknife sets are fed in the fitting routine one by one.
The square root of the eigenvalues  play the role of the error bars on these
linear combinations. 
%
\chapter{Regge theory in 2+1 dimensions\la{ap:regge2d}}
The theory of the $S$-matrix can be developed in an entirely analogous way 
to the 3+1 dimensional case (see chapter~\ref{ch:hehr}).
We denote by $A$ the $a+b\rightarrow c+d$ amplitude. 
The partial wave expansion in the $s$-channel reads (cf. Eqn.~\ref{eq:part_wave})
\be
A(s,t)= a_0(s)+2\sum_{\lam\ge1}a_\lam(s)C_\lam(1+\frac{2t}{s}).
\ee
Here $C_\lam(\cos{\theta})=\cos{\lam\theta}$ is a Chebyshev polynomial.
 The absence of a factor 2 in the first term
originates from the geometric difference between the spin 0 and the other
partial waves. If we define a parity axis along the axis of the collision, 
then  while the left- and right-winding spin $\lam$ wave functions 
add up to $2\cos{\lam\theta}$, in the spin 0 sector 
only the $0^+$ state contributes as a partial wave. This separation of the 
spin 0 sector is necessary in order to carry out the analytic continuation 
in $\lam$ through the Sommerfeld-Watson transform. 
We now write the amplitude for the $t$-channel process and then 
assume that it can be analytically continued to the region $s>0,~t<0$.
In 2+1 dimensions, the complications due to the signature $\eta=\pm1$
also appears since the wave functions of spin $\lam$ are associated with a 
phase $(-1)^\lam$ under a rotation by $\pi$. Thus we have to introduce two
analytic functions $a^+(\lam,t)$ and $a^-(\lam,t)$, so that
\be
A(s,t)=a_0(t)+i\int_C d\lam~\sum_\eta \frac{\eta+e^{-i\pi \lam}}{2} 
\frac{a^{(\eta)}(\lam,t)}{\sin{\pi \lam}}~C(\lam,1+\frac{2s}{t})
\ee
We now want to deform the contour as is done in 3+1 dimensions.
However because 
\be
C_\lam(z)\sim z^{|\lam|}\qquad (|z|\rightarrow\infty),
\ee
we cannot reduce the `background integral' by pushing it to
 $\re \lam=-\frac{1}{2}$. Therefore we integrate along the imaginary axis
and  arrive at the  following expression:
\ba
A(s,t)&=&a_0(t)
+i\int_{\epsilon-i\infty}^{\epsilon+i\infty}d\lam~
\sum_\eta \frac{\eta+e^{-i\pi \lam}}{2}
\frac{a^{(\eta)}(\lam,t)}{\sin{\pi\lam}}~
C(\lam,1+\frac{2s}{t})+\nonumber\\
&&\sum_{\eta}\sum_{n}\frac{\eta+e^{-i\pi \alpha_{n_\eta}}}{2} \frac{2\pi
\beta_{n_\eta}(t)}{\sin{\pi\alpha_{n_\eta}(t)}}~C(\alpha_{n_\eta}(t),1+\frac{2s}{t})
\ea
$\beta_n(t)$ is the residue of Regge pole $\alpha_n(t)$.
So unless  $a_0(t)$  and the background integral vanish, we obtain 
\be
A(s,t)\sim s^{\mathrm{max}({\bar\alpha}(t),0)}
\ee
where  ${\bar\alpha}(t)$ is the pole with the largest real part (`leading
Regge pole'). Using the optical theorem at high energies (Eqn.~\ref{eq:opt_thm}),
we obtain the prediction, for scattering driven by  Regge-pole-exchange,
\be
\sigma_{tot}\sim s^{\bar\alpha(0)-1}
\ee
If all Regge trajectories have negative intercept, 
the background term prevails at high-energy.
In the case of potential scattering, $\lam=0$ is an accumulation point of 
Regge poles at threshold.
It is for that reason that we kept the 
background integral along $\re \lam=\epsilon$.
\paragraph{Potential scattering $\&$ bound states in the plane\\}
\noindent The Ansatz 
\be
\psi(r,\varphi)=\sum_{\lam=-\infty}^\infty \frac{\phi_\lam(r)}{\sqrt{r}}~ e^{i\lam\varphi}
\ee
plugged into the Schr\"odinger equation
leads to the following radial equation for $\phi_\lam$:
\be
-\phi_\lam(r)''+\left(\frac{(\lam^2-\frac{1}{4})}{r^2}+V(r)\right)\phi_\lam(r)=E\phi_\lam(r)
\ee
Thus there is a trivial correspondence between scattering in 3d and 2d
via the substitution:
\be
\ell=\lambda-\frac{1}{2}\qquad \Rightarrow \qquad \ell(\ell+1)=\lam^2-\frac{1}{4}
\ee
This effective shift in the angular momentum has important consequences. Regge originally
showed for a large class of potentials 
in 3d scattering that the partial wave amplitudes are meromorphic in $\ell$ in the 
region $\re \ell > -\frac{1}{2}$ ; this corresponds to the region $\re \lam > 0$ in 2d.
It was already known in the 1960's that at threshold $E\rightarrow 0$,
there is an accumulation of an infinite number of Regge poles around $\lam=0$.
Any attractive potential, however weak, will create a bound
state at $\lam=0$. A heuristic calculation can be found 
in~\cite{Landau:1987gn} showing that the binding energy is a non-perturbative 
expression in the potential:
\be
E\simeq \exp{-\frac{1}{\int V(r) rdr}}
\ee
The same peculiarity is seen in low-energy potential scattering.
It was shown in~\cite{Chadan:1998qm} that under 
very general conditions, the $s$-wave amplitude vanishes logarithmically 
at threshold. This can be interpreted as a branch point 
singularity in the complex $\lam$ 
plane:
\be
a_0\sim \frac{\pi}{2\log{k}}=\frac{\pi}{2} \int\frac{d\lam}{2\pi i} k^\lam 
\log{\lam}\qquad (k\rightarrow 0).
\ee
%
%
%
\chapter{Regge trajectories in the Isgur-Paton model\la{ap:ft}}
To quantize the Hamiltonian~(\ref{eq:ft_hamil}), 
it is convenient to change the variable to
\be
x=(\rho\sqrt{\sigma})^{3/2},
\ee
after what the kinetic term takes the standard 
form $\frac{9}{16\pi}p_x^2$. The Schr\"odinger equation to be solved
is thus
\be
\left\{-\frac{d^2}{dx^2} + \frac{8}{9}\frac{J(J+1)-\Lambda^2}{x^2}
+V(x)\right\} \psi(x) = \epsilon\psi(x)
\ee
where $\epsilon=\frac{16\pi E}{9\sqrt{\sigma}}$ and 
\be
V(x)=\frac{16\pi}{9}\left(2\pi x^{2/3} +\frac{M-\gamma}{x^{2/3}}\right).
\ee
%
\section{The phononic trajectory}
Since we are interested in the leading trajectory, we want to find the states
with largest angular momentum for a given energy. Therefore we select the
case where $\Lambda=M$ (see Eqn.~\ref{phonon_e},~\ref{phonon_j}), 
i.e. when there is no cancellation of angular momentum between the phonons.
In that case the effective potential for the one-dimensional 
quantum mechanics problem reads
\be
V_{\rm eff}(x)=\frac{8}{9}\left[\frac{M}{x^2} + 
2\pi\left(2\pi x^{2/3} + \frac{M-\gamma}{x^{2/3}}\right)\right]
 \la{eq:eff_pot_phon}
\ee
When $M$ is large, the minimum of the potential energy is situated at 
\be
x_0=\left(\frac{M-\gamma+3+ {\cal O}\left(\frac{1}{M}\right)}{2\pi}\right)^{3/4} 
\ee
and evaluates to
\be
\epsilon_{\rm pot} = V_{\rm eff}(x_0)=\frac{32\pi}{9}\sqrt{2\pi M}
\left(1+\frac{1-\gamma}{2M}+ {\cal O}\left(\frac{1}{M^2}\right)\right)~.
\ee
Because this energy grows with $M$, the kinetic term turns out to be 
subleading - we are in a regime where semi-classical methods are expected
to perform well. We evaluate the effect of the kinetic piece and the residual
part of the potential in the harmonic approximation:
\be
\epsilon_{\rm quad}=\frac{ \omega_{\rm eff}}{2}= 
\sqrt{\frac{V''_{\rm eff}(x=x_0)}{2}} = \frac{16\pi^{3/2}}{9\sqrt{M}}.
\ee
Thus 
\be
\epsilon \simeq \epsilon_{\rm quad} + \epsilon_{\rm pot} =
 \frac{64\pi^2}{9}\sqrt{\frac{ M}{2\pi}}
\left(1+\frac{1+1/\sqrt{2}-\gamma}{2M}+ {\cal O}\left(\frac{1}{M^2}\right)\right)
\ee
and the Regge trajectory of Eqn.~(\ref{eq:regge_phon}) is obtained.

\section{The orbital trajectory}
In this case we consider $M=\Lambda=0$ (no phonons)
and therefore solve for the 
fundamental state of the one-dimensional effective potential
\be
V_{\rm eff}(x)=\frac{8}{9} \left[\frac{J(J+1)}{x^2} +
2\pi\left(2\pi x^{2/3} - \frac{\gamma}{x^{2/3}}\right)\right]
 \la{eq:eff_pot_orbi}
\ee
We can proceed in a similar fashion as for the phonons.
The minimum of the potential is located at
\be
x_0=\left(\frac{\sqrt{3}}{2\pi}\right)^{3/4}
\left(J+\frac{1}{2}-\frac{\gamma}{2\sqrt{3}}+
{\cal O}\left(\frac{1}{J}\right)\right)^{3/4},
\ee
leading to
\be
\epsilon_{\rm pot}=V_{\rm eff}(x_0)=
 \frac{8}{9}~\left(\frac{2\pi}{\sqrt{3}}\right)^{3/2} ~\sqrt{J}~
\left(4-\frac{\sqrt{3}\gamma-1}{J}+{\cal O}\left(\frac{1}{J^2}\right)\right).
\ee
The quadratic piece evaluates this time to 
\be
\epsilon_{\rm quad}=  \frac{8}{9}\left(\frac{2\pi}{\sqrt{3}}\right)^{3/2}~~ 
\sqrt{\frac{3}{J}}~~
\left(1+{\cal O}\left(\frac{1}{J}\right)\right)
\ee
The total energy $\epsilon=\epsilon_{\rm quad}+\epsilon_{\rm pot}$
then leads to the Regge trajectory of Eqn.~(\ref{eq:regge_orbi}).